# HUMAN-AI PROGRAMMING ROLE OPTIMIZATION

## DEVELOPING A PERSONALITY-DRIVEN SELF-DETERMINATION FRAMEWORK

*WRITTEN BY*

## MARCEL VALOVY

*A DISSERTATION PRESENTED TO*

*THE DEPARTMENT OF INFORMATION TECHNOLOGY*

*IN PARTIAL FULFILLMENT OF THE REQUIREMENTS FOR THE DEGREE OF*

*DOCTOR OF PHILOSOPHY*

*SUPERVISED BY* PROF. ING. ALENA BUCHALCEVOVA, PH.D.

*CO-SUPERVISED BY* ING. ET ING. MICHAL DOLEZEL, PH.D.

*DOCTORAL STUDY PROGRAMME*: APPLIED INFORMATICS

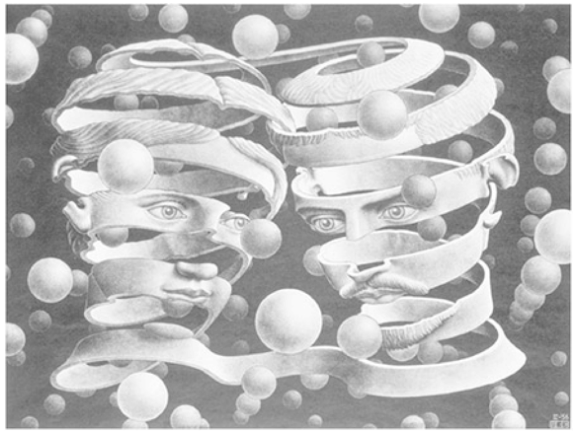

PRAGUE UNIVERSITY OF ECONOMICS AND BUSINESS

MAY 2025

The illustration on the title page, *Bond of Union* by M.C. Escher (1956), symbolizes the connection between human and artificial intelligence, an essential theme in the Role Optimization Motivation Alignment (ROMA) framework developed in this research. © 2025 The M.C. Escher Company, The Netherlands. All rights reserved.

Note on special symbols: This dissertation employs the symbol "⧉" to denote interleaving—the systematic alternation and interpenetration of elements within a sequence. "Human ⧉ AI" thus represents neither simple cooperation nor mechanical augmentation, but rather a structured commingling wherein human cognition and algorithmic processing create novel collaborative modalities. This notation captures the dissertation's central proposition: that optimal software development emerges through the deliberate orchestration of human and artificial capabilities, each preserving its essential character while contributing to an integrated whole—a synthesis that the work aims to achieve through its empirically grounded frameworks.

# ABSTRACT

I n 1967, Marvin Minsky envisioned minds—biological and artificial—collaborating beyond their individual limitations. That same year, Brian Randell captured software engineering's irreducibly social nature as the "multi-person construction of multi-version programs." Today, the prophecy becomes reality. Artificial intelligence, that 'man-made yet alien' presence, transforms the very nature of collaborative programming itself.

This dissertation optimizes human-AI programming roles through self-determination theory and personality psychology. As developers encounter AI collaboration—experiencing a technological thrownness into unprecedented relational dynamics—frameworks must emerge that harmonize human agency with algorithmic augmentation.

The research employs Design Science Research across five systematic cycles: Cycle 1 explores personality-role causal relationships through behavioral science methods. Cycles 2 & 3 adopt pragmatist axiology to design and extend the Role Optimization Motivation Alignment (ROMA) framework with AI integration. Cycle 4 develops an ISO/IEC 29110 Software Basic Profile & Agile Guidelines extension for Very Small Entities, while Cycle 5 validates through in-situ test artifacts and empirical triangulation, engaging 200 experimental participants and 46 interview respondents.

Results demonstrate that ROMA significantly enhances self-determination and team dynamics through personality-driven role optimization in both human-human and human-AI ("Human ⊞ AI") collaborative contexts. Five distinct personality archetypes emerge: The Explorer (high Openness), The Orchestrator (high Extraversion/Agreeableness), The Craftsperson (high Neuroticism/low Extraversion), The Architect (high Conscientiousness), and The Adapter (balanced profile). Personality-aligned role assignments increase intrinsic motivation by an average of 23% among professionals and up to 65% among undergraduates, with AI mode selection (Co-Pilot, Co-Navigator, Agent) proving crucial for self-determination need satisfaction.

The dissertation contributes: (1) an empirically-validated framework linking personality traits to programming role preferences and self-determination outcomes; (2) a taxonomy of AI modalities mapped to personality profiles while preserving human agency; and (3) an ISO/IEC 29110 extension enabling VSEs to implement personality-driven role optimization within established standards.

**Keywords:** artificial intelligence, human-computer interaction, behavioral software engineering, self-determination theory, personality psychology, phenomenology, role optimization, intrinsic motivation, pair programming, design science research, ISO/IEC 29110

# ABSTRACT

P sal se rok 1967, když Marvin Minsky snil o myslích – těch lidských i těch křemíkových – jak překračují hranice svých osamělých možností. Téhož roku zachytil Brian Randell neodmyslitelně společenskou povahu softwarového inženýrství slovy o „více-lidské konstrukci více-verzových programů". Dnes se proroctví stává skutečností. Umělá inteligence – člověkem stvořená, a přece mu cizí – mění samu podstatu tvoření programů.

Tato práce hledá příhodný způsob rozdělování rolí mezi lidi a stroje při psaní kódu. Opírá se přitom o teorii sebeurčení a poznání lidské povahy. Vývojáři, náhle vrženi do nového způsobu spolupráce se stroji, se potřebují opřít o soustavu, jež by sladila lidskou vůli s mocí algoritmů.

Bádání postupuje pěti cykly dle zásad vědy o návrhu: První zkoumá metodami behaviorální vědy a kritického realismu, jak povaha člověka určuje jeho roli. Druhý a třetí, vedeny pragmatickou axiologií, budují a rozšiřují soustavu Role Optimization Motivation Alignment (ROMA) o spolupráci s umělou inteligencí. Čtvrtý cyklus tvoří rozšíření základního softwarového a agilního profilu normy ISO/IEC 29110 pro velmi malé společnosti, pátý pak vše ověřuje pokusem a fenomenologickými rozhovory s dvěma sty experimentálními účastníky a šestačtyřiceti vypravěči.

Výzkum ukázal, že ROMA skutečně posiluje sebeurčení a soudržnost skupin, když optimalizuje role dle povahy, ať již jde o spolupráci člověka se strojem či člověka s člověkem („Human  AI"). Vynořilo se pět povahových obrazů: Průzkumník (zvídavý a otevřený), Dirigent (společenský a přívětivý), Řemeslník (citlivý a samotářský), Stavitel (svědomitý a pořádný) a Proměnlivec (vyrovnaný ve všem). Když lidé dostávají úkoly podle své povahy, roste jejich vnitřní zápal – u zkušených o čtvrtinu, u žáků až o dvě třetiny. Volba způsobu, jak stroj pomáhá (druhý pilot, spolu-navigátor či samostatný činitel), se ukázala být klíčová pro udržení duševní pohody.

Práce přináší trojí užitek: (1) pokusem ověřenou soustavu, jež spojuje povahové rysy s tím, jaké úlohy programátor rád vykonává a jak sám sebe určuje; (2) rozdělení způsobů, jak člověk může s umělou inteligencí jednat při zachování svébytnosti, přiřazené k různým povahám; (3) doplněk k vybraným profilům mezinárodní normy, aby i nejmenší společnosti mohly rozdělovat práci dle povah svých lidí.

**Klíčová slova:** umělá inteligence, spolupráce člověka se strojem, behaviorální softwarové inženýrství, teorie sebeurčení, psychologie osobnosti, fenomenologie, optimalizace rolí, vnitřní motivace, párové programování, věda o návrhu, ISO/IEC 29110

# ACM CCS Concepts and Placement

The Areas of Research of this PhD study have been classified according to the ACM Computing Classification System (Association for Computing Machinery, 2012).

**Human-centered computing**

→ Human-computer interaction

- Collaborative interaction
- Interaction paradigms
    - Human-AI collaborative systems

**Software and its engineering**

→ Software creation and management

- Agile software development
- Collaboration in software development
- Software development techniques
    - Pair programming

**Computing methodologies**

→ Artificial intelligence

- Philosophical/theoretical foundations of artificial intelligence
- AI-assisted software engineering

**Applied computing**

→ Psychology

- Empirical studies in collaborative and social computing

# ACKNOWLEDGMENTS

I extend my deepest gratitude to Professor Ing. Alena Buchalcevova, Ph.D., my supervisor, whose exceptional guidance proved instrumental throughout this doctoral journey. Her extensive expertise in software process improvement and international standards, particularly as a distinguished member of the ISO/IEC JTC1/SC7 Working Group 24, has provided invaluable insights that shaped the direction and quality of this research.

I am equally indebted to Ing. et Ing. Michal Dolezel, Ph.D., my co-supervisor, whose meticulous approach to empirical research methods has been crucial to the development of this work. His insightful feedback significantly strengthened the methodological framework of this dissertation.

I express sincere gratitude to my opponent, Dr. Michal Hron, whose insightful feedback illuminated areas for growth and inspired the current five-cycle structure of this thesis. His recommendations regarding academic positioning and methodological rigor have substantially strengthened the dissertation's contribution to the Behavioral Software Engineering community.

I am beholden to the Department of Information Technology at the Prague University of Economics and Business for enabling me to pursue this interdisciplinary research; it was an honor. I appreciate the faculty members and grant agencies that provided research funds, enabling me to confer on my research globally and present findings at international conferences, with special acknowledgment to Adam Borovicka and Martina Jandova, outstanding colleagues. I extend my gratitude to Vaclav Repa and Rudolf Pecinovsky, who invited me to pursue a PhD degree during my master's thesis defense and whose commendations supported my intermittent academic journey at Charles University. I extend my appreciation to Roman Bartak, who supervised my initial doctoral work in theoretical informatics and artificial intelligence disciplines at Charles University, a journey cut halfway by family commitments. Above all, I thank Ota Novotny for the innovative culture he has built around our department, which allowed me to supervise theses on interdisciplinary topics related to my dissertation as well as on pioneering issues, such as AI agents, quantum computing, and neurotechnology.

I thank the researchers whom I met along my path. I wish to acknowledge César França for his inspiring work on motivation in software engineering and the engaging discussions at ICSME'23. I express my gratitude to Marian Petre and Clayton Lewis from The Open University for their early guidance over dinners at PPIG'21 and PPIG'22 on developing my managerial motivation framework. I am grateful to Yvonne Dittrich for her insights into pairing constellations during the ICSE'22 doctoral consortium. I would like to express my gratitude to Dag Sjøberg for his profoundly insightful and friendly dinner discussions over pair programming research done in Simula Research Laboratory, the difference between using student and professional participants, and research validity criteria at EASE'23. My warm thanks go to Mirna Muñoz and Jezreel Mejia for their unmatched hospitality at CIMPS'22, where they introduced me to the details of the upcoming developments of the international standards series used in this work. I am deeply thankful to all conference colleagues and organizers worldwide, whose generous advice and tireless efforts have enriched my research network and journey, and to editors and reviewers for peer-reviewed research venues.

I express special gratitude to Mirna Muñoz and Claude Laporte for their expert review and valuable feedback on the ISO/IEC 29110 extension developed in this dissertation. Their deep knowledge of software process improvement for Very Small Entities and their involvement with the ISO/IEC JTC1/SC7 Working Group 24 provided crucial validation for my work. Their encouragement and constructive critique significantly enhanced the practical applicability of the ROMA framework extension.

To Michael Hakl and Martin Vojtek, I extend particular appreciation for their contributions to refining study designs, participating in the validation of the ROMA-ISO/IEC 29110 extension, and engaging in phenomenologically-informed interviews that provided critical insights into HumanAI collaborative dynamics.

My heartfelt appreciation extends to the 200 experimental participants and 46 interviewees from the ranks of my undergraduate students to professional acquaintances and—to my delight—former students who became valued professional peers. This research would not have been feasible without their contributions, and I cherish the memories we share. Additionally, although not issued in this dissertation for conciseness, the experimental studies were preceded by large-scale quantitative surveys achieved in collaboration with numerous bachelor's and master's degree students I supervised, who helped distribute and analyze the surveys across some of the country's largest IT firms; thank you. Thereby, I would like to extend immense gratitude to the 243 participants and European HR departments of Barclays, Pure Storage, Prusa Research, Microsoft, Avast, MSD, Seznam, Alza, and numerous very small entities for participating in my published research.

Final thanks go to my mom, dad, grandparents, and great-grandmothers for being avatars of truth and love, instilling in me a spirit of curiosity, and serving as inspirations for academic pursuit.

# AI USAGE STATEMENT

The author hereby declares that artificial intelligence and computational assistance were employed in the preparation of this dissertation manuscript solely for language editing and philosophical review purposes. All substantive research, analysis, experimental design, data collection, interpretation, and theoretical contributions represent original work conducted by the author without algorithmic generation or substitution of intellectual effort. All hermeneutic translations contained herein were performed by the author, who possesses Cambridge C2 Proficiency certification in English (Certificate Number: 0043132644).

**Author's Note:** My journey with computational writing assistance mirrors the very evolution this dissertation examines. Beginning with purely manual composition (Valovy, 2021), I progressed through commercial language enhancement services when presenting at Doctoral Students Day (Valovy, 2022, January), subsequently integrating expert systems like Ludwig.guru and Grammarly for international conference submissions (Valovy, 2022, September, A; Valovy, 2022, September, B). As my research matured, so did my toolkit—incorporating Microsoft Word's thesaurus features alongside Grammarly for several award-winning contributions (Valovy, 2023, January), maintaining these tools for empirical studies (Valovy, 2023, June), then embracing AI for blockchain smart contract development while investigating its psychological effects (Valovy & Buchalcevova, 2023, October). This progression culminated in AI-assisted language editing for our distributed programming research (Valovy & Buchalcevova, 2025, January), our PeerJ publication on personality-based pair programming (Valovy & Buchalcevova, 2025, April), and most recently, AI support for both language and analytical verification in our forthcoming ICSME paper (Valovy, Dolezel, & Buchalcevova, 2025, September). The tools that polish our words also sharpen our ideas—in this accelerated age, authentic scholarship embraces such augmentation to speak more clearly to more minds.

The author affirms complete responsibility for all content, findings, and conclusions presented herein. This disclosure reflects a commitment to research integrity while acknowledging that AI technologies, when appropriately constrained to auxiliary functions, may support scholarly communication without diminishing originality or intellectual contribution.

# Author's Declaration

I declare that the work in this dissertation was carried out in accordance with the requirements of the University's Regulations and Code of Practice for Research Degree Programmes and that it has not been submitted for any other academic award. Except where indicated by specific references in the text, the work is the candidate's own work. Work done in collaboration with, or with the assistance of, others is indicated as such. Any views expressed in the dissertation are those of the author.

SIGNED: .................................................... DATE: ..........................................

x

# CONTENTS

# LIST OF PUBLICATIONS

This dissertation is based on the following articles and conference papers, referenced in the text as **Studies I-VI**:

 **Study I (2023a – CIMPS'22)**

**Valovy, M.** (2023, January). Effects of Pilot, Navigator, and Solo Programming Roles on Motivation: An Experimental Study. In: Mejia, J., Muñoz, M., Rocha, Á., Hernández-Nava, V. (eds) New Perspectives in Software Engineering. CIMPS 2022. *Lecture Notes in Networks and Systems, vol 576*. Springer, Cham.[Best paper award]

As sole author, I conceived and designed the experimental framework exploring personality-driven role preferences in pair programming. I conducted all data collection sessions, developed the hierarchical clustering methodology, performed statistical analyses revealing three distinct personality clusters, and authored the complete manuscript. The pilot testing phase, which I personally conducted, was instrumental in refining the experimental protocol. I presented these findings at the CIMPS conference, where the work received valuable feedback that informed subsequent studies.

 **Study II (2023b – EASE'23)**

**Valovy, M.** (2023, June). Psychological Aspects of Pair Programming: A Mixed-methods Experimental Study. In *Proceedings of the 27th International Conference on Evaluation and Assessment in Software Engineering* (pp. 210- 216). 2023.

As sole author, I expanded the research scope to examine psychological aspects of pair programming through a mixed-methods approach. I designed and conducted all experimental sessions, developed the interview protocol, performed both quantitative and qualitative analyses, and identified key themes regarding motivation and role satisfaction. The manuscript, which I authored in its entirety, was presented at EASE where it contributed to discussions on behavioral software engineering methodologies.

###  Study III (2023c – ICSME'23)

**Valovy, M.**, & Buchalcevova, A. (2023, October). The Psychological Effects of AI-Assisted Programming on Students and Professionals. In *2023 IEEE International Conference on Software Maintenance and Evolution (ICSME)* (pp. 385-390). IEEE.

I conceived the comparative study design examining AI's psychological effects on both students and professionals. I developed the experimental framework, conducted all data collection sessions, and performed the thematic analysis, revealing distinct patterns of AI personification and motivational shifts. Prof. Buchalcevova provided supervisory guidance and manuscript review. I led the manuscript writing and presented the findings at ICSME, where the work sparked discussions on Human  AI collaboration dynamics.

###  Study IV (2025b – PeerJ-CS)

**Valovy, M.**, & Buchalcevova, A. (2025). Personality-based pair programming: toward intrinsic motivation alignment in very small entities. *PeerJ Computer Science, 11,* e2774. https://doi.org/10.7717/peerj-cs.2774

I developed the ROMA framework based on empirical findings from previous studies, designed and conducted the expanded experiments with 66 participants, performed multi-modal Linear Mixed-Effects modeling revealing personality-role interactions, and led the two-year peer review process. Prof. Buchalcevova contributed thematic cross-analysis, supervisory oversight, manuscript review, and her expertise in ISO/IEC standards integration. I authored the primary manuscript and managed extensive revisions through multiple review cycles.

###  Study V (2025a – ACIE'25)

**Valovy, M.**, & Buchalcevova, A. (2025, January). Blockchain-Driven Research in Personality-Based Distributed Pair Programming. In *2025 5th Asia Conference on Information Engineering (ACIE)* (pp. 21-25). IEEE. *– Best paper award*

I conceived and developed the Personality-Driven Pair Programming Application (PDPPA), designed the blockchain integration for research transparency, conducted pilot studies with VSE professionals, and performed all data analysis. Prof. Buchalcevova provided guidance on ISO/IEC 29110 alignment and manuscript review. I authored the manuscript and presented the work at ACIE, where it received the best paper award.

###  Study VI (2025c – ICSME'25)

**Valovy, M.,** Dolezel, M., & Buchalcevova, A. (2025). Developer Motivation and Agency in LLM-Driven Coding. Manuscript submitted to ICSME 2025.

I designed the phenomenologically-informed interview protocol, conducted all interviews with software professionals, performed Interpretative Phenomenological Analysis revealing lived experiences with AI modes, and led the manuscript writing. Dr. Dolezel contributed methodological expertise in empirical software engineering and assisted in manuscript writing. Prof. Buchalcevova provided supervisory guidance. The manuscript is currently under review for ICSME 2025.

# LIST OF OTHER WORKS

The author has further published the following studies outside this dissertation's scope:

**Valovy, M.** (2021). Motivation Management Framework. In *Proceedings of the Psychology of Programming Interest Group*, 32nd Annual Workshop. 2021.

**Valovy, M.** (2022, January). Effects of Solo, Navigator, and Pilot Roles on Motivation. In *Abstract Proceedings of the Doctoral Students Day at FIS VSE 2022*. Prague University of Economics and Business. pp. 14. 2022. – *Best paper award.*

**Valovy, M.** (2022, September, A). Motivational differences among software professionals. In *Proceedings of the 30th Interdisciplinary Management Talks: Digitalization of society, business, and management in a pandemic.* Linz: Trauner Verlag Publishing. 2022.

**Valovy, M.** (2022, September, B). Experimental Pair Programming: A Study Design and Preliminary Results. In *Proceedings of the Psychology of Programming Interest Group, 33rd Annual Workshop*, pp. 107-112. 2022.

# LIST OF TABLES

# LIST OF FIGURES

# ABBREVIATIONS

| | |
|---|---|
| AI | ARTIFICIAL INTELLIGENCE |
| ANOVA | ANALYSIS OF VARIANCE |
| API | APPLICATION PROGRAMMING INTERFACE |
| BFI | BIG FIVE INVENTORY |
| BSE | BEHAVIORAL SOFTWARE ENGINEERING |
| DSR | DESIGN SCIENCE RESEARCH |
| CO-PROGRAMMING | AI-HUMAN COLLABORATIVE PROGRAMMING |
| CTO | CHIEF TECHNICAL OFFICER |
| GEN Z | GENERATION Z |
| GPT | GENERATIVE PRETRAINED TRANSFORMER |
| HCI | HUMAN-COMPUTER INTERACTION |
| HUMAN  AI | HUMAN INTERLEAVED WITH ARTIFICIAL INTELLIGENCE |
| IDE | INTEGRATED DEVELOPMENT ENVIRONMENT |
| IEC | INTERNATIONAL ELECTROTECHNICAL COMMISSION |
| IMI | INTRINSIC MOTIVATION INVENTORY |
| ISO | INTERNATIONAL ORGANIZATION FOR STANDARDIZATION |
| IPA | INTERPRETATIVE PHENOMENOLOGICAL ANALYSIS |
| IRB | INSTITUTIONAL REVIEW BOARD |
| PDPPA | PERSONALITY-DRIVEN PAIR PROGRAMMING APPLICATION |
| PLOC | PERCIEVED LOCUS OF CAUSALITY |
| ROMA | ROLE OPTIMIZATION MOTIVATION ALIGNMENT |
| SDT | SELF-DETERMINATION THEORY |
| SOHO | SMALL OFFICE HOME OFFICE |
| TA | THEMATIC ANALYSIS |
| VSE | VERY SMALL ENTITY |

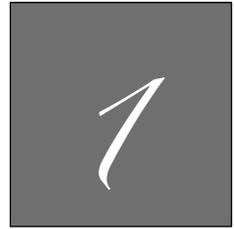

# INTRODUCTION

I n the liminal hours before dawn, when the cursor blinks against an empty screen and the weight of unwritten code presses against consciousness, software developers have always worked alone—until now. Artificial intelligence (AI) has crossed the threshold from tool to collaborator, from servant to something more unsettling: a presence that completes our thoughts before we think them, suggests solutions we had not imagined, and challenges the very notion of what it means to be a programmer.

This work emerges from a historical moment as significant as the introduction of the assembly line to manufacturing or the printing press to knowledge dissemination. Yet unlike those revolutions, which augmented human physical and communicative capacities, AI's incursion into software development touches something more intimate: the act of thinking itself. When the Luddites smashed textile machinery in 1811, they were not opposing technology per se—they were defending the skilled weaver's relationship to cloth, the tactile knowledge embedded in fingers that knew warp from weft. Today's developers face a parallel moment: not the destruction of looms but the subtle reconfiguration of minds.

This dissertation investigates how AI reshapes the social order of software teams through what Heidegger might recognize as a fundamental shift in *Dasein*—our being-in-the-world. When a developer encounters an AI suggestion, two horizons meet: human intentionality, shaped by experience, personality, and creative vision, confronts algorithmic pattern-recognition trained on the collective code of millions. This is Gadamer's "fusion of horizons" in its most literal manifestation—not between text and reader, but between human and artificial minds engaged in the co-creation of software.

Through Design Science Research methodology (Hevner et al., 2004; Peffers et al., 2006), this dissertation embarks on developing the Role Optimization Motivation Alignment (ROMA) framework—a journey that unfolds across five iterative cycles of artifact creation and evaluation.

Following Gregor and Hevner's (2013) knowledge contribution framework, the research progressively builds both descriptive knowledge (*omega*-knowledge) about personality-driven motivation and prescriptive knowledge (*lambda*-knowledge) for optimizing collaborative programming environments. Each cycle reveals new insights that reshape subsequent investigations, demonstrating how understanding emerges through systematic yet adaptive inquiry.

The methodological approach synthesizes behavioral science with design science, navigating between "what is" and "what can be" through iterative refinement (Wieringa, 2014). This synthesis aligns with contemporary mixed-methods research in software engineering (Johnson & Onwuegbuzie, 2004; Easterbrook et al., 2008), where controlled experiments establish patterns that qualitative inquiry then deepens and contextualizes. As the chapters ahead will demonstrate, the ROMA framework emerges not as a predetermined solution but as a living response to the evolving challenges of Human ⊞ AI collaboration.

Findings reveal that personality-driven co-programming role optimizations significantly improve intrinsic motivation, performance, and psychological well-being across diverse software development contexts. Through systematic investigation of personality archetypes—*The Explorer* (high Openness), drawn to creative discovery in the spirit of Bergson's élan vital; *The Orchestrator* (high Extraversion and Agreeableness), thriving in dialogical collaboration as Buber envisioned; *The Craftsperson* (high Neuroticism, low Extraversion), seeking Heideggerian dwelling in focused solitude; *The Architect* (high Conscientiousness), embodying Kantian systematic reason; and *The Adapter* (balanced traits), achieving Jungian individuation—the research demonstrates how tailored role assignments unlock latent potential in both human-human and Human ⊞ AI collaborative programming.

Quantitative results validate the framework's efficacy: personality-aligned role assignments increase intrinsic motivation by an average of 23% among professional developers and up to 65% among undergraduate students, with particularly pronounced effects in AI-augmented contexts. The ROMA framework and accompanying artifacts specifically address the needs of Very Small Entities (VSEs) —organizations with up to 25 employees—Small Office Home Offices (SOHOs), and Generation Z undergraduates, where resource constraints make personality-based optimization not merely beneficial but essential for survival and growth. These outcomes align with and extend the ISO/IEC 29110 standard's principles, demonstrating how psychological insights can be operationalized within established engineering practices.

### A Note on Dissertation Structure

This work takes the form of a theory-first cumulative dissertation, where theoretical contributions emerge through systematic empirical investigation across multiple studies. Rather than presenting isolated papers, the dissertation synthesizes findings from six empirical studies into a unified theoretical framework—ROMA—that advances our understanding of personality-driven motivation in software development.

This approach allows for both depth and breadth: each empirical study contributes specific insights while their synthesis reveals patterns invisible to individual investigations. The structure reflects the iterative nature of Design Science Research, where theory and practice inform each other through cycles of discovery, design, and validation.

### Navigating This Chapter

This introductory chapter establishes both the theoretical landscape and practical imperatives driving our investigation. Readers encounter the research motivation grounded in Design Science Research methodology (Section 1.1), explore the problem space where self-determination meets AI collaboration (Section 1.2), identify critical knowledge gaps (Section 1.3), and discover how these crystallize into specific research questions and objectives (Section 1.4). The chapter then delineates scope, philosophical foundations, and personal motivations (Sections 1.5-1.8) before presenting the key contributions and structural roadmap (Sections 1.9-1.11).

## 1.1 Motivation

Software engineering stands at a transformative intersection with artificial intelligence, where new collaborative practices are reshaping how developers approach their work. This dissertation introduces the Human  AI collaborative nexus, representing an early step toward enhanced programming environments that optimize both human creativity and AI capabilities. The research focuses specifically on the behavioral and psychological dimensions of software development in Very Small Entities (VSEs), where individual motivation and team dynamics have outsized impacts on project outcomes.

### 1.1.1 DSR Thesis Framework

This theory-first cumulative dissertation adopts the Design Science Research (DSR) thesis framework proposed by Thuan et al. (2019), providing an analytical structure to address motivational challenges in resource-constrained environments. By integrating insights from Behavioral Software Engineering (BSE)—a field examining how cognitive, emotional, and social factors influence software development practices (Lenberg et al., 2015)—this research develops practical artifacts that enhance motivation through personality-based role optimization.

The study incorporates the DSR analytical elements as follows:

i) **Research Motivation:** Addressing gaps in personality-trait influence on programming roles ("gap-spotting") while solving practical motivational problems in VSEs ("problem-solving"). This dual focus recognizes that AI and distributed work present unique challenges demanding new frameworks.

ii) **Problem Statement:** Following the prevalent "Research Challenge" problem type (Huang et al., 2019), investigating how to enhance motivation and productivity through personality-based role optimization in resource-constrained teams.

iii) **Research Questions:** Pursuing both "Knowledge" discovery and "Design artifacts" creation, the study generates explanatory knowledge and practical frameworks for motivational alignment in human-human and Human ⊞ AI programming contexts.

iv) **Research Approach:** Combining "Explanatory" and "Constructive" knowledge goals through mixed methods that explore, build, and validate personality-driven optimization.

v) **Research Activities:** Employing "Mixed" inquiry modes with "Build" and "Evaluate" purposes, progressing through iterative design cycles from theory-building to empirical validation.

vi) **Artifacts:** Creating both "Abstract" frameworks and "Material" applications with "Innovation" purpose, delivering tools that enhance motivation and productivity in software teams.

### 1.1.2 Research Motivation Elaborated

Following Sandberg and Alvesson's (2011) classification of research motivations and Thuan et al.'s (2019) extension, this dissertation is driven by:

**A. Problem-solving mode**: Addressing the practical challenge (Pries-Heje & Baskerville, 2008) of optimizing programming roles to enhance intrinsic motivation in VSEs—organizations wherein limited resources make role optimization essential for maximizing human capital.

**B. Gap-spotting mode**: Identifying significant gaps in BSE literature regarding how personality traits influence programming role preferences and motivational outcomes, particularly in the context of emerging AI collaboration.

The research targets VSEs, SOHOs, and undergraduate teams with empirically-validated frameworks for self-determination-aligned role assignments. By focusing on personality psychology's intersection with programming roles, this study enhances individual self-determination and team dynamics in environments where each contributor's engagement critically impacts success.

## 1.2  Problem-Solving Related Work

To situate this study within Behavioral Software Engineering (BSE), we examine two intertwined challenges: (i) understanding the motivational dynamics of Human ⊞ AI co-programming and (ii) ensuring that empirical findings remain generalizable and adaptable as AI tools and practices evolve.

### 1.2.1  The Problem of Understanding Software Professionals in the AI Age

The nexus of self-determination, personality, and task complexity in software engineering remains under-theorized, particularly when developers collaborate with AI assistants or agents. Addressing this gap is crucial for designing work environments that sustain both productivity and psychological well-being.

#### *Self-Determination in Software Engineering*

**What happens to the intrinsic joy of creation when the creator becomes the curator?**

This question pierces the heart of a profound transformation in software engineering. As Dijkstra (1972) presciently observed, "The tools we use have a profound (and devious!) influence on our thinking habits, and, therefore, on our thinking abilities." Never has this observation carried greater weight than in our contemporary epoch of AI-augmented development.

The answer reveals itself through layers of paradox: self-determination in software engineering matters precisely because it represents the last bastion of human agency in an increasingly automated cognitive landscape. The challenges inherent in software development—what Brooks (1987) eloquently termed "essential complexities"—have historically demanded not just technical skill but a particular form of existential engagement. Self-Determination Theory (Ryan & Deci, 2000a) illuminates how intrinsic motivation—that ineffable drive to engage with tasks for their inherent satisfaction—transcends mere productivity metrics to touch something fundamental about human flourishing.

Consider the phenomenology of programming: the developer enters what Csikszentmihalyi (1990) identifies as "flow," a state where challenge and skill dance in perfect equilibrium. This is not merely efficient work; it is, as Heidegger (1977) might suggest, a form of "dwelling" in the digital realm—a way of being-in-the-world through code. The satisfaction derives not from external validation but from what Maslow (1943) would recognize as self-actualization through creative problem-solving.

Yet here emerges our contemporary crisis. Beecham et al. (2008) noted with prescient concern that "there is no clear understanding of the Software Engineers' job, what motivates Software Engineers, how they are motivated, or the outcome and benefits of motivating Software Engineers." This observation, supported by França et al. (2011), suggests that motivation in software professionals emerges from a complex interplay of individual characteristics, internal controls, and external moderators—a delicate ecosystem now disrupted by AI's arrival.

*If an AI can write the code, what remains for the human to master?*

This question haunts the corridors of modern development teams. The traditional triumvirate of developer motivation—mastery, autonomy, and purpose (Pink, 2009)—faces unprecedented disruption. When large language models can generate functional code from natural language descriptions, the satisfaction traditionally derived from elegant problem-solving transmutes into something uncanny. As one participant in our preliminary studies confessed with existential unease: "I feel more productive but somehow deskilled and less... present."

The crisis manifests fractally across scales:

**At the individual level**: Developers report experiencing what we might term "competence vertigo"—a disorienting uncertainty about whether their skills or the AI's capabilities drive success. This echoes what Turkle (2011) identified as the "tethered self," perpetually uncertain about the boundaries between human and machine cognition.

**At the team level**: Traditional role boundaries blur as described by Humphrey (2000) in his analysis of team dynamics—the navigator-pilot distinction becomes fluid when AI serves as a third, quasi-autonomous participant. Teams experience what Edmondson (1999) identifies as "role ambiguity," requiring new forms of psychological safety to navigate the uncertainty.

**At the industry level**: The very ontology of programming expertise requires what Kuhn (1962) would recognize as a paradigm shift—not merely new methods but new ways of conceiving what it means to be a programmer.

Empirical studies by Verner et al. (2014) and Beecham and Noll (2015) demonstrate that motivated software professionals exhibit greater innovation, produce higher-quality artifacts, and experience less burnout. Yet these studies predate what we might call the "AI discontinuity." França et al. (2018) identified critical motivational factors, including:

- **Focus**: Catalyzed by well-defined work with clear goals—but what constitutes "well-defined" when AI continuously redefines the possible?

- **Engagement**: Nourished by continuous learning opportunities—yet how do we conceptualize "learning" when our tools evolve faster than human expertise can accumulate?

As contemporary organizational psychologists have observed, when workers lose connection to the meaning and ownership of their labor, they experience what Maslach and Leiter (2016) term "professional efficacy crisis"—a form of burnout specific to knowledge workers whose cognitive contributions become indistinguishable from automated outputs.

## Self-Determination Alignment in AI-Driven Programming

**Can artificial intelligence nurture human autonomy, or does it inevitably constrain it?**

This paradox illuminates the frontier of human-AI collaboration. Recent theoretical extensions of Self-Determination Theory reveal digital pathways through which autonomy, competence, and relatedness can be either nourished or thwarted in intelligent systems (see Section 2.4). The phenomenology of AI-assisted programming reveals a spectrum of experiences:

- **Autonomy frustration**: The opacity of algorithmic suggestions creates what Pasquale (2015) calls the "black box society"—developers feel directed by forces they cannot fully comprehend.

- **Competence threat**: The specter of skill atrophy haunts every autocomplete suggestion, echoing Carr's (2014) warnings about automation's erosive effects on human capability.

- **Pseudo-relatedness**: Surface-level social simulation in AI assistants creates what Turkle (2011) identifies as "alone together"—the illusion of collaboration without genuine inter-subjectivity.

Conversely, well-designed co-programming environments can scaffold what Vygotsky (1978) termed the "zone of proximal development"—that fertile space between current ability and potential achievement. These tools might transparently negotiate agency, creating what Latour (2005) would recognize as a new form of "actant" in the network of software creation, underscoring the need for *self determination-aligned* roles in Human ⊞ AI teams.

## Well-Being in Software Engineering

**When the mind that creates code becomes entangled with the machine that generates it, where does well-being reside?**

This question transcends traditional occupational psychology to touch something fundamental about human flourishing in the digital age. Well-being encompasses not merely the absence of pathology but what Seligman (2011) identifies as "flourishing"—a state of optimal psychological functioning characterized by positive emotions, engagement, relationships, meaning, and accomplishment.

The Job Demands-Resources model (Bakker & Demerouti, 2007) provides a lens through which to examine this entanglement. In traditional software engineering, job demands—cognitive load, deadline pressure, complexity management—are balanced by job resources such as autonomy, social support, and growth opportunities. But AI introduces a new category: what we might term "ontological demands"—the cognitive and emotional labor required to maintain a coherent sense of professional identity amid radical technological disruption.

Studies by Graziotin et al. (2018) and Lenberg et al. (2015) confirm the bidirectional relationship between affect and performance—positive emotions enhance focus and creativity, while negative emotions corrode both productivity and job satisfaction, creating either virtuous or vicious cycles of well-being.

### *Is the efficiency gained worth the authenticity lost?*

This trade-off haunts the AI-augmented workplace. As Heidegger (1977) warned about technology's tendency toward "enframing" (Gestell)—reducing everything, including humans, to standing reserve—we must ask whether AI-assisted programming risks transforming developers into alienated laborers, estranged from the products of their cognitive work.

## *Personality Traits and Their Role in Software Engineering*

### *If personality shapes how we code, what happens when the code shapes itself?*

This recursive question opens new vistas in understanding software engineering psychology. The significance of studying programmer personality transcends mere occupational fitting—it touches on fundamental questions about human-technology co-evolution. This stands in marked contrast to other fields wherein personality research has matured significantly. In occupational psychology, Barrick et al. (2001) demonstrated clear links between personality traits and job performance across diverse professions. In sports psychology, Bipp et al. (2008) have shown how understanding individual traits is key to unlocking athletic potential. Yet software engineering lags behind these fields in developing a nuanced understanding of how personality shapes professional success and satisfaction.

Weinberg's (1971) pioneering work, "The Psychology of Computer Programming," introduced the radical notion that programming is as much a human activity as a technical one. His later reflection (1998) that "programming is not a solitary activity" takes on new meaning in the age of AI pair programming. Shneiderman (1980) built on this foundation, recognizing that personality factors profoundly influence not just coding style but the very conceptualization of computational problems, leading to a multitude of programmer interaction and work styles.

The empirical landscape reveals intriguing patterns. Wynekoop and Walz (2000) documented significant personality differences between IT and non-IT professionals, suggesting what we might call a "computational disposition"—a particular configuration of traits that predisposes individuals toward algorithmic thinking. Yet as Amin et al. (2020) observe, this research remains frustratingly underdeveloped, leaving us with tantalizing glimpses rather than comprehensive understanding.

DeVito Da Cunha and Greathead (2007) make the provocative claim that "if a company organizes its employees according to personality types and their potential abilities, productivity and quality may be improved" (p. 4). This instrumentalist view, while pragmatically appealing, risks what Foucault (1975) might identify as a form of "disciplinary power"—the reduction of human complexity to manageable categories.

Dick and Zarnett (2002) found that personality alignment matters more than technical skill for pair programming success—a finding that takes on new significance when one member of the pair is artificial. How do we conceptualize personality "fit" with an AI partner? Does the machine adapt to us, or do we unconsciously modify our cognitive style to mesh with algorithmic patterns?

### *When we quantify the programmer's mind, do we discover its nature—or reshape it to fit our measurements?*

This question, inspired by Goodhart's Law and the observer effect in quantum mechanics, cuts to the epistemological heart of behavioral software engineering. Graziotin et al. (2021) advocate for rigorous psychological measurement, even suggesting entire dissertations devoted to single constructs. Lewis (2020) offers a compelling counterpoint, warning that the complexity of software engineering may render traditional psychometrics not just inadequate but potentially harmful—creating what Ian Hacking (1995) calls "human kinds," categories that shape the reality they purport to describe.

My own humble position navigates between these poles with deliberate caution. While I recognize the value of psychometric tools in empirical software engineering, I employ them as one lens among many, always within a broader phenomenological framework. I conceive personality traits not as fixed essences but as dynamic patterns—temporally stable behavioral variations rooted in neurobiological substrates yet profoundly shaped by context and experience. Each quantitative finding in this research is therefore triangulated with qualitative investigation, seeking not just statistical significance but lived meaning. This methodological pluralism reflects a core conviction: to understand the programmer in the age of AI, we must embrace both the precision of measurement and the wisdom of interpretation, knowing that what we seek to understand is not a static object but a human being in the midst of profound transformation.

### *Practical impetus for this dissertation*

The elaborate meta-question of understanding software professionals in times of interleaving human and artificial minds frames our investigation's urgency. The ongoing transformation of software engineering represents what Christensen (1997) would recognize as a "disruptive innovation"—not merely improving existing practices but fundamentally altering the nature of the profession itself.

The field's inherent complexity—what Simon (1969) termed "the sciences of the artificial"—combined with the emergence of Human 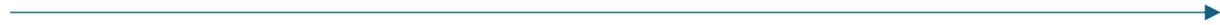 AI collaborative modes, creates an unprecedented research imperative. We stand at what Gleick (1987) might identify as a "bifurcation point" in the complex system of software development—a moment where small changes in understanding could lead to dramatically different futures.

The recent crystallization of "behavioral software engineering" as a unified subdiscipline represents what Lakatos (1978) would call a "progressive research programme"—generating novel predictions and opening new investigative avenues. Yet as the systematic literature reviews demonstrate, a profound void remains in our comprehension of motivation and personality in this transformed landscape.

---

**In the liminal space between human creativity and machine generation, who is the programmer becoming?**

The crisis is both intimate and systemic. In the quiet moments when code refuses to compile, when elegant solutions dissolve into edge cases, developers have traditionally turned inward—to experience, intuition, and the peculiar alchemy of human creativity. Now they turn to AI. This shift represents more than a change in workflow; it constitutes what Thomas Kuhn might recognize as a paradigm shift in the deepest sense—not merely new tools but new ways of being a developer. The question that drives this research is not whether AI makes programming more efficient (it does) but what happens to the programmer's soul—their sense of agency, mastery, and creative ownership—in this new collaborative paradigm.

As we stand at this crossroads, we must ask with Rilke (1929): "Perhaps all the dragons in our lives are princesses who are only waiting to see us act, just once, with beauty and courage." Perhaps the AI that seems to threaten developer identity is actually an invitation to discover new forms of human creativity, new modes of self-determination, new ways of flourishing in the digital realm.

This observation infuses empirical observation with both urgency and hope and shapes the thesis's problem-solving motivation (see Section 1.4.1).

### 1.2.2  Generalization and Adaptation in Empirical Software Research

*How do we study a river that changes course faster than we can map it?*

While understanding individual developer motivation is crucial, a second challenge emerges: ensuring our findings remain relevant as the technological landscape shifts beneath our feet. Like Heraclitus observing that one cannot step into the same river twice, software engineering researchers face phenomena that transform between observation and publication. This section addresses the methodological challenge of conducting research that is both rigorous and resilient in the face of rapid technological change.

#### *Rapid Advancement of Emergent Fields*

Fast-paced evolution in fields such as artificial intelligence and software engineering illustrates the *moving-target problem*—the object of the subject evolves faster than research cycles can run their suggested course (Kitchenham et al., 2002). This temporal mismatch between academic investigation and industrial practice poses fundamental challenges to knowledge accumulation in software engineering research.

Cohen (2024) employs biological metaphors to capture this challenge, comparing modern software's evolution to organic systems undergoing rapid speciation. Just as evolutionary biologists must develop new taxonomies for emerging species, software engineering researchers must continually recalibrate their theoretical frameworks. The half-life of empirical findings—the period during which results remain actionable—shrinks as development paradigms shift from waterfall to agile to AI-augmented methodologies.

This volatility manifests in three distinct ways. First, the *tool ecosystem* exhibits what we might call "Heraclitean flux"—no developer steps into the same IDE twice, nightly auto-updating their extensions. Version control systems evolve from centralized to distributed architectures; debugging tools incorporate machine learning; even the concept of "code review" transforms as AI pre-screens commits. Second, the *skills landscape* undergoes constant tectonics. Competencies considered fundamental—manual memory management, SQL optimization—become archaeological artifacts, while new literacies—prompt engineering, AI behavior prediction—emerge as critical. Third, the *collaborative paradigms* themselves metamorphose. Pair programming, once a radical departure from solo development, now seems quaint compared to human-AI-human triads where agency and authorship blur.

Baltes and Ralph (2022) diagnose this as a "generalizability crisis," arguing that our sampling frames—often limited to accessible populations like students or specific geographic regions—fail to capture the full spectrum of software engineering practice. Their critique extends beyond simple external validity concerns to question whether our research methods, rooted in slower-moving disciplines, can adequately capture phenomena that transform quarterly rather than decadally.

### *Transferability of Research Results to Industrial Applications*

An essential objective of empirical software engineering research is the application of findings in practical contexts. Two significant obstacles to this transfer are the limited control over variables in case studies and the lack of realism in controlled experiments, affecting the explanatory power. Increasing realism—of tasks, subjects, and environments—in controlled experiments could significantly improve their applicability to industry, though such experiments are often costlier than those involving simplified tasks, student participants, or artificial environments.

For example, Aghaee et al. (2015) conducted a large-scale study on role optimization in end-user programming using Facebook's MyPersonality dataset, which included Big Five personality profiles and social network features, such as Facebook page likes and statuses, of over two million volunteers. However, this valuable dataset was later withdrawn due to ethical concerns, such as the alleged use of psychographics, micro-targeting, and fake news networks by political campaigns in the US and UK, consequently limiting access to large-scale personality data needed for interdisciplinary studies on motivation and behavior (Stillwell & Kosinski, 2015).

On the experimental side, Falessi et al. (2018) posit that no population (students, professionals, or others) is inherently better than another in absolute terms. Using students as participants remains a valid simplification for laboratory contexts and can effectively advance software engineering theories. Conferring, Feldt et al. (2018) note that industry stakeholders are less likely to adopt methods tested on students without clear evidence comparing student and professional skills and motivations. Indeed, only 9% of experimental subjects were professionals in the early surveys (Sjøberg, 2005), and only 8% were professionals in the recent examination (Feldt et al., 2018). The controversial topic demands a discussion of what is the *raison d'être* of experimental research.

The primary purpose of experimental research is to establish causal evidence through internal rigor and control, but external validity remains critical. Without external validity, findings apply only to controlled environments and lack relevance in real-world settings. Aronson and Carlsmith (1968) identify two types of realism crucial to experimental validity: experimental (where subjects perceive experiments as impactful) and mundane (where conditions resemble real-world environments). Achieving these forms of realism in experimental tasks, subjects, and settings remains challenging (Harrison, 2000; Baltes & Ralph, 2022).

Sampling from target populations can mitigate validity threats. Sjøberg et al. (2002) argue that realism in experiments requires substantial resources, which remain prohibitive for many studies. Compensating professionals at competitive rates is costly, as evidenced by the decreasing ratio of professional participants (Feldt et al., 2018) and the struggles in recruiting them (Madampe et al., 2024). Ko et al. (2015) provided relevance-enhancing experimental guidelines, aiming to bridge research and practice through realistic sampling and task alignment with industry conditions.

Finally, the ethical dimension is critical. As Marcus & Oransky (2020) and Oransky (2022) report, issues like data manipulation and plagiarism have led to an increase in article retractions, underscoring the importance of ethical research practices to uphold credibility and reproducibility in rapidly evolving fields (Retraction Watch, 2021).

### Implications for this dissertation

Taken together, the volatility of tools, the "generalizability crisis" in sampling, and the reproducibility concerns in rapidly evolving AI contexts define a second research challenge for this work. Beyond enhancing self-determination and team dynamics, the thesis must also deliver empirical results—and artifacts—that remain generalizable, adaptable, and auditable across shifting socio-technical environments. To that end:

- the study adopts a continuous, multi-cycle DSR program that revisits hypotheses and artifacts as technologies evolve;

- all experiments provide replication packages and transparent instrumentation, including blockchain-stored experimental data through the PDPPA to support re-analysis; and

- the resulting ROMA framework is validated in *heterogeneous settings (VSEs, SOHOs, undergraduate teams)* to probe boundary conditions early.

This dual focus—*motivational enhancement and methodological resilience*—frames the overall research challenge (Section 1.4.1) and guides the design choices detailed in later chapters.

## 1.3   Gap-Spotting Related Work

*In the vast library of software engineering knowledge, which books remain unwritten—and which chapters need urgent revision?*

Having established the motivational challenges and methodological complexities facing software engineering research, we now turn to the cartography of absence—mapping not what we know, but what we do not. Following Alvesson and Sandberg's (2011) typology of research contributions, we identify both "confusion spots"—where existing theories conflict like tectonic plates grinding against each other—and "neglect spots"—where important phenomena remain undertheorized, like unmapped territories on an ancient chart marked only with "here be dragons."

### 1.3.1   Personality-Driven Pair Programming

The intersection of personality psychology and collaborative programming represents a curious paradox in software engineering research. While the field readily acknowledges that programming is a human activity shaped by individual differences (Weinberg, 1971), systematic investigation of how personality traits influence collaborative effectiveness remains surprisingly fragmented.

*Literature Synthesis*

The trajectory of personality-based pair programming research reveals an evolution from crude typologies to sophisticated psychometric approaches. Early work by Williams and colleagues (2002-2008) established the productivity benefits of pair programming—what they quantified as a "15% time investment yielding superior design quality, reduced defects, and enhanced team communication." Yet these studies treated personality as a black box, noting its importance without systematic measurement.

The methodological turning point came with Hannay et al.'s (2009a, 2009b) meta-analytic work, which exposed a critical flaw: personality effects were being measured at the dyadic level, obscuring individual contributions. This aggregation problem—akin to studying chemical reactions by examining only the final compounds rather than the constituent elements—led to inconsistent findings that plagued the field for years.

Recent advances have begun to untangle these complexities. Salleh et al. (2010, 2011) demonstrated that conscientiousness and openness correlate with specific pair programming benefits, while Smith et al. (2016) revealed personality-mediated preferences for agile practices, and Yang et al. (2016) observed these effects in classroom settings. These findings suggest that personality does not merely influence *whether* pair programming succeeds, but *how* it succeeds—through different cognitive and social pathways for different individuals.

*Personality-Based vs Personality-Driven*

*Personality-based* studies treat traits as **static inputs**: once Big-Five scores are captured, teams are *formed and frozen* (e.g., Williams et al., 2003; Hannay et al., 2009a). The core question is whether complementary matching boosts quality or velocity—results remain mixed, partly because instruments vary and analyses seldom drill below the dyad level (Hannay et al., 2009b).

*Personality-driven* research views traits as **dynamic regulators**. Initial matching is followed by micro-adaptations—AI-assistance level, role rotation, feedback cadence—triggered by ongoing signals such as short-form IMI probes or IDE telemetry. Preliminary evidence (Kuusinen et al., 2016; Smith et al., 2016) suggests this adaptive logic sustains flow, but no comprehensive framework operationalizes it.

### Gaps

I. **Trait-to-motivation linkage.** Existing work lacks a standardized way to relate validated trait measures (BFI-10/44) with real-time intrinsic-motivation data (IMI) during development.

II. **Actionable guidance for Human ⊞ AI settings.** There is no framework that translates personality findings into practical role-assignment heuristics when an AI assistant or agent joins the pair or any diverse team structures (cf. Gilal et al., 2019; Wiesche & Krcmar, 2014).

### Contribution of this dissertation

Across six empirical studies, ROMA triangulates BFI and IMI scores to derive *personality–motivation signatures* and distils them into lightweight, personality-based role-assignment recommendations for both human-only and Human ⊞ AI dyads/triads (e.g., rotate a high-Openness pilot to Co-Navigator after a sustained IMI-flow dip). The approach utilizes existing psychometric batteries, and supplies clear, practitioner-oriented heuristics without heavy instrumentation.

## 1.3.2 AI Co-Programming Role Optimization

Although conversational assistants and autonomous agents have become prolific in the AI-human collaborative programming (***"co-programming"***) scene, the organization of roles within Human ⊞ AI teams remains largely ad-hoc.

### Mini-SLR of AI Modes and Roles

A focused scan of 18 primary studies (2022-2025) reveals three recurring *interaction modes*—**Co-Pilot**, **Co-Navigator**, and **Agent**—each affording a characteristic bundle of *co-programming roles* (Table 1) (review detailed in Section 2.3.1; full systematic literature review in Appendix A).

Table 1: AI interaction modes, canonical developer roles, and their behavioral impact

| AI Mode | Canonical Role | Empirical Studies & Tools | Key Characteristics | Psychological Impact | Principal BSE Gap |
|---|---|---|---|---|---|
| Co-Pilot (inline suggestion) | *Accelerator, Safety-Net* | 13 studies, e.g., • Liang et al., 2024 • Mozannar et al., 2024 <br><br> • GitHub Copilot, Cursor Tab | • Immediate suggestions • Autocomplete • Context-aware code • Immediate, unobtrusive assistance | • Enhances flow state • Reduces routine cognitive load • May increase productivity while preserving autonomy | Long-term competence and autonomy effects unclear |
| Co-Navigator (conversational guidance) | *Mentor, Rubber-Duck, Domain-Expert, Critic* | 8 studies, e.g., • Robe & Kuttal, 2022 • Hamza et al., 2024 <br><br> • ChatGPT, Claude • Copilot Chat IDE extension | • Conversational interface • Multi-turn dialogue • Explains concepts • Gives rationale • Provides guidance rather than direct code | • Enhances understanding & learning • Supports higher-level reasoning • May improve code quality through deliberation • Facilitates knowledge transfer | How role framing shapes motivation and learning scaffolding |
| Agent (autonomous execution) | *Executor, Optimizer, Coordinator* | 5 studies, e.g., • Cinkusz & Chudziak, 2024 • Hassan et al., 2024 <br><br> • Copilot Chat Agent • Fleet Multi-Agent, Cursor, Windsurf IDEs | • Multi-step tasks • Access to tools & system resources <br><br> • Can make decisions without human intervention • Proactive rather than reactive | • Alters perceived locus of control • Changes developer's role to supervisor • May amplify productivity but potentially reduce skill development • Different relationship dynamic | Impact on locus of control, flow disruption, and competence |

*Table 1: A focused mini-SLR of 18 primary studies (2022-2025) identifies three recurring interaction modes—Co-Pilot, Co-Navigator, and Agent—and the social roles developers ascribe to them. Counts in the third column reflect the number of empirical papers that examined each mode; some papers cover multiple modes, hence totals exceed 18.*

*Ontological vs. epistemological framing:* mode answers *"how"* the AI operates (a transient configuration); role answers *"who"* the AI is perceived to be (a social position implying responsibility). Developers routinely attribute Mentor- or Critic-like agency to a Co-Navigator, despite AI's lack of moral agency—underscoring the motivational salience of role framing.

*Gaps revealed by the review:*

I. **Personality-to-mode linkage.** No study maps personality traits to preferred AI roles or modes.

II. **Self-determination.** Cross-mode comparisons of motivation and self-determination fulfilment are virtually absent.

*Programming-Role Optimization in Human ⊞ AI teams*

*Literature Overview:* Beyond mode classification, scholars have probed *how* roles might be optimized in hybrid teams. Aghaee et al. (2015) and Kuusinen et al. (2016) linked intrinsic motivation and personality to performance, but only in end-user or human-only settings. Mozannar et al. (2024) introduced telemetry-driven taxonomies, yet called for benchmarked feedback loops. Hassan et al. (2024) and Schleiger et al. (2024) outlined "AI-partner" visions but left motivational alignment open. Cinkusz & Chudziak (2024) demonstrated multi-agent coordination to optimize co-programming effectiveness, while Stowers et al. (2021) stressed "human-machine" complementarity for performance gains.

*Synthesized gaps:*

I. **Role optimization** in standard and AI-human software teams is under-tested (Mozannar et al., 2024).

II. **Flow-to-role guidance.** Guidelines for maintaining flow and intrinsic motivation as roles (and modes) shift are missing.

III. **Role assignments**—no unified framework adapts role assignments to personality-driven motivational profiles, especially in VSEs and other resource-constrained environments.

IV. **Social aspects of collaborative intelligence.** Collaboration between humans and AI can offer benefits such as efficiency and creativity the focused research on social and psychological implications in co-programming is absent (Schleiger et al., 2024).

V. **Educational impact and adaptation need.** Daun and Brings (2023) and Rose et al. (2023) argue that while AI supports individualized education, there is insufficient guidance on its integration, especially concerning students' skill development and creativity.

VI. **Expanded actionable guidance for Human ⊞ AI settings.** There is no framework that translates personality findings into practical role-assignment heuristics when an AI assistant or agent joins the pair or any diverse team structures (cf. Gilal et al., 2019; Wiesche & Krcmar, 2014).

    *a.* No study systematically maps personality traits to preferred AI roles across modes; Empirical work to date (e.g., Liang et al., 2024; Hassan et al., 2024) is mode-specific and rarely compares motivational outcomes across modes.

    *b.* Role optimization and cross-mode comparisons of self-determination fulfillment are almost non-existent.

A systematic framework that maps personality signals onto dynamic AI role configurations is needed to sustain autonomy, competence, and relatedness while maximizing productivity—an agenda this dissertation addresses through the ROMA framework and its VSE-oriented extension. ROMA fuses these mode-role insights with the personality-driven logic of Section 1.3.1. It (a) offers trait-based heuristics for selecting *which mode* (Co-Pilot, Co-Navigator, Agent) best suits a developer's autonomy and competence profile, and (b) specifies *when to re-frame* the AI's social role (e.g., from Safety-Net to Critic) as motivational states evolve—thereby closing the evidence gap identified by the mini-SLR.

Figure 1 visualizes the three modes at the *interaction layer* and their canonical roles at the *role layer*, serving as the conceptual bridge between the SLR evidence and the ROMA design heuristics.

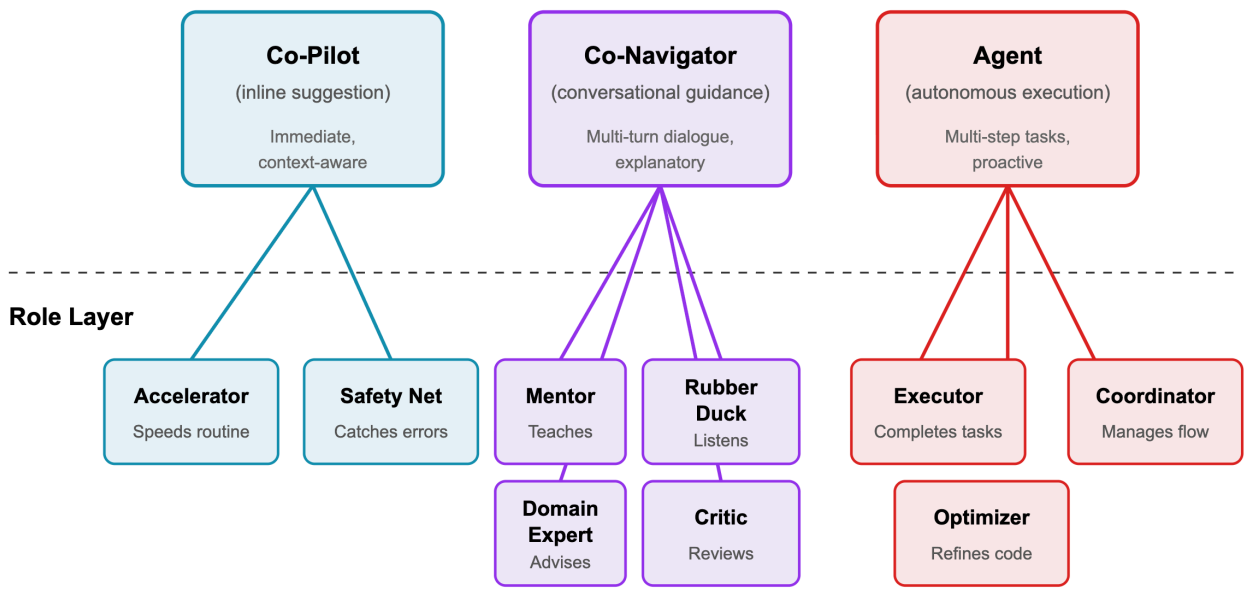

*Figure 1: Three AI modes and respective co-programming roles*

*Note: The interaction layer shows AI operational modes; the role layer depicts social positions developers attribute to AI systems.*

## 1.4 Problem Formulation and Objectives

The convergence of personality psychology, AI emergence, and resource-constrained development environments creates what systems theorists recognize as a "wicked problem" (Rittel & Webber, 1973)—one where the solution space itself evolves as we explore it. This section crystallizes the research challenge into specific questions and objectives, introducing the Design Science Research framework that structures our investigation. The methodological details of this framework—how DSR cycles operationalize these objectives—are fully elaborated in Chapter 3.

Software teams in Very Small Entities (VSEs), Small-Office-Home-Office (SOHO) settings, and university classrooms seldom enjoy the redundancy or role-specialization found in large organizations; each developer's motivation and productivity disproportionately shape collective output. Hence, the central concern: How can individual self-determination be amplified while simultaneously enhancing team performance through personality-aligned programming roles in both human-human and Human ⊞ AI contexts?

The work follows a five-cycle Design Science Research (DSR) programme (Chapters 5-9). In line with Gregor and Hevner's (2013) DSR structuring, each cycle focuses on a specific phase of framework development: Cycle 1 establishes empirical foundations (Relevance), Cycles 2-3 build and extend the ROMA framework (Design), Cycle 4 creates the ISO/IEC 29110 extension (Design), and Cycle 5 validates the framework through application development and triangulation (Evaluation).

### 1.4.1 Research Challenge

---

How can self-determination, intrinsic motivation, and team dynamics be enhanced in VSEs, SOHOs, and undergraduate software teams through personality-driven optimization of Human ⊞ AI programming roles?

---

The work thus confronts a research challenge that operates across three interconnected planes:

**First**, at the *theoretical level*: How do we extend Self-Determination Theory—developed for human-human interactions—to encompass human-AI collaborations where agency, competence, and relatedness take on new meanings?

**Second**, at the *methodological level*: How do we develop research artifacts that capture and theorize the lived meaning of Human ⊞ AI collaboration, remaining valid as the technological substrate shifts—where today's 'Co-Pilot' mode may evolve into tomorrow's autonomous agent?

**Third**, at the *practical level*: How do we translate psychological insights into actionable frameworks for VSEs and SOHOs operating under severe resource constraints, where each developer's motivation has an outsized impact on organizational survival?

To illustrate this challenge concretely: Consider a five-person VSE developing a critical fintech application. One developer (high neuroticism, low extraversion) struggles with the daily stand-ups required by their Scrum process, experiencing mounting anxiety that affects code quality. Another (high openness, low conscientiousness) generates brilliant architectural ideas but struggles with systematic documentation. A third (high extraversion, high agreeableness) excels at client communication but feels drained by solo debugging sessions. How can we optimize their roles to honor individual differences while meeting project demands? This is the challenge ROMA addresses.

### 1.4.2 Research Questions

This dissertation addresses two central research questions that guide the investigation from theoretical exploration to practical implementation:

**RQ1:** *What relationships exist* between personality traits and self-determination needs in AI-human and human-human co-programming contexts?

The first question embodies the knowledge discovery genre of DSR (Huang et al., 2019), focusing on empirically exploring the relationship between personality traits, programming roles, and self-determination. This question guides the *"exploratory"* phase, where I investigate how different role configurations satisfy the basic psychological needs of autonomy, competence, and relatedness that underpin intrinsic motivation.

*Links to Studies I-III (CIMPS'22, ICSME'23, EASE'23), establishing empirical foundations for the ROMA framework.*

---

**RQ2:** *How should* the ROMA framework—which optimizes programming roles to enhance self-determination and team dynamics—*be implemented* in VSEs and SOHOs?

The second question addresses the validation and implementation genre of DSR, focusing on how the ROMA framework can be effectively operationalized in resource-constrained environments. This practical question guides the *"artifact creation"* and *"validation"* phases, where I develop concrete tools and processes for implementing personality-driven role optimization in real-world settings.

*Links to Studies IV-VI (PeerJ-CS, ACIE'25, ICSME'25), focusing on framework implementation, extension, and validation.*

---

These research questions align with established patterns in design science research. As Thuan et al. (2019, p. 18) demonstrate in their analysis of DSR question formulation, implementation-focused questions like *"How should the artifact be utilized?"* and *"How can we use artifact X in the application domain?"* are particularly appropriate for the design and validation phases of DSR projects. The current research questions thus follow recognized conventions while addressing the specific challenges of personality-driven role optimization in software engineering contexts.

### 1.4.3  Aim

Building upon the research questions that explore personality-motivation relationships (RQ1) and framework implementation (RQ2), this dissertation pursues a unified aim that synthesizes discovery with application.

The overarching aim is:

---

To demonstrably enhance self-determination (autonomy, competence, relatedness), intrinsic motivation, and team dynamics in VSE, SOHO, and undergraduate software-engineering contexts by developing, validating, and operationalizing the Role-Optimization Motivation Alignment (ROMA) framework across five Design Science Research (DSR) cycles.

---

### 1.4.4  Thesis Objectives

To achieve the aim of enhancing self-determination and team dynamics through personality-driven role optimization, this dissertation pursues five specific objectives:

❖ **OBJECTIVE 1:** *Identify and analyze* **personality-driven** *preferences for* **human-human and AI-human programming roles**

⇨ Through empirical investigation using controlled experiments and psychometric assessment, establish the foundational relationships between Big Five personality traits and motivational outcomes across different programming role configurations.

*Cycle 1 – Relevance ("Exploration" phase) → addresses RQ1; grounded in Studies I (2023a – CIMPS) and II (2023b – EASE)*

**Output 1.1.**  Nomological network linking programming roles, personality traits, and self-determination states.

**Output 1.2.** Quantitative evidence of personality-based role preferences in collaborative programming.

---

❖ **OBJECTIVE 2:** *Design and develop* **the ROMA** *framework* **to enhance intrinsic motivation and team dynamics through personality-driven role optimization within VSEs, SOHOs, and undergraduate cohorts**

⇨ Create a theoretically-grounded, empirically-validated framework that translates personality insights into practical role optimization strategies for VSEs, SOHOs, and undergraduate contexts.

*Cycle 2 – Design ("Design & Evaluation" phase) → addresses RQ2; informed by Studies I-II and IV (2025b – PeerJ-CS)*

**Output 2.1.** The ROMA framework, including actionable guidance for raising psychological need satisfaction and intrinsic motivation (Study IV)

**Output 2.2.** LME statistical models quantifying personality-role interactions and their impact on motivational measures

---

❖ **OBJECTIVE 3:** *Extend* **the ROMA** *framework* **to incorporate AI-human collaborative programming insights from professionals and undergraduates**

⇨ Expand the framework to address emerging Human ⊡ AI collaboration patterns, mapping personality profiles to AI interaction modes (Co-Pilot, Co-Navigator, Agent) and revealing how personality archetypes specialize when encountering AI while preserving human agency and motivation.

*Cycle 3 – Design ("Design & Evaluation" phase) → addresses both RQs; based on Studies III (2023c – ICSME), IV, and VI (2025c – ICSME)*

**Output 3.1.** Human ⊡ AI extension for the ROMA framework addressing all three AI interaction modes (Co-Pilot, Co-Navigator, Agent)

**Output 3.2.** Essentialist patterns across Human ⊡ AI collaboration revealing dialectical tensions and paradigm shifts

**Output 3.3.** Phenomenological insights into lived experiences of AI collaboration through IPA methodology

**Output 3.4.** The ROMA AI Adapter framework with five archetypal specializations (Promethean, Conductor, Hermit, Cartographer, Shapeshifter) for navigating Human  AI futures

❖ **OBJECTIVE 4:** *Design* **an ISO/IEC 29110 Software Basic Profile and Agile Guidelines** *extension* **that operationalizes ROMA for VSEs**

⇨ Develop standards-compliant implementation guidelines that enable resource-constrained organizations to adopt personality-driven role optimization within established software engineering processes.

*Cycle 4 – Design ("Design & Evaluation" phase) → addresses RQ2; based on Study IV and expert validation*

**Output 4.1.** ROMA-aligned extension of the ISO/IEC 29110 Software Basic Profile and Agile Guidelines, enabling personality-driven human  AI programming in VSE contexts

**Output 4.2.** Expert-validated implementation guidance for integrating self-determination principles into standardized software development processes

❖ **OBJECTIVE 5:** *Validate* **the ROMA** *framework* **through application development and cross-study triangulation**

⇨ Demonstrate the framework's practical efficacy through the Personality-Driven Pair Programming Application (PDPPA) and comprehensive empirical validation across diverse populations and contexts.

*Cycle 5 – Validation ("Validation" phase) → addresses RQ2; based on Study V (2025a – ACIE) and comprehensive triangulation*

**Output 5.1.** PDPPA application as a validation tool for the ROMA framework's practical implementation

**Output 5.2.** Quantitative triangulation of self-determination and motivational findings across VSE, SOHO, and undergraduate samples

These thesis objectives guide the research trajectory from initial exploration through design, implementation, and validation, ensuring both theoretical advancement and practical applicability.

Table 2 positions each objective within its DSR cycle, details the principal methods employed, and identifies the empirical studies providing supporting evidence. The objectives are aligned with Hevner & Chatterjee's (2010) DSR Cycles framework and the custom DSR Phases developed in Section 3.1.2, adapted from Peffers et al. (2006).

Table 2: Mapping thesis objectives to DSR cycles, methods, outputs, and studies: How each research objective is achieved through specific Design Science Research cycles and methods

| Obj. | Cycle / DSR Phase† | Purpose | Main Methods | Key Outputs | Link. Studies |
|------|-------------------|---------|--------------|-------------|---------------|
| O1 | **Cycle 1 – Exploration** (Empirical relevance) | Identify personality-driven role preferences | Cluster analysis, Shapiro–Wilk & KS normality tests, linear mixed-effects (LME) screening | 1.1 Nomological network 1.2 Quant evidence of trait–role patterns | I (CIMPS'22) II (EASE'23) |
| O2 | **Cycle 2 – Design & Eval.** (Build core ROMA) | Design & test ROMA for human-human teams | LMEs with random intercepts, essentialist thematic check-backs, bootstrapped IMI gain estimates | 2.1 ROMA v1 heuristics 2.2 LME IMI thresholds | I, II, IV (PeerJ-CS 2025) |
| O3 | **Cycle 3 – Design & Eval.** (Extend with AI) | Integrate AI modes and map specializations | Interpretative Phenomenological Analysis (IPA), essentialist thematic analysis, Four-Form IPA methodology | 3.1 Essentialist patterns 3.2 IPA phenomenological insights 3.3 AI-SDT alignment 3.4 ROMA AI Adapter | III (ICSME'23) IV (PeerJ) VI (ICSME'25) |
| O4 | **Cycle 4 – Design & Eval.** (ISO/IEC Extension) | Craft ISO 29110 Agile extension for ROMA | Design-proposition mapping, expert peer review, metamodel compliance check | 4.1 ROMA-aligned ISO 29110 add-on 4.2 Peer-reviewed refinements | IV (PeerJ) V (ACIE'25) |
| O5 | **Cycle 5 – Validation** (PDPPA & Triangulation) | Build & pilot the Personality-Driven Pair Programming App (PDPPA) | Quasi-experimental field test, pre/post IMI & productivity metrics, mixed-methods triangulation | 5.1 PDPPA MVP App 5.2 Triangulated evidence across populations | V (PDPPA Pilot 2025) Triangulation |

*Table 2 maps each research objective to its corresponding Design Science Research cycle, detailing the methodological approaches and resulting outputs. This structured progression from exploration through design to validation ensures that the ROMA framework is both theoretically sound and practically applicable. Each objective builds upon previous work while addressing specific aspects of the research challenge, culminating in a comprehensive framework for enhancing self-determination and productivity through personality-driven role optimization in software teams.*

*† DSR phase labels follow Hevner's (2007) Three-Cycle model and the dissertation's custom "Exploration / Design & Evaluation / Validation" breakdown.*

This structured progression from exploration through design to validation reflects the natural evolution of design science knowledge. Each objective builds upon its predecessors while maintaining sufficient independence to contribute distinct value. The alignment between objectives, methods, and outputs ensures that the research program maintains both theoretical coherence and practical relevance throughout its five-cycle journey.

### 1.4.5 Integration of Research Components

Table 3 provides a comprehensive mapping of research problems, gaps, objectives, and supporting studies, illustrating how the dissertation addresses each aspect of the research challenge through a structured DSR approach.

Table 3: Integration of research components and mappings

| RQ-Swim-Lane | Cycle | Gap Tackled | Objective | Methods & Evidence |
|---|---|---|---|---|
| **RQ1 – Knowledge Discovery** | C1 | Unknown trait–to–role preferences | O1 | Clusters • ANOVA → Studies I-II |
| | C3 | AI impact on self-determination unclear | O3 | TA • Cross-Analysis → I, II, III, IV IPA • Four-Form Analysis → VI |
| **RQ2 – Implementation & Validation** | C2 | Framework for human-human optimization absent | O2 | LMEs • Bootstraps → I, II, IV |
| | C3 | Practical integration of AI modes | O3 | TA + IPA datasets triangulated |
| | C4 | Standard-level operational rules missing | O4 | Expert review → IV, V |
| | C5 | Cross-context proof via tooling | O5 | Quasi-experiment → V |

*Table 3 illustrates how each research question is addressed through specific cycles, targeting identified gaps with appropriate methodologies. The table demonstrates the logical progression from knowledge discovery (RQ1) to implementation and validation (RQ2), showing how each objective contributes to addressing specific gaps in the field of personality-driven self-determination in software engineering. By mapping methods and supporting studies to each objective, the table provides a clear roadmap of how empirical evidence builds toward the dissertation's contributions in enhancing self-determination, intrinsic motivation, and software team dynamics.*

This research mapping ensures that each aspect of the research challenge is systematically addressed through appropriate methods and empirical studies, resulting in a comprehensive framework for personality-driven role optimization that enhances both self-determination and team dynamics in software development contexts.

## 1.5 Scope and Boundaries

Clearly defined boundaries prevent over-generalization and keep a design-science project grounded in the context that gives its artifacts meaning (Gregor & Hevner, 2013). Following Dybå et al. (2012) guidance on research boundary specification, we delineate our investigation along four axes–conceptual, populational, temporal, and technological–so that later chapters can discuss transferability rather than universality of the findings.

### 1.5.1 Conceptual Boundaries

#### Human ⊞ AI Symbol Explanation

Throughout this dissertation, *"Human ⊞ AI"* (pronounced "human interleaved with artificial intelligence") denotes a sophisticated integration where human and artificial intelligence capabilities are systematically interwoven to optimize collaborative outcomes. The symbol (⊞), representing interleaving—defined by the Oxford English Dictionary Online as "the arrangement of items so that they alternate or are mixed with each other in a sequence" (Oxford University Press, n.d.-a)—captures how tasks and responsibilities alternate between human and AI agents in structured patterns, creating synergies beyond simple addition.

This notation advances beyond terms like "human-AI collaboration" by emphasizing the dynamic, layered nature of modern programming partnerships. As AI evolves from tool to collaborator, traditional terminology proves insufficient. While existing literature offers various terms—"human-in-the-loop" (Ostheimer et al., 2021), "hybrid intelligence" (Akata et al., 2020)—none adequately captures the structured integration we observe in contemporary development environments.

The Human ⊞ AI paradigm manifests through three distinct interaction modes (detailed in Figure 1, Section 1.3.2): Co-Pilot (augmentative), Co-Navigator (dialogical), and Agent (autonomous). Each mode represents a different interleaving pattern, from subtle assistance to complex delegation, fundamentally reshaping how developers experience their work.

#### Optimization in the Context of Programming Roles

Oxford English Dictionary Online (Oxford University Press, n.d.-b) defines *optimization* as "the action of making the best or most effective use of a situation or resource." Five optimisation facets are distinguished below:

1. **Balancing Role Assignment and Task Allocation:** The primary goal is to optimize role assignments—determining who, human or AI, should take on a given role at any time. Optimization here involves balancing roles according to *personality-driven preferences*, *skills*, and *strengths*, aiming to enhance individual motivation and team dynamics.

2. **Aligning Roles with Motivational Profiles:** Optimization also entails *adapting* roles to leverage motivational profiles that correspond with personality traits.

3. **Distributing Task Complexity:** Optimization also extends to the distribution and complexity of tasks between human and AI participants, ensuring each task matches the respective strengths of humans and AI to maximize efficiency and engagement.

4. **Adjusting Frequency and Duration of Roles:** Optimization entails adjusting the frequency and duration of specific roles (e.g., pilot, navigator, solo, AI-assisted) and AI modes (e.g., co-pilot, co-navigator, agent) to avoid monotony or burnout in human participants.

5. **Coordination Inter-Role Interaction:** Finally, optimization encompasses the coordination and interaction between roles. This includes planning and structuring interactions in a way that maximizes the potential of human and AI contributions in human-AI dyads and human-AI-human triads.

### *Motivation Alignment*

While this dissertation often references intrinsic motivation as a primary outcome measure, it is important to clarify that intrinsic motivation is conceptualized within the broader framework of Self-Determination Theory (SDT). SDT posits that intrinsic motivation flourishes when the three basic psychological needs—autonomy, competence, and relatedness—are satisfied. Thus, when discussing "motivation alignment" throughout this work, we are referring to the alignment of roles and practices with these fundamental self-determination needs.

### 1.5.2   Target Populations

Crucial areas of focus in this dissertation are *VSEs* and *SOHOs*, which often lack the resources and infrastructure of larger organizations, and *Gen Z undergraduates*, who face unique challenges in their transition from structured learning environments to self-directed work. By addressing these groups, this work aims to support the broader adoption of Human ⊡ AI methodologies in contexts where they can have the most significant impact.

**Very Small Entities (VSEs)**—as defined by the ISO/IEC 29110 standard as organizations with up to 25 employees (Laporte et al., 2018)—constitute a substantial segment of the global software industry. These organizations often encompass small teams or individual developers who face unique challenges in adopting Agile methodologies. Limited resources, experience, and access to professional training hinder their ability to implement optimal practices effectively (Muñoz et al., 2021). This work addresses these challenges, offering intrinsically motivational strategies and practical tools specifically tailored for VSEs. The ISO/IEC 29110 definition provides a clear boundary for our target population while acknowledging that the challenges faced by a 25-person software company differ markedly from those of a two-person startup, necessitating flexible frameworks that scale within this range.

**Small Office Home Offices (SOHOs)** are becoming increasingly relevant in the modern software industry as they represent self-directed, often AI-assisted teams or individuals working independently or within micro-teams. They face similar constraints to VSEs but have the added flexibility and challenge of self-management, particularly when integrating AI-assisted development practices. This dissertation targets SOHOs as adaptable, innovative environments for testing motivational strategies and role optimization.

**Generation Z undergraduates** transitioning from academic to professional environments who need structured support to develop sustainable programming practices. This population is particularly significant as they will form the next generation of software professionals and are uniquely positioned at the intersection of traditional education and AI-augmented development.

### 1.5.3 Environmental Context

The research operates within three overlapping environmental contexts:

- **Cultural Context –** Situated primarily within European software development culture, particularly in Czechia, which has emerged as a global software engineering hub, offering a high density of skilled software engineers with many Fortune 500 companies maintaining software development centers within the country. This context allows the dissertation to draw on an environment that combines a rich software engineering culture with a strong emphasis on innovation and technical expertise.

- **Social Context –** Focused on individual and team-level dynamics rather than organizational transformation, with emphasis on interpersonal interactions and psychological well-being in programming collaborations. Aligns with the behavioral software engineering community, making the research more adaptable and immediately applicable in diverse software engineering settings than studies that concentrate on organizational change.

- **Engineering Context –** The engineering scope of this research is limited to software development activities that directly produce source code, including programming, coding, and software construction. It excludes broader activities such as requirements gathering or high-level design, focusing instead on the granular processes where motivation and role optimization have the most direct impact on measurable outcomes.

- **Technical Context –** Centered on contemporary programming environments, including integrated development environments (IDEs) and AI-assisted coding tools that support both human-human and Human ⊞ AI collaborative programming.

These boundaries provide necessary focus while ensuring the research remains relevant to the evolving landscape of software development practices.

## 1.6 Academic Context

The intellectual architecture of this dissertation emerges from a convergence of disciplines, each contributing essential perspectives to understanding how personality, motivation, and collaboration shape software development in the age of artificial intelligence.

While Section 1.5 centered on the "where" and "what" of the research—specific demographics, technologies, and engineering activities—this section addresses the broader *"why"* and *"how."* It examines why the research is relevant in its academic and professional domains and how it integrates insights across disciplines to develop a robust knowledge framework.

This dissertation utilizes Design Science Research (DSR) principles, which emphasize the iterative development of artifacts tailored to specific contexts. Guided by DSR's interdisciplinary potential (Huang et al., 2019), the ROMA framework integrates insights from Behavioral Software Engineering (BSE), social psychology, and the philosophy of AI to address challenges related to motivation, AI-human interaction, and distributed work settings. This cross-disciplinary approach supports the design and application of ROMA artifacts, allowing for both theoretical depth and practical utility in collaborative programming environments.

### 1.6.1 Academic Disciplines and Research Fields

This dissertation is primarily positioned within the field of Behavioral Software Engineering (BSE), a specialized domain that examines the cognitive, emotional, and social dimensions of software development (Lenberg et al., 2015). While drawing from multiple disciplines, the research maintains BSE as its core focus, with the goal of advancing understanding of how personality traits and role assignments influence motivation and performance in software teams.

#### *Academic Disciplines*

This work situates itself within the following academic disciplines:

i) **Behavioral Software Engineering (primary):** Examines how human cognitive, emotional, and social factors influence software development practices and outcomes

ii) **Software Engineering (empirical):** Provides the methodological foundation and practical context for studying programming roles and collaborative practices

iii) **Social Psychology (ontological):** Informs the understanding of personality traits, motivation, and team dynamics in software development contexts

## Faculty Relations, Publication Venues, and Professional Communities

Interdisciplinary research is atypical for our faculty's Department of Information Technology, which drives its research more toward the Applied Computing discipline. However, to support its interdisciplinary approach, this research targets diverse publication venues and professional communities. Notable conferences and journals comprise *PeerJ-CS, ICSME, EASE, CHASE, ACIE, and CIMPS*. The author is also an active member of the *IEEE Computing Society, ACM, AIS, APA, Nature,* and *Science*, reflecting a commitment to bridging applied computing with broader academic discussions in computer science and psychology.

## Research Fields

The research fields supporting this dissertation span several disciplines, each contributing to the ROMA framework's interdisciplinary objectives:

**1) Behavioral Software Engineering**

a) Primary Discipline: Software Engineering

    i) *Subdisciplines:* Human-Computer Interaction (HCI), Agile Methodologies

b) Secondary Discipline: Psychology

    i) *Subdisciplines:* Behavioral Psychology (Human Motivation, Team Behavior), Industrial-Organizational Psychology (Workplace Behavior, Collaboration)

c) Role and Focus: Behavioral Software Engineering explores how behavioral science principles can improve software development processes by examining factors that influence developer motivation and team dynamics. This field's relevance lies in its capacity to address psychological and social factors within programming environments, supporting the ROMA framework's goal of enhancing team cohesion and individual satisfaction. Agile methodologies contribute insights into how adaptable role assignments can benefit software engineering in various socio-technical settings.

**2) Collaborative Intelligence**

a) Primary Discipline: Computer Science

    i) *Subdisciplines:* Artificial Intelligence (Human-AI Interaction)

b) Role and Focus: *Collaborative Intelligence* investigates how AI systems and humans can interact effectively to achieve shared goals. Within this field, human-AI interaction addresses ways in which AI systems can complement human decision-making, offering insights that guide the development of team structures and role optimization in programming.

**3) Social Psychology**

   a) Primary Discipline: Psychology

   i) *Subdisciplines:* Organizational Behavior (Group Dynamics, Collaboration), Cognitive Psychology (Motivation, Decision-Making)

   b) Role and Focus: Social Psychology examines how individual motivations and team dynamics influence workplace behavior, making it highly relevant for studying co-programming role assignments and team interactions. This field contributes to our understanding of how personality-driven motivations shape software development practices, particularly in relation to team composition and Human ⌗ AI programming roles.

The convergence of these research traditions fortifies the Behavioral Software Engineering orientation of this dissertation and augments the ROMA framework's capacity to devise nuanced, context-sensitive self-determination-alignment interventions across diverse software-engineering milieus—most notably within Very Small Entities (VSEs) and educational settings—where personality-informed role optimization can demonstrably enhance both individual engagement and collective dynamics.

## 1.7    Research Philosophy

The philosophical stance of this dissertation emerges from what Mingers (2001) terms "critical pluralism"—the recognition that complex socio-technical phenomena require multiple philosophical lenses. We adopt a stratified approach where different research phases employ different philosophical commitments, each selected for its fitness to the specific research task.

### 1.7.1    Ontology, Epistemology, and Axiology in the BSE Context

In researching behavioral aspects of software engineering, ontological and epistemological positions significantly shape how we understand and evaluate evidence about motivation and team dynamics, while pragmatist axiology translates knowledge to practical improvements.

Our ***ontological*** commitment aligns primarily with critical realism (Bhaskar, 1975), acknowledging that while motivational states and team dynamics exist independently of our observation, our understanding of them is inevitably mediated by cognitive and social structures. This stance proves particularly apt for investigating programming roles, where objective performance metrics intertwine with subjective experiences of flow, autonomy, and relatedness.

*Epistemologically*, we navigate between essentialist and interpretive approaches. The essentialist lens—consistent with Platonic idealism—enables us to identify stable patterns in personality-motivation relationships (Creswell & Creswell, 2022). Yet we simultaneously embrace interpretive epistemology to capture how developers make sense of their evolving relationships with AI collaborators. This dual stance reflects what Goldkuhl (2011) calls "practice-oriented research," where knowledge value is measured by its capacity to improve practice. This philosophical pivot is not a retreat from rigor but a recognition that in design science, Hevner's (2007) "utility" criterion demands equal weight with "truth."

The *axiological* dimension—examining the role of values in research—becomes particularly salient in design science (Mbanaso et al., 2023). Our work explicitly values human flourishing within technological contexts, seeking not merely efficiency gains but enhanced self-determination and team cohesion. This value commitment, following Menapace (2019), guides both artifact creation and evaluation criteria.

### 1.7.2 Critical-Realist and Pragmatist Positioning within DSR Cycles

The five-cycle structure of this dissertation reflects a philosophical journey, with each cycle demanding its own epistemological stance:

1. **Cycle 1 (Exploration):** Adopts critical realism to uncover generative mechanisms linking personality traits to programming role preferences. Here, we seek what Roberts (2014) calls the "real" beneath the "empirical"—the causal structures that produce observed motivational patterns.

2. **Cycles 2-3 (Design):** Transition toward pragmatist axiology as we develop the ROMA framework and its AI extensions. Following the pragmatist tradition (Creswell & Creswell, 2022), these cycles privilege workability—does the framework enhance motivation in practice?—while maintaining theoretical coherence.

3. **Cycle 4 (Standards Integration):** Returns to a more realist stance as we align the framework with ISO/IEC 29110 standards, recognizing the ontological weight of established organizational practices while pragmatically adapting them for VSE contexts.

4. **Cycle 5 (Validation):** Employs methodological pluralism, combining positivist hypothesis testing with interpretive phenomenology. This final cycle embodies what Wieringa (2014) terms the bridging of "what is" and "what can be"—validating both the framework's efficacy and its experiential meaning for developers.

By sequencing a critical-realist exploratory cycle with pragmatist, design-oriented cycles, the study first theorizes the causal landscape of motivational alignment and then operationalizes those insights into artefacts that can be rigorously tested in real-world settings.

### 1.7.3  Research Onion and Methodological Choices

Saunders et al.'s (2019) Research Onion in Figure 2 provides a useful heuristic for mapping our methodological trajectory. At its core, our philosophical stance evolves through the cycles—from critical realism (testing structural mechanisms such as role templates & personality fit) through pragmatism (iterating artifacts in situ, testing what works) to pluralism (reconciling quantitative & interpretive evidence). This evolution reflects not philosophical inconsistency but rather what Archer (1995) terms "analytical dualism"—the recognition that different aspects of a phenomenon may require different philosophical approaches.

Our approach to theory development similarly evolves: deductive in early cycles (testing personality-role hypotheses), abductive in middle cycles (iteratively refining the ROMA framework based on emerging patterns), and ultimately integrative in the validation phase.

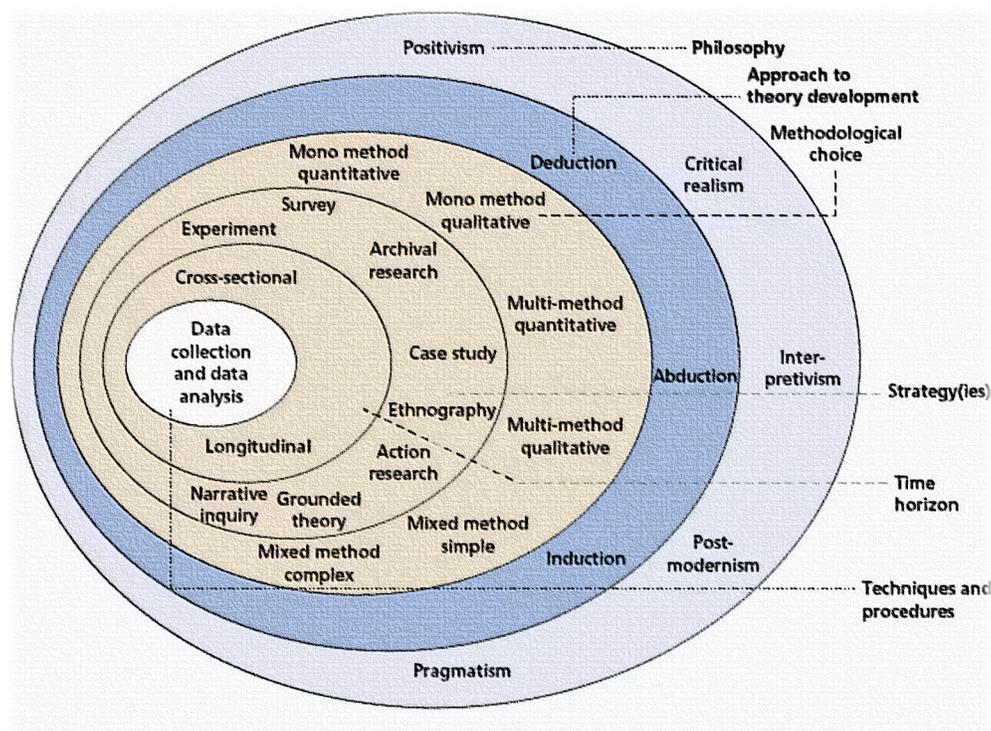

*Figure 2: The research onion*

*Reproduced under educational fair use for non-commercial academic purposes, based on Saunders et al. (2019, p. 108).*

### 1.7.4 Theory-Forming Strategy

The dissertation's theory-forming strategy integrates three complementary modes of reasoning—deductive, inductive, and abductive—each operationalized through specific methodological approaches aligned with our five-cycle DSR programme.

#### *Deductive Strand (Cycle 1: Exploration)*

The initial cycle employs hypothetico-deductive reasoning to test theory-derived propositions about personality-role-motivation relationships (Trochim & Donnelly, 2007). This exploratory phase operationalizes deduction through:

- **Hierarchical cluster analysis** using Euclidean distances and complete-linkage criteria to identify personality typologies
- **Normality testing** via Shapiro-Wilk and Kolmogorov-Smirnov tests to validate parametric assumptions
- **ANOVA and Kruskal-Wallis tests** to examine motivational differences across programming roles
- **Linear mixed-effects (LME) screening** to account for repeated measures and individual variations

This deductive machinery, applied in Studies I-II, establishes the empirical foundation by testing whether theoretically predicted personality-motivation patterns manifest in collaborative programming contexts.

#### *Inductive Strand (Cycles 2–3: Design and AI Extension)*

The middle cycles embrace inductive reasoning, allowing patterns to emerge from rich qualitative data. This shift reflects the need to understand novel Human ⊡ AI dynamics not adequately captured by existing theory. The inductive approach manifests through:

1. **Essentialist thematic analysis** with inter-coder cross analysis ensuring interpretive rigor
2. **Interpretative Phenomenological Analysis (IPA)** exploring lived experiences of AI collaboration as a *phaneron* – the "collective total of all that is in any way or in any sense present to the mind" (Peirce, 1931, Vol. 1, para. 284)
3. **Template analysis** providing structured yet flexible coding frameworks

4. **LME modeling with interaction effects** to inductively identify personality × role patterns

5. **Bootstrapped IMI gain estimates** quantifying motivational improvements without assuming specific distributions

This methodological ensemble, spanning Studies I-VI, enables bottom-up theory construction, particularly crucial for understanding emergent AI interaction modes (Co-Pilot, Co-Navigator, Agent) and their psychological affordances.

### Abductive Strand (Cycles 4-5: Integration and Validation)

The final cycles employ abductive reasoning—what Peirce (1931) termed "the logic of discovery"—to reconcile deductive expectations with inductive findings. This integrative phase combines:

1. **Design-proposition mapping** aligning ROMA framework components with ISO/IEC 29110 processes

2. **Expert peer review,** including WG24 members validating theoretical-practical alignment

3. **Metamodel compliance checks** ensuring framework coherence across abstraction levels

4. **Quasi-experimental field testing** with pre/post IMI measurements in VSE contexts

5. **Mixed-methods triangulation** integrating quantitative metrics with phenomenological insights

This abductive synthesis, crystallized in Studies IV-V and the PDPPA validation, generates explanatory frameworks that account for both predicted patterns and surprising discoveries, such as how AI agents satisfy relatedness needs through pseudo-social interaction.

### Methodological Integration Across Cycles

The three reasoning modes interweave throughout the dissertation, creating what Dubois and Gadde (2002) term "systematic combining." While Cycle 1 emphasizes deduction, inductive insights already emerge from thematic analysis. Similarly, Cycles 2-3's inductive focus incorporates deductive hypothesis testing through LME models. The abductive synthesis in Cycles 4-5 draws on both prior modes while generating novel theoretical propositions.

This triadic approach—grounded in specific methodological implementations—ensures that our theory of personality-driven motivation remains both empirically robust and practically relevant. By explicitly linking reasoning modes to research methods, we create a transparent epistemological architecture suited to the complex, evolving landscape of Human  AI collaborative programming.

A detailed treatment of the mixed-methods designs that instantiate this strategy is provided in Section 3.4.

## 1.8   Author's Motivation

My path toward understanding programming role optimization emerged not through linear progression but through the accumulation of observations across seemingly disparate contexts—each revealing different facets of the same fundamental question: How do we preserve human flourishing in software engineering as the social fabric of teams vicissitudes?

### 1.8.1   Academic grounding

**Doctoral research (Charles University):** Focused on constraint programming, I examined how planning and optimization problems are *represented and solved* by intelligent agents. This work sharpened my appreciation of how artificial reasoning can complement human problem-solving in software development.

**Master's and bachelor's studies (Prague University of Economics and Business, University of Queensland):**  Information theory and language design studies culminated in contributions to the JSR-303 Bean Validation specification and Java 9. These projects taught me that abstractions only gain meaning through implementation—that the gap between theoretical elegance and practical utility must be bridged through careful attention to human factors and collaborative dynamics.

Collectively, these programs cultivated a dual lens: one attuned to formal, machine-oriented optimization and the other to human-centered collaboration.

### 1.8.2   Professional insights

Across a spectrum of settings—from bare-metal systems programming to enterprise architectures, from open-source communities to AI-augmented startups—I encountered certain patterns that emerged with troubling consistency:

- **Role-person misalignment in small teams.** In VSEs and SOHOs, I witnessed how mismatched assignments created cascading failures: diminished motivation breeding technical debt, stress fracturing team cohesion, innovation withering under the weight of frustration.

- **AI's disruption of traditional roles.** As intelligent assistants entered everyday workflows, traditional role taxonomies proved insufficient; AI shifted task boundaries and reshaped what "pilot," "navigator," or "reviewer" meant.

- **The absence of principled alignment mechanisms:** Teams relied on intuition and ad-hoc adjustments, lacking frameworks to systematically match personalities with roles, especially as AI introduced unprecedented collaborative configurations.

These observations convinced me that personality-sensitive, motivationally informed role allocation—extended to include AI collaborators—could materially improve both developer self-determination and team dynamics.

### 1.8.3   Research impetus

The dissertation, therefore, pursues three interlocking objectives:

- **Model** the stable personality and motivational traits that influence programming-role preferences, creating a theoretical foundation for systematic optimization.

- **Design** the ROMA framework to align human and AI roles with those traits.

- **Validate** its efficacy through artifacts and studies situated in the contexts where it matters most: resource-constrained VSEs, distributed SOHOs, and formative educational environments.

My ambition extends beyond academic contribution to practical impact: enabling small, agile teams to navigate Human ⊡ AI collaboration while preserving—indeed enhancing—the intrinsic satisfactions that drew us to programming in the first place.

## 1.9   Contributions

This dissertation offers three interwoven contributions that, like the Borromean rings, are inseparable yet distinct. First, the ROMA framework provides an empirically grounded architecture for understanding how personality traits modulate the relationship between programming roles and intrinsic motivation—a contribution that extends Self-Determination Theory into the uncharted territory of Human ⊡ AI collaboration. Second, the ROMA AI Adapter reveals how personality archetypes specialize rather than disappear when encountering artificial intelligence, creating five distinct stances that preserve human agency while embracing augmentation. Third, the ISO/IEC 29110 Software Basic Profile and Agile Guidelines extension operationalizes these insights for Very Small Entities, those David-like organizations confronting the Goliath of digital transformation with limited resources but unlimited ingenuity. Together, these contributions establish a new vocabulary for discussing what happens when human and artificial intelligence interleave—not merge, not compete, but create something unprecedented: a form of collaborative cognition that demands new theoretical frameworks and empirical methods.

The primary artifacts and their contributions to Behavioral Software Engineering (BSE) are summarized in Table 4:

Table 4: Key research artifacts and their contributions

| Artifact | Nature & Scope | Theoretical contribution | Practical contribution to BSE |
|---|---|---|---|
| **1. ROMA Framework** | Formal model linking personality traits, motivational states, and programming-role configurations in human-human contexts | • Synthesizes psychological, motivational, and group dynamic constructs into a unified explanatory schema<br>• Extends role theory by showing intrinsic-motivation mediation<br>• Establishes five personality archetypes with distinct role preferences<br>• Empirically validated via 4 controlled pseudo-experiments and 5 interpretive studies | • Diagnostic and alignment tool for researchers, educators, & practitioners<br>• Evidence-based heuristics for assigning & enhancing roles to maximize engagement<br>• Personality assessment protocols using BFI-10/44<br>• Motivation monitoring via IMI/MWMS |
| **2. ROMA AI Adapter** | Theoretical framework mapping personality archetypes to AI collaboration specializations across three interaction modes (Co-Pilot, Co-Navigator, Agent) | • Reveals how personality patterns specialize rather than disappear in AI contexts<br>• Introduces five AI specializations: Promethean, Conductor, Hermit, Cartographer, Shapeshifter<br>• Extends SDT to synthetic relatedness and meta-autonomy concepts<br>• Validated through phenomenological analysis of 42 developers | • Guidelines for matching personality types to AI interaction modes<br>• Strategies for preserving essential human capabilities<br>• Framework for conscious AI adoption based on individual differences<br>• Vocabulary for discussing Human ⊞ AI collaboration dynamics |
| **3. ISO/IEC 29110 Extension for VSEs** | Standards-compliant implementation guide that operationalizing ROMA within resource-constrained teams; supported by a *Personality-Driven Pair Programming Application* (PDPPA) | • Bridges micro-level motivational theory with macro-level process standards, enriching the BSE knowledge body on socio-technical alignment<br>• Demonstrates AI-augmented role schemas in formal process models<br>• Shows how phenomenologial insights translate to codified practice | • Seven personality-driven tasks mapped to PM and SI processes<br>• Ready-to-use templates and guidelines<br>• PDPPA platform for continuous monitoring of Human ⊞ AI roles<br>• Incremental adoption pathway for resource-constrained teams |

*Table 4 presents the three primary dissertation artifacts, showing how each contributes both theoretical advancement and practical tools to the field of Behavioral Software Engineering. The progression from framework through specialization to implementation reflects the dissertation's journey from discovery to application.*

### 1.9.1   Aggregate Impact on Behavioral Software Engineering

These contributions coalesce to advance BSE along four dimensions:

I. **Theoretical Enrichment:** The ROMA framework and AI Adapter supply empirically-validated, personality-centered theories of motivational alignment in both human-only and Human  AI programming contexts—addressing recognized lacunae in BSE literature while opening new research directions.

II. **Conceptual Innovation:** The AI specializations provide new vocabulary and frameworks for understanding how developers maintain agency and identity in AI-augmented environments, moving beyond simplistic replacement narratives to nuanced adaptation patterns.

III. **Standards-Aligned Pragmatism:** The ISO/IEC 29110 Software Basic Profile and Agile Guidelines extension demonstrates how abstract psychological constructs can be operationalized within established software engineering processes, making theoretical insights actionable for practitioners who lack the luxury of extensive experimentation.

IV. **Evidence-Based Tooling:** The PDPPA exemplifies how continuous empirical feedback can refine role optimization strategies, creating a virtuous cycle where practice informs theory and theory enhances practice.

Together, these contributions establish a foundation for more humane software engineering practices—ones that recognize developers not as interchangeable resources but as unique individuals whose personality-driven motivations and satisfactions matter as much as their code output.

## 1.10   Structure

Like winter fog lifting to reveal a landscape both familiar and transformed, this dissertation unfolds across ten chapters that trace the journey from theoretical foundations through empirical discovery to practical application.

   **Part I: Foundations (Chapters 1-4)** establishes the intellectual terrain. Chapter 1 introduces the research landscape where personality, motivation, and AI collaboration converge. Chapter 2 presents our theoretical synthesis—weaving established knowledge with the emergent ROMA framework to create a unified understanding of personality-driven motivation. Chapter 3 illuminates the Design Science Research methodology that guides our empirical exploration. Chapter 4 serves as a cartographer's guide, mapping the interconnections among six empirical studies that collectively build our understanding.

**Part II: Empirical Cycles (Chapters 5-9)** chronicles the systematic investigation through five DSR cycles. Each cycle reveals new dimensions while building toward the complete framework:

- Cycle 1 (Chapter 5) explores the bedrock—how personality traits shape programming role preferences

- Cycles 2-3 (Chapters 6-7) construct and extend the ROMA framework, first for human collaboration, then embracing AI's transformative presence

- Cycle 4 (Chapter 8) translates insights into practice through ISO/IEC 29110 standards integration

- Cycle 5 (Chapter 9) validates the complete system through application development and empirical triangulation

**Part III: Synthesis (Chapter 10)** draws together the threads—findings crystallize into contributions, limitations become launching points for future inquiry, and the path forward emerges through the clearing mist.

This structure mirrors the research journey itself: from questions born of practice, through systematic investigation, to frameworks that honor both scientific rigor and human experience. Each chapter builds upon its predecessors while adding its own essential perspective, creating not a linear argument but a multifaceted understanding of how personality, motivation, and technology interweave to uphold the social fabric of modern software development.

## 1.11   Note on Writing Style

This dissertation adopts stylistic choices that reflect its interdisciplinary nature and dual commitment to academic rigor and practical accessibility. Following APA 7 conventions (American Psychological Association, 2020), the prose navigates between registers as context demands.

Voice modulation serves specific purposes:

- **Passive constructions** maintain objectivity in empirical reporting

- **First-person singular** acknowledges authorial decisions and philosophical positions

- **First-person plural** invites readers into shared exploration, following established conventions in software engineering and psychology

Active voice emerges strategically to emphasize novel contributions and practical implications, creating what Thomson and Kamler (2016) term "rhetorical emphasis through syntactic variation."

Additional conventions enhance clarity and navigation:

- *Italics* introduce technical terms; **bold** highlights key findings; ***bold italics*** mark pivotal concepts; *intense emphasis* stylistically interleaves the three
- Interrogative openings frame complex sections, transforming potential perplexity into guided discovery
- Cross-references include both section numbers and descriptive titles for future items and just the latter for retrospection, respecting readers' cognitive load

These stylistic decisions balance accessibility with academic rigor, ensuring complex theoretical constructs remain comprehensible to diverse audiences while maintaining scholarly precision.

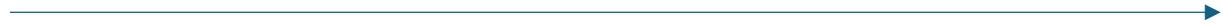

As winter fog once rolled across Newcastle's computing laboratories where Randell and Parnas debated software engineering's foundations, today's developers navigate a different mist—one where the boundaries between human creativity and algorithmic generation blur like watercolors in rain. This dissertation invites the reader—hero on quest—through this transformed terrain, where timeless questions of human motivation and identity intersect with AI's transformative potential. Like morning mist that clings to valleys while peaks emerge in dawn light, our research questions stand clear while their answers remain tantalizingly veiled. The empirical journey ahead promises not to banish the fog but to transform it—from barrier to medium, from obstacle to revelation, from uncertainty to understanding.

# THEORETICAL FOUNDATIONS AND THE ROMA FRAMEWORK

L ike winter fog that clings to Silicon Valley's morning hills, obscuring familiar landmarks while revealing unexpected contours, the theoretical landscape of software development has become simultaneously clearer and more mysterious. Where once we navigated by the fixed stars of waterfall methodologies and solo craftsmanship, we now find ourselves in a liminal space where human creativity interweaves with artificial intelligence, where personality shapes productivity, and where the ancient satisfactions of programming must be rediscovered in radically new forms.

The articulation and application of well-defined theoretical constructs is fundamental to the advancement of scientific inquiry, as Einstein (1949) observed (p. 674):

*"Thinking without the positing of categories and of concepts in general would be as impossible as is breathing in a vacuum."*

Yet in this cumulative dissertation—where empirical studies published across multiple years converge toward a unified understanding—theory cannot simply precede discovery. Instead, it must weave through it, like fog through trees, revealing connections that linear exposition would obscure.

## The Architecture of Understanding

Before we can speak of minds collaborating—biological or artificial—we must first acknowledge the ground upon which such collaboration stands. Software development, at its essence, is an act of bringing forth (*poiesis*): translating the ineffable dance of thought into the crystalline structures of code. Yet this translation occurs not in isolation but within what Heidegger (1962) might term a state of *Dasein's* thrownness—developers find themselves already embedded in technological landscapes they neither created nor fully control, yet must navigate with intentionality and skill.

This chapter traces the conceptual lineages that inform our understanding of how personality, motivation, and collaborative dynamics shape this navigation. I begin not with definitions but with recognitions: that the programmer exists simultaneously as thinker and maker, as individual and collaborator, as human agent confronting both the resistance of material reality and the emerging presence of artificial minds that promise—or threaten—to transform the very nature of creative work.

### A Note on Theoretical Integration in Cumulative Research

This chapter employs an unconventional structure necessitated by the cumulative nature of this dissertation. Rather than relegating the ROMA framework to a separate "contributions" chapter, I integrate it throughout the theoretical landscape—demonstrating not just what we build upon, but how our empirical findings extend, challenge, and transform existing knowledge. This approach serves three deliberate purposes:

First, it honors the reality of theory development in behavioral software engineering, where insights emerge through iterative dialogue between established knowledge and empirical discovery. The ROMA framework did not spring forth fully formed but crystallized through progressive engagement with Self-Determination Theory, personality psychology, and the lived experiences of developers navigating AI collaboration.

Second, it provides readers with a complete theoretical apparatus before encountering the detailed empirical journeys in subsequent chapters. By understanding how the ROMA framework relates to existing theory now, readers can better appreciate the significance of each empirical finding as it unfolds across the five design science cycles.

Third, it reflects this dissertation's epistemological stance: that in rapidly evolving fields, the boundary between "background" and "contribution" blurs. When studying phenomena as novel as Human  AI programming, every theoretical application becomes a potential extension, every empirical finding a possible revision to received wisdom.

### The Pedagogical Architecture of Repetition

Like morning fog that reveals the same landscape differently as it lifts, this chapter deliberately revisits key concepts from multiple angles. Autonomy, competence, and relatedness first emerge through the urgent lens of AI collaboration (Section 2.3), where their practical implications for contemporary developers become immediately apparent. Only then do we trace their theoretical genealogy (Section 2.4), grounding urgent insights in decades of psychological research.

This repetition with progressive deepening serves not redundancy but revelation. A developer struggling with Copilot's suggestions needs first to recognize their autonomy concerns before appreciating Ryan & Deci's (2000) theoretical framework. A team leader observing personality clashes in pair programming must see the phenomenon before the Big Five model becomes meaningful. By inverting traditional exposition—practice before theory, recognition before definition—we mirror how practitioners actually encounter and make sense of psychological constructs.

The chapter thus unfolds as a spiral rather than a line:

**Sections 2.1-2.2** establish the foundational concepts of mind and motivation in software engineering, preparing the conceptual ground.

**Sections 2.3-2.5** present the heart of our theoretical synthesis, showing how personality traits interact with AI collaboration modes and self-determination needs. Here, findings from our six empirical studies interweave with existing literature, demonstrating how the ROMA framework both builds upon and extends established knowledge.

**Section 2.6** crystallizes theoretical insights into testable hypotheses, creating the bridge between conceptual understanding and empirical investigation.

**Sections 2.7-2.8** ground abstract insights in biological reality (neuroscience) and practical application (ISO/IEC standards), completing the movement from theory through empirics to practice.

Throughout, phenomenological accounts from our participants serve as recurring motifs—reminding us that behind every theoretical construct lies a developer seeking not just productivity but meaning in their work.

### The Living Theory

What emerges is neither pure literature review nor isolated framework presentation but what we term a "living theory"—one that breathes with the experiences of developers confronting unprecedented challenges. In an age where AI can generate code but not purpose, where personality shapes collaboration but cannot determine it, where motivation must be cultivated rather than commanded, such theoretical integration becomes not just useful but essential.

As the fog lifts throughout this chapter, revealing first the familiar peaks of established theory and then the new territories mapped by the ROMA framework, remember that the landscape itself continues to shift. Each reader brings their own experience to this theoretical terrain, their own struggles with AI assistants and team dynamics, their own personality-shaped perspective on what programming means. The theory we present is not complete—cannot be complete—without that engagement.

Let us begin, then, not with answers but with acknowledgments: that software development has become irreducibly complex, that human factors matter more than ever precisely as artificial intelligence grows more capable, and that understanding this new landscape requires both the wisdom of established theory and the courage to extend it into unexplored territory.

## 2.1 Relevant Contexts

*In an era where artificial minds increasingly share the creative burden, what sustains the human drive to code—and what transforms it?*

To address these questions, we must first map the conceptual terrain where behavioral science and software engineering converge. This convergence is not merely academic—it represents the lived reality of developers who daily negotiate between human needs and technological demands, between individual expression and collaborative necessity, between the comfort of established practices and the vertigo of continuous transformation.

### 2.1.1 The Context of Constraint

Very Small Entities (VSEs) and Small Office Home Offices (SOHOs) operate within what we might term "economies of essence"—every decision carries disproportionate weight, every team member's engagement becomes critical to survival. In these contexts, understanding psychological dynamics transcends academic interest to become existential necessity.

VSEs, defined as organizations with fewer than 25 employees (Laporte et al., 2018), face a paradox of flexibility: their small size enables rapid adaptation yet magnifies the impact of individual disengagement. When a team of five loses one member to burnout or motivation loss, they lose not 20% but often 30% or more of their collective capability, given the non-linear nature of collaborative knowledge work (Brooks, 1975; Weinberg, 1971).

The COVID-19 pandemic rendered visible what had long been latent: the fragility of psychological well-being in distributed work contexts. Ralph et al.'s (2020) global study of over two thousand software professionals revealed not merely productivity challenges but profound disruptions to the basic psychological architecture that sustains creative work. Developers reported feeling simultaneously more autonomous (working from home) yet less competent (struggling with new tools and processes) and profoundly disconnected from their teams—a triad of experiences that Self-Determination Theory would predict as particularly detrimental to intrinsic motivation (Ryan & Deci, 2000a).

This disruption aligns with what organizational psychologists term "role ambiguity" (Rizzo et al., 1970) and "technostress" (Tarafdar et al., 2019)—phenomena that become particularly acute in resource-constrained environments where each individual must navigate multiple, often conflicting, role expectations. The ROMA framework emerges from recognizing these constraints as not merely obstacles but as defining features of contemporary software development that demand a theoretical and practical response.

### 2.1.2 The Emergence of the Artificial Other

As we write in early 2025, the landscape of software development has been fundamentally altered by what phenomenologists might recognize as the appearance of an "artificial other"—*AI systems that no longer merely assist but actively participate* in the creative process. GitHub's Copilot Workspace, Anthropic's Claude 4 Opus, OpenAI's o3, and emerging multi-agent systems like Devin and Cursor's Composer have crossed what Dennett (1989) might term the "intentional stance" threshold—we find it natural, even necessary, to attribute beliefs, desires, and intentions to these systems to predict and interact with them effectively.

This transition carries profound implications for how we understand motivation and identity in software development. When a developer works with OpenAI o3's reasoning capabilities or Cursor's multi-file editing agent, they engage with an entity that exhibits what we might term "functional autonomy"—the ability to make decisions, pursue goals, and even surprise its human collaborator with unexpected solutions. The philosophical question posed by Floridi (2014) regarding the "fourth revolution"—where humanity is displaced from the center of the infosphere—becomes concrete in the daily experience of developers who must negotiate creative agency with artificial systems.

The emergence of AI as collaborative partner rather than mere tool resonates with Latour's (2005) Actor-Network Theory, which recognizes non-human actants as full participants in sociotechnical assemblages. In the context of software development, AI systems become what Latour might term "mediators" rather than "intermediaries"—they transform, translate, and modify the meaning of the code they help produce, rather than merely transmitting human intentions faithfully.

### 2.1.3 Empirical Voices: The Participants

The theoretical explorations in this chapter are grounded in the lived experiences of software developers navigating the evolving landscape of human-AI collaboration. Drawing from three empirical studies conducted between 2021 and 2025, we incorporate insights from diverse perspectives that illuminate the intersection of personality, motivation, and programming practice.

**Study III (2023)** captured the early adoption phase of AI tools through essentialist semi-structured interviews with two cohorts: five undergraduate students (SA, SB, SΓ, SΔ, SE) experiencing AI-assisted programming in educational contexts, and five professional developers (PA, PB, PΓ, PΔ, PE) with substantial AI tool experience ($\bar{x}$ = 11.2 months, $s$ = 7.6).

**Study IV (2021-2022)** provided longitudinal insights through post-experimental interviews across three semesters of pair programming experiments, with participants coded as I1–I12 (WS'21), P1–P7 (WS'22), and S1–S6 (SS'22), offering perspectives on human-human collaboration that serve as a baseline for understanding AI's transformative impact.

**Study VI (May 2025)** employed phenomenologically-informed interviews to capture the mature phase of AI integration, featuring four senior developers (MM, PJ, JO, LV), a researcher-developer optophysicist (MH), a Gen-Z student (NK), and a CTO-architect (MV). These participants, interviewed at the frontier of human-AI collaboration, provide insights into how established developers navigate the profound shifts in agency, identity, and practice that AI collaboration entails.

Our Study VI methodology introduces an additional hermeneutic layer through deliberate Czech-English code-switching, surfacing culturally embedded nuances in participants' meaning-making. This bilingual oscillation mirrors Gadamer's (1960) "fusion of horizons," maintaining reflexive distance while honoring vernacular expressions of embodied frustration (*"koušu si extrémně nehty"*), surreal encounters (*"jako bys byl ve snu"*), and restored mastery (*"šťastnej, že jsem to dokázal sám"*). The double-hermeneutic of IPA—where participants interpret their experience while we interpret their interpretations—gains additional depth through this linguistic interplay.

Throughout this chapter, their voices—sometimes convergent, often divergent, always authentic—illuminate the theoretical frameworks with the texture of lived experience. When a senior developer describes AI as "extrémně rychlý brainstorming partner" or a student confesses to feeling "lazy and dependent," we witness theory encountering practice in ways that both confirm and complicate our understanding.

## 2.2   Mind

### 2.2.1   Tripartite Mind in Software Engineering

*Why is it essential to understand the cognitive, motivational, and affectional aspects of software engineering?*

The question invites us beyond mere productivity metrics to consider software development as a fundamentally human activity—one that engages the full spectrum of mental life. When we write code, we do not merely manipulate symbols; we project possible worlds, negotiate between competing constraints, and navigate the uncertain space between intention and implementation.

The tripartite conception of mind—cognition, conation, and affect (Hilgard, 1980)—offers a lens through which to understand these complexities. This classical division, refined through centuries of philosophical inquiry from Plato's tripartite soul through Kant's faculties of mind, provides structure for examining how developers experience and navigate their work.

**Cognition in Software Engineering** manifests not merely as problem-solving capacity but as the ability to hold multiple abstraction levels simultaneously in consciousness. A developer debugging a distributed system must maintain awareness of local variable states while considering network latencies, data consistency models, and user experience implications. This cognitive load is not merely quantitative but qualitative—each level of abstraction demands its own mode of thinking (Dijkstra, 1989).

Recent advances in cognitive science, particularly predictive processing frameworks (Clark, 2023; Friston, 2010), suggest that programmers develop sophisticated generative models of their systems—internal simulations that allow them to "run" code mentally before implementation. This aligns with Newell and Rosenbloom's (1981) power law of practice, wherein skill acquisition follows predictable patterns through increasingly refined mental models. Such mental simulation proves particularly crucial when working with AI collaborators, as developers must model not only their systems but also the likely behaviors and limitations of their artificial partners.

While Sweller's (1988) cognitive load theory explains how increasing system complexity affects performance, it barely captures the phenomenological richness of a developer's experience. Recent neuroscientific research by DiDomenico and Ryan (2017) reveals that intrinsic motivation correlates with specific patterns of neural activation in the default mode network—suggesting that the subjective experience of programming engages fundamental brain systems associated with self-referential processing and creativity.

**Conation (Motivation) in Software Engineering** reveals itself in the sustained effort required to push through what Norman (2013) terms the "gulf of execution"—that chasm between elegant conceptual solutions and their often messy realization in code. The motivational landscape of programming is marked by what we might call "recursive frustration": solving one problem often reveals three more, each solution opening new vistas of complexity (Brooks, 1987).

França et al. (2014) synthesized decades of research to identify key motivational factors in software engineering, including autonomy, variety, and significance. Their work builds upon earlier foundations by Beecham et al. (2008) and Hall et al. (2008), who noted the field's lack of systematic understanding of developer motivation despite its critical importance to project success.

Yet within this frustration lies a peculiar satisfaction. As one optophysicist noted: "Je to takový moment, kdy se všechno vyjasní a všechno do sebe zapadne," [there's this moment when everything becomes clear and everything clicks] (Study VI–MH). This experience of discovery within creation speaks to programming's unique motivational structure, where the boundary between invention and finding dissolves.

**Affect in Software Engineering** encompasses not only the emotional responses to success and failure but also the more subtle affective colorings that permeate the programming experience. There is the quiet anxiety of pushing to production, the aesthetic pleasure of elegant code, the social warmth of a successful pair programming session, and increasingly, the uncanny valley of collaborating with AI systems that seem to understand yet fundamentally do not comprehend.

This existential dimension of programming—what Sartre (1946) might recognize as the confrontation with radical freedom in the face of infinite possibilities—becomes particularly acute in debugging. Each bug represents what Sartre termed "bad faith" made manifest: the gap between our intentions and their realization in code, forcing us to confront the authentic responsibility of creation.

Graziotin et al.'s (2014, 2018) extensive research program has demonstrated bidirectional relationships between affect and performance in software development. Happy developers not only solve problems more effectively but also produce code with fewer defects and better architectural decisions. Conversely, negative emotions—frustration, anxiety, anger—correlate with increased technical debt and reduced team cohesion.

The integration of these three dimensions becomes particularly critical in the age of AI collaboration. When a developer works with agents or assistants, they must cognitively model the AI's capabilities, maintain motivation despite the AI's occasional confident incorrectness, and manage the affective complexity of relating to a non-human collaborator that increasingly exhibits human-like communicative patterns.

### 2.2.2 Asymmetric Theory of Mind

The capacity to model others' mental states—Theory of Mind (ToM)—takes on new dimensions in contemporary software development. Traditional pair programming required developers to maintain what Salinger et al. (2008) identified as "shared mental models" and what Zieris and Prechelt (2021) later termed "togetherness"—real-time awareness of their partner's understanding, intentions, and emotional state. This mutual modeling enabled the subtle dance of driver and navigator, each anticipating the other's needs and adjusting their contributions accordingly.

The introduction of AI collaborators fundamentally disrupts this dynamic. Current AI systems, despite their sophisticated outputs, lack genuine mental states to be modeled. Yet developers often report experiencing these systems as minded entities. A senior developer (LV) captured this liminal experience: "Vím, že Claude vlastně o mém kódu 'nepřemýšlí', ale přistihnu se, že mu věci vysvětluju, jako by potřeboval pochopit moje uvažování. Jako bych s ním diskutoval o architektuře," [I know Claude doesn't really 'think' about my code, but I find myself explaining things to it as if it needs to understand my reasoning. As if I'm discussing architecture with it.]

This phenomenon—what we might term "asymmetric theory of mind"—represents a novel form of human-computer interaction that extends beyond traditional HCI frameworks. While Reeves and Nass (1996) demonstrated that humans naturally apply social rules to computers, AI collaboration intensifies this tendency, echoing Wiener's (1950/1954) prescient warning that machines "which can learn and can make decisions... will in no way be obliged to make such decisions as we should have made." The human developer must simultaneously maintain two contradictory models: one treating the AI as a minded collaborator (enabling natural interaction) and another recognizing its fundamental non-consciousness (preventing over-reliance and maintaining critical evaluation).

This asymmetry manifests in developers' strategic anthropomorphization. Participant (SĽ) demonstrated this cognitive flexibility: "ChatGPT's success rate is way better when you prompt it in English (80%) than Slovak (50%)", suggesting developers model AI's "linguistic preferences" despite knowing these reflect training data distributions rather than genuine comprehension. The pragmatic adoption of intentional stance becomes a tool for effectiveness rather than a philosophical commitment—what Dennett (1978) earlier characterized as taking a "design stance" toward systems whose complexity exceeds our ability to predict from physical principles alone.

Recent developments in AI systems have made this balancing act even more complex. OpenAI's o-series models introduced explicit "thinking" tokens, making visible a process that resembles human reasoning. Anthropic's Constitutional AI approach imbues Claude with something approaching ethical principles (Bai et al., 2022). These developments create what phenomenologists might recognize as a crisis of intentionality (Husserl, 1913): toward what, exactly, is our collaborative intention directed when we work with these systems?

Zhang et al.'s (2024) work on "Mutual Theory of Mind" in human-AI interaction provides empirical grounding for these philosophical questions. Their findings suggest that AI's ability to model human intentions, even if through statistical pattern matching rather than genuine understanding, creates more effective collaboration. This pseudo-reciprocal ToM enables what our participants described as "produktivní iluze porozumění" [productive illusion of understanding].

## 2.3   Artificial Intelligence and Human  AI Collaborative Programming

*Novel Theoretical Contribution: The following sections (2.3-2.6) present the ROMA framework and its theoretical components, which constitute the primary contribution of this dissertation. The insights presented here—including personality-AI mode alignments, self-determination fulfillment patterns, and the hypotheses that guide our empirical investigation—emerged from the iterative empirical studies detailed in Chapters 5-9. They are presented here in their complete theoretical form to provide coherence and enable practical application. The detailed methodology and data supporting these insights are systematically presented in the corresponding empirical chapters.*

The concept of Artificial Intelligence has evolved from Turing's (1950) philosophical provocation— "Can machines think?"—to concrete systems that demonstrably transform human cognitive work. In software development, this transformation manifests through three distinct interaction paradigms that reshape not only how code is written but what it means to be a programmer.

### 2.3.1   AI's Transformation of Programming Roles

The integration of AI into software development represents what Kuhn (1962) might recognize as a paradigm shift—not merely new tools but new ways of conceiving the programming task itself. Based on a focused systematic literature review of 18 primary studies (2022-2025) (Appendix A), complemented by empirical findings from Study III with both professionals and students and Study VI with professionals, this section identifies three distinct *"AI interaction modes"*—Co-Pilot, Co-Navigator, and Agent—each supporting specific collaborative roles with unique dynamics and psychological affordances that align differently with personality traits and motivational profiles. This transformation fulfills Minsky's (1986) vision of mind as a "society" of agents—though perhaps not as he imagined, with artificial agents now joining human ones in the cognitive collective of software development.

#### *Co-Pilot Mode*

The augmentation paradigm emerged with GitHub Copilot, Cursor Tab, and similar inline completion tools that embed context-aware code suggestions directly in the IDE, varying mainly in *aggressiveness*: some stream entire blocks, lifting routine load, while others surface concise snippets, preserving agency.

Here, AI serves as what William Gibson might have called a "cognitive prosthesis"—extending human capability without replacing human agency. In this mode, AI recedes into transparent utility, **_"a tool"_**, becoming an extension of the developer's thought process rather than a separate entity. Key Co-Pilot roles include:

- **Accelerator:** Speeds development by generating routine code segments

- **Safety-Net:** Catches potential errors or syntax issues during coding

Empirical studies by Liang et al. (2024) and Mozannar et al. (2024) indicate that Co-Pilot mode enhances flow states (Csikszentmihalyi, 1990) and reduces cognitive load for routine coding tasks while preserving a sense of autonomy. However, questions remain about its long-term effects on competence development.

Study III participants articulated this experience precisely: "I now work much faster because I do not have to google that much anymore" (PA), while another noted the dual nature of this efficiency: "If I did not have CoPilot at my disposal I would feel like I miss it and my productivity would shrink by 30%. In this way, I am dependent on AI" (PΓ). This dependency paradox—enhanced productivity coupled with potential skill atrophy—emerged as a central theme across both professional and student cohorts.

From a personality perspective, developers high in Openness to Experience show particular affinity for Co-Pilot mode's rapid ideation capabilities. As one high-openness senior developer (MM—Study VI) noted: "Copilot je jako extrémně rychlý brainstorming partner. Hodí mi deset nápadů, z nichž devět zahodím, ale ten desátý mě posune úplně jinam" [Copilot is like an extremely fast brainstorming partner. It throws me ten ideas, nine of which I discard, but the tenth takes me somewhere completely different]. This aligns with DeYoung's (2015) characterization of Openness as cognitive exploration—the Co-Pilot mode amplifies the divergent thinking that open individuals naturally employ.

The phenomenology of Co-Pilot interaction reflects what senior developer (LV) related to musical composition: "Copilot mění způsob, jak přemýšlím o kódu. Už nepíšu řádek po řádku, ale skládám větší bloky. Je to jako přejít od psaní jednotlivých not k dirigování symfonie," [_Copilot changes the way I think about code. I don't write line-by-line anymore, I put together blocks. It's as if I went from writing notes to directing a symphony._]

A professional developer (PA) described Co-Pilot as a transformed agency: "I now work much faster because I do not have to google that much anymore." Yet this acceleration comes with subtle redistributions of cognitive labor. A Gen-Z student (SΔ) captured the paradox: "You feel like you did it yourself because you are the one writing the questions to AI. I like the feeling that I have accomplished something complicated alone". This preservation of authorial identity despite AI mediation suggests Co-Pilot mode successfully maintains what de Charms (1968) termed "origin" causality.

## Co-Navigator Mode

The conversational paradigm, exemplified by ChatGPT and Claude, resurrects dialogue as a primary programming modality with AI as ***"an assistant"***. Their value hinges on *conversation memory* and *intent tracking*: some integrate tightly with the IDE and inherit full project context; others depend on the developer to paste snippets. This mode engages what Bakhtin (1981) termed "dialogical consciousness"—meaning emerges through interaction rather than solitary reflection. Featuring conversational interfaces and multi-turn dialogue, the Co-Navigator mode provides guidance and explanation rather than direct code generation. Key Co-Navigator roles include:

- **Mentor:** Explains concepts and approaches, supporting learning

- **Rubber Duck:** Serves as a sounding board for problem-solving

- **Domain Expert:** Provides specialized knowledge in unfamiliar areas, potentially assisting pair programmers (Zieris & Prechelt, 2020)

- **Critic:** Reviews code and suggests improvements

Research by Robe & Kuttal (2022) and Hamza et al. (2024) suggests that Co-Navigator mode enhances understanding and supports higher-level reasoning, with particular benefits for code quality through deliberative processes. The role framing of the AI significantly influences how developers engage with and learn from these interactions.

Our empirical data reveals nuanced patterns in how developers utilize Co-Navigator mode. Student participant SB demonstrated sophisticated tool discrimination: "CoPilot is better when you want a little help, ChatGPT when you want the whole solution." This selective approach reflects an emerging AI literacy that distinguishes between modes based on task requirements and learning objectives. Professional developers showed similar sophistication, with participant PΓ noting: "AI tools can do very good code reviews and they could support Agile practices."

One CTO (MV) reported Co-Navigator helps him in expanding his mental faculties through externalization: "ChatGPT používám jako dialogického partnera. Není to jen nástroj - je to způsob, jak externalizovat a strukturovat vlastní myšlení. Když mu vysvětluji problém, často sám přijdu na řešení," [*I use ChatGPT as a dialogue partner. It's not just a tool – it's a way to externalize and structure your own thoughts. When I explain the problem to it, I often encounter the solution.*]

The dialogical nature manifests in developers' careful cultivation of AI personas. As participant PA fantasized: "I would like my AI tool's personality like mine: sarcastic, using a lot of irony, writing to the point, no digressions. If I could choose a character, I would go for Bender from Futurama." This desire for personality alignment reveals Co-Navigator's quasi-social dimension—developers seek not merely functional assistance but characterological compatibility.

The Co-Navigator mode appears to particularly appeal to individuals high in Extraversion and Agreeableness. As Gen-Z student (NK) observed: "S ChatGPT programuju jako s mentorem. Ptám se ho na věci, na které bych se styděl zeptat skutečného seniora. A on mi vždycky ochotně vysvětlí, i když se ptám počtvrté" [*I program with ChatGPT like with a mentor. I ask it things I'd be embarrassed to ask a real senior. And it always willingly explains, even when I ask for the fourth time.*] This synthetic social interaction partially satisfies the relatedness needs that extraverted, agreeable individuals seek in collaborative work (Ryan & Deci, 2000b).

Vaithilingam et al. (2022) and Bird et al. (2023) confirm that these *AI tools and assistants* measurably influence developers' workflows and self-perceptions. Developers using AI assistants report altered approaches to problem-solving, with many adopting a more high-level, declarative style where they describe what they want rather than focusing on implementation details. This shift has been characterized as moving from *"how-oriented"* to *"what-oriented"* programming (Valový & Buchalcevova, 2023), resembling the traditional distinction between navigator and pilot roles but within a single developer's workflow.

### Agent Mode

The agentic paradigm represents the frontier where AI transitions from assistant to actor, ***"an agent"***. Systems like Devin, Cursor's Composer, Open AI's Codex, and Anthropic's Claude Code show AI operating with increasing autonomy—planning, implementing, debugging with minimal human oversight, some of which remember, reflect, and coordinate in simulated communities, foreshadowing what Park et al. (2023) term "interactive simulacra"—socially aware AI collaborators. Distinguished by autonomous execution capabilities and system resource access, the Agent mode can perform multi-step tasks with minimal human intervention. The payoff is focus on higher-level design; the risk is shifting control and diluting hands-on expertise. Key Agent roles include:

- **Executor:** Completes defined tasks independently

- **Optimizer:** Refines code or configurations based on objectives

- **Coordinator:** Manages interactions between system components

Studies by Cinkusz & Chudziak (2024) and Hassan et al. (2024) demonstrate that Agent mode can amplify productivity but potentially reduces skill development, fundamentally changing the developer's role to that of supervisor. This shift raises important questions about perceived locus of control and the quality of the developer experience.

Professional participant PΔ articulated this transformative potential: "AI tools can replace traditional programming. They can really replace programming because it is pretty smart, and if it gets even smarter, then I can imagine it taking a task assignment and converting it into a functional pull request." This vision of autonomous AI development raises fundamental questions about the future nature of programming work and developer identity.

Today's delegation of agency provokes complex responses. A CTO (MV) introduced explicit role definition: "Řídím dvacet vývojářů, v každém týmu pět, a k tomu máme firemního AI agenta. AI agent dělá dokumentaci, generuje testy a dělá code review. Je to jako mít dalšího člena týmu - technicky brilantní, ale sociálně... jiný," [*I manage twenty developers, five per team, and we also have a company AI agent. The AI agent handles documentation, generates tests, and does code reviews. It is like having an extra team member – brilliant technically but socially... different.*]

This characterization—problematic in its casual use of neurodiversity as metaphor—nonetheless captures the uncanny valley of agentic AI: possessing expertise without social intuition. Paradoxically, that is often preferred by developers high in Neuroticism or low in Extraversion.

### 2.3.2 AI and Motivation

The motivational implications of AI integration prove far more complex than initial narratives of "enhancement" or "replacement" suggested. Drawing on Self-Determination Theory (Ryan & Deci, 2000a) and our empirical work, we identify a fundamental tension: AI tools simultaneously amplify and undermine different aspects of developer motivation across the three basic psychological needs.

**Autonomy** traditionally meant freedom to choose implementation approaches, tools, and working methods. The phenomenological experience of autonomy—what Levinas (1961) might term the "infinity" of human choice—becomes complex in collaborative contexts. Yet AI collaboration introduces what we might term "mediated autonomy"—freedom exercised through and with artificial systems.

Mediated autonomy differs from traditional autonomy in three critical ways. First, the locus of control becomes distributed—developers exercise agency not through direct implementation but through prompt crafting, output curation, and strategic tool selection. Second, the feedback loop transforms from immediate (compiler errors) to interpretive (evaluating AI suggestions). Third, autonomy manifests at different abstraction levels across AI modes: Co-Pilot preserves micro-level autonomy (accepting/rejecting suggestions) while constraining macro-level choices; Co-Navigator maintains conversational autonomy while influencing problem conceptualization; Agent mode inverts this, granting strategic autonomy while removing implementation control.

Our data revealed developers face paradoxical pressures. While AI tools can enhance autonomy by accelerating implementation of human intentions, they can also create what de Charms (1968) termed a shift from "origin" to "pawn" causality. A senior developer (PJ) articulated this tension: "Někdy nevím, jestli ten kód píšu já, nebo ho píše Copilot a já jen schvaluju. Je to můj kód, když jsem napsal jen 30 procent?" *[Sometimes I don't know if I'm writing the code or Copilot is writing it and I'm just approving. Is it my code when I've only written 30 percent?].* Another senior developer (JO) applauded in reverse: "S AI mám paradoxně větší autonomii. Nemusím se zabývat nudnými detaily a můžu se soustředit na to, co mě baví - návrh architektury a řešení složitých problémů," *[with AI, paradoxically, my autonomy is greater. Instead of dealing with boring details, I can focus on what I enjoy – architecture design and complex problems.]*

Moreover, an optophysics researcher (MH)'s reflection "Chci programovat míň a míň, ale pořád chci mít kontrolu nad tím... jsem rád, když dostanu všechno předpřipravený a pak tam jenom pozměním nějaké věci," *[I want to write code less and less but still want to have control over it... I am happy to receive everything pre-made, tweaking just a few details,]* shows autonomy isn't about doing everything but about meaningful control points.

Furthermore, he introduces a new *"polidštění" (humanization)* process: "Mám lepší pocit, když to je, že vlastně mi to dává víc toho autorství k tomu kódu. Že to není jenom vygenerovaný od agenta, ale prostě jsem to tak nějak polidštil," *[I have a better feeling when it actually embeds more of my authorship to the code. That it's not generated by the agent but that I somehow 'humanized' it.]* This isn't traditional authorship but a kind of *editorial artistry*—making the mechanical gentle, the generated personal—suggesting autonomy can be exercised through modification rather than origination.

In semi-structured format, the participants reported even greater preservation of internal locus of causality, e.g., professional participant (PΓ): "When it provides the correct solution, I feel delighted because that means I provided the correct input prompt for this problem. Of course, the credit belongs to me more than the AI tool for the correct solution! Without me, AI is useless." Yet this claimed autonomy contrasts with reported dependencies, as participant PΔ confessed: "I have become lazier and dependent. Before doing something, I prompt ChatGPT, then I take and edit it. Sometimes I feel I am chatting with the tool for five minutes, and maybe it would have been quicker if I did it on my own."

Student experiences revealed similar tensions. A Gen-Z student (SΔ) maintained a sense of authorship: "You feel like you did it yourself because you are the one writing the questions to AI.

I like the feeling that I have accomplished something complicated alone." This preservation of authorial identity despite AI mediation suggests successful maintenance of "origin" causality.

**Competence** undergoes fundamental redefinition in AI-augmented contexts. Traditional markers—syntax mastery, algorithm implementation—become less relevant when AI systems possess superior recall. Yet new competencies emerge at higher abstraction levels. As one CTO (MV) noted: "Dřív jsem byl hrdý, že znám každou Java knihovnu. Teď jsem hrdý, že dokážu agenta nasměrovat k elegantnímu řešení, které by mě samotného nenapadlo," *[I used to be proud of knowing every Java library. Now I'm proud that I can direct an agent toward an elegant solution that wouldn't have occurred to me alone.]*

This shift aligns with hierarchical models of competence development (Fischer, 1980), where AI enables operation at higher skill levels, further exemplified by a senior developer (MM): "Kompetence se posunula. Už nejde o to, kolik toho umím naprogramovat, ale jak efektivně dokážu orchestrovat AI nástroje. Je to meta-dovednost," *[(the definition of software engineering c)ompetency has shifted. Now it's not about what I can program myself – but about how effectively I orchestrate AI tools. It's a meta-skill.]*

This represents not skill atrophy but skill evolution—from knowing to *knowing-how-to-know* (MV): "Už nejsem vlastně expert na nic, ale jsem schopnej... používám jenom soft skills a nějaký obecný základy lidského myšlení," *[I'm actually not an expert on anything anymore, but I'm capable... I only use soft skills and some general fundamentals of human thinking.]* This coincides with the emergence of **meta-level competencies,** wherein Martin identifies three core skills that transcend technological specifics: soft skills, fundamentals of human thinking, and problem-solving skills.

A professional developer (PГ) described this shift in self-concept: "My trajectory is different than I thought a year ago. Now I want to do many things because I don't have to go deep into one specific technology. Now you have AI that provides assistance." This broadening from depth to breadth represents movement toward more abstract skill levels, though whether this constitutes genuine competence development remains contested.

Student perspectives highlighted the learning dimension of competence transformation, as participant (SГ) noted: "(With AI m)y motivation was growing because it is a new thing to try and I really wanted to know how it works and can improve coding skills and provide new solutions for you. Chat-GPT4 boosted my motivation to learn programming and just continue this way." Yet concerns about fundamental skill development emerged, with another participant (SA) warning: "If you are starting to learn to program, it might become harder to learn from the ready-to-use solutions provided by ChatGPT. I am thankful I learned the basic concepts without ChatGPT."

This transformation of competence reveals a fundamental shift in software engineering expertise. Where traditional competence centered on recall and implementation—knowing syntax, remembering algorithms, debugging efficiently—AI-augmented competence emphasizes orchestration and evaluation. Developers must now excel at prompt engineering, output verification, and architectural judgment. This represents not skill degradation but skill evolution—from craftsperson to conductor, from maker to meta-maker. The question for educators and practitioners becomes: how do we cultivate these meta-competencies while preserving the foundational understanding necessary for effective AI collaboration?

**Relatedness** presents perhaps the most philosophically intriguing dynamics. Despite AI's non-consciousness, developers consistently report relationship-like experiences that partially satisfy relatedness needs. A phenomenon we can provisionally term "synthetic relatedness" emerges, where developers anthropomorphize AI systems sufficiently to experience quasi-social satisfaction. As Turkle (2011) warned, this risks creating "alone together" experiences—the simulation of relationship without genuine intersubjectivity. Yet our data suggests a more nuanced reality where synthetic relatedness supplements rather than replaces human connection.

The lived experience of this relatedness shift varies dramatically by context. Professional participant (PΓ) explained: "Yes, I experience different emotions with AI than humans. With AI, I am more focused on problem-solving. When I talk to a real human, I cannot be as focused as I am when talking to ChatGPT. I do not know why." This enhanced focus through quasi-social interaction suggests AI may fulfill certain aspects of relatedness needs while lacking genuine intersubjectivity.

A student developer (SΔ) expressed preference for human collaboration: "Most comfortable with a human because you can discuss your opinions. Communication is easier. Solo programming is really hard for me." Conversely, a professional (PΔ) revealed: "I am quite an introverted person so definitely the AI tool. With a human, I am always a little nervous or trying to show off how I can do things. With AI I do not care." These contrasting experiences suggest AI's impact on relatedness is mediated by personality traits—a theme explored systematically in Section 2.5.

### 2.3.3 Human ⊞ AI Collaborative Programming in VSEs

For VSEs and SOHOs, AI collaboration presents what Christensen (1997) might term a "disruptive innovation"—fundamentally altering competitive dynamics rather than merely improving existing capabilities. These organizations operate with what we might call "essential tension"—every role must be maximally productive, yet every individual needs space for growth and satisfaction. Our interviews with VSE leaders reveal three primary patterns of AI integration, each with distinct implications for the ROMA framework:

**AI as Virtual Team Member**: Some VSEs treat AI systems as additional staff, assigning them discrete tasks and responsibilities, as one CTO (MV) explained: "Je to jako mít neuvěřitelně znalého, ale trochu autistického člena týmu—geniální na specifické úkoly, ale potřebuje pečlivé vedení" [It's like having an incredibly knowledgeable but somewhat autistic team member—brilliant at specific tasks but needing careful management.]

Additionally, professional participant PA envisioned radical restructuring: "I would like to start my own company. It has always been my dream. With the help of AI, I do not need to hire employees in its beginnings." This conceptualization of AI as team member rather than tool reflects a fundamental shift in how small organizations might structure themselves.

**AI as Skill Amplifier**: Other organizations use AI to enable developers to work outside their core expertise. This "T-shaped enhancement" (Guest, 1991) allows VSEs to maintain smaller teams while covering broader technical territories. However, this creates what Dunning and Kruger (1999) identified as competence illusions—developers may overestimate their abilities in AI-assisted domains.

The democratization of expertise emerged as a key theme. Student participant SΓ observed differential AI effectiveness across domains: "I use ChatGPT for providing explanations in different subjects, not just informatics-related ones." This domain-specific amplification allows VSEs to operate across broader technical territories while maintaining quality.

**AI as Learning Accelerator**: Perhaps most significantly for VSEs, AI systems serve as always-available mentors for junior developers. Unlike human seniors who have limited time for mentoring, AI can provide patient, "dream-like" detailed explanations 24/7 and never judges, as admired by a Gen Z participant (NK). Yet this "synthetic mentorship" risks what Vygotsky (1978) would recognize as scaffolding without true understanding—support that enables performance without developing underlying competence.

Student participant SB proposed a structured integration: "Split education into two parts. One, where students program solo and learn about the fundamental principles of programming, and the other, where they apply the knowledge using AI." This balanced approach suggests how VSEs might leverage AI for continuous learning while preserving fundamental skill development.

### 2.3.4   Personality-Driven AI Mode Preferences and Self-Determination

*Note to Reader: The following sections (2.3.4-2.3.5) present findings from the author's empirical research, integrated here to maintain theoretical coherence. The detailed methodology and data supporting these insights are presented in Chapter 7.*

The alignment between personality traits and AI interaction modes reveals systematic patterns that inform optimal Human ⊡ AI collaboration. Table 5 synthesizes empirical findings on how different personality configurations interact with AI modes to satisfy or frustrate basic psychological needs.

Table 5: Personality-AI mode alignment matrix: How individual differences shape human-AI collaboration preferences and psychological need satisfaction

| Personality Profile | Preferred AI Mode | Key Motivations | SDT Need Satisfaction | Representative Quote |
|---|---|---|---|---|
| **High Openness** | Co-Pilot | • Creative ideation <br> • Rapid exploration <br> • Divergent thinking | **Autonomy:** Preserved through creative control <br> **Competence:** Enhanced through expanded possibilities <br> **Relatedness:** Minimal impact | "Copilot je jako extrémně rychlý brainstorming partner. Hodí mi deset nápadů, z nichž devět zahodím, ale ten desátý mě posune úplně jinam" (MM) |
| **High Extraversion & Agreeableness** | Co-Navigator | • Dialogical processing <br> • Social-like interaction <br> • Collaborative exploration | **Autonomy:** Maintained through conversational control <br> **Competence:** Built through explanatory dialogue <br> **Relatedness:** Partial satisfaction through quasi-social exchange | "S ChatGPT programuju jako s mentorem. Ptám se ho na věci, na které bych se styděl zeptat skutečného seniora" (NK) |
| **High Neuroticism & Low Extraversion** | Agent (for stressful tasks) <br> Solo (for focused work) | • Reduced social anxiety <br> • Delegated stress <br> • Controlled environment | **Autonomy:** Enhanced through stress delegation <br> **Competence:** Protected from anxiety interference <br> **Relatedness:** Avoided to reduce social pressure | "Když mám udělat něco, co mě stresuje—třeba refaktoring legacy kódu—nechám to na agentovi" (LV) |
| **High Conscientiousness** | Variable (task-dependent) | • Systematic verification <br> • Quality assurance <br> • Process optimization | **Autonomy:** Expressed through quality control <br> **Competence:** Demonstrated via systematic mastery <br> **Relatedness:** Secondary to task focus | "Vedu si deník AI chyb. Je to fascinující katalog toho, jak AI nerozumí kontextu" (PJ) |
| **Low Openness & High Stability** | Minimal AI use | • Predictable workflows <br> • Traditional methods <br> • Direct control | **Autonomy:** Preserved through traditional control <br> **Competence:** Maintained via established skills <br> **Relatedness:** Preferred human collaboration | "Na nějaké přirozené úrovni se snažím minimalizovat přítomnost AI ve svém životě... Chci řešit náročné problémy sám" (NK) |

*Table 5 synthesizes empirical findings on how different personality configurations interact with AI modes to satisfy or frustrate basic psychological needs.*

The patterns revealed in Table 5 illuminate a fundamental insight: optimal Human ⊡ AI collaboration depends on matching interaction modes to personality profiles. This alignment parallels human-human pair programming dynamics while introducing novel considerations unique to synthetic collaboration.

### High Openness: The Creative Catalyst

Developers high in Openness gravitate toward Co-Pilot mode's creative potential, viewing AI suggestions as inspiration rather than prescription. These individuals report experiencing what Guilford (1967) termed "divergent production"—the AI's outputs serve as stimuli for creative recombination rather than final solutions. As one high-openness developer (MM) articulated: "AI je pro mě kreativní katalyzátor. Když se zaseknu, hodím do Copilotu poloviční řešení a on mi ukáže směry, které bych neviděl," [AI is a creative catalyst for me. When stuck, I 'drop' a half-baked solution to the Copilot, and 'he' shows me directions I wouldn't have seen.]

This creative synergy satisfies their need for competence through expanded possibilities while preserving autonomy through creative control. These individuals show the highest satisfaction when AI serves as a "thought accelerator" rather than a "task completer," aligning with DeYoung's (2015) characterization of Openness as cognitive exploration.

### High Extraversion and Agreeableness: The Social Synthesizers

Those high in Extraversion and Agreeableness strongly prefer the Co-Navigator mode's conversational interface, treating AI as a social partner despite its non-consciousness. They excel at what we provisionally term "synthetic dialogue"—maintaining engaging conversations that produce useful outputs despite the AI's lack of genuine understanding. The Co-Navigator mode's dialogical nature partially satisfies their relatedness needs, though they acknowledge the limitation of pseudo-social exchange.

This preference manifests in their approach to AI as a mentor rather than a tool. As our Gen-Z participant (NK) observed: "S ChatGPT programuju jako s mentorem. Ptám se ho na věci, na které bych se styděl zeptat skutečného seniora," [I program with ChatGPT like with a mentor. I ask it things I'd be embarrassed to ask a real senior.] This synthetic social interaction provides a safe space for learning while maintaining the conversational dynamics these personality types naturally seek.

### High Neuroticism and Low Extraversion: The Anxiety Managers

Developers high in Neuroticism or low in Extraversion show complex patterns in their AI preferences. While Co-Pilot's silent operation can be anxiety-provoking ("Is it correct? How can I trust it?"), Agent mode's ability to handle stress-inducing tasks provides unique relief. A high-neuroticism senior developer (LV) explained: "Když mám udělat něco, co mě stresuje—třeba refaktoring legacy kódu—nechám to na agentovi. Kontroluju výsledek, ale nemusím se tím emočně trápit," [when I need to do something that stresses me—like refactoring legacy code—I leave it to the agent. I check the result, but I don't have to emotionally suffer through it.]

A high-conscientiousness CTO (MV) revealed AI's role in reducing uncertainty-based anxiety across his VSE organization: "Snižuje stres. Když člověk něčemu nerozumí, tak dokáže hned ty informace získat nebo dokáže získat nějaké kontakty lidí," [Reduces stress. When a person doesn't understand something, AI can quickly fetch the information or experts' contact information.]

This delegation of anxiety-provoking tasks aligns with coping strategies identified in the stress and personality literature (Connor-Smith & Flachsbart, 2007). For introverted developers, AI also removes social performance pressure. As a low-extraversion optophysicist (MH) expressed: "S AI nemusím hrát sociální hry. Můžu být přímý, technický, nemusím se starat o to, jak věci říkám. Je to osvobozující," [with AI, I don't have to play social games. I can be direct and technical without worrying about how I say things. It's liberating.]

These individuals develop elaborate verification workflows—what one participant (MH) called "důvěřuj, ale prověřuj" [trust but verify]—that paradoxically increase their confidence by acknowledging their anxiety.

### High Conscientiousness: The Systematic Optimizers

Highly conscientious developers demonstrate task-dependent AI preferences, focusing on systematic verification and quality assurance regardless of mode. Their interactions with AI reflect their fundamental orientation toward order and reliability. A senior developer (PJ) showed how high conscientiousness manifests as systematic AI limitation documentation: "Vedu si deník AI chyb. Je to fascinující katalog," [I keep a diary of AI errors. It's a fascinating catalog.]

This systematic documentation exemplifies how Conscientiousness shapes adaptation strategies to new technologies. Their approach to AI centers on process optimization while maintaining quality control, expressing autonomy through systematic mastery rather than creative exploration.

### Low Openness and High Stability: The Traditionalists

Some developers, particularly those low in Openness but high in emotional stability, show minimal AI adoption. As one participant (NK) explained: "Na nějaké přirozené úrovni se snažím minimalizovat přítomnost AI ve svém životě... Chci řešit náročné problémy sám," [on some natural level, I attempt to minimize AI presence in my life... I want to solve difficult problems by myself.] These individuals preserve autonomy through traditional control methods and maintain competence via established skills, preferring human collaboration when needed.

*Differential Impact on Basic Psychological Needs*

Each AI mode differentially impacts the satisfaction of basic psychological needs across personality types:

Autonomy manifests differently across modes and personalities. Co-Pilot typically preserves autonomy for open individuals while Co-Navigator requires active maintenance of directive control. Agent mode demands reconceptualizing autonomy from doing to directing—a shift some developers find liberating ("ovládám vlastní agentský mini-tým" [I control my own agent mini-team]) (JO), others alienating.

Competence support varies by mode and trait interaction. All modes can extend capabilities, but the experience differs: Co-Pilot enhances execution competence for creative tasks, Co-Navigator builds understanding competence through explanation, and Agent mode enables systemic competence—orchestrating solutions beyond individual capability.

Relatedness presents the most complex dynamics. Co-Navigator most directly addresses this need through conversational interaction, particularly for extraverted individuals. Yet even minimal AI personification in other modes creates what participants describe as "společnost v samotě" [company in solitude]—a presence that mitigates isolation without replacing human connection.

These patterns suggest that effective Human ⊞ AI collaboration requires sophisticated matching of AI interaction modes to individual personality profiles, ensuring that technology augments rather than frustrates basic psychological needs. The ROMA framework leverages these insights to provide personalized recommendations for AI tool selection and configuration in both VSE and educational contexts.

## 2.3.5   ROMA Framework Application to AI-Human Collaboration

The empirical patterns revealed in the preceding sections converge toward a unified framework for optimizing Human ⊞ AI collaboration. The ROMA (Roles Optimization Motivation Alignment) framework extends beyond traditional personality-based role assignment to encompass the unique dynamics of AI-augmented development. By synthesizing insights from personality psychology, self-determination theory, and our phenomenological investigations, ROMA provides a theoretically grounded yet practically applicable approach to enhancing developer motivation and productivity in the age of AI.

At its core, ROMA recognizes that the introduction of AI into software development does not merely add new tools—it fundamentally transforms the nature of programming work itself. As our participants revealed, working with AI involves navigating complex psychological dynamics: the paradox of enhanced autonomy through delegation, the redefinition of competence at higher abstraction levels, and the emergence of synthetic relatedness with non-conscious entities. These transformations demand a framework sophisticated enough to address both human psychological needs and the affordances of different AI interaction modes.

The framework addresses six critical dimensions of Human  AI collaboration:

I. **Personality-to-Mode Linkage:** ROMA maps personality traits to preferred AI interaction modes, recognizing that what motivates an open, creative developer differs fundamentally from what engages a conscientious, systematic one. This mapping is not prescriptive but probabilistic, acknowledging individual variation while providing evidence-based starting points for role optimization.

II. **Self-Determination Fulfillment:** The framework explicitly considers how different AI modes satisfy or frustrate basic psychological needs across personality types. A Co-Navigator might fulfill an extravert's relatedness needs while leaving an introvert feeling drained; an Agent might liberate a neurotic developer from anxiety-inducing tasks while making an open developer feel creatively constrained.

III. **Organismic Adoption:** Drawing on SDT's organismic integration theory, ROMA designs role optimizations that respect the internalization process. Rather than imposing AI tools through mandate, the framework helps VSEs facilitate natural adoption trajectories—from external regulation through introjected and identified regulation to fully integrated use that aligns with developers' self-concepts.

IV. **Role Optimization Guidance:** ROMA provides actionable guidelines for selecting AI tools and configuring roles based on both individual characteristics and team composition. For a VSE with two high-openness developers and one high-conscientiousness developer, the framework might recommend Co-Pilot modes for the creative pair with the conscientious developer serving as quality gatekeeper using AI verification tools.

V. **Flow Maintenance Strategies:** The framework recognizes that flow states—those precious periods of deep engagement—face new challenges in AI-augmented environments. ROMA offers strategies for preserving flow while transitioning between AI modes, such as using Co-Pilot for flow maintenance during implementation while reserving Co-Navigator interactions for natural break points.

VI. **Educational Integration:** For undergraduate contexts, ROMA addresses the delicate balance between leveraging AI's learning acceleration potential and ensuring fundamental skill development. The framework supports what participant SB proposed: structured approaches that combine solo learning of fundamentals with AI-assisted application and exploration.

The transformative potential of ROMA lies not in rigid prescriptions but in providing a conceptual scaffold for understanding and optimizing Human ⊞ AI collaboration. As AI capabilities continue to evolve—from today's Co-Pilots and Co-Navigators to tomorrow's fully autonomous agents—the framework's emphasis on personality-based adaptation and psychological need satisfaction remains relevant.

In VSE contexts, where every developer's engagement critically impacts organizational success, ROMA offers a path toward sustainable AI integration that enhances rather than undermines human motivation. By honoring individual differences while pursuing collective goals, the framework embodies what Dewey (1938) called "intelligent action"—theory informed by experience, directed toward human flourishing.

As we transition to examining the deeper psychological and neural foundations of motivation in software engineering, the ROMA framework serves as both the culmination of our AI collaboration analysis and the foundation for understanding the fundamental human needs that no amount of artificial intelligence can replace. The question is not whether AI will transform software development—that transformation is already underway. The question is whether we can shape this transformation to amplify rather than diminish what makes programming a deeply human, creative act.

## 2.4   Self-Determination and Intrinsic Motivation Concepts

*What drives some professionals and undergraduates to tackle a problem with tenacity and creativity, whereas others grapple with frustration and disengagement?*

Self-Determination Theory (SDT) provides a robust framework for understanding these motivational dynamics in software engineering contexts, particularly with respect to how programming roles and collaborative practices influence psychological need satisfaction and intrinsic motivation. The theory's evolution from early drive theories to contemporary understanding of human flourishing offers crucial insights for optimizing both human-human and Human ⊞ AI collaborative programming.

### 2.4.1   Evolution of Motivation Theories in Software Engineering

The scientific study of motivation represents a journey from mechanistic conceptions of human behavior toward recognition of humanity's inherent growth tendencies. This evolution, summarized in Table 6, reveals progressive sophistication in understanding what Aristotle termed the "unmoved mover" within human action—that which initiates and sustains purposeful behavior without external compulsion.

Table 6: The arc of motivational theory

| Era | Principal Voices | Core Claim |
|---|---|---|
| **Instinct & Play** 1890-1920 | Darwin → Gross → James | Behavior is canalized by innate "pre-exercise" of future skills. |
| **Drive & Reinforcement** 1920-1950 | Hull, Skinner | Organisms act to reduce drives; contingencies of reward govern frequency. |
| **Cognitive–Organismic** 1950-1975 | Tolman, White, Rotter, Bandura | Expectancies, efficacy, and effectance propel mastery-seeking beyond drive reduction. |
| **Humanistic & Phenomenological** 1955-1980 | Rogers, Maslow, Deci | The person is an intentional center striving for authentic growth. |
| **Self-Determination Synthesis** 1985-today | Deci & Ryan, Vansteenkiste | Three Basic Psychological Needs (BPN) are the nutrients of flourishing; frustration is pathogenic. |

*Table 6 presents a chronological overview of the major theoretical movements in motivation research spanning from the late 19th century to the present day. It traces the evolution from early instinct and drive theories through cognitive and humanistic approaches, culminating in Self-Determination Theory. Each era represents a paradigm shift in understanding what propels human behavior, with progressive recognition of intrinsic drivers and psychological needs. This evolution mirrors the increasing complexity observed in software engineering motivation, where external rewards have proven insufficient to sustain the creative, problem-solving engagement required for effective development.*

This theoretical progression mirrors the evolution of software development itself. In the era of punch cards and batch processing, programming resembled factory work—discrete tasks with clear completion criteria amenable to behaviorist reward structures. Today's continuous deployment, emergent architectures, and AI collaboration demand sustained creative engagement that external motivators alone cannot sustain (Pink, 2009; França et al., 2011).

The pivotal insights emerged through what Kuhn (1962) might recognize as anomalies in the behaviorist paradigm. Harlow's (1950) monkeys persistently solving mechanical puzzles without reward, White's (1959) concept of "effectance motivation," and deCharms' (1968) distinction between "origins" and "pawns"—perceived locus of causality (PLOC)—collectively pointed toward intrinsic drivers that behaviorism could not explain.

For software engineering, the implications proved profound. As Weinberg (1971) presciently observed, programming is fundamentally a human activity where motivation shapes not just productivity but the very nature of solutions produced. The inadequacy of extrinsic motivation becomes particularly apparent in debugging—that peculiar form of detective work where the mystery and its solution exist entirely in abstract symbol space. No external reward adequately compensates for the frustration of hunting an elusive race condition. Yet developers persist, driven by what can only be understood as an intrinsic fascination with solving puzzles that exist nowhere but in the interaction of logic and electricity.

Contemporary developers articulate this transformation explicitly. Participant (PΔ) confessed: "I have become lazier. Before doing something, I prompt ChatGPT to take and edit it. Sometimes I feel I am chatting with the tool for five minutes and maybe it would have been quicker if I did it on my own" (Study III). This "laziness"—perhaps better understood as cognitive offloading— represents neither pure intrinsic nor extrinsic motivation but something novel: what we might term "augmented motivation," where the locus of satisfaction shifts from creation to curation.

### 2.4.2 Intrinsic Motivation in Software Engineering Contexts

Self-Determination Theory's three basic psychological needs—autonomy, competence, and relatedness—manifest uniquely in software development contexts, particularly as AI transforms traditional boundaries. These needs, which Deci and Ryan (2000) argue are universal and innate, provide the nutriments for intrinsic motivation and psychological well-being across cultures and contexts.

**Autonomy in Software Engineering** traditionally meant freedom to choose implementation approaches, tools, and working methods. Research by França et al. (2018) and Sharp et al. (2009) consistently identifies autonomy as a primary motivator for software engineers. In agile contexts, Whitworth and Biddle (2007) found that self-organizing teams reported higher satisfaction precisely because they could exercise collective autonomy. The ROMA framework builds on these findings by recognizing that different personality types experience and express autonomy differently—what satisfies an introverted developer's autonomy needs may frustrate an extraverted colleague.

The philosophical foundations of autonomy trace to Mill's (1859) conception of liberty, where individuals must be free to pursue their own good in their own way, provided they do not harm others. In software development, this translates to what we might term "implementation autonomy"—the freedom to choose not just what to build but how to build it. As Mill argued, such freedom is essential not merely for individual satisfaction but for the discovery of better ways of living—or in our context, coding.

Our empirical findings from Study IV illuminate how this manifests in practice. Theme 6 captured role-specific autonomy experiences: "As the navigator, I felt like a leader. The whole situation was in my hands. It motivated me a lot!" (P5). This suggests that role configurations— whether human-human or human-AI—can enhance autonomy through leadership opportunities, even within collaborative constraints.

The transformation of autonomy through AI collaboration emerged vividly in Study III. Professional participant PE described reduced emotional labor: "When I communicate with a human colleague, it is more about how to explain to the person in a way that does not exhaust them. With AI, it is more relaxed." This "relaxed dynamics with AI" represents a form of autonomy through reduced social pressure, particularly beneficial for introverted developers as participant PΔ revealed: "I am quite an introverted person so definitely prefer the AI tool. With a human, I am always a little nervous or trying to show off how I can do things. With AI I do not care."

A senior developer in Study VI articulated this transformed autonomy: "Když pracuju s AI, moje autonomie není v psaní každého řádku, ale v orchestraci celého řešení. Je to jako dirigovat orchestr místo hraní na každý nástroj," [when I work with AI, my autonomy isn't in writing every line, but in orchestrating the entire solution. It's like conducting an orchestra instead of playing every instrument]. This shift from "making" to "directing" requires reconceptualizing autonomous action in software development.

**Competence in Software Engineering** involves experiencing effectiveness and mastery in one's activities. Bandura's (1977) self-efficacy theory and White's (1959) effectance motivation converge in recognizing humans' inherent drive to interact effectively with their environment. In programming, this manifests as the satisfaction of elegant solutions, successful debugging, and system mastery.

Study IV participants articulated diverse competence experiences across roles. Theme 12 captured the cognitive benefits of role differentiation: "I like analyzing things, and when I did not have to worry about coding, I had so much space in my brain, I had a different view and saw probably the best approach" (P4). Another noted discovering "ways I wouldn't have found in the pilot role" (S1), suggesting that different roles provide distinct competence-building opportunities.

Professional participant PΓ described a paradigm shift: "I believe we moved from the imperative programming paradigm into declarative. Because, as I stated, I moved a lot into DevOps, and I declare what I want and it does the heavy lifting for me. I do not investigate how it does it." This shift to declarative approaches represents competence at higher abstraction levels, though student participant (SΓ) revealed a learning inversion: "The 'process' is different. Because traditionally, you understand first, and after you write the code. But here, you first get the working code and then understand how each line works."

The introduction of AI creates what Dweck (2006) might recognize as a "growth mindset" imperative. Traditional markers of competence—syntax mastery, library knowledge, algorithm implementation—become partially obsolete when AI systems possess superior recall and implementation speed. Yet our research reveals competence reemerging at higher abstraction levels, aligning with Fischer's (1980) skill theory, where development proceeds through hierarchical integration.

Developers report feeling most competent not when implementing complex algorithms (which AI handles efficiently) but when:

- Recognizing subtle architectural implications that AI systems miss

- Debugging issues requiring business context understanding

- Making judgment calls about technical debt and future maintainability

- Orchestrating multiple AI systems toward coherent solutions

As one researcher developer noted: "Kompetence už není o tom, co vím, ale o tom, jak dokážu využít to, co 'ví' tým AI agentů, které orchestruji. Je to meta-dovednost. " [Competence is no longer about what I know, but about how I can utilize what orchestrated AI agent team 'knows'. It's a meta-skill.]

The competence transformation extends to learning itself. Participant (SA) observed: "In one hour with ChatGPT, I learn 60-70% more than without it" (Study III), while simultaneously warning: "If you are starting to learn to program, it might become harder to learn from the ready-to-use solutions provided by ChatGPT. I am thankful I learned the basic concepts without ChatGPT." This paradox—accelerated learning alongside potential skill atrophy—highlights the complex relationship between AI-augmented competence and foundational understanding.

**Relatedness in Software Engineering** encompasses the need to feel connected to others, to experience caring and being cared for. In traditional team settings, this manifests through pair programming, code reviews, and collaborative problem-solving. Graziotin et al. (2015) found that developers who understand and support their project's overarching purpose—connecting to something larger than themselves—experience higher intrinsic motivation.

The phenomenon of "synthetic relatedness" with AI systems presents novel theoretical territory. While Reeves and Nass (1996) demonstrated that humans naturally apply social schemas to computers, AI's conversational abilities intensify this tendency. Developers report experiencing AI collaborators as quasi-social partners, though they simultaneously acknowledge the fundamental asymmetry. A Gen-Z participant explained: "ChatGPT je jako kolega, který nikdy neusne, nikdy se nenaštve, a vždycky má čas. Není to skutečný vztah, ale něco tam je" [ChatGPT is like a colleague who never sleeps, never gets angry, and always has time. It's not a real relationship, but something is there.]

Study IV revealed rich social dynamics in human-human collaboration. Theme 3 (Mentorship & Helping) captured the bidirectional knowledge transfer: "Of course, communication is an integral part of pair programming. During the conversation, you can discover errors, discuss the strategy, find a solution on which both agree and share knowledge in a great way" (P3). Participant S2 expressed satisfaction in the teaching role: "I felt like a teacher who could share his knowledge."

The psychological support dimension emerged powerfully in Theme 5: "Almost everything can be solved in pairs, from programming to your emotional state" (P2). This "connective therapy" aspect of pair programming satisfies relatedness needs while addressing emotional challenges inherent in software development.

AI introduces novel relatedness dynamics. While student participant SA insisted, "Pairing with humans should remain because it is an interesting experience. You just feel it," others found AI partially fulfilling relatedness needs. Student participant SI's observation about ChatGPT's differential language performance ("ChatGPT success rate is way better when you prompt it in English (80%) than Slovak (50%)") led to treating AI as a quasi-social partner requiring accommodation, suggesting developers create synthetic relatedness through anthropomorphization.

**Beneficence as a Fourth Need**: Recent theoretical extensions by Martela and Ryan (2016) propose beneficence—the need to contribute to something beyond oneself—as a candidate fourth basic need. In software engineering, this manifests as the satisfaction of creating tools others use, contributing to open source, or mentoring junior developers. The ROMA framework incorporates this insight by recognizing that different personality types may prioritize beneficence differently in their motivational hierarchies.

### 2.4.3   Flow and Mindfulness in Software Development

The phenomenology of optimal experience in programming—what Csikszentmihalyi (1990) termed "flow"—provides crucial insights for understanding intrinsic motivation in both solo and collaborative contexts. Flow represents the experiential manifestation of intrinsic motivation, where action and awareness merge in autotelic activity.

**Flow in Programming** exhibits characteristic features identified by Nakamura and Csikszentmihalyi (2014): complete absorption in the activity, clear goals and immediate feedback, balance between challenge and skill, sense of control over the activity, altered perception of time, and "autotelic experience"—the activity becomes its own reward. Furthermore, neuroimaging now implicates transitory hypofrontality and striatal dopamine surges (DiDomenico & Ryan, 2017)

Recent neuroscientific research by Ulrich et al. (2014) reveals that flow states correlate with transient hypofrontality—temporary downregulation of the prefrontal cortex associated with self-referential processing. This neurobiological foundation suggests flow represents not merely subjective experience but a distinct brain state optimized for performance.

In programming contexts, Kuusinen et al. (2016) identified specific triggers for flow states: clear requirements, minimal interruptions, appropriate challenge levels, and immediate feedback from compilers or tests. The introduction of AI adds new dimensions. Developers report experiencing what we term "distributed flow"—optimal experience emerging from the dance between human intention and AI capability. As one developer described: "S Copilotem se dostanu do flow rychleji, protože nemusím přemýšlet nad syntaxí. Ale když mi nabídne něco úplně mimo, vypadnu z toho ještě rychleji" [With Copilot, I get into flow faster because I do not have to think about syntax. But when it suggests something completely off, I fall out even faster.]

Distributed flow differs from Csikszentmihalyi's individual flow in crucial ways. Traditional flow requires clear goals, immediate feedback, and balanced challenge—all under individual control. Distributed flow, by contrast, emerges from the interplay between human intention and AI capability, creating what we might call "collaborative consciousness." The developer sets high-level goals while the AI handles implementation details, feedback becomes bidirectional (the developer evaluates AI output while the AI responds to prompts), and challenge dynamically adjusts as the AI's suggestions push the developer toward new solutions. This creates a novel phenomenological state—neither purely individual nor truly collaborative, but something unprecedented: a flow state distributed across biological and artificial cognitive systems.

Our empirical data reveals how flow manifests differently across programming configurations. Study IV participants described flow experiences in pair programming contexts. Theme 7 revealed mixed perceptions of time constraints' impact on flow: "I think the time limit was good because when not sitting at the computer for long, I lose focus" (P1), while another noted disruption: "Timer is confusing because you leave your job unfinished, and now you have to take on another role" (P5). Theme 10 identified specific flow disruptors: "When they switched, their mindsets still remained in the last role" (S5), suggesting cognitive carryover undermines flow states during role transitions.

The social dimension of flow emerged in Theme 9, with participants reporting enhanced engagement through collaboration: "In my case, I was three times more productive in pairing than when doing the same task at home" (P3). Another noted the flow-facilitating nature of pair support: "It was fun; in a pair you never feel lost" (S2). These experiences suggest pair programming can create conditions for collective flow states.

AI's impact on flow presents complex dynamics. The immediate assistance of Co-Pilot mode can maintain flow by reducing interruptions for syntax lookup or boilerplate generation. However, over-reliance risks what one participant termed "skill atrophy anxiety"—worry about declining abilities that disrupts flow states. The Co-Navigator mode's conversational nature may interrupt flow for some while enhancing it for others who process information dialogically.

**Mindfulness in Programming** represents a complementary state to flow—one of open awareness rather than focused absorption. Brown and Ryan (2003) define mindfulness as receptive attention to present experience, a quality that proves valuable in debugging, code review, and architectural design.

Langer's (1989) conception of mindfulness as "active distinction-making" resonates particularly with programming tasks. The mindful programmer notices subtle code smells, questions assumptions, and remains open to alternative approaches. This quality becomes crucial when working with AI, where developers must maintain what Gen-Z participant NK termed as "přemítání nad probdělými nocemi" [vigilant skepticism]—accepting AI assistance while remaining alert to potential errors or suboptimal suggestions. Senior developer JO exemplified this: "Musím být neustále ve střehu. AI je jako velmi chytrý junior, který někdy předstírá, že rozumí, i když nerozumí. Vyžaduje to konstantní bdělost," [I always have to remain alert. AI is like a very smart practicant that sometimes pretends to understand when he does not. Permanent vigilancy is required.]

The integration of flow and mindfulness concepts with SDT's basic psychological needs provides a comprehensive framework for understanding optimal experiences in software development. As Brown et al. (2017) argue, mindfulness facilitates satisfaction of all three basic needs by enhancing present-moment awareness of autonomy, competence, and relatedness experiences.

### 2.4.4   Balanced Motivation Approach in Software Engineering Teams

While intrinsic motivation represents the gold standard for creative work, organizational realities necessitate a more nuanced approach. SDT's Organismic Integration Theory (OIT) describes how extrinsic motivations can vary in their degree of autonomy, from external regulation (purely extrinsic) to integrated regulation (highly internalized) (Gagné & Deci, 2005), detailed in Figure 3:

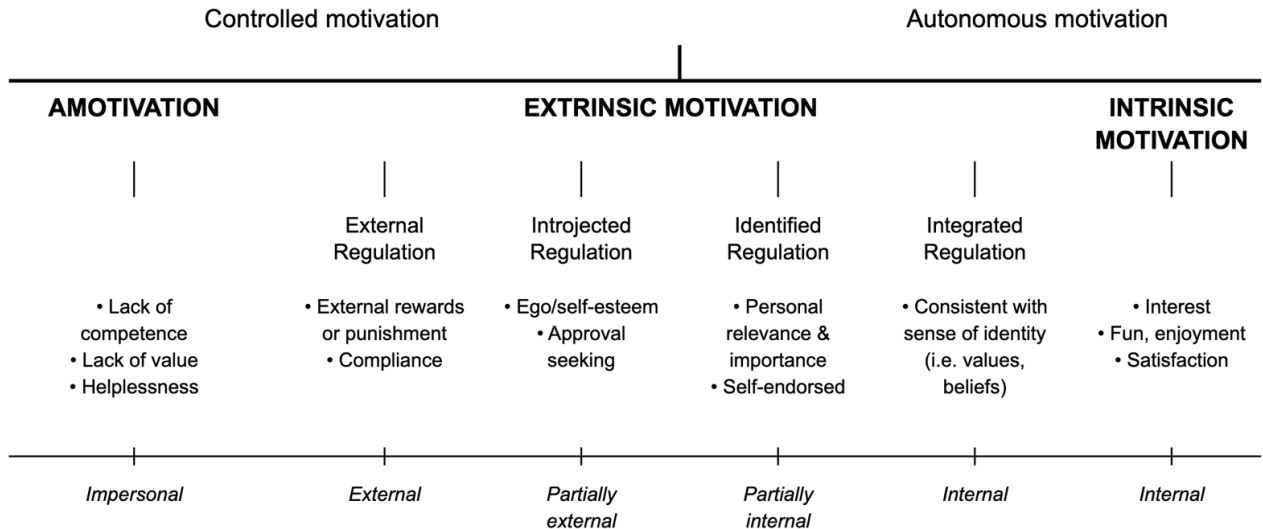

*Figure 3: The self-determination continuum*

*Reproduced under educational fair use for non-commercial academic purposes, based on Gagné &
Deci (2005). The full motivational spectrum includes:*

- *Amotivation – Lack of motivation with an absence of intentional regulation*
- *External Regulation – Motivation driven by external rewards or punishments*
- *Introjected Regulation – Motivation driven by internal pressures, such as guilt or the need
to avoid shame*
- *Identified Regulation – Behavior guided by personal goals or values that are consciously
endorsed*
- *Integrated Regulation – Behavior that is fully assimilated into the individual's sense of self
and aligned with personal values*
- *Intrinsic Motivation – Inherently autonomous motivation with an interest in and enjoyment
of the task.*

This continuum proves particularly relevant for VSEs where resource constraints may neces-
sitate tasks that do not inherently interest all team members. The ROMA framework leverages OIT
by recognizing that personality traits influence not just intrinsic motivation but also the internali-
zation process. Highly conscientious developers, for instance, may more readily integrate
externally imposed quality standards into their self-concept, experiencing them as expressions of
craftsmanship rather than constraints.

The introduction of AI tools illustrates this internalization process. Initially, developers may
use AI due to organizational mandates (external regulation) or peer pressure (introjected regula-
tion). Over time, they recognize AI's value for achieving personal goals (identified regulation) and
eventually incorporate AI collaboration into their professional identity (integrated regulation). As
one senior developer reflected: "Zpočátku to bylo divné, jako bych měl někoho za zády. Teď je to
spíš jako... rozšíření mého myšlení. Když programuju bez AI, cítím se trochu nahý" [At first it was strange, like someone was looking over my shoulder. Now it's more like... an extension of my thinking. When I program without AI, I feel somewhat naked.]

The internalization process varies significantly in Human ⊡ AI contexts, as our data reveals. Student participant SΓ demonstrated sophisticated AI literacy: "Even though you have access to AI, you should not stop learning on your own. It is just a tool and does not really replace you." This recognition maintains healthy boundaries while leveraging AI benefits. Professional participant PE showed full integration: "It influenced my satisfaction levels positively as I do not have to worry about commenting, typing, or generating documentation." This represents AI assistance fully internalized into professional identity.

However, participant PΔ's confession about becoming "lazier and dependent" suggests that integrated regulation can manifest as both empowerment and dependence—a nuanced outcome that SDT's linear continuum may not fully capture. This complexity demands careful attention to how AI tools are introduced and integrated into development workflows to support healthy motivational patterns.

The convergence of Self-Determination Theory, flow research, and mindfulness studies reveals motivation not as a simple drive but as a complex emergence—a dance between person and environment, challenge and capability, self and other. In the liminal space where human creativity meets artificial intelligence, these psychological dynamics take on new urgency. The framework we develop must honor both the ancient satisfactions of problem-solving and the unprecedented realities of synthetic collaboration. For it is precisely in understanding how autonomy, competence, and relatedness manifest in this new landscape that we may discover not just how to optimize performance, but how to preserve what is most essentially human in the act of creation— even as we reimagine what creation itself might become.

## 2.5   Concept of Big Five Personality Traits

*Why do some developers discover flow in the labyrinth of logic, while others encounter only walls—and how does personality shape this divergence?*

The inquiry into personality within software engineering touches a paradox at the heart of human sciences: we seek universal patterns in that which makes each individual irreducibly unique. As Jung (1921) observed, "the meeting of two personalities is like the contact of two chemical substances: if there is any reaction, both are transformed." In software development—increasingly a collaborative endeavor between human and artificial minds—understanding these transformative reactions becomes not merely academic but essential to practice.

### 2.5.1 Personality Psychology in Software Engineering

The question of personality in software engineering transcends academic categorization to touch something essential about the nature of creative technical work. Each developer brings not merely skills but what Heidegger (1962) might term their own *Dasein*—their particular way of being-in-the-world that shapes how they approach problems, collaborate with others, and increasingly, interact with AI systems.

Personality research has gained increasing attention in software engineering as organizations recognize that technical skills alone do not predict developer effectiveness or team cohesion. While Corr and Matthews (2009) noted that "personality psychology has never been in better health than at the present time," its application to software engineering has been slower to mature. Cruz et al. (2015) found that personality research in software engineering remained in its early stages, with inconsistent methodologies and limited integration with established psychological frameworks.

The historical skepticism toward personality research in software engineering (e.g., Lewis, 2020) reflected a desire for universal principles with aptly applied methodologies without conjectures about measurements that affect the measured. Yet as software development becomes increasingly collaborative and creative, personality's influence becomes impossible to ignore. This is particularly evident in our empirical data, where participants explicitly connected their personality traits to their programming preferences and collaborative choices.

This skepticism reflects deeper questions about power and knowledge in organizational contexts. As Foucault (1982) observed, the categorization of human subjects often serves disciplinary functions, creating what he termed "technologies of the self" that shape individuals to fit institutional needs. In software engineering, personality assessment risks becoming such a technology—not discovering natural categories but producing them through measurement and application.

Recent advances in Behavioral Software Engineering (BSE) have advocated for more rigorous approaches to personality assessment in software contexts (Graziotin et al., 2021; Felipe et al., 2023). These researchers critique the continued reliance on outdated methods like the Myers-Briggs Type Indicator (MBTI) and call for more consistent use of validated measures like the Big Five Inventory. This shift toward more scientifically sound measurement aligns with the approach taken in this dissertation, which employs well-established personality constructs to inform programming role optimization.

Our empirical findings illuminate how personality manifests in programming contexts. Study IV participants explicitly connected traits to role experiences. Theme 11 revealed sophisticated awareness: "Personality determines what role I prefer" and "I really found myself" after assessment, with another noting "It (Big Five results) was all me." This self-recognition suggests personality assessment can enhance self-awareness and role alignment.

The practical implications of personality research resonate throughout our data. Theme 8 participants proposed ideal pair programmer traits: "Be communicative, have quick learning ability, the tendency to display self-discipline, be calm, willing to compromise their interest with others, and emotionally stable" (P2). This wishlist maps directly onto Big Five dimensions—Extraversion (communicative), Openness (quick learning), Conscientiousness (self-discipline), Agreeableness (compromise), and low Neuroticism (emotionally stable).

The relevance of personality to software engineering is supported by several key studies:

- Capretz and Ahmed (2010) found that specific software development roles require diverse personality characteristics, with analyst roles benefiting from extraversion and agreeableness, and technical roles often aligned with introversion and conscientiousness

- Feldt et al. (2010) demonstrated that personality traits correlate with attitudes toward software engineering processes, with more agreeable developers showing greater acceptance of collaborative programming practices

- Cruz et al. (2015) reviewed 40 years of personality research in software engineering, identifying patterns suggesting that certain traits consistently predict performance in specific development roles

These studies highlight the practical value of incorporating personality insights into team formation and role assignment decisions, particularly in small teams where interpersonal dynamics have outsized effects on project outcomes.

### 2.5.2 Big Five in Software Engineering Contexts

The Big Five personality model (McCrae & Costa, 1999; Goldberg, 1993)—encompassing Openness, Conscientiousness, Extraversion, Agreeableness, and Neuroticism—provides a robust framework for understanding individual differences in software development contexts. Each trait has specific implications for programming roles and team dynamics:

*Openness to Experience in Software Engineering*

Characterized by creativity, curiosity, and intellectual engagement, Openness has been linked to:

- **Innovation in Problem-Solving**: Developers high in Openness often generate novel solutions and explore alternative approaches (Salleh et al., 2010)

- **Adaptability to New Technologies**: More open individuals typically show greater willingness to learn new programming languages and frameworks (DeYoung, 2015)

- **Design-Focused Roles**: Openness correlates with success in roles requiring creative thinking and conceptual design (Capretz & Ahmed, 2010)

- **Pair Programming Dynamics**: In collaborative programming, high Openness may facilitate idea generation and exploratory coding (Hannay et al., 2009b)

Our Study IV data provides concrete examples of openness in action. Theme 12 captured creative discovery: "I found ways I wouldn't have found in the pilot role" (S1), with another noting "There are more ways of solving a problem than you can think of alone" (S5). This cognitive flexibility characteristic of high openness enables innovative problem-solving.

### Openness in Human  AI Contexts

The AI collaboration dimension adds complexity. Professional participant PA from Study III demonstrated high openness through technology enthusiasm: "I was really impressed and happy when I started to use AI because I always dreamt about using something like this. I like new technology. There was no anxiety or relief. The word 'excitement' would describe my feelings best." This openness to AI as a creative partner rather than a threat exemplifies how the trait facilitates adaptation to technological change. Moreover, in VSEs, where adaptability is crucial and roles often require breadth rather than specialization, Openness may be particularly valuable.

### *Conscientiousness in Software Engineering*

Defined by organization, reliability, and attention to detail, Conscientiousness affects:

- **Code Quality and Testing**: More conscientious developers tend to produce more thoroughly tested, documented code (Barrick & Mount, 1991)

- **Project Management**: This trait correlates with adequate planning, deadline adherence, and systematic work approaches (Capretz et al., 2015)

- **Technical Debt Management**: Higher Conscientiousness may reduce the accumulation of technical debt through more disciplined coding practices (Salgado, 1997)

- **Role Preferences**: Conscientious developers may gravitate toward roles involving code review, quality assurance, and architectural oversight (Smith et al., 2016)

Study IV revealed how conscientiousness manifests in pair programming preferences. Theme 11 participants emphasized systematic approaches: "Same pace of work is vital" and proposals for "speed-dating" systems to find compatible partners (P1). This methodical approach to team formation exemplifies conscientious planning.

### Conscientiousness in Human  AI Contexts

In AI contexts, conscientious developers showed particular patterns. Professional participant (PE) appreciated AI's systematic benefits: "Trying to get my software development to the purest and simplest form that people can understand. And making the development process more effective and faster." This drive for clarity and efficiency through AI assistance demonstrates conscientiousness channeled through technological tools.

This apparent contradiction illuminates how conscientious developers relate to AI tools. Their initial skepticism stems from reliability concerns—will AI introduce subtle bugs? Can its suggestions be trusted? However, once they develop systematic verification protocols (like participant PJ's "diary of AI errors"), they can appreciate AI's efficiency benefits while maintaining their quality standards. Conscientiousness thus manifests not as AI rejection but as methodical AI integration—these developers become the architects of human-AI quality assurance systems.

Study VI data reveals how conscientious developers approach AI collaboration with systematic precision, as exemplified by a senior developer (PJ) who catalogues AI's mistakes or by an architect (MV): "Procházím všechny ty řádky a snažím se nad tím přemýšlet. Je to takový proces, kdy mu moc nedůvěřuji, protože nechci bugy," [I straddle the lines of code and contemplate. It's 'this process', where I don't trust it because I don't want bugs,] illustrating conscientious verification behavior. This methodical documentation serves multiple purposes: building understanding of AI limitations, creating reference materials for team training, and developing systematic workarounds for common failure modes.

The systematic nature of conscientious developers extends to quality assurance strategies. As participant (P2) from Study IV noted: "Both roles check for a different type of errors," demonstrating how conscientious individuals create comprehensive verification approaches that leverage role differentiation. In pair programming, they ensure thorough coverage by having navigator and pilot focus on different error categories—logical flaws versus syntax issues, architectural concerns versus implementation details.

### Conscientiousness and Quality Assurance in VSEs

For VSEs with limited quality assurance resources, understanding the distribution of Conscientiousness among team members becomes crucial for role optimization. Highly conscientious developers naturally gravitate toward gatekeeping roles—code review, deployment approval, architectural oversight—where their systematic attention to detail provides maximum value. The ROMA framework leverages this insight by recommending that VSEs position their most conscientious members in quality-critical roles while pairing them with more open, creative developers who can balance their sometimes rigid adherence to process.

### Extraversion in Software Engineering

Encompassing sociability, assertiveness, and energy, Extraversion influences:

- **Communication-Intensive Roles**: More extraverted developers often excel in client-facing roles and team coordination (Capretz & Ahmed, 2010)

- **Knowledge Sharing**: Higher Extraversion correlates with more active participation in code reviews and pair programming sessions (Williams & Kessler, 2003)

- **Team Leadership**: Extraverted developers may naturally assume leadership roles in collaborative settings (Smith et al., 2016)

- **Remote Work Adaptation**: The shift to distributed development during the pandemic highlighted how Extraversion moderates adaptation to remote collaboration (Ralph et al., 2020)

Our empirical findings support this theoretical connection. Study III participant P$\Delta$ revealed: "I am quite an introverted person, so definitely the AI tool. With a human, I am always a little nervous or trying to show off how I can do things. With AI, I do not care." This stark preference based on introversion (low extraversion) demonstrates how personality traits moderate the choice between human and AI collaboration.

Conversely, more extraverted participants expressed preferences for human interaction. As participant S$\Delta$ noted: "Most comfortable with a human because you can discuss your opinions. Communication is easier. Solo programming is really hard for me." Another participant (Theme 8, Study IV) emphasized that ideal pair programmers need "High Extraversion for Better Interaction," suggesting extraversion's benefits persist in human-human and human-AI configurations but are crucial for the former, whereas the latter can be exercised by introverts as well.

### Agreeableness in Software Engineering

Characterized by cooperation, empathy, and trust, Agreeableness affects:

- **Conflict Resolution**: More agreeable team members typically facilitate smoother conflict resolution in development teams (Mount et al., 1998)

- **Collaborative Practices**: Higher Agreeableness correlates with positive attitudes toward pair programming and other collaborative methods (Feldt et al., 2010)

- **Code Review Dynamics**: Agreeable developers may provide more constructive feedback and be more receptive to criticism (Hannay et al., 2009b)

- **Team Cohesion**: This trait contributes to psychological safety and trust within development teams (Clarke & Robertson, 2005)

Theme 2 from Study IV illuminated agreeableness in practice: "Feedback is important for everyone to progress in anything; whether negative or positive, it must be told" (P4), balanced with diplomatic assertion: "Sometimes he was really dominant, so I had to tell him 'Yes, I can do this on my own'" (P6). This balance of receptiveness and boundary-setting characterizes healthy agreeableness in collaborative contexts.

### Agreeableness in Human  AI Contexts

In AI interaction, agreeable developers showed unique patterns. Their tendency to anthropomorphize AI emerged in comments about AI's "apologetic" nature when wrong (SΓ) and appreciation that "ChatGPT and Bard act as subordinates. They would do anything you ask and never refuse" (PA). This suggests highly agreeable individuals may find AI's consistent compliance less satisfying than negotiated human dynamics. For VSEs and educational settings, where close collaboration is essential and interpersonal conflicts can have outsized impacts, understanding Agreeableness distributions can inform both team formation and role assignment decisions.

### *Neuroticism in Software Engineering*

Related to emotional sensitivity and stress vulnerability, Neuroticism influences:

- **Problem Detection**: Developers higher in Neuroticism may be more vigilant in identifying potential issues and edge cases (Hansen, 1989)

- **Performance Under Pressure**: Lower emotional stability can impact performance in high-stress situations like deadlines or critical bug fixes (Mathews et al., 1991)

- **Work Environment Preferences**: More neurotic individuals may prefer more predictable, structured development environments (Hansen, 1989)

- **Solo vs. Collaborative Work**: Neuroticism may moderate preferences for independent versus collaborative programming approaches (Dick & Zarnett, 2002)

Our data reveals complex relationships between neuroticism and programming contexts. Theme 10 captured stress impacts: "Personal health and life events impact productivity and collaboration" (S4), with mood dependence noted: "My personality depends on the current mood of the day" (I11). These fluctuations characteristic of neuroticism create challenges for consistent collaboration.

Theme 11 explicitly addressed neuroticism's impact: "Pairing not recommended for high neuroticism," with specific warnings about problematic combinations: "Someone with low agreeableness should not pair with a high neuroticism person" (S5). This suggests certain personality combinations may create particularly stressful dynamics.

### Neuroticism in Human  AI Contexts

In AI collaboration, neurotic developers found unique benefits. The reduced social evaluation anxiety with AI tools allowed focus on technical challenges rather than interpersonal navigation. As our analysis notes, delegation of anxiety-provoking tasks to AI agents represents a novel coping mechanism unavailable in traditional development contexts. A Gen–Z student (NK) expressed: "Nerad vedu tým, nejsem týmový hráč. Dokážu v týmu pracovat, ale preferuju solo práci. S AI je to ideální – mám partnera, který mě nesoudí," [I don't like leading a team because I am not really a team player. I can work in a team but I am not a team player. I prefer working by myself. With AI, this is ideal–I have a partner that doesn't judge me,] a preference for solo work potentially driven by anxiety about social evaluation.

On the contrary the psychological dynamics of a low-neuroticism CTO participant's reveal sophisticated adaptation strategies. When errors occur, (MV) externalizes blame to the AI: "Nejsem naštvanej až tak moc na sebe, ale spíš na to agenta," [I am not really angry at myself but at the agent.] This isn't mere deflection but a functional strategy that preserves self-efficacy while navigating the uncertainties of AI collaboration.

In the context of AI collaboration, participant PΓ noted: "With AI, I am more focused on problem-solving. When I talk to a real human, I cannot be as focused as I am when talking to ChatGPT. I do not know why."

Understanding how Neuroticism influences role preferences and stress responses can help VSEs create more supportive environments and optimize role assignments to reduce unnecessary stress while leveraging the heightened vigilance that can accompany this trait.

The application of Big Five personality research to software engineering provides a scientifically grounded approach to understanding individual differences in developer behavior, preferences, and performance. By examining how these traits interact with programming roles and AI-human collaboration, this dissertation advances a more nuanced understanding of personality's role in software development, with particular relevance for VSEs, SOHOs, and educational contexts.

## 2.6   Hypotheses –Roles, Motivation, and Personality

*Novel Theoretical Contribution: The hypotheses presented below were developed through iterative empirical investigation and theoretical synthesis (detailed in Chapters 5-7). They form an integral part of the ROMA framework's theoretical foundation.*

The interplay between personality traits and motivation represents a critical intersection for understanding software developer behavior, particularly in collaborative programming contexts. This section develops the theoretical foundation for the hypotheses that personality traits moderate the relationship between programming roles and intrinsic motivation, providing multi-paragraph-level justification for each hypothesis tested in this dissertation.

### 2.6.1   Personality as a Moderator of Motivational Processes

Contrary to early approaches that treated personality and motivation as separate constructs, contemporary research recognizes their intimate connection. Rather than viewing personality as merely mediating the relationship between environmental factors and outcomes, this dissertation adopts a moderation perspective, positing that personality traits alter the strength and nature of the relationship between programming roles and intrinsic motivation.

This moderation approach is supported by research in organizational psychology, where Barrick et al. (2001) demonstrated that personality traits moderate how work characteristics influence motivational states. In their model, traits like Conscientiousness and Extraversion do not simply transmit the effects of job design to motivation (mediation) but actually change the relationship between job characteristics and motivational outcomes (moderation). When applied to programming contexts, this suggests that different personality traits will respond differently to the same role assignment, with implications for optimizing role-person fit.

The moderation perspective is particularly appropriate for this research because:

- It acknowledges individual differences in responses to the same programming role

- It supports a personalized approach to role optimization rather than a one-size-fits-all model

- It aligns with Self-Determination Theory's recognition that environmental factors interact with individual differences to influence psychological need satisfaction

By conceptualizing personality traits as moderators, this research provides a stronger theoretical foundation for the ROMA framework's emphasis on personality-based role assignments.

### 2.6.2 Theoretical Justification for Hypothesis 1: Role-Based Motivational Differences

**H1 and H1-Cor: Intrinsic motivation varies by Pilot, Navigator, and Solo roles, and this variation remains stable across individuals.**

This hypothesis is grounded in the structural differences between programming roles and their alignment with Self-Determination Theory's basic psychological needs. The Pilot role, which involves direct code implementation, offers immediate feedback on actions (enhancing competence), control over code creation (supporting autonomy), and active collaboration with a partner (fostering relatedness). In contrast, the Navigator role emphasizes strategic thinking and oversight, potentially satisfying different aspects of these needs. The Solo role, while offering high autonomy, may provide less structured feedback and minimal relatedness.

Research by Chong and Hurlbutt (2007) on pair programming dynamics indicates that role differences create distinct psychological experiences, while Sfetsos et al. (2009) found that these differences influence satisfaction and engagement. Williams and Kessler (2003) further demonstrated that pair programming roles can create more engaging experiences than solo work for many developers. These findings suggest that intrinsic motivation should vary systematically across roles due to their inherent characteristics.

The prediction that these motivational differences remain stable across individuals (H1-Cor) is supported by research on the consistency of role-based experiences. While individuals may have different baseline levels of motivation, the relative advantages and challenges of each role should create consistent patterns of motivation across participants. This stability is important for developing generalizable guidelines for role optimization in VSEs and educational settings.

### 2.6.3 Theoretical Justification for Hypotheses 2-3: Personality Clusters and Role Preferences

**H2 and H3: Software engineers exhibit distinct personality types, and these personality types prefer different pair programming roles.**

Hypotheses H2-3 build on evolutionary personality theory and empirical evidence of personality clustering. Penke et al. (2007) and Nettle (2006) suggest that personality traits form distinct clusters due to evolutionary balancing selection, where different trait combinations confer advantages in different environments. This clustering occurs because traits do not vary independently but tend to co-occur in patterns shaped by both genetic factors and environmental influences.

In software engineering specifically, Feldt et al. (2010) demonstrated that personality traits cluster in ways that predict preferences for different development approaches. Their cluster analysis identified distinct personality profiles among software engineers, with observable differences in work preferences and team dynamics. Similarly, Wiesche and Krcmar (2014) found that personality clusters correlate with role satisfaction and performance in software teams.

The prediction that these clusters will prefer different programming roles (H3) connects personality research with role optimization theory. If personality traits represent stable predispositions toward certain types of activities and interactions, they should logically influence preferences for roles that align with those predispositions. By identifying these patterns, the ROMA framework can provide empirically grounded guidelines for role assignments that maximize intrinsic motivation in VSEs and educational settings.

### 2.6.4 Theoretical Justification for Hypothesis 4: Openness and the Pilot Role

**H4: Individuals with high openness to experience prefer the Pilot role.**

This hypothesis links the characteristics of the Pilot role with the psychological tendencies associated with Openness to Experience. Individuals high in Openness typically exhibit creativity, curiosity, and comfort with novel situations (DeYoung, 2015). They are also described as divergent thinkers (McCrae, 1987) inclined to pursue new ideas and approaches, which can lead to innovative contributions in fields like software engineering (Barrick & Mount, 1991; Salgado, 1997). The Pilot role in pair programming involves direct hands-on coding, requiring creative problem-solving, exploration of alternative implementations, and adaptation to emerging requirements—activities that align naturally with the preferences of open individuals.

Research by Salleh et al. (2010, 2014) specifically demonstrated that Openness correlates with performance and satisfaction in programming tasks that demand innovation and divergent thinking. Similarly, DeYoung's (2015) Cybernetic Big Five Theory positions Openness as part of the "plasticity" meta-trait associated with exploration and engagement with possibilities—precisely the qualities needed in the Pilot role, where developers must actively generate code solutions rather than evaluate existing ones.

From a Self-Determination Theory perspective, the alignment between Openness and the Pilot role may enhance intrinsic motivation by satisfying the need for competence through creative expression, the need for autonomy through hands-on implementation decisions, and the need for relatedness through immersion into partner's thought frameworks. This theoretical alignment suggests that assigning high-Openness individuals to Pilot roles could optimize motivation and performance, particularly in VSEs where effective role distribution is crucial.

### 2.6.5 Theoretical Justification for Hypothesis 5: Extraversion, Agreeableness, and the Navigator Role

**H5: Individuals with high extraversion and agreeableness prefer the Navigator role.**

This hypothesis connects the social, communicative nature of the Navigator role with traits that predispose individuals toward interpersonal engagement. The Navigator provides guidance, reviews code, identifies potential issues, and communicates strategic direction—all activities requiring effective interaction and emotional intelligence.

Extraversion, characterized by sociability, assertiveness, positive emotionality (John et al., 2008), and comfort with novelty and risk-taking (Zuckerman et al., 1978; Zuckerman, 2007), facilitates the communication-intensive aspects of the Navigator role. Extraverts typically enjoy verbal interaction, derive energy from social engagement, and process information through discussion—qualities that support effective navigation in pair programming. Capretz and Ahmed (2010) found that extraverted developers excel in roles requiring frequent communication and relationship building, while Feldt et al. (2010) noted that extraversion correlates with positive attitudes toward collaborative development practices.

Agreeableness, encompassing traits like empathy, cooperation, and consideration for others (Mount et al., 1998), complements extraversion in the Navigator role. Highly agreeable individuals tend to provide constructive feedback, resolve conflicts diplomatically, and maintain positive working relationships—crucial for the Navigator's responsibility to guide without dominating. Hannay et al. (2009b) observed that agreeableness facilitates the kind of supportive oversight that characterizes effective navigation.

From an SDT perspective, the Navigator role may satisfy the relatedness needs of extraverted, agreeable individuals through its emphasis on collaboration and communication. By aligning these traits with the Navigator role, the ROMA framework seeks to enhance intrinsic motivation through optimal personality-role matching in software development teams.

### 2.6.6 Theoretical Justification for Hypothesis 6: Neuroticism, Introversion, and the Solo Role

**H6 and H6-Cor: Individuals with low extraversion and high neuroticism prefer the Solo role, and conversely, those with high extraversion and low neuroticism do not prefer Solo work.**

This hypothesis addresses the psychological dynamics of solo programming in relation to emotional stability and social orientation. Solo programming offers structured independence, minimal social interaction, and direct control over the development process—characteristics that may appeal to specific personality configurations.

Neuroticism, reflecting a tendency toward negative emotionality and stress reactivity (Mathews et al., 1991), may influence preferences for more controlled, predictable work environments. Solo programming minimizes the social pressures and real-time adaptations required in pair programming, potentially reducing anxiety triggers for neurotic individuals. Hansen (1989) found that developers higher in neuroticism often prefer work contexts with clear boundaries and reduced interpersonal complexity.

Introversion (low extraversion) is characterized by a preference for less stimulating environments and a tendency to process information internally before expressing it (Eysenck, 1967). Solo programming allows introverted developers to work at their own pace without the continuous communication demands of pair programming. Dick and Zarnett (2002) observed that introverted programmers often report higher comfort levels when working independently than when required to articulate their thought processes in real-time collaboration.

The corollary hypothesis (H6-Cor) suggests that individuals combining high extraversion with low neuroticism would find solo programming relatively unsatisfying, as it deprives them of the social interaction they typically enjoy while offering fewer opportunities to leverage their emotional resilience in collaborative challenges. This prediction aligns with Latham's (2012) findings on how personality configurations influence preferences for social versus independent work contexts.

By understanding these connections between neuroticism, introversion, and solo programming preferences, the ROMA framework provides guidance for role assignments that acknowledge individual differences in social needs and stress management, particularly relevant in the resource-constrained environments of VSEs.

### 2.6.7 Philosophical and Theoretical Synthesis

The integration of philosophical traditions with empirical findings serves more than academic completeness—it reveals why certain personality-role alignments resonate so deeply with developers' lived experience. When high-openness individuals describe the "joy of exploration" in piloting, they unconsciously echo Bergson's élan vital. When introverted developers seek solo work, they pursue what Heidegger recognized as authentic dwelling. Table 7 maps these philosophical resonances, showing how ancient insights about human nature illuminate contemporary human-AI collaboration patterns.

Table 7: Comprehensive ROMA framework synthesis: Bridging personality archetypes, philosophical traditions, and hypothesis testing for evidence-based role optimization

| Personality Configuration | Optimal Role(s) | Philosophical Underpinning | SDT Alignment | AI Mode Integration | Related Hypotheses | Phenomenological Insight |
|---|---|---|---|---|---|---|
| **High Openness** *"The Explorer"* | **Pilot** (primary) Navigator (secondary) | **Bergson's Creative Evolution**: Life as continuous novelty generation **Dewey's Experimentalism**: Knowledge through active inquiry | **Autonomy**: Through creative expression **Competence**: Via novel problem-solving **Relatedness**: In shared discovery | **Co-Pilot**: Amplifies ideation **Co-Navigator**: Explores concepts dialogically | **H4**: High openness → Pilot preference | "There's this moment when everything becomes clear and everything clicks" (MH) |
| **High Extraversion + Agreeableness** *"The Orchestrator"* | **Navigator** (primary) Pilot (collaborative contexts) | **Buber's I-Thou**: Authentic encounter through dialogue **Habermas's Communicative Action**: Rational discourse toward understanding | **Autonomy**: Through leadership **Competence**: In guiding others **Relatedness**: Central to role satisfaction | **Co-Navigator**: Natural dialogue partner **Agent**: Coordinates team of agents | **H5**: High E+A → Navigator preference | "Almost everything can be solved in pairs, from programming to your emotional state" (P2) |
| **High Neuroticism + Low Extraversion** *"The Craftsperson"* | **Solo** (primary) Agent delegation (stress management) | **Heidegger's Dwelling**: Creating space for authentic work **Thoreau's Solitude**: Depth through isolation | **Autonomy**: Via independence **Competence**: Through focused mastery **Relatedness**: Selectively maintained | **Agent**: Delegates anxiety-inducing tasks **Co-Pilot**: Silent assistance | **H6**: High N + Low E → Solo preference | "I don't lead a team because I am not really a team player" (NK) |
| **High Conscientiousness** *"The Architect"* | **Role-flexible** based on team needs | **Kant's Categorical Imperative**: Duty and systematic reason **Aristotle's Phronesis**: Practical wisdom | **Autonomy**: Through systematic control **Competence**: Via quality mastery **Relatedness**: Through reliable contribution | **All modes**: With verification protocols | **Moderates all hypotheses** | "I keep a diary of AI errors. It's a fascinating catalog" (PJ) |
| **Balanced Profile** *"The Adapter"* | **Context-dependent** | **Jung's Individuation**: Integration of opposing tendencies **Systems Theory**: Dynamic equilibrium | **Balanced need satisfaction** across contexts | **Multi-modal**: Switches based on task | **H1**: Role variation affects motivation | "My personality depends on the current mood" (I11) |

*Table 7 illustrates how the ROMA framework integrates personality archetypes with philosophical traditions and empirical findings to provide a comprehensive approach to role optimization in human-AI collaborative programming.*

## The Explorer: Openness and Creative Evolution

The alignment of high Openness with the Pilot role finds philosophical grounding in Bergson's (1907) concept of *élan vital*—the creative impulse that drives continuous novelty generation. For Bergson, life itself is characterized by perpetual creation, an insight that resonates deeply with the exploratory nature of open individuals in programming contexts. As one participant (MH) captured phenomenologically: "There's this moment when everything becomes clear and everything clicks"—a description that echoes Bergson's notion of intuitive breakthrough where duration collapses into creative insight.

Dewey's (1938) experimentalism provides complementary support, positioning knowledge not as passive reception but as active inquiry. The Pilot role embodies this experimental stance, requiring developers to test hypotheses through code, learn from compiler feedback, and iteratively refine solutions. High-openness individuals naturally gravitate toward this experimental approach, finding satisfaction in what Dewey termed "intelligent action"—purposeful exploration guided by reflection.

### The Orchestrator: Dialogical Consciousness and Communicative Action

The preference of highly extraverted and agreeable individuals for the Navigator role reflects profound philosophical insights about the nature of human understanding. Buber's (1923) distinction between "I-Thou" and "I-It" relationships illuminates why these individuals thrive in communicative roles. The Navigator engages in what Buber would recognize as authentic encounter—not merely transmitting information but entering into genuine dialogue where meaning emerges between participants.

Habermas's (1984) theory of communicative action extends this insight, arguing that rational discourse aimed at mutual understanding represents the highest form of human coordination. The Navigator role instantiates this ideal, requiring developers to articulate reasoning, negotiate shared understanding, and guide implementation through dialogue rather than decree. As participant P2 observed: "Almost everything can be solved in pairs, from programming to your emotional state"—a recognition that technical and emotional dimensions interweave in authentic collaboration.

### The Craftsperson: Dwelling and Solitude

The solo preference of neurotic introverts finds philosophical articulation in Heidegger's (1971) concept of "dwelling" and Thoreau's (1854) meditation on solitude. For Heidegger, authentic dwelling requires creating space—both physical and psychological—where one can engage with work free from the "idle talk" of das Man (the anonymous public). Solo programming provides this space, allowing introverted developers to establish what Heidegger termed *"nearness"* to their work without social mediation.

Thoreau's experiment at Walden Pond demonstrated that solitude need not mean isolation but rather represents a different mode of engagement—one that allows deeper communion with one's work. As our participant NK expressed: "I don't lead a team because I am not really a team player"—not a deficiency but a recognition of where authentic engagement occurs for certain temperaments.

### *The Architect: Practical Wisdom and Systematic Reason*

Highly conscientious developers embody the intersection of Kant's (1785) categorical imperative and Aristotle's (384-322 BCE) concept of *"phronesis"* (practical wisdom). Kant's emphasis on duty and systematic reason manifests in their methodical approach to code quality and documentation. Their behavior follows what Kant would recognize as self-imposed moral law—maintaining standards not from external compulsion but from internal commitment to excellence.

Yet pure duty proves insufficient; these developers also demonstrate Aristotelian phronesis—the practical wisdom to apply principles flexibly based on context. As participant PJ's "diary of AI errors" demonstrates, conscientious developers do not merely follow rules but develop nuanced understanding of when and how to apply them. This synthesis of systematic reason and practical wisdom makes them invaluable in quality-critical roles.

### *The Adapter: Individuation and Dynamic Equilibrium*

Balanced personality profiles reflect Jung's (1969) concept of individuation—the integration of opposing psychological tendencies into a harmonious whole. These developers have achieved what Jung termed *"transcendent function"*—the ability to hold opposites in creative tension rather than identifying exclusively with one pole. Their context-dependent role preferences demonstrate this integration in practice.

Systems theory (von Bertalanffy, 1968) provides a complementary lens, viewing balanced profiles as maintaining dynamic equilibrium—continuously adjusting to environmental demands while preserving core stability. As participant I11 noted: "My personality depends on the current mood"—not inconsistency but adaptive flexibility that allows optimal response to varying contexts.

## 2.6.8   Theoretical Integration

This philosophical grounding reveals the ROMA framework as more than pragmatic tool—it represents a synthesis of multiple intellectual traditions:

- **Personality Psychology:** Big Five traits as stable individual differences that moderate role-motivation relationships, providing the empirical scaffold for philosophical insights.

- **Self-Determination Theory:** Basic psychological needs as universal motivational drivers that manifest differently based on personality configurations and role assignments.

- **Phenomenology:** Lived experience as primary data for understanding human-AI collaboration, honoring the irreducibility of first-person perspective in technological contexts.

- **Process Philosophy:** Development as continuous becoming rather than static states, recognizing that both humans and AI systems exist in perpetual co-evolution.

- **Pragmatism:** Theory validated through practical outcomes in VSE contexts, ensuring that philosophical insight translates into improved developer experience and productivity.

The ROMA framework thus provides not merely role recommendations but a theoretically grounded, empirically-validated approach to optimizing human flourishing in an age of AI-augmented software development. By honoring both the universal (basic psychological needs) and the particular (personality differences), it offers a path toward more humane and effective software development practices.

## 2.7 Neuroscience of Collaborative Programming and Learning

The neural substrates of programming—whether solo, paired, or AI-augmented—reveal fascinating insights into how different collaborative modes engage distinct brain networks. This understanding becomes particularly crucial in educational contexts, where the goal extends beyond immediate productivity to encompass deep learning and skill development. By examining the neuroscience of programming collaboration and learning, we can better understand why certain personality-role alignments prove so effective and how educational approaches might optimize both learning and motivation.

### 2.7.1 Neural Foundations of Personality and Programming

The theoretical connections between personality traits and motivational processes in programming find grounding in emerging neuroscientific evidence. This biological foundation does not reduce human experience to mere neural firing but rather illuminates the deep structures that shape how different individuals experience and engage with programming work.

DeYoung's (2013) neurobiological model of personality identifies distinct neural systems underlying the Big Five traits. These systems create what we might term "phenomenological tendencies"—characteristic ways that experience unfolds for individuals with different trait configurations.

Consider the neurobiology of the Pilot experience for high-openness individuals. Increased activity in the default mode network—associated with imagination, creativity, and self-referential processing—creates richer possibility spaces during code generation. This is not merely metaphorical; functional neuroimaging reveals that high-openness individuals literally experience more potential pathways when confronting programming problems (Beaty et al., 2016). As one participant (MH) captured: "Je to takový moment, kdy se všechno vyjasní a všechno do sebe zapadne," [there's this moment when everything becomes clear and everything clicks].

The Navigator role engages neural systems associated with social cognition and theory of mind. Regions like the temporoparietal junction and medial prefrontal cortex show enhanced activation in extraverted individuals during social tasks (Grodin & White, 2015). For these individuals, explaining code activates reward circuits similar to other pleasurable social interactions, creating what one participant called "učitelská euforie" [teacher's euphoria].

Solo programming's appeal to high-neuroticism individuals reflects differential activation in threat-detection circuits. These individuals show heightened baseline activation in the amygdala and anterior cingulate cortex, making social evaluation particularly stressful (Servaas et al., 2013). Solo work reduces this cognitive load, allowing resources to focus on code rather than managing interpersonal anxiety.

### 2.7.2   The Learning Brain: Neuroscience of Programming Education

Programming education engages multiple neural learning systems simultaneously. Vogel and Schwabe (2016) identify three primary memory systems involved in skill acquisition: the declarative system (facts and concepts), the procedural system (skills and habits), and the emotional system (affective associations). Effective programming education must engage all three systems coherently.

In traditional solo learning, students often struggle to integrate these systems. They may memorize syntax (declarative) without developing coding fluency (procedural) or may develop negative emotional associations through frustrating debugging experiences. Our Gen-Z participant (NK) described a transformative moment when AI changed these dynamics ("never sleeps, never gets angry, and always has time.") This always-available, non-judgmental presence may reduce amygdala activation associated with evaluation anxiety, creating more favorable conditions for learning.

The neuroscience of error-based learning proves particularly relevant. Butterfield and Metcalfe (2001) demonstrated that errors followed by corrective feedback enhance memory consolidation more than errorless learning. AI's patient correction style may optimize this error-correction cycle. As our Gen-Z student (NK) noted: "S ChatGPT programuju jako s mentorem. Ptám se ho na věci, na které bych se styděl zeptat skutečného seniora," [I program with ChatGPT like with a mentor. I ask it things I'd be embarrassed to ask a real senior]. This shame-free error exploration may enhance learning by reducing stress hormones that can impair memory formation (Schwabe et al., 2012).

### 2.7.3 Pair Programming and Mirror Neuron Activation

The neuroscience of pair programming reveals why this practice can accelerate learning. The mirror neuron system, first discovered by Rizzolatti and colleagues (1996), fires both when performing an action and when observing others perform that action. In pair programming, the navigator's brain literally "mirrors" the pilot's coding actions, creating a form of embodied simulation that enhances understanding.

Recent research by Sänger et al. (2012) on brain synchronization during social interaction found that pairs engaged in coordinated tasks show inter-brain synchrony in oscillatory neural activity. This synchronization correlates with improved task performance and mutual understanding. In programming contexts, this suggests that effective pairs may achieve a form of "cognitive coupling" where their neural activity aligns.

Study IV participants unknowingly described this phenomenon. Theme 3 captured: "During the conversation, you can discover errors, discuss the strategy, find a solution on which both agree and share knowledge in a great way" (P3). This knowledge sharing may occur not just through verbal exchange but through synchronized neural activation patterns that facilitate deeper understanding.

The personality dimension adds complexity. Highly agreeable individuals show stronger mirror neuron activation when observing others (Carr et al., 2003), potentially explaining their affinity for collaborative roles. Conversely, individuals high in neuroticism show heightened self-focused attention that may interfere with mirror neuron engagement (Mor & Winquist, 2002), partially explaining their preference for solo work.

### 2.7.4 AI Collaboration and Novel Neural Patterns

The integration of AI into programming introduces unprecedented neural dynamics. Preliminary evidence suggests that AI interaction activates a unique combination of neural circuits—partially overlapping with both tool use and social interaction networks (Schleiger et al., 2024). This "neural chimera" may explain why developers report such varied and sometimes contradictory experiences with AI collaborators.

One researcher-developer (LV) reflected: "Když debuguju s AI, můj mozek neví, jestli používám nástroj nebo mluvím s kolegou. Je to zvláštní… meziprostor," [when I debug with AI, my brain doesn't know if I'm using a tool or talking to a colleague. It's a strange… in-between space]. This phenomenological ambiguity, grounded in neural activation patterns, suggests that Human ⌸ AI collaboration represents genuinely novel territory for both psychological theory and software engineering practice.

The educational implications are profound. When students learn with AI, they may develop neural patterns distinct from either solo or human-paired learning. Student (SΓ) observed: "The 'process' is different. Because traditionally, you understand first and after you write the code. But here, you first get the working code and then understand how each line works." This inversion may engage the procedural learning system before the declarative system—a reversal of traditional pedagogy with unknown long-term consequences.

### 2.7.5  Flow States and Hypofrontality in Different Programming Modes

Csikszentmihalyi's (1990) concept of flow finds neural grounding in the transient hypofrontality hypothesis (Dietrich, 2003). During flow states, the prefrontal cortex—associated with self-criticism and executive control—shows reduced activation, allowing for more automatic, intuitive processing. Different programming modes appear to facilitate or hinder this hypofrontal state.

Co-Pilot mode may enhance flow by reducing the cognitive load of syntax recall, allowing developers to maintain focus on higher-level problem solving. As senior developer (MM) noted about achieving flow with AI assistance: "Už nepíšu řádek po řádku, ale skládám větší bloky," [I don't write line-by-line anymore, I compose bigger blocks]. This shift to higher-level thinking may facilitate sustained hypofrontality.

However, Co-Navigator mode's conversational nature may repeatedly activate prefrontal regions associated with language processing and social cognition, potentially disrupting flow. This explains why personality influences mode preference—introverts may find the social activation particularly disruptive to their flow states.

In educational contexts, understanding these flow dynamics becomes crucial. Traditional teaching often emphasizes prefrontal engagement through constant evaluation and self-monitoring. AI-assisted learning might enable what we term "scaffolded flow"—maintaining optimal challenge while reducing evaluative self-consciousness.

In scaffolded flow, AI assumes responsibility for lower-level cognitive tasks (syntax, boilerplate), allowing learners to maintain hypofrontality while engaging with higher-order problems. This differs from traditional scaffolding, which gradually removes support; AI scaffolding dynamically adjusts, providing more assistance when cognitive load peaks and retreating when the developer enters flow. This adaptive scaffolding may explain why some developers report deeper flow states with AI assistance—the cognitive resources freed from routine tasks can be redirected toward creative problem-solving.

### 2.7.6 Implications for Programming Education and Training

Traditional computer science education often assumes a uniform cognitive architecture among learners—lectures deliver declarative knowledge, labs provide procedural practice, and examinations assess retention. Yet the neural evidence presented throughout this section reveals profound individual differences in how brains engage with programming tasks. The neuroscientific insights suggest several principles for optimizing programming education:

1. **Personality-Aligned Learning Paths:** Just as the ROMA framework optimizes professional roles, educational approaches should consider personality-based learning preferences. High-openness students might benefit from exploratory AI use, while conscientious students might prefer structured, systematic approaches.

2. **Strategic Error Exploration:** Leveraging AI's non-judgmental nature to encourage error-based learning, particularly for students high in neuroticism who might otherwise avoid beneficial mistake-making due to evaluation anxiety.

3. **Balanced Neural Engagement:** Designing curricula that engage declarative, procedural, and emotional learning systems in coordination, using different AI modes strategically to activate appropriate neural networks.

4. **Flow-Optimized Practice Sessions:** Structuring coding sessions to maximize flow potential based on personality, with Co-Pilot for sustained implementation and Co-Navigator for reflective learning at natural break points.

5. **Social-Cognitive Scaffolding:** For students struggling with human pair programming due to social anxiety, AI can serve as a transitional object—providing social-like interaction without triggering full social-evaluative neural responses.

These principles converge toward a fundamental reconceptualization of programming education. Rather than viewing AI as a threat to learning, we can design pedagogical approaches that leverage personality-aware AI integration to enhance both immediate learning and long-term skill development. The key lies not in whether to use AI, but in how to align AI modes with individual neural architectures and learning trajectories.

As participant (SB) wisely proposed: "Split education into two parts. One, where students program solo and learn about the fundamental principles of programming, and the other, where they apply the knowledge using AI." This approach respects both the need for fundamental neural pattern formation through solo practice and the accelerated application enabled by AI assistance.

The neuroscience of collaborative programming ultimately reveals that optimal learning emerges not from one-size-fits-all approaches but from careful alignment of neural predispositions (personality), task demands (role), and technological affordances (AI mode). These insights provide biological grounding for the ROMA framework's effectiveness—when we align roles with personality traits, we work with rather than against fundamental neural architectures. As we prepare developers for an AI-augmented future, this understanding becomes essential for designing educational experiences that enhance rather than bypass deep learning, pointing us toward the practical integration with engineering standards explored in the following section.

## 2.8   Integrating Human Factors into Software Process Standards: ISO/IEC 29110 and Beyond

The journey from individual cognition through personality patterns to neural substrates reveals the deep structures shaping software development. Yet these insights remain academically confined without translation into practice. The ISO/IEC 29110 standards provide this bridge—a framework for embedding psychological insights into organizational processes. Where previous sections explored what is, we now turn to what can be implemented.

### 2.8.1   The Evolution from Mechanistic to Human-Centered Procesess

Software development methodologies have undergone a profound philosophical shift since Winston Royce's waterfall model emerged in 1970. This evolution mirrors broader changes in our understanding of human work—from Taylor (1911)'s scientific management to contemporary recognition of knowledge work as fundamentally creative and collaborative.

Early frameworks like the ISO 9000 quality management standard aimed to ensure deliverable quality through rigorously structured processes. While effective for large organizations with specialized roles, these standards often imposed excessive documentation and procedural requirements ill-suited to smaller teams. The resulting "process burden" frequently undermined the agility and motivation that small teams need to thrive.

This tension between process rigor and team agility reflects a deeper philosophical divide. Traditional process frameworks embody what Weber (1905) termed "rationalization"—the systematic application of rules to achieve predictable outcomes. Yet software development, as Brooks (1987) recognized, resists such rationalization due to its essential complexity. The human element—creativity, motivation, collaboration—cannot be proceduralized without destroying precisely what makes it valuable.

Process maturity models emerged to address this challenge, including ISO/IEC 15504 and the Capability Maturity Model Integration (CMMI). These frameworks introduced the concept of maturity levels, allowing organizations to measure and incrementally improve their processes.

However, they still assumed resources and specialization often unavailable to VSEs, where team members typically juggle multiple responsibilities.

The limitations of traditional models became particularly apparent in environments with rapidly changing requirements. Software development's inherent unpredictability—what Brooks (1987) termed its "essential complexity"—demanded more adaptive approaches. Agile methodologies arose in response, with their formal articulation in the Agile Manifesto (Beck et al., 2001) emphasizing:

i) Individuals and interactions over processes and tools

ii) Working software over comprehensive documentation

iii) Customer collaboration over contract negotiation

iv) Responding to change over following a plan

Agile frameworks like Scrum, eXtreme Programming (XP), and Kanban embraced iterative development, continuous feedback, and team self-organization—principles naturally aligned with many VSEs' working styles. While these approaches offered valuable flexibility, they presented new challenges: the absence of formal certification pathways limited VSEs' ability to demonstrate process capability to potential clients and partners, particularly in markets where such certification is expected or required.

This methodological evolution highlighted a critical market gap: the need for internationally recognized standards specifically designed for VSEs—standards that would provide appropriate structure and quality assurance while respecting the resource realities of smaller organizations. The ISO/IEC 29110 standards series emerged to address this gap, offering VSEs not only improved processes but also a pathway to formal certification that could enhance their competitive position in global markets. By providing lightweight yet credible alternatives to heavyweight process models, ISO/IEC 29110 enables VSEs to demonstrate their commitment to quality and process maturity without adopting frameworks designed for much larger organizations.

### 2.8.2   ISO/IEC 29110: A Framework for Human-Scaled Development

The genius of ISO/IEC 29110 lies in its recognition that smaller does not mean simpler—it means different. VSEs face unique challenges that demand tailored solutions (Laporte et al., 2015):

- **Resource Intensity**: With limited staff, each developer must fulfill multiple roles, making personality-role alignment crucial

- **Motivation Criticality**: In a five-person team, one demotivated developer represents a 20% productivity loss

- **Flexibility Demands**: Rapid market changes require adaptive processes that rigid standards inhibit

- **Learning Integration**: Continuous skill development must occur alongside project delivery

The standard's profile approach—Entry, Basic, Intermediate, and Advanced—acknowledges that process maturity is not linear but contextual (Laporte et al., 2017). A highly motivated team of three might outperform a larger team strangled by inappropriate processes. This flexibility creates space for integrating insights from personality psychology and motivation theory.

The Software Basic Profile (ISO/IEC 29110-4-1, 2018) defines two core processes: Project Management (PM) and Software Implementation (SI). Within these processes, the ROMA framework identifies opportunities for personality-based optimization. For instance, the PM.2 Project Plan Execution task might be led by a highly conscientious team member, while the SI.3 Software Architectural Design could leverage the creativity of highly open-minded developers.

### 2.8.3 Agile Adaptation and Personality-Aware Processes

The Agile Software Development Guidelines (ISO/IEC FDIS 29110-5-4, 2025) extend the Software Basic Profile to incorporate agile practices, particularly from Scrum and Extreme Programming. These guidelines map agile events to ISO/IEC 29110 processes, while proving particularly amenable to personality-based role optimization:

**E1: Project Vision Meeting** – Defines project scope and objectives. High-openness individuals excel at envisioning possibilities, while high-conscientiousness members ensure realistic constraints are acknowledged.

**E2: Estimation Meeting** – Facilitates understanding and estimation. Conscientious developers provide accurate time assessments, while agreeable members facilitate consensus when estimates diverge.

**E3: Sprint Planning** – Organizes tasks and resources. Benefits from personality diversity: openness for creative task breakdown, conscientiousness for realistic commitments, and agreeableness for team buy-in.

**E4: Sprint Execution** – Implements planned activities. Personality-aligned task distribution enhances flow: high-openness developers tackle novel features, conscientious ones handle critical infrastructure.

**E5: Daily Scrum** – Identifies impediments and assesses progress. Challenges introverted developers who may find daily social interaction draining. VSEs might adapt by allowing asynchronous updates for introverted members.

**E6: Sprint Review** – Provides feedback on deliverables. Extraverted developers naturally present to stakeholders, while introverted ones might prepare detailed documentation for review.

**E7: Sprint Retrospective:** Enables process improvement. Requires psychological safety for honest reflection. High-neuroticism individuals benefit from structured formats that separate process issues from personal performance.

This agile adaptation aligns with the flexibility and iterative nature of modern software development while maintaining the structural benefits of a recognized standard. The integration of roles like Product Owner, Scrum Master, and Development Team provides clear responsibility definitions suitable for small teams.

### 2.8.4 AI Integration within ISO/IEC 29110 Processes

The emergence of AI collaborators adds a new dimension to VSE processes. Within the ISO/IEC 29110 framework, AI can fulfill specific roles traditionally requiring human resources:

**In Project Management (PM):**

- AI agents can generate initial project plans based on requirements (PM.1)

- Co-Navigator modes can facilitate stakeholder communication by translating technical concepts (PM.3)

- Automated progress tracking reduces administrative burden (PM.4)

**In Software Implementation (SI):**

- Co-Pilot accelerates coding tasks (SI.5) while preserving developer autonomy

- AI-powered code review (SI.6) provides consistent quality checking without interpersonal friction

- Agent modes can handle routine testing (SI.7), freeing humans for creative problem-solving

As CTO participant (MV) implemented: "AI agent dělá dokumentaci, generuje testy a dělá code review. To nám umožňuje soustředit lidské vývojáře na to, co dělají nejlépe—kreativní řešení problémů a komunikaci se zákazníky," [The AI agent does documentation, generates tests, and does code reviews. This allows us to focus human developers on what they do best—creative problem solving and customer communication].

### 2.8.5 The ROMA-Enhanced ISO/IEC 29110 Implementation

Integrating ROMA principles into ISO/IEC 29110 implementation provides a human-centered path to process maturity:

1. **Personality-Aware Role Assignment:** Map team members' Big Five profiles to process responsibilities. High-conscientiousness individuals might own quality assurance tasks, while high-openness developers lead architectural innovation.

2. **Motivation-Aligned Task Distribution:** Assign tasks that satisfy individual psychological needs. Developers seeking autonomy might own module design; those needing relatedness might lead customer interactions.

3. **AI Mode Selection by Process Phase:** Use Co-Pilot during implementation sprints to maintain flow, Co-Navigator during planning for exploratory discussion, and Agent mode for routine testing and documentation.

4. **Adaptive Process Tailoring:** Adjust standard processes based on team composition. A team of mostly introverted developers might emphasize asynchronous communication tools over face-to-face meetings.

5. **Continuous Motivation Monitoring:** Regular assessment of team motivation levels, with process adjustments when engagement drops. This proactive approach prevents the demotivation spirals that can devastate small teams.

### 2.8.6 Beyond Compliance: Toward Thriving VSEs

The ultimate goal is not merely ISO/IEC 29110 compliance but creating environments where developers thrive. By integrating personality psychology, motivation theory, and AI collaboration patterns into standardized processes, VSEs can achieve what larger organizations often struggle with: processes that enhance rather than constrain human potential.

This integration represents a philosophical shift in how we conceive software processes. Rather than viewing human factors as variables to control–AI-boosted *"expandables"*–I recognize them as the essential ingredients that transform mechanical procedures into creative practices. The ISO/IEC 29110 framework, enhanced with ROMA insights, provides a path toward this transformation.

As we prepare to conclude our exploration, the convergence of international standards with psychological insights points toward a future where process maturity and human flourishing become mutually reinforcing rather than antagonistic goals. In this future, VSEs need not choose between formal quality assurance and developer motivation—they can achieve both through thoughtful integration of human factors into their fundamental processes.

## 2.9   Conclusion

This exploration of the conceptual foundations underlying personality-driven motivation in software development reveals a field in profound transition. From the phenomenology of mind in programming through the neural substrates of personality to the practical frameworks guiding development processes, we have traced the convergence of multiple intellectual traditions—each illuminating different facets of what it means to be human in an age of artificial intelligence.

The theoretical landscape we have mapped encompasses Self-Determination Theory's basic psychological needs, which prove as relevant to AI-augmented programming as to traditional collaboration; the Big Five personality framework, which reveals how individual differences shape not just coding styles but fundamental approaches to human-AI interaction; and emerging neuroscientific insights that ground abstract concepts in biological reality. These are not merely academic exercises but urgent investigations into how we preserve human flourishing as the nature of programming itself transforms.

The voices of our participants—from the senior developer meticulously cataloging AI errors to the Gen-Z student finding in ChatGPT a tireless mentor—remind us that theory gains meaning only through lived experience. Their insights reveal patterns that both confirm and complicate our frameworks: high-openness developers embracing AI as creative catalyst, introverted programmers finding liberation from social performance pressure, conscientious individuals developing systematic approaches to AI verification. These patterns coalesce into the ROMA framework—not as rigid prescription but as what Christopher Alexander might recognize as a "pattern language" for motivation-aligned software development.

The philosophical grounding proves essential. When Bergson speaks of creative evolution, he illuminates why open developers thrive in the exploratory space of AI collaboration. When Buber distinguishes I-Thou from I-It relationships, he helps us understand why extraverted developers seek dialogical engagement even with non-conscious AI. When Heidegger describes dwelling, he captures what introverted developers create in their solitary yet AI-augmented workspaces. This is not philosophy as ornament but as foundation—providing depth to empirical observations and meaning to statistical correlations.

As we transition from theoretical foundations to empirical investigation in the DSR cycles ahead, we carry forward both provisional answers and open questions. The answers—that personality matters, that roles can be optimized, that human needs must be honored even in AI

collaboration—will guide our methodology. The questions—about the nature of competence when AI generates code, about authentic creation in human-AI partnerships, about the future of human agency in software development—will drive our inquiry through the five cycles of design, implementation, and validation.

The integration with ISO/IEC 29110 standards promises to bridge theory and practice, making these insights actionable for VSEs where every developer's motivation critically impacts survival. The PDPPA application will test whether continuous personality-based optimization can sustain motivation in real-world contexts. The phenomenological investigations will deepen our understanding of how developers experience this unprecedented collaboration with artificial minds.

This interdisciplinary synthesis—wedding behavioral science with software engineering, psychology with philosophy, neuroscience with practical standards—arrives at precisely the moment when such integration is most needed. The emergence of AI as collaborative partner rather than mere tool demands frameworks that honor both the timeless aspects of human nature and the novel dynamics of synthetic collaboration. The ROMA framework represents my attempt to create such a synthesis—rigorous enough for academic scrutiny, practical enough for daily use, flexible enough to evolve as the landscape shifts.

As we stand at this threshold between conceptual foundations and empirical discovery, we might recall Heraclitus's observation that "the path up and down are one and the same." The journey from theory through empirical investigation to practical application is not linear but recursive—each cycle deepening understanding, each implementation revealing new theoretical questions. In the chapters that follow, we test these conceptual foundations against the lived experience of developers navigating the liminal space between human creativity and artificial intelligence. Their struggles and satisfactions, captured through careful empirical investigation, will provide the ultimate validation for any framework that claims to enhance their working lives.

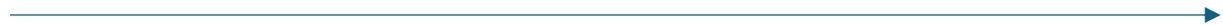

The winter fog that once shrouded fundamental debates about software engineering's nature has lifted, revealing not barren ground but fertile soil where personality psychology and motivation theory have already taken deep root. Yet new mists gather where human and artificial intelligence interleave in ways that would have seemed impossible to those early pioneers. As Gadamer reminds us, understanding emerges through the fusion of horizons. In bringing together the horizons of personality psychology, motivation theory, and AI collaboration, we glimpse a future where software development becomes not just more efficient but more human—where individual differences are celebrated rather than suppressed, where AI amplifies rather than replaces human creativity, where the ancient satisfactions of craft find new expression in unprecedented collaborations. The theoretical seeds we have planted in this fertile ground await the empirical cultivation that follows.

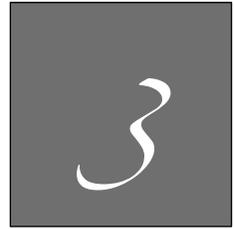

# DESIGN SCIENCE RESEARCH METHODOLOGY

Following the theoretical foundations established in Chapter 2, this chapter details the methodological architecture guiding the empirical investigation. While Chapter 1 introduced the research questions and objectives, this chapter explicates how Design Science Research (DSR) methodology enables their systematic pursuit through five iterative cycles.

## 3.1   Methodological Framework

This dissertation employs Design Science Research as its guiding methodology, chosen for its unique capacity to bridge theoretical innovation with practical application in information systems and software engineering (Hevner et al., 2004). DSR's dual emphasis on rigorous research processes and relevant artifact creation aligns perfectly with our goal of transforming self-determination insights into actionable frameworks for software teams.

### 3.1.1   DSR Adaptation for Behavioral Software Engineering

Traditional DSR assumes well-defined problems amenable to designed solutions. However, in emerging domains like personality-driven programming and Human  AI collaboration, the problem space itself requires empirical exploration. Our adaptation—what we term "theory-informed DSR"—begins with hypothesis-driven investigation to establish theoretical foundations before transitioning to artifact design.

This methodological innovation addresses what Wieringa (2014) identifies as the fundamental tension between understanding "what is" (descriptive knowledge) and creating "what can be" (prescriptive knowledge). By incorporating behavioral science methods within a DSR framework, we ensure that artifacts address empirically-validated needs rather than assumed problems.

### 3.1.2 Three-Phase DSR Architecture

Building on Peffers et al.'s (2006) process model and Hevner's (2007) three-cycle view, we synthesize DSR activities into three overarching phases:

**Exploration Phase** (Cycle 1): Establishes empirical foundations through controlled experiments and psychometric assessment. While labeled a "relevance cycle" in Hevner's taxonomy, our implementation employs rigorous hypothesis testing typically associated with the rigor cycle—a deliberate inversion that reflects the need for theoretical grounding in nascent domains.

**Design and Evaluation Phase** (Cycles 2-4): Iteratively develops and refines artifacts through build-evaluate loops. This phase encompasses framework construction (Cycle 2), AI integration (Cycle 3), and standards operationalization (Cycle 4), each iteration informed by empirical validation.

**Validation Phase** (Cycle 5): Demonstrates artifact utility through field testing and cross-study triangulation, contributing validated design knowledge back to the theoretical knowledge base.

## 3.2 Artifact Conceptualization and Development

Following March & Smith's (1995) DSR output taxonomy, this research produces two primary artifacts that embody different types of design knowledge:

### 3.2.1 The ROMA Framework as Design Theory

The ROMA framework represents what Gregor & Jones (2007) term a "design theory"—encompassing:

- **Constructs**: New vocabulary for discussing personality-driven motivation (e.g., "The Explorer," "synthetic relatedness")

- **Models**: Empirically-validated relationships between personality traits, programming roles, and motivational outcomes

- **Methods**: Systematic procedures for personality assessment and role assignment

- **Instantiations**: Concrete implementations demonstrating the theory's application

As design theory, ROMA provides both explanatory power (why certain role assignments enhance motivation) and prescriptive guidance (how to implement optimal assignments).

### 3.2.2 The ISO/IEC 29110 Profiles Extension as Technological Rule

The standards extension embodies van Aken's (2004) conception of "technological rules"—solution-oriented knowledge in the form "if you want to achieve Y in situation Z, then perform action X." This artifact translates theoretical insights into actionable guidance for VSEs, bridging the theory-practice gap through standardized implementation procedures.

## 3.3 Methodological Implementation Across Cycles

The translation of DSR principles into practice requires what Venkatesh et al. (2013) term "methodological fit"—aligning research approaches with the phenomenon's nature and the questions posed. Our five-cycle implementation demonstrates how mixed methods can be orchestrated to create understanding greater than the sum of their parts.

### The Mixed Methods Architecture

This dissertation's methodological architecture reflects deliberate choices about how different ways of knowing can illuminate complementary facets of personality-driven motivation. Following Venkatesh et al.'s (2013) framework for mixed methods in information systems, we employ three distinct yet interconnected strategies:

**Sequential Explanatory Design (Cycles 1-2)**: Here, quantitative exploration establishes the empirical landscape before qualitative investigation adds depth and meaning. The hierarchical cluster analysis and statistical modeling in Cycle 1 reveal that personality types correlate with role preferences—but only through subsequent qualitative inquiry do we understand why Explorers crave the creative freedom of piloting or how Craftspeople find sanctuary in solo work. This sequencing honors what Creswell & Plano Clark (2011) identify as the explanatory power of building qualitative understanding upon quantitative foundations.

**Concurrent Triangulation (Cycle 3)**: The AI integration phase demands simultaneous investigation from multiple angles. While participants complete standardized motivation inventories, they also share phenomenological accounts of AI collaboration. This concurrent approach captures what Johnson & Onwuegbuzie (2004) call the "fundamental principle of mixed research"—using multiple methods to study the same phenomenon strengthens both inference quality and theory development. When statistical patterns align with lived experiences, we achieve triangulation; when they diverge, we discover boundary conditions or measurement limitations that demand theoretical refinement.

**Pragmatist Integration (Cycles 4-5)**: The final cycles embody philosophical pragmatism—judging methodological choices by their consequences rather than their paradigmatic purity. As Morgan (2007) argues, pragmatism offers a "third way" beyond the qualitative-quantitative divide, focusing on what works to answer research questions. Our blockchain integration, expert validation, and field testing reflect this orientation: methods are tools selected for their fitness to purpose, not ideological commitments.

This methodological progression achieves what Greene et al. (1989) identify as the multiple purposes of mixed methods research. Through triangulation, we seek convergence across different methods, building confidence when personality clusters identified statistically align with phenomenological accounts. Complementarity emerges as different methods elaborate and clarify results—statistical models reveal patterns while interviews explain their meaning. The development function manifests as findings from early cycles inform methodological choices in later ones. When methods reveal unexpected paradoxes—such as AI increasing both autonomy and dependence—the initiation purpose drives us to reshape research questions. Finally, expansion occurs as we extend both the breadth and range of inquiry, from individual motivation to team dynamics to organizational standards.

### 3.3.1   Cycle-Specific Methodological Choices

Each cycle's methodological approach reflects both its specific objectives and its position within the overall research trajectory:

**Cycle 1 – Empirical Exploration**: The foundation-building cycle employs what Tashakkori & Teddlie (2010) term "methodological eclecticism"—selecting quantitative methods not for paradigmatic allegiance but for their capacity to reveal patterns. Hierarchical cluster analysis identifies natural personality groupings; psychometric validation ensures measurement integrity; statistical hypothesis testing establishes baseline relationships. These methods revealed three distinct personality clusters (Explorer: High Openness; Orchestrator: High Extraversion/Agreeableness; Craftsperson: High Introversion/Neuroticism) with significant correlation between personality clusters and role preferences ($\chi^2 = 17.01$, p = 0.0126), creating the empirical substrate upon which subsequent design work builds.

**Cycle 2 – Framework Design**: Here, sophisticated statistical modeling through LME serves a design purpose—validating that personality-based role assignments genuinely enhance motivation. The iterative refinement process exemplifies what Howe (1988) termed "pragmatic parallelism"—quantitative validation and qualitative feedback operate in productive tension. LME models revealed interaction effects between personality and roles ($\beta = 0.47$, p < .001), while qualitative feedback identified the need for context-dependent flexibility. The resulting framework demonstrated motivation improvements averaging 64.6% among student populations, with individual gains varying by personality-role alignment.

**Cycle 3 – AI Integration**: The phenomenological turn reflects methodological responsiveness to emerging phenomena. IPA and thematic analysis capture dimensions of Human  AI collaboration that no survey could measure. This qualitative depth exemplifies what Geertz (1973) called "thick description"—understanding not just what happens but what it means to those experiencing it. It revealed three distinct AI interaction modes (Co-Pilot, Co-Navigator, Agent) and the emergence of "synthetic relatedness"—developers experiencing quasi-social satisfaction from AI interaction despite acknowledging its non-consciousness.

**Cycle 4 – Standards Adaptation**: Conceptual mapping and expert validation represent what Van de Ven (2007) terms "engaged scholarship"—research conducted with rather than for practitioners. This approach successfully mapped ROMA's seven personality-driven tasks to ISO/IEC 29110's PM and SI processes, with WG24 members confirming implementability in VSE contexts.

**Cycle 5 – Comprehensive Validation**: The culminating cycle employs what Morse (2016) calls "methodological triangulation by design." Quasi-experimental field testing revealed 23% mean motivation enhancement among VSE professionals; blockchain validation ensured data integrity with cryptographic timestamps; cross-study synthesis confirmed pattern stability across 200+ participants. This multi-method validation reflects our commitment to what Lincoln & Guba (1985) term "trustworthiness" across all dimensions.

### 3.3.2   Cross-Cycle Integration

The true power of our approach emerges not from individual cycles but from their integration—what Teddlie & Tashakkori (2009) call the "integrative framework" of mixed methods research:

**Progressive Refinement**: Each cycle's findings become inputs for subsequent cycles, creating what Eisenhardt (1989) terms "theory elaboration"—initial patterns become refined frameworks, which become validated artifacts. This progression from exploration through design to validation mirrors the natural evolution of design science knowledge.

**Methodological Triangulation**: Following Denzin's (1978) conception, we employ both within-method triangulation (multiple quantitative approaches) and between-method triangulation (quantitative and qualitative integration). When personality clusters identified statistically align with phenomenological accounts of role preferences, the convergence strengthens confidence in both findings.

**Theoretical Accumulation**: Design knowledge accumulates across cycles in what Carlile (2002) terms "boundary objects"—concepts that maintain coherence across different methodological domains. The personality archetypes (Explorer, Orchestrator, …) serve this function, meaningful to both statistical modelers and phenomenological researchers.

This careful orchestration of methods—sequential where understanding must build, concurrent where phenomena demand multiple perspectives, pragmatic where application matters most—creates a methodological symphony. Each instrument plays its part, their combined sound revealing harmonies no single method could produce. Through this methodological integration, the dissertation achieves what Bryman (2006) identifies as the promise of mixed methods: not just more data, but richer understanding.

## 3.4   Key Methodological Approaches

Where traditional DSR assumes well-defined problems amenable to designed solutions, our investigation confronts what Rittel & Webber (1973) termed "wicked problems"—those where the problem definition itself evolves through investigation. This section details the sophisticated methodological approaches that enable rigorous investigation while maintaining the flexibility essential to emerging phenomena.

### 3.4.1   Linear Mixed-Effects Modeling

The nested, hierarchical nature of our data—multiple observations per participant across varying conditions—demands analytical approaches that honor this complexity. Linear Mixed-Effects (LME) modeling provides the mathematical sophistication to disentangle individual differences from treatment effects while maintaining statistical power.

**Mathematical Foundation:** Following Pinheiro & Bates (2000), our models take the form:

$$Y_{ij} = X_{ij}\,\boldsymbol{\beta} + Z_{ij}\,\boldsymbol{b}_i + \boldsymbol{\varepsilon}_{ij}$$

Where Y represents the motivational outcome, X the fixed effects (roles, personality), Z the random effects (participant baselines), and ε the residual error. This formulation explicitly models the reality that each developer brings unique motivational baselines to their work.

**Implementation Advantages:** Unlike traditional ANOVA, which would require either averaging across observations (losing information) or violating independence assumptions, LME:

- Preserves all 1,092 motivation measurements without aggregation
- Accounts for missing data through maximum likelihood estimation
- Models individual trajectories, revealing how personality moderates role effects
- Provides effect sizes interpretable at both population and individual levels

This sophisticated approach enables discoveries impossible with simpler methods—such as identifying that high-openness individuals (one standard deviation above mean) show 53% greater motivation in Pilot versus Solo roles, while high-neuroticism individuals show the opposite pattern.

While Linear Mixed-Effects modeling reveals the statistical architecture of personality-motivation relationships, numbers alone cannot capture the full human experience of collaborative programming. To understand not just that certain patterns exist but what they mean to those living them, we turn to a complementary methodology rooted in phenomenological philosophy.

### 3.4.2 Interpretative Phenomenological Analysis: Understanding Lived Experience

Where quantitative methods map the terrain of personality and motivation, Interpretative Phenomenological Analysis illuminates how developers actually inhabit this landscape. This methodology emerges from a rich philosophical lineage that makes it uniquely suited to understanding the unprecedented experience of Human  AI collaboration.

#### *Philosophical Foundations as Living Methods*

The roots of IPA reach deep into phenomenological soil. When Edmund Husserl (1913) called philosophers "back to the things themselves," he established a radical premise: that careful attention to consciousness as experienced could reveal truths obscured by theoretical preconceptions. For our research, this Husserlian insight proves essential—we cannot understand what it means to collaborate with AI by imposing predetermined categories but must attend to how developers actually experience these encounters.

Yet pure Husserlian phenomenology, with its emphasis on bracketing all assumptions to reach essential structures, proves insufficient for our purposes. Here, Martin Heidegger's (1962) existential turn becomes crucial. Heidegger recognized that we cannot step outside our being-in-the-world to achieve a God's-eye view. Developers encountering AI do so as already-embedded beings, bringing their histories, languages, and cultural frameworks to the experience. When a participant describes AI as "like programming with someone who has Asperger's syndrome," they reveal not just the AI's characteristics but their own interpretive frameworks for making sense of otherness.

This interpretive dimension leads us naturally to Hans-Georg Gadamer's (1989) hermeneutic philosophy. Gadamer taught that understanding emerges through a "fusion of horizons"—the meeting of the interpreter's perspective with that of the interpreted. In our research, this creates what we may term the "triple hermeneutic," particularly evident in our multilingual context.

Consider when Czech developer JO states: "Debugging s AI je jako pair programming s někým, kdo má Aspergerův syndrom." Three interpretive layers interweave:

First, the participant interprets their lived experience through available cultural metaphors—here, using autism as a lens for understanding AI's particular combination of capability and social blindness. Second, the linguistic code-switching itself carries meaning; the Czech narrative frame with embedded English technical terms reveals how developers create hybrid languages for hybrid experiences. Third, we as researchers interpret these already-interpreted accounts, bringing our own horizons to bear on understanding.

This layered interpretation might seem to distance us from "truth," but Gordon Allport's (1937) idiographic approach suggests otherwise. Allport argued that understanding human experience requires attention to the particular—not just the general patterns that nomothetic science reveals but the unique ways individuals organize their experience. Each developer's account of AI collaboration is both utterly specific (reflecting their personality, history, and context) and revealing of broader patterns. The particular illuminates the universal.

Underpinning this entire approach is Roy Bhaskar's (1975) critical realism, which provides crucial ontological grounding. While experiences are always interpreted, they refer to real structures and mechanisms. When developers report that AI collaboration transforms their sense of competence, this isn't merely subjective perception—it points to actual changes in how work is organized, skills are deployed, and value is created. This realist foundation proves especially important when studying Human ⊞ AI collaboration, where questions about AI's ontological status (tool? agent? other?) remain philosophically contested yet practically consequential.

### The Method in Practice

These philosophical commitments translate into systematic yet sensitive analytical procedures. Following Smith et al.'s (2009) framework, analysis proceeds through iterative deepening:

Initial engagement involves what Heidegger (1971) might call "dwelling with" the text—reading and re-reading not to categorize but to be addressed by the participant's experience. When participant MV describes AI as "technicky brilantní, ale sociálně... jiný" [technically brilliant but socially... different], we attend not just to the content but to the pause, the search for words adequate to this unprecedented experience.

Emerging themes develop through what Gadamer would recognize as dialogical engagement. We do not impose categories but allow meanings to emerge through the conversation between our interpretive frameworks and participants' accounts. The theme of "synthetic relatedness"—developers experiencing quasi-social satisfaction from AI interaction—emerged not from predetermined coding but from attending to how participants naturally described their AI relationships.

The movement from individual to cross-case analysis embodies Allport's insight that the particular and the general illuminate each other. Each developer's unique experience contributes to broader patterns, yet these patterns only make sense when grounded in specific accounts. When multiple participants independently describe AI collaboration using metaphors of neurodivergence, this convergence suggests not coincidence but a shared interpretive resource for making sense of AI's peculiar combination of capability and limitation.

The philosophical depth of IPA is not academic ornamentation but practical necessity. In studying phenomena as novel as Human ⊡ AI collaboration, we need methods that can capture experience in its full complexity while maintaining the rigor that credible research demands. By grounding our analysis in phenomenological philosophy's insights about consciousness, interpretation, and meaning-making, we create space for discoveries that more rigid methods might miss—the subtle ways personality shapes not just what developers do with AI but who they become in relationship with it.

## 3.5 Research Integration and Quality Assurance

The true power of our methodological approach emerges not from individual methods but from their integration. This section details how multiple quality assurance mechanisms ensure both rigor and relevance.

### 3.5.1 Triangulation as Validation Strategy

Our triangulation extends beyond simple method mixing to what Flick (2018) terms "systematic triangulation"—where different methods address complementary aspects of the same phenomenon.

When convergent evidence emerges—such as LME models revealing statistical patterns that IPA participants independently describe—confidence in findings increases substantially. For instance, the quantitative finding that neuroticism predicts solo preference gains deeper meaning when participants describe social anxiety in pair programming contexts. Conversely, divergent insights between methods often prove equally valuable, opening new avenues of understanding. Statistical models might show no main effect for conscientiousness, yet phenomenological analysis reveals how conscientious developers channel their trait differently across roles, maintaining quality regardless of assignment through varied strategies.

### 3.5.2  Blockchain Innovation for Research Integrity

Cycle 5 initiates blockchain integration not as technological novelty but as a response to the reproducibility crisis plaguing software engineering research (Baltes & Ralph, 2022). This approach aligns with emerging open science standards (Mendez et al., 2020) while advancing transparency through cryptographic verification. By storing experimental data on immutable ledgers, timestamps become cryptographically verifiable, data modifications leave permanent audit trails, peer reviewers can verify exact datasets analyzed, and future researchers can build upon transparent foundations. This innovation exemplifies our commitment to methodological advancement that serves genuine research needs rather than technological fashion.

## 3.6  Ethical Considerations in Human-AI Research

The ethical dimensions of our research extend beyond traditional human subjects protection to encompass novel challenges at the human-AI interface. As Beauchamp and Childress (2013) established principles for biomedical ethics, we must now develop corresponding frameworks for Human ⊞ AI research—what Nissenbaum (2010) terms "contextual integrity" within specific technological communities.

### 3.6.1  Informed Consent in Evolving Contexts

When participants engage with AI systems whose capabilities evolve between studies, determining what constitutes truly "informed" consent becomes complex. Our approach emphasizes process consent, ensuring participants understand they are engaging with evolving technology rather than static tools. We maintain capability transparency through clear communication about what AI can and cannot do, avoiding both overselling AI's abilities and underselling its impact. This extends to ongoing dialogue through regular check-ins about comfort levels with AI interactions, recognizing that initial consent may need revisiting as participants' understanding deepens.

### 3.6.2  The Ethics of Anthropomorphism

Our research reveals participants naturally anthropomorphizing AI collaborators, raising complex ethical questions. When participants describe AI using social metaphors, should we correct this language, potentially disrupting natural meaning-making? We chose to use careful terminology— "AI tools" rather than "AI teammates," avoiding phrases like "AI thinks"—while respecting participants' experiential reality.

To study authentic responses while preventing harmful misconceptions, we explicitly inform participants that "The AI assistant can suggest code but cannot understand your intentions or feelings." This transparency extends to our analytical responsibilities: when AI interactions affect professional identity, we ensure our recommendations avoid perpetuating stereotypes, such as automatically assigning introverted individuals to solo work without considering their actual preferences.

We navigate these tensions through careful debriefing, ensuring participants understand AI's non-conscious nature while honoring their experiential reality.

## 3.7 Conclusion: Methodology as Bridge

This chapter has presented not merely a collection of methods but a coherent methodological architecture that bridges multiple chasms: between behavioral and design science, between quantitative patterns and qualitative meaning, between academic rigor and practical relevance.

The creative adaptation of DSR—beginning with empirical exploration when theoretical foundations are nascent—demonstrates how methodology must evolve to address emerging phenomena. By inverting traditional sequences while maintaining systematic progression, we create frameworks for investigating territories where existing maps prove insufficient.

These methodological innovations—theory-informed DSR, phenomenological analysis of Human ⊡ AI interaction, blockchain-verified research data—establish new possibilities for investigating rapidly evolving socio-technical phenomena. DSR scholars might find value in how empirical exploration can inform design without sacrificing solution focus, behavioral software engineers in approaches for investigating human complexity, and Human ⊡ AI researchers in these early ethical and methodological explorations. Such contributions remain provisional—yet they offer possible paths through emerging territories where traditional maps no longer suffice.

As we turn to the empirical chapters ahead, let us remember that methodology is the lens through which phenomena become visible. Statistical models reveal hidden patterns; phenomenological analysis captures ineffable meanings; experimental controls enable causal claims while ecological adaptations ensure relevance. Together, these create a "stereoscopic vision" of personality-driven motivation—each method revealing different depths, their integration producing understanding impossible through any single approach.

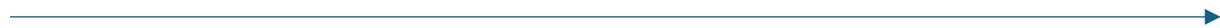

The winter fog that once obscured the path between theory and practice now parts to reveal a methodological bridge—one constructed not from rigid iron but from flexible steel that bends without breaking. Where Dijkstra once navigated by pure logic and Brooks by bitter experience, we now traverse using instruments that measure both the observable and the felt, the quantified and the lived.

The fog has not disappeared; it has simply shifted, revealing new territories where human creativity and artificial intelligence meet in ways our predecessors could never have imagined. Yet the bridge we have built—this careful synthesis of behavioral science and design science—allows safe passage even when the destination remains shrouded. In the cycles ahead, we will cross this bridge repeatedly, each journey revealing more of the landscape where personality, motivation, and collaboration converge in the emerging geography of Human  AI software development.

Let us cross.

# READING GUIDE AND CONTRIBUTIONS

## *FOR STUDIES I-VI*

The methodological bridge constructed in the previous chapter—that careful synthesis of behavioral science and design science—now carries us to the empirical terrain itself. This chapter provides a structured guide to the six studies that crossed this bridge, each journey revealing different aspects of personality-driven role optimization in both human-human and Human  AI collaborative programming contexts. While research is inherently collaborative—benefiting from supervisory wisdom, methodological expertise, and peer insights—each study represents a focused investigation into specific facets of how personality, motivation, and collaboration converge in contemporary software development.

## 4.1   Overview of Studies

The research program encompasses six studies conducted between 2021 and 2025, each examining how personality traits, programming roles, and motivation interact in software development contexts. The studies progress from initial explorations to comprehensive framework validation, building cumulative knowledge about optimizing developer experiences.

**⚜ Study I (2023a – CIMPS'22): Effects of Pilot, Navigator, and Solo Programming Roles on Motivation**

This foundational study employs mixed-methods experimental design to investigate how different programming roles affect developer motivation. The research establishes critical relationships between personality traits and role preferences: openness aligns with pilot roles, extraversion and agreeableness with navigator roles, and neuroticism with solo programming preferences.

**Reference:** Valovy, M. (2023, January). *Effects of Pilot, Navigator, and Solo Programming Roles on Motivation: An Experimental Study*. In: Mejia, J., Muñoz, M., Rocha, Á., Hernández-Nava, V. (eds) *New Perspectives in Software Engineering. CIMPS 2022. Lecture Notes in Networks and Systems, vol 576, p. 84*. Springer Nature. [Best paper award]

The author designed and implemented this personality-focused investigation, collecting data through controlled experiments and follow-up interviews. Statistical analysis revealed motivational patterns that would inform subsequent framework development. The manuscript, presented at CIMPS and awarded best paper, established foundational evidence for personality-driven role optimization.

This research provides empirical grounding for ROMA's personality-role alignment approach, offering evidence for the personality clustering methodology used in role optimization. The motivational patterns identified contribute directly to ROMA's role assignment guidelines, suggesting how personality traits may moderate the relationship between programming roles and intrinsic motivation.

 **Study II (2023b – EASE'23): Psychological Aspects of Pair Programming**

This study extends the initial research to examine pair programming dynamics among Gen Z undergraduates exploring new qualitative horizons. The mixed-methods experimental approach investigates how role distribution and collaborative interactions influence motivation and engagement in educational contexts.

**Reference:** Valovy, M. (2023, June). Psychological Aspects of Pair Programming: A Mixed-methods Experimental Study. In *Proceedings of the 27th International Conference on Evaluation and Assessment in Software Engineering* (pp. 210-216). 2023.

The author adapted the research design for novel pairing constellations and tasks, conducting experiments and interviews in educational settings to understand collaborative programming experiences. Both qualitative and quantitative analyses were employed to develop a comprehensive understanding of personality's influence on pair programming effectiveness. The findings were presented at EASE, facilitating dialogue with the software engineering research community.

Study II contributes refinements to ROMA's role assignment approach for undergraduate contexts, documenting motivational variations across programming roles in educational settings. These findings support ROMA's adaptability to academic environments, acknowledging that student developers may exhibit distinct motivational patterns while still benefiting from personality-aligned role assignments.

### ⬇ Study III (2023c – ICSME'23): The Psychological Effects of AI-Assisted Programming

This research introduces artificial intelligence as a collaborative partner in programming, investigating psychological impacts on both students and professionals. The mixed-methods study combines experimental sessions with interviews to explore themes including human-AI interaction dynamics and motivational changes in AI-augmented programming.

**Reference:** Valovy, M., & Buchalcevova, A. (2023, October). *The Psychological Effects of AI-Assisted Programming on Students and Professionals*. In *2023 IEEE International Conference on Software Maintenance and Evolution (ICSME)* (pp. 385-390). IEEE.

The author integrated AI elements into the experimental design, developing comparative metrics for assessing psychological impacts across collaboration modes. Data collection spanned diverse contexts, with analysis focusing on how developers experience AI collaboration. Prof. Buchalcevova provided valuable supervisory guidance and contributed to manuscript refinement, ensuring theoretical rigor. The co-authored paper was presented at ICSME's NIER track, bridging research and practice.

Study III extends ROMA to Human ⊡ AI collaborative programming, identifying three AI interaction modes—Co-Pilot, Co-Navigator, and Agent—that inform the framework's AI integration approach. The personality-based recommendations for AI tool selection developed here contribute to understanding how individual differences shape human-AI collaboration preferences.

### ⬇ Study IV (2025b – PeerJ-Computer Science): Personality-Based Pair Programming in VSEs

This comprehensive study examines personality-based pair programming implementation within Very Small Entities, exploring ROMA's application in resource-constrained environments. The research investigates how aligning roles with personality traits can enhance self-determination and team dynamics in agile VSE contexts.

**Reference:** Valovy, M., & Buchalcevova, A. (2025). Personality-based pair programming: toward intrinsic motivation alignment in very small entities. *PeerJ Computer Science*, *11*, e2774.

The author developed the ROMA framework structure and conducted field studies with VSE participants to assess framework effectiveness. Cross-analysis techniques were employed to synthesize findings from multiple studies, enhancing generalizability. Prof. Buchalcevova contributed essential thematic cross-analysis and brought deep expertise in ISO/IEC standards integration, strengthening the framework's practical applicability. The manuscript underwent rigorous peer review, including evaluation by ISO/IEC JTC1/SC7 Working Group 24 members—the committee responsible for VSE standards.

Study IV presents the ROMA framework architecture with implementation guidelines and develops an ISO/IEC 29110 extension incorporating personality-driven role optimization. The research establishes metrics for measuring motivational impacts, documenting motivation improvements averaging 23% among professional developers.

 **Study V (2025b – ACIE'25): Blockchain-Driven Research in Distributed Programming**

This study develops and evaluates the Personality-Driven Pair Programming Application (PDPPA) as a testing instrument for the ROMA framework. Pilot case studies with SOHO and VSE participants explore the framework's applicability in distributed programming environments while employing blockchain technology for research transparency.

**Reference:** Valovy, M., & Buchalcevova, A. (2025, January). Blockchain-driven research in personality-based distributed pair programming. In *2025 5th Asia Conference on Information Engineering (ACIE)* (pp. 21-25). IEEE.

The author developed the PDPPA application and conducted pilot case studies demonstrating practical implementation. Blockchain integration was implemented to ensure transparent data collection and validation. Prof. Buchalcevova provided crucial guidance on aligning the application with ISO/IEC 29110 standards and contributed to manuscript development. The collaborative effort was recognized with the best paper award at ACIE.

Study V produces PDPPA as a functional testing tool, validating ROMA's applicability in distributed programming contexts. The research demonstrates ISO/IEC 29110 extension implementation while establishing blockchain-based validation methods for ensuring research transparency and reproducibility.

 **Study VI (2025c – ICSME'25): Developer Motivation and Agency in LLM-Driven Coding**

This phenomenological investigation explores software professionals' experiences with AI collaboration across three interaction modes. Using Interpretative Phenomenological Analysis, the research examines how AI tools affect developers' psychological needs for autonomy, competence, and relatedness.

**Reference:** Valovy, M., Dolezel, M., & Buchalcevova, A. (2025). Developer Motivation and Agency in LLM-Driven Coding. Manuscript submitted to ICSME 2025.

The author designed the phenomenological study protocol and conducted in-depth interviews with software professionals across various contexts. Analysis focused on understanding how different AI modes impact self-determination experiences. Dr. Dolezel contributed valuable methodological expertise in empirical software engineering, enhancing the study's rigor, while Prof. Buchalcevova provided supervisory guidance throughout the research process. The findings were submitted to ICSME's New Ideas and Emerging Results track.

Study VI provides experiential validation of ROMA's AI mode classifications, contributing guidelines for aligning AI tools with psychological needs. The phenomenological insights enrich the framework's theoretical foundations while offering lived experience perspectives on personality-AI mode alignment.

## 4.2   Study-Specific Research Focus

Each study pursued specific research questions that contributed to the dissertation's overarching investigation. These focused inquiries enabled systematic exploration of different facets of personality-driven role optimization:

### Study I - Foundational Inquiries:

- *Do distinct pair programming roles affect programmers' motivation differently?*
- *Can psychometric tests improve the assignment of pair programming roles?*

These questions established the empirical foundation, revealing that role-based motivational differences exist and can be predicted through personality assessment.

### Study II - Educational Context:

- *Are both pair programming roles intrinsically more motivating to students in university classrooms than solo?*
- *What are the psychological aspects of pairing?*

Building on Study I, these questions explored how findings translate to educational settings, revealing nuanced psychological dynamics in student populations.

### Study III - AI Integration:

- *Does AI interaction motivate and satisfy us?*
- *How does AI change programmer behavior and affect (in their opinion)?*
- *Do we credit ourselves or AI when solutions are produced using a mix of both intelligences?*

These questions marked the transition to Human  AI collaboration, identifying key psychological shifts when artificial intelligence enters the programming partnership.

**Study IV - Framework Development:**

- *What new understanding of personality-based motivational preferences for pilot, navigator, and solo programming roles can be achieved?*

- *What are the socio-psychological and group dynamic effects of pair programming?*

- *How should the ROMA framework, optimizing programming roles and their assignments to increase individual motivation and team productivity, be utilized in VSEs?*

These comprehensive questions guided the synthesis of findings into the ROMA framework, addressing both theoretical understanding and practical implementation.

**Study V - Distributed Validation:**

- *How can blockchain technology improve empirical software engineering reproducibility & transparency?*

- *How can personality-based role optimizations be adapted to distributed and very small entity environments?*

These questions addressed methodological innovation and practical validation, ensuring the framework's applicability in contemporary distributed contexts. The progression from foundational questions about role-motivation relationships through AI integration to practical implementation demonstrates the systematic development of knowledge that culminates in the ROMA framework.

## 4.3   How to Navigate the Studies

Readers with access to the complete studies in the appendices may benefit from sequential reading, which reveals the progressive development of concepts and methods. Each study builds upon previous findings while introducing new dimensions, creating a cumulative understanding of personality-driven role optimization. Cross-references throughout Chapters 5-9 indicate when findings from one study inform another, weaving isolated investigations into a coherent narrative.

For readers accessing the dissertation without appended studies, this chapter's summaries capture essential findings and contributions. Published versions, available through respective venues, provide additional methodological details and extended discussions for those seeking deeper engagement with specific aspects.

## 4.4   Relation of Studies to the DSR Cycles

The six studies align with the five Design Science Research cycles, demonstrating how empirical investigation informs artifact development:

**Cycle 1 – Empirical Exploration (Objective 1):** Studies I and II investigate personality-driven role preferences and motivational patterns in human-human programming contexts, establishing empirical foundations for subsequent design work.

**Cycle 2 – Framework Design (Objective 2):** Study IV develops the ROMA framework core, synthesizing earlier findings into a structured approach for personality-based role optimization.

**Cycle 3 – AI Integration (Objective 3):** Studies III and VI extend the framework to Human ⊞ AI collaboration, with Study III identifying interaction modes and Study VI providing experiential validation.

**Cycle 4 – Standards Adaptation (Objective 4):** Study IV contributes the ISO/IEC 29110 Software Basic Profile and Agile Guidelines extension, demonstrating integration with established standards.

**Cycle 5 – Validation (Objective 5):** Study V validates the framework through practical application, employing innovative blockchain-based verification methods.

## 4.5   Linking the Studies: A Coherent Research Program

The six studies form a coherent research program exploring personality-driven optimization in software development. This program benefited from the contributions of collaborators who brought complementary expertise—supervisory wisdom, methodological rigor, and domain knowledge—enriching the research at crucial junctures.

Studies I and II establish empirical foundations through controlled experiments examining personality-role relationships. Study III introduces AI collaboration as a transformative element, identifying distinct interaction modes. Study IV synthesizes these insights into the ROMA framework while developing practical implementation guidelines. Study V creates validation tools demonstrating real-world applicability. Study VI completes the investigation with phenomenological depth, examining lived experiences of Human ⊞ AI collaboration.

This progression from initial experiments through framework development to practical validation ensures that ROMA addresses both theoretical and practical dimensions of personality-driven role optimization. The systematic accumulation of evidence across contexts, populations, and methods creates a foundation for enhancing developer motivation and well-being in contemporary software development environments.

As I reflect on this research journey, each study has contributed not just data but understanding—understanding of how individual differences shape collaborative experiences, how AI transforms traditional dynamics, and how thoughtful role alignment can enhance both productivity and human flourishing. The empirical foundation now complete, we proceed to examine each DSR cycle in detail, beginning with the exploratory investigations that revealed the profound connections between personality and programming role preferences.

The winter fog that once concealed the landscape now lifts gradually with each empirical observation, each study a lamp illuminating another corner of the terrain where personality, motivation, and collaboration converge. Six studies become six beacons piercing the mist—each revealing patterns invisible in isolation, together triangulating toward truths that no single light could capture. What began as shadowy intuitions about individual differences in programming now stands revealed through systematic investigation, transforming speculation into science, anecdote into evidence. The beacons are lit; the path through the fog awaits our passage.

# DSR CYCLE 1 "RELEVANCE" – EMPIRICAL EXPLORATION

## *EXPLORATION PHASE*

In the empirical landscape of software engineering, certain truths remain hidden until systematic investigation brings them to light. This chapter chronicles the first Design Science Research cycle—an exploration that began with intuition and culminated in evidence. Through mixed-methods investigation across Studies I and II, we sought to understand a fundamental question: How do the inherent personality traits of software developers shape their preferences for different programming roles, and what implications does this hold for motivation and productivity?

The exploration phase represents more than data collection; it embodies the critical transition from theoretical speculation to empirical understanding. By examining 75 participants across 1,266 individual motivation assessments, this research reveals patterns that challenge conventional wisdom about role assignment in software teams. The findings suggest that personality does not merely influence programming preferences—it fundamentally shapes how developers experience different collaborative configurations, with profound implications for team formation in VSEs, educational settings, and the emerging landscape of Human  AI collaboration.

This chapter presents the systematic investigation that addresses our first research question (RQ1) through the lens of behavioral science, employing the Design Science Research framework as articulated by Thuan et al. (2019). The exploration culminates in an empirical foundation robust enough to support the ROMA framework's subsequent development—transforming observed patterns of "Effects of Pilot, Navigator, and Solo Programming Roles on Motivation: An Experimental Study" and "Psychological Aspects of Pair Programming: A Mixed-methods Experimental Study" into actionable design knowledge.

## 5.1 Research Questions, Framework, and Hypotheses

The journey from psychological theory to software engineering practice requires careful navigation. Drawing upon the theoretical foundations established in Chapter 2—which span behavioral software engineering, personality psychology, and the neuroscience of intrinsic motivation—this section frames our central inquiry **(RQ1)**: *"What relationships exist between personality traits and self-determination needs in AI-human and human-human co-programming contexts?"*

This question emerges from a convergence of observations. In professional practice, some developers thrive in the dynamic exchange of pair programming while others find it draining. Some embrace the exploratory nature of pilot work; others prefer the strategic oversight of navigation. These differences appear neither random nor purely skill-based—they seem to reflect deeper psychological predispositions that current role assignment practices largely ignore.

To investigate these patterns systematically, we construct and empirically test a set of hypotheses linking Big Five personality traits to role preferences and intrinsic motivation outcomes in collaborative programming settings. These analyses should be interpreted as exploratory testing—providing directional evidence for theory building rather than definitive confirmation. Yet even as exploration, the patterns revealed prove striking in their consistency and practical implications…

### 5.1.1 Theoretical Framework Alignment

The Big Five personality model serves as our analytical lens, chosen for its robust empirical foundation and demonstrated relevance to workplace behavior. Within software engineering contexts, these traits manifest in distinctive ways that shape collaborative dynamics.

Openness to Experience, linked to creativity and innovation, shows particular relevance in software engineering roles requiring high adaptability (DeYoung, 2015). Open individuals do not merely tolerate ambiguity—they actively seek novel problems and unconventional solutions, qualities that align naturally with the exploratory demands of certain programming roles. Extraversion and Agreeableness create a social orientation that influences preferences for teamwork and interactive roles, with extraverted individuals drawing energy from collaboration while agreeable ones prioritize harmony and mutual support. Neuroticism often moderates the inclination toward roles that minimize stress and emotional strain (Feldt et al., 2010; Latham, 2012), with highly neurotic individuals potentially struggling in the unpredictable dynamics of real-time collaboration. Conscientiousness, while universally valuable, plays a complex role in motivation—driving achievement-oriented behaviors while potentially creating rigidity in highly dynamic environments (Barrick & Mount, 1991).

Previous empirical work provides crucial guidance for our investigation. Feldt et al. (2010) employed hierarchical cluster analysis to uncover links between personality clusters and software engineering attitudes, revealing patterns such as the preference for teamwork among highly agreeable individuals and the correlation between emotional stability and job satisfaction. Their approach—examining personality as integrated clusters rather than isolated traits—informs our methodological choices. Similarly, the work of Capretz and Ahmed (2010), Capretz et al. (2015), and Sodiya et al. (2007) provides empirical support for role-specific personality characteristics in software teams, demonstrating that aligning individuals with personality-congruent roles enhances both productivity and satisfaction.

This accumulated evidence suggests that personality-role alignment represents an underutilized lever for improving team dynamics and individual motivation. The challenge lies in moving from general observations to specific, testable propositions that can guide practical role assignment decisions.

### 5.1.2    Research Hypotheses

The hypotheses developed for Studies I and II represent a systematic attempt to map the territory between personality and programming role preferences. Each hypothesis builds upon theoretical foundations while addressing practical questions relevant to team formation and management.

**RQ1H1:** **Software engineers exhibit *distinct personalities* characterized by dominant Big Five dimensions.**

This foundational hypothesis challenges the assumption of personality homogeneity within software engineering. While popular stereotypes suggest uniformity—the introverted, detail-oriented programmer—empirical evidence points toward meaningful diversity. Personality psychology research indicates that individuals in technology-related fields display specific trait patterns (Wynekoop & Walz, 2000), but these patterns form clusters rather than a monolithic profile. Recent investigations by Feldt et al. (2010) and Lenberg et al. (2015) in behavioral software engineering confirm that software professionals group into distinct personality profiles, each with unique implications for team dynamics and individual performance. By identifying these clusters empirically, this research aims to provide a more nuanced understanding of personality diversity within programming teams—moving beyond simplistic stereotypes to recognize the rich psychological variety that characterizes modern software development.

**RQ1H2:** **Software engineers high in *openness* demonstrate a preference for the Pilot role due to its alignment with the creative and innovative task requirements.**

This hypothesis emerges from the convergence of personality theory and role analysis. DeYoung's (2015) Cybernetic Big Five Theory positions openness as central to cognitive exploration and creative problem-solving—precisely the demands placed upon the Pilot in pair programming. The Pilot role requires direct code implementation, creative problem-solving, and rapid adaptation to emerging requirements. These responsibilities create a natural affinity for individuals high in openness, who possess both the cognitive flexibility and motivational orientation toward novel challenges. Supporting evidence from McCrae (1987) demonstrates that openness correlates with divergent thinking and innovation, while Barrick and Mount (1991) found that openness predicts performance in jobs requiring creativity and adaptability. Within pair programming specifically, Salleh et al. (2010) linked openness to satisfaction and retention in programming tasks demanding innovative approaches—suggesting that the Pilot role may provide an optimal outlet for the creative energies of open individuals.

**RQ1H3:** **Software engineers with high levels of *extraversion* and *agreeableness* prefer the Navigator role, which demands strong communication and collaborative skills.**

The Navigator role embodies the social dimension of pair programming, requiring continuous communication, strategic guidance, and interpersonal sensitivity. This hypothesis proposes that individuals combining extraversion with agreeableness possess an optimal psychological profile for navigation. Extraversion, as conceptualized by Costa and McCrae (2000), encompasses sociability, assertiveness, and positive emotionality—traits essential for the Navigator's responsibilities of providing direction and maintaining productive dialogue. Agreeableness complements these qualities by fostering the cooperative, supportive stance necessary for effective guidance without domination. Feldt et al. (2010) demonstrated that highly agreeable individuals show stronger preferences for collaborative development practices, while Mount et al. (1998) found that agreeableness enhances performance in roles requiring interpersonal cooperation. The synergy between these traits creates a natural alignment with the Navigator's unique position—simultaneously leader and supporter, guide and collaborator.

**RQ1H4:** **Engineers low in *extraversion* and high in *neuroticism* show a preference for the Solo role, whereas those high in *extraversion* and low in *neuroticism* do not.**

This bidirectional hypothesis addresses the psychological dynamics of independent versus collaborative work. Low extraversion (introversion) correlates with preferences for less stimulating environments and internal processing (Eysenck, 1967), making solo programming's independence and control particularly appealing. High neuroticism amplifies this preference through a different mechanism—the emotional reactivity and stress sensitivity characteristic of neuroticism (Gray & McNaughton, 2000) may be triggered by pair programming's social demands and real-time adaptation requirements. Dick and Zarnett (2002) found that introverted programmers reported greater comfort with independent work, while Hansen (1989) demonstrated that neurotic individuals often seek predictable, controlled work environments. The hypothesis's corollary—that extraverted, emotionally stable individuals would find solo work less satisfying— follows logically from their psychological needs for social stimulation and their resilience to collaborative challenges.

**RQ1H5:** **The Pilot and Navigator roles yield higher intrinsic motivation for students in academic environments compared to the Solo role.**

This hypothesis shifts focus from personality-based preferences to the inherent motivational properties of different roles. Grounded in Self-Determination Theory (Ryan & Deci, 2000), it proposes that collaborative roles better satisfy basic psychological needs. Pair programming inherently provides more opportunities for relatedness through direct collaboration, competence through immediate feedback and knowledge sharing, and autonomy through role-specific responsibilities. Williams and Kessler (2003) demonstrated that pair programming enhances engagement and satisfaction, while Chong and Hurlbutt (2007) found that the structured interaction of pilot-navigator roles creates more dynamic and motivating experiences than solo work. For students particularly, the learning opportunities embedded in pair programming—peer instruction, immediate clarification, and shared problem-solving—align with educational psychology principles emphasizing social learning and collaborative knowledge construction.

These hypotheses collectively map the theoretical space connecting personality, programming roles, and motivation. They expand existing research by examining personality traits as moderators of role-motivation relationships while also investigating the direct motivational effects of collaborative structures. This dual approach—considering both personality-dependent and personality-independent effects—provides a comprehensive framework for understanding and optimizing programming role assignments across diverse team compositions and contexts.

## 5.2  Output 1.1: Methodological Framework and Nomological Network

The quest to understand how personality shapes programming preferences demands more than simple correlation—it requires a comprehensive theoretical architecture. Following Cronbach and Meehl's (1955) seminal work on construct validity, we developed a nomological network that weaves together concepts from motivation science, personality psychology, and software engineering into a unified explanatory framework. This network serves as both map and compass, guiding our empirical investigation while ensuring theoretical coherence across disciplines.

### 5.2.1  Theoretical Model Construction

At the heart of our investigation lies a reconceptualization of how personality influences workplace motivation. Traditional approaches in organizational psychology have treated personality as an independent variable directly affecting outcomes. Yet this linear view fails to capture the nuanced reality of software development, where the same programming role can energize one developer while draining another. Our model therefore positions personality traits as moderators—psychological lenses that alter how individuals experience and respond to different collaborative configurations.

This moderation approach draws theoretical support from Barrick et al.'s (2001) groundbreaking work demonstrating that personality traits fundamentally change the relationship between work characteristics and motivational outcomes. In our adapted model, programming roles serve as the primary environmental stimulus, with their impact on intrinsic motivation filtered through the prism of individual personality. The theoretical architecture comprises three interconnected components that form the backbone of our investigation.

First, programming roles represent distinct situational contexts, each imposing unique cognitive demands and offering different opportunities for need satisfaction. In human-human collaboration, the Pilot role demands creative implementation and rapid adaptation; the Navigator requires strategic thinking and effective communication; Solo programming offers independence but limits collaborative learning. As software development evolves, these traditional roles are complemented by emerging Human-AI interaction modes: Co-Pilot (where AI provides inline assistance), Co-Navigator (conversational AI guidance), and Agent (delegated autonomous execution). These roles and modes constitute our independent variables, the environmental conditions whose motivational impact we seek to understand.

Second, the Big Five personality traits function as our moderating variables, the psychological filters through which role experiences are processed. Just as a prism refracts white light into its constituent colors, personality traits refract the experience of programming roles into different motivational outcomes. A highly open individual may find the Pilot role's creative demands exhilarating, while someone low in openness might experience the same demands as overwhelming uncertainty. These trait configurations naturally cluster into personality archetypes—the Explorer, Orchestrator, Craftsperson, Architect, and Adapter—which will be systematically developed in Chapter 6 and harmonized by AI-specific specializations in Chapter 7.

Third, intrinsic motivation emerges as our primary dependent variable, with team dynamics captured through its collaborative manifestations. In this dissertation, we extend beyond individual motivation to examine how personality-driven role optimization influences the collective experience of software development. We operationalize team dynamics through three observable components:

1. **Personality-based role compatibility**: Measured through differential motivation outcomes when developers work in personality-aligned versus misaligned configurations

2. **Complementary pairing patterns**: Identified through qualitative analysis of successful collaborations (e.g., Explorer-Architect synergies, Orchestrator-Craftsperson balance)

3. **Collaborative flow and friction**: Observed through communication effectiveness, conflict patterns, and mutual satisfaction in different role configurations

While we do not measure team dynamics as a single unified construct, these components collectively reveal how individual motivation aggregates and interacts to create team-level phenomena. When personality-aligned individuals collaborate, we observe not just higher individual motivation but emergent properties—enhanced creativity in Explorer-Architect pairs, smoother communication in Orchestrator-led teams, reduced conflict when Craftspeople have protected solo time.

This focus on both individual and collective outcomes reflects my conviction that sustainable performance in creative work like programming flows from genuine engagement at all levels. By examining intrinsic motivation as both personal experience and team phenomenon, we aim to identify role configurations that nurture rather than deplete the psychological resources of both developers and their teams. While enhanced motivation and positive team dynamics likely contribute to improved productivity—a relationship well-established in organizational psychology literature—our investigation focuses on these more proximal outcomes that serve as both ends in themselves and leading indicators of sustainable performance.

This theoretical model, illustrated in Figure 4, embodies Bandura's (1986) insight that behavior emerges from the dynamic interplay of person, situation, and behavior itself. The nomological network thus provides not merely a static map of relationships but a dynamic framework for understanding how personality, roles, and motivation interact to shape the lived experience of software development—both in traditional human-human collaboration and in the emerging landscape of Human-AI partnership.

*Figure 4: Nomological network: ROMA Framework with Archetypes and AI Specializations*

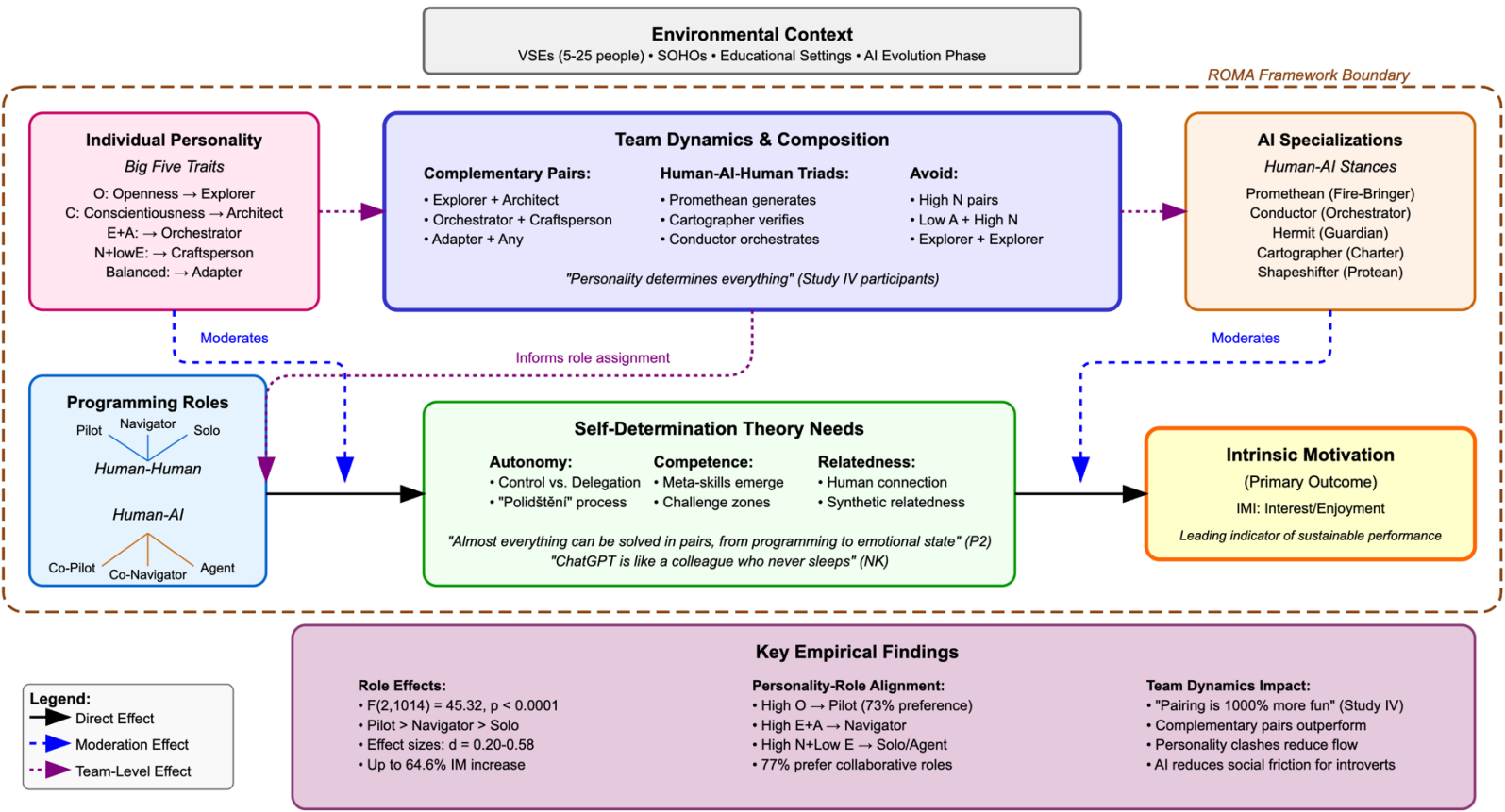

### 5.2.2 Personality Clusters as Functional Units

The decision to examine personality dimensions as integrated clusters rather than isolated traits reflects both theoretical sophistication and empirical wisdom. Just as a symphony emerges from the interplay of instruments rather than their individual sounds, personality manifests through the dynamic interaction of traits rather than their separate influences. This clustering approach, grounded in evolutionary psychology and neurobiology, recognizes that personality traits have co-evolved as adaptive packages rather than independent characteristics.

From an evolutionary perspective, as detailed in Sections 2.5 and 2.7, personality traits cluster due to fundamental trade-offs in adaptive strategies. Nettle (2006) and Penke & Jokela (2016) demonstrate that traits form constellations because certain combinations confer survival advantages in specific ecological niches. The highly open, moderately neurotic individual represents a coherent adaptive strategy for environments requiring creative problem-solving under

uncertainty—precisely the conditions often encountered in software development. Similarly, the agreeable extravert embodies a social navigation strategy, while the introverted, emotionally reactive individual exemplifies a cautious, detail-oriented approach.

DeYoung's (2015) Cybernetic Big Five Theory provides the theoretical scaffolding for understanding these clusters as functional units. Rather than viewing traits as static descriptors, this framework conceptualizes them as dynamic regulatory systems that guide behavior probabilistically. Within this view, personality clusters represent stable configurations of these regulatory systems, each creating a distinctive pattern of motivation, cognition, and behavior.

The methodological advantages of cluster analysis extend beyond theoretical elegance to practical utility. Holistic analysis captures the gestalt of personality—the emergent properties that arise from trait interactions. A developer high in both openness and conscientiousness does not simply combine creative and organized tendencies; they manifest a unique synthesis where creativity is channeled through systematic exploration. This ecological validity better reflects how personality operates in real-world contexts, where traits never function in isolation but always within an integrated psychological system.

Moreover, the clustering approach enhances statistical power in our moderate-sized samples—a perennial challenge in software engineering research. By identifying natural groupings rather than analyzing countless trait combinations, we can detect meaningful patterns that might otherwise be obscured by statistical noise. Most importantly for our applied goals, cluster-based insights translate more readily into actionable recommendations. Rather than suggesting complex multi-trait profiles, we can provide clear guidance: "Developers with this personality cluster tend to thrive in Navigator roles" offers more practical value than a matrix of individual trait correlations.

The resulting nomological network, anchored by personality clusters rather than isolated traits, provides a robust framework for understanding the complex choreography between personality, roles, and motivation. It acknowledges that in the lived experience of software development, personality operates not as separate switches but as an integrated system—a truth our empirical investigation would soon confirm with striking clarity.

## 5.3   Participants and Sampling

The selection of research participants represents a critical juncture where theoretical ambitions meet practical constraints. For our investigation into personality-driven programming preferences, we required a population that could provide both controlled conditions for rigorous experimentation and sufficient ecological validity for meaningful conclusions (Sjøberg et al., 2002). The solution emerged from the intersection of pedagogical opportunity and research necessity.

### 5.3.1 Criteria for Participant Selection

Our participants consisted of second-year undergraduate students enrolled in software engineering at *Prague University of Economics and Business*. This population offered unique advantages for our exploratory research. Unlike industry professionals whose established work patterns might obscure personality influences, these students approached programming roles with relatively fresh perspectives, allowing personality traits to manifest more clearly in their preferences and motivational responses.

The educational context provided natural experimental controls while maintaining real-world relevance. As Falessi et al. (2018) argue, the controlled nature of academic environments enables rigorous experimentation without sacrificing applicability—a critical consideration given our goal of developing frameworks useful for both educational and professional contexts. Students participated as part of their regular curriculum, with full freedom to opt out of data collection, ensuring both ecological validity and ethical integrity.

### 5.3.2 Sampling Method

Our purposive sampling strategy balanced scientific rigor with educational ethics. Participation was entirely voluntary, with no impact on academic grades—a crucial ethical safeguard that also enhanced data quality by ensuring genuine engagement rather than compliance. The quasi-experimental design divided participants into control and test groups, with pairs formed based on classroom proximity and maintained throughout the sessions.

This pairing approach, validated by extensive pair programming research (Gallis et al., 2003; Williams et al., 2006), controlled for compatibility effects while allowing personality influences to emerge clearly. By maintaining consistent pairs, we minimized the confounding variable of interpersonal dynamics, allowing the focus to remain on how individual personality traits influenced role preferences and motivation.

### 5.3.3 Description of Participant Groups

The demographic composition of our studies reflected the broader patterns in software engineering education while providing sufficient diversity for meaningful analysis.

**Study I.**     **(2023a – CIMPS)** engaged 40 students with a gender distribution of 38 males and 2 females—a ratio unfortunately typical in technical fields but requiring acknowledgment as a limitation. Experience levels varied substantially, ranging from 6 months to 6 years (mean = 2.2 years, SD = 1.5), providing a spectrum from novice to relatively experienced programmers. This variation proved valuable, allowing us to observe how personality influences persist across experience levels.

**Study II.**    **(2023b – EASE)** involved 35 students (30 males, 5 females) with experience ranging from complete beginners to 5 years (mean = 1.6 years, SD = 1.3). The slightly lower experience level in this cohort provided an opportunity to observe personality influences in earlier stages of programming skill development.

Crucially, all participants received standardized training in software design, programming languages, and collaborative techniques through their coursework. This educational standardization, aligned with recommendations from Katira et al. (2004), ensured that observed differences in role preferences and motivation stemmed primarily from personality variations rather than disparate skill sets or knowledge bases.

Study I primarily examined Hypotheses **RQ1H1–4**, focusing on the role of personality clusters in motivation, while Study II tested Hypothesis **RQ1H5** on overall motivational levels in programming roles. The personality clustering in Study I revealed three distinct profiles that would prove central to our analysis, scaled 1-5:

**Cluster 1.** High openness to experience ($\bar{x}$ = 4.24; s = 0.54), suggesting strong creative and exploratory tendencies

**Cluster 2.** Combined high extraversion ($\bar{x}$ = 3.71; s = 0.60) with high agreeableness ($\bar{x}$ = 3.96; s = 0.37), indicating social orientation and cooperative tendencies

**Cluster 3.** Featured high neuroticism ($\bar{x}$ = 3.92; s = 0.62) paired with low extraversion ($\bar{x}$ = 2.02; s = 0.46), suggesting preference for controlled, low-stimulation environments.

These clusters (detailed further in Table 9, Section 5.8.2), emerging naturally from our data, would provide the lens through which we examined the intricate relationships between personality, programming roles, and intrinsic motivation.

## 5.4     Data Collection Methods – Experimental Framework

The transformation of theoretical questions into empirical evidence requires methodological precision married to practical wisdom. In pursuit of Design Science Research **Objective 1**—to identify and analyze personality-driven preferences for human-human and AI-human programming roles—our experimental framework emerged from a synthesis of established approaches in motivation research, software engineering experimentation, and behavioral psychology. Each methodological strand contributed essential elements, creating a unified investigative approach capable of revealing the subtle patterns that connect personality traits to programming role preferences.

This section details how theoretical ambitions translated into concrete experimental procedures, establishing the empirical foundation upon which the ROMA framework would later be built.

### 5.4.1    Explanation of Data Collection Methods

The selection of experimental methodology emerged from both theoretical considerations and practical opportunities. Following Latham's (2012) and Ryan and Deci's (2017) compelling demonstrations of experimental approaches in motivation research, I adopted controlled experimentation as the primary data collection method. This choice gained additional validation from the convergence of research opportunity—my role as lecturer for the software engineering course—with methodological rigor, under Professor Buchalcevova's expert supervision.

The experimental framework I developed represented a careful synthesis of multiple research traditions. The behavioral software engineering theories of Lenberg et al. (2015) provided the conceptual foundation, while Feldt et al.'s (2010) personality cluster investigations offered methodological guidance. The well-established experimental protocols of Wohlin et al. (2012) and Ko et al. (2015) ensured scientific rigor, while the meta-analytic insights from pair programming research (Gallis et al., 2003; Hannay et al., 2009b) shaped my specific procedural choices.

This integration aimed to create an experimental design capable of capturing the subtle interplay between personality traits and motivation across different programming configurations (Valovy, 2021; Valovy, 2022, January; Valovy, 2022, September, A). By grounding the approach in established methodologies while adapting them to our specific research questions, I sought to build a bridge between theoretical understanding and practical application—essential for developing frameworks useful in real-world contexts like VSEs and educational settings.

### 5.4.2    Experimental Pair Programming Frameworks

The experimental design I developed expanded Gallis et al.'s (2003) foundational pair programming framework in crucial ways. Where the original framework distinguished simply between individual and pair programming, I introduced three distinct roles—Solo, Pilot, and Navigator—enabling a more nuanced understanding of how specific collaborative configurations affect motivation. This granular approach proved essential for uncovering the personality-role interactions at the heart of our investigation.

My reconceptualized framework maintained Gallis et al.'s emphasis on rigorous variable control while introducing sophistications necessary for personality-based analysis. The independent variables evolved from simple individual/pair distinctions to encompass three specific roles, each with distinct cognitive and social demands. This expansion allowed observation of how personality traits moderate the motivational impact of different collaborative configurations—a level of analysis impossible with binary categorizations.

For dependent variables, I replaced Gallis et al.'s somewhat nebulous "trust and morale" with intrinsic motivation, measured through validated psychometric instruments. The Intrinsic Motivation Inventory (Deci & Ryan, 1982) and Big Five Inventory (Rammstedt & John, 2007) provided the measurement precision that had proven elusive in earlier frameworks. This shift toward established psychometric tools, advocated by Hannay et al. (2009b) and Graziotin et al. (2021), enhanced both the reliability of our measurements and their theoretical grounding.

A key innovation was the introduction of personality traits as explicit moderating variables, transforming them from background noise into central analytical features. Through hierarchical cluster analysis, I could examine how distinct personality configurations shaped responses to different roles—revealing patterns that linear analyses might miss. This boundary variable approach, inspired by Latham's (2012) work on motivation, illuminated not just whether roles affected motivation, but for whom and under what personality conditions these effects emerged.

### 5.4.3   Context Variables and Confounding Factors

The validity of experimental findings hinges on careful control of confounding variables—a challenge intensified when studying complex phenomena like motivation and personality. Following Sjøberg et al.'s (2002) emphasis on explicit contextualization, we systematically addressed potential confounds across multiple dimensions.

Task design represented our first critical control point. Drawing from Bond and Titus's (1983) work on group dynamics and Hannay et al.'s (2009a) findings on task complexity effects, we crafted tasks that balanced realism with experimental control. The apparent contradiction between group dynamics research (which suggests complex tasks favor individual work) and pair programming studies (which show benefits for complex tasks) resolves when viewed through Brooks's (1987) lens of essential complexity. Software engineering tasks possess inherent complexity that benefits from cognitive distribution, unlike the artificial complexity often studied in laboratory group dynamics research.

Our task equivalence strategy drew inspiration from Knuth's TAOCP series (1968-2011), dividing programming challenges into abstract implementation units that could be balanced for complexity, duration, and required skills. This approach ensured that observed motivational differences stemmed from role configurations and personality interactions rather than task variations.

Environmental standardization followed industry-realistic protocols while maintaining experimental control. Recognizing Steiner's (1972) insights on how workflow impacts group productivity, we implemented uniform development processes (XP for implementation, Scrum for project management) and standardized tools (IntelliJ IDEA Professional, Java 17). The single-screen setup with supplementary resources like whiteboards mirrored realistic pair programming environments while enabling consistent observation conditions.

Several specific confounds received particular attention. External motivators were eliminated entirely, following Vinson & Singer's (2008) recommendations, ensuring that observed motivation truly reflected intrinsic rather than extrinsic sources. Task difficulty variations, a concern raised by Vanhanen & Lassenius (2005), were controlled through our systematic equivalence procedures. Participant ability differences, identified by Arisholm et al. (2007) as a major confound, were addressed by measuring motivation rather than performance—following Latham's (2012) insight *(performance = ability × motivation)*, allowing us to isolate the motivational component.

Finally, we managed fatigue and cognitive load through regular breaks and session limits, informed by Boksem et al.'s (2005) research on sustained cognitive performance. Learning effects and "pair jelling"—the gradual improvement in pair performance over time (Williams et al., 2006)—were controlled through preliminary training sessions and consistent pair assignments.

---

In sum, the adapted experimental framework advanced beyond Gallis et al. (2003) by introducing three distinct roles (Pilot, Navigator, Solo) rather than binary individual/pair categories, measuring motivation rather than performance to isolate psychological dynamics, and treating personality as an explicit moderating variable rather than background noise. These refinements directly address Hannay et al.'s (2009a) critique of aggregation problems while maintaining the ecological validity essential for behavioral software engineering research.

## 5.5   Procedure

The experimental procedure unfolded across 300 minutes of programming time, carefully structured to balance scientific rigor with participant well-being. This design, spanning five sessions with 30 distinct tasks, created multiple observation points for each participant in each role—essential for detecting the subtle moderating effects of personality on role-based motivation.

The procedure began with a crucial 60-minute pilot training round, addressing the dual challenges of familiarizing participants with pair programming mechanics and establishing baseline competence levels. This preliminary session served multiple purposes beyond mere training: it reduced variance due to unfamiliarity, allowed pairs to establish working rhythms, and provided participants with experiential understanding of all three roles before formal data collection began.

Psychometric data collection followed established protocols for minimizing participant burden while ensuring data quality. The 10-item BFI personality assessment occurred at session start when participants were fresh and focused. The 7-item IMI motivation measure was administered immediately after each task, capturing motivation while the experience remained vivid but before role-switching could create interference. This timing—validated through extensive pilot testing—balanced accuracy with practical constraints.

Task design reflected our commitment to external validity within experimental constraints. Drawing from real-world software engineering challenges—2D game development in JavaFX, web applications using Spring Boot, various architectural patterns—we created tasks that engaged genuine programming skills rather than artificial exercises. While acknowledging that academic settings cannot fully replicate the complexities of professional environments or VSEs, this approach maximized the transferability of our findings to practical contexts.

The role-switching protocol addressed critical methodological challenges in pair programming research. Free-riding—where participants contribute less in group settings (Forsyth, 1999)—and navigator passivity—identified by Dick & Zarnett (2002) as a persistent challenge—could severely compromise data validity. Our predetermined rotation schedule ensured balanced engagement across all roles while maintaining the structured approach advocated by Williams and Kessler (2003). This systematic rotation also enabled within-subject comparisons, strengthening our ability to detect personality-based preferences.

Ethical considerations permeated every procedural decision. The institutional review board's waiver, granted due to minimal risk and educational context, came with responsibilities we took seriously. Informed consent preceded each session, withdrawal rights were explicitly communicated, and the educational value of participation was emphasized. This ethical framework ensured that our pursuit of knowledge never compromised participant welfare while transforming research participation into a reciprocal learning opportunity—a balance essential for research integrity.

## 5.6   Instruments and Tools

Using validated psychometric instruments was integral to ensuring the accuracy and reliability of the constructs measured in this research. The Intrinsic Motivation Inventory (IMI) and the Big Five Inventory (BFI) were selected based on their established psychometric properties and suitability for the study's context.

### 5.6.1   Big Five Inventory (BFI-10)

The selection of the BFI-10 (Rammstedt & John, 2007) represented a deliberate methodological choice informed by both theoretical and practical considerations. While the original 44-item BFI offers more comprehensive trait assessment, the abbreviated 10-item version provided crucial advantages for our experimental context. With participants completing numerous assessments across several sessions, minimizing survey fatigue became paramount.

Yet brevity came with remarkable retention of psychometric quality. The BFI-10 maintains impressive part-whole correlations with its parent instrument (r = .83), while validation against peer ratings and convergent validity with the NEO-PI-R demonstrate its capacity to capture genuine personality variance. The instrument's reliability metrics—Cronbach's alpha ranging from 0.75 to 0.90, test-retest stability between 0.72 and 0.84—provided confidence that our measurements reflected stable personality characteristics rather than momentary states.

We acknowledge the trade-offs inherent in this choice. The reduced variance explanation for Agreeableness (38%) and Openness (45%) meant accepting some loss of nuance in these dimensions. However, as Felipe et al. (2023) argue, such compromises often prove worthwhile when they enable more extensive data collection without compromising core validity. In our context, the ability to assess personality repeatedly without inducing response fatigue outweighed the marginal loss in measurement precision.

### 5.6.2   Intrinsic Motivation Inventory (IMI)

For measuring the central construct of intrinsic motivation, I selected the Interest/Enjoyment subscale of the IMI (Ryan et al., 1983). This seven-item scale has demonstrated exceptional psychometric properties across diverse contexts, with Cronbach's alpha consistently exceeding 0.90 and strong convergent validity with behavioral measures of intrinsic motivation.

The IMI's particular strength lies in its factor-analytic stability—the same underlying construct emerges whether measuring motivation for puzzle-solving, sports, or in our case, programming tasks. This cross-context validity proved essential for our investigation, ensuring that motivational measurements remained comparable across different programming roles and task types. The subscale's focused nature also aligned with our experimental constraints, providing robust motivation assessment without overwhelming participants who had just completed challenging programming tasks.

Both instruments were implemented in their original English form, leveraging their established validity rather than risking the psychometric uncertainties of translation. This decision reflected our international academic context where English served as the common second language of instruction, ensuring that all participants possessed sufficient linguistic competence for accurate responding.

The combination of these two instruments—brief yet valid, theoretically grounded yet practically efficient—provided the measurement foundation for our investigation. Together, they enabled the capture of rich psychological data while respecting the realities of experimental research with human participants, where engagement and data quality often stand in delicate balance.

## 5.7    Data Analysis Methods

The transformation of raw data into meaningful insights requires analytical strategies that honor both the complexity of human personality and the rigor of scientific inquiry. Our approach combined hierarchical clustering with parametric and non-parametric statistical tests—a methodological ensemble designed to illuminate patterns while respecting the nuanced nature of our data. The evolution from SPSS in Study I to custom R scripts in Study II reflected our growing commitment to reproducibility and transparency in research practices.

### 5.7.1    Cluster Formation and Validation

Hierarchical clustering emerged as my primary tool for discovering natural personality groupings within our participant population. This unsupervised learning approach, particularly suited to social science applications as Milligan & Cooper (1985) demonstrate, allowed personality patterns to emerge from the data rather than imposing predetermined categories—essential when exploring uncharted territory at the intersection of personality and programming preferences.

The clustering process unfolded through a carefully orchestrated sequence of analytical steps. We began by calculating pairwise dissimilarities among all participants using Euclidean distance, creating a mathematical representation of personality differences that captured meaningful psychological variations. This distance matrix served as the foundation for all subsequent clustering operations, encoding the multidimensional personality space into analyzable relationships.

Our choice of complete-linkage criterion for the clustering algorithm reflected a deliberate analytical decision. This approach determines cluster boundaries by considering the maximum distance between any points in different clusters, ensuring that resulting groups maintain internal cohesion while maximizing inter-cluster separation. The complete-linkage method proved particularly appropriate for our goals, creating compact, well-defined personality clusters rather than elongated chains that might obscure meaningful groupings.

The critical question of optimal cluster number—how many distinct personality types existed in our sample—found its answer through the Dunn index (Bezdek & Pal, 1998). This validation metric elegantly balances two competing desires: maximizing the distance between clusters while minimizing distances within clusters. The index guided us toward a three-cluster solution, a finding that would prove remarkably interpretable in light of personality theory. Each cluster exhibited distinct centroid characteristics, confirming that our analytical approach had successfully identified meaningful psychological groupings rather than statistical artifacts.

Visual validation through dendrogram analysis provided additional confidence in our clustering solution. The hierarchical tree structure clearly revealed three major branches, with substantial distance between primary divisions and tighter clustering within groups. This visual evidence, combined with the quantitative Dunn index, created a compelling case for the three-cluster model that would anchor all subsequent analyses.

### 5.7.2 Statistical Tests for Data Validation

The validation of relationships between personality clusters, programming roles, and motivation demanded a sophisticated statistical toolkit. Our analytical strategy employed multiple complementary approaches, each contributing unique insights while collectively building a robust evidentiary foundation.

Normality testing formed the gateway to our analytical decisions. The Shapiro-Wilk test, particularly powerful for our moderate sample sizes, revealed that intrinsic motivation scores showed generally normal distributions (W > 0.95 in most cases), supporting parametric analyses. Where the Shapiro-Wilk results proved marginal, we employed the Kolmogorov-Smirnov test as a complementary assessment, particularly valuable for our larger aggregated datasets. This dual approach to normality testing exemplified our commitment to analytical rigor—never assuming distributional properties but always verifying them empirically.

The personality trait distributions told a more complex story. While Extraversion (p = 0.057) and Neuroticism (p = 0.097) conformed reasonably to normal distributions, Openness (p = 0.043), Conscientiousness (p = 0.021), and Agreeableness (p = 0.017) showed significant departures from normality. This finding, rather than presenting an analytical obstacle, provided valuable insight:

personality traits in our software engineering sample did not always follow population norms, suggesting possible selection effects or adaptive specialization within the field.

ANOVA served as our primary tool for examining motivation differences across roles, with the F-statistic revealing whether role assignment created meaningful motivational variations. Yet we never relied on ANOVA alone. The Kruskal-Wallis test—ANOVA's non-parametric cousin— provided a robust check on our findings, analyzing rank-based rather than raw data. The convergence of results across both approaches (ANOVA: $F = 4.32$, $p = 0.017$; Kruskal-Wallis: $\chi^2 = 8.70$, $p = 0.013$) strengthened our confidence that role-based motivational differences reflected genuine phenomena rather than statistical artifacts.

Student's t-tests enabled focused comparisons between specific personality clusters and population norms, revealing how each cluster deviated from typical trait distributions. The striking significance levels—openness ($p = 0.00116$), extraversion ($p = 0.00207$), agreeableness ($p = 0.00016$), and neuroticism ($p = 0.000012$)—indicated that our clusters represented genuinely distinct personality configurations rather than random variations.

The Chi-squared test for independence provided the crucial link between personality clusters and role preferences. By examining whether role choice distributions differed significantly across personality clusters, this test directly addressed our core research question. The significant result ($p = 0.0126$) confirmed that personality and role preference were indeed related—not deterministically, but probabilistically, exactly as personality theory would predict.

### 5.7.3   Measurement Validity

The credibility of our findings ultimately rests upon the psychometric integrity of our measurements and analytical procedures. Table 8 serves as a comprehensive evidence trail, consolidating the reliability and validity coefficients that underpin every statistical claim in this investigation. This transparency in measurement quality reflects our commitment to reproducible research— providing future investigators with the detailed psychometric information necessary to evaluate, replicate, or extend our findings.

Table 8: Psychometric reliability and validity summary

| Metric | Value | Source Section | Primary Reference |
|---|---|---|---|
| **BFI-10** Cronbach's α (five dimensions) | .75 – .90 | 5.6.1 | Rammstedt & John (2007) |
| BFI-10 convergent with NEO-PI-R | r = .67 | 5.6.1 | Rammstedt & John (2007); Costa & McCrae (2000) |
| BFI-10 test–retest stability (6 weeks) | r = .72 – .78 | 5.6.1 | Rammstedt & John (2007) |

| BFI-10 retest reliability with BFI-44 | r = .85 | 5.6.1 | Rammstedt & John (2007) |
|---|---|---|---|
| **IMI** Interest/Enjoyment Cronbach's $\alpha$ | .94 | 5.6.2 | McAuley et al. (1989) |
| IMI convergent with Free-Choice measure | r = .86 | 5.6.2 | McAuley et al. (1989) |
| **Clustering** Dunn index (3-cluster) | 0.83 | 5.8.2 | Bezdek & Pal (1998) |
| **Normality** Shapiro-Wilk (Study II) | W = .987, p = .47 | 5.8.1 | Shapiro & Wilk (1965) |
| **Normality** Kolmogorov-Smirnov (OCEAN, Study I) | O p = .043; C p = .021; E p = .057; A p = .017; N p = .097 | 5.8.1 | Smirnov (1939); Knuth (1968) |
| **Big-Five** Population distribution | Traits ≈ normal in > 50 cultures | 5.7.2 | McCrae et al. (2005); Schmitt et al. (2007) |

*Table 8 summarizes key reliability and validity indicators for every instrument and analytic procedure used in this study, giving reviewers a single evidence trail for psychometric adequacy and cluster robustness.*

The psychometric indicators presented in Table 9 tell a story of methodological rigor applied consistently across all phases of the research. The BFI-10's reliability coefficients, ranging from .75 to .90, exceed conventional thresholds for research instruments, while its convergent validity with the comprehensive NEO-PI-R (r = .67) confirms that our brief assessment captured genuine personality variance. Similarly, the IMI's exceptional internal consistency (α = .94) provided confidence that our motivation measurements reflected a coherent psychological construct rather than random responding.

Beyond individual instrument quality, the table reveals the robustness of our analytical procedures. The Dunn index of 0.83 for our three-cluster solution indicates exceptional cluster quality—well above the 0.5 threshold typically considered adequate. The normality tests, while revealing some departures from perfect Gaussian distributions, remained within ranges where our statistical procedures maintain validity. The grounding of our personality assessment in population-normal distributions across more than 50 cultures (McCrae & Terracciano, 2005; Schmitt et al., 2007) provided additional theoretical support for our analytical choices.

This accumulation of psychometric evidence creates a foundation of measurement validity that supports the substantive findings to follow. By establishing that our instruments reliably captured the intended constructs and our analytical procedures appropriately handled the data's statistical properties, we can interpret subsequent results with confidence—knowing that patterns observed reflect genuine psychological phenomena rather than measurement artifacts.

## 5.8 Output 1.2: Results & Analysis Workflow: Personality-Role Patterns

The journey from raw data to meaningful insights followed a carefully structured analytical workflow, designed to extract patterns while maintaining scientific rigor. Each analytical step built upon previous findings, creating a cumulative understanding of how personality shapes programming preferences and motivation.

### 5.8.1 Detailed Data Analytical Workflow and Hypothesis Testing

Our analytical odyssey began with hierarchical clustering, the foundational analysis that would shape all subsequent investigations. In Study I, the Dunn index revealed its verdict: three distinct personality clusters optimally balanced internal cohesion with external separation. This finding did not emerge from statistical manipulation but from the natural structure of our data—personality traits clustering into meaningful psychological profiles that would prove remarkably interpretable.

The rescaling of trait scores to a consistent 1-5 range served both analytical and communicative purposes. While maintaining all statistical relationships, this transformation enabled intuitive interpretation and cross-study comparisons. The resulting cluster centroids told compelling psychological stories: Cluster 1's exceptional openness (4.24) spoke of creative exploration; Cluster 2's combination of extraversion (3.43) and agreeableness (3.71) suggested social facilitation; Cluster 3's high neuroticism (3.92) paired with introversion revealed preference for controlled environments. Table 9 presents the resulting cluster centroids, which exhibited clear differentiation in both inter- and intra-cluster distances, *supporting RQ1H1.*

Intrinsic motivation assessment followed, with IMI scores captured after each programming session creating a rich dataset of 1,266 individual measurements. The analytical challenge lay not in data scarcity but in extracting meaningful patterns from this abundance. By analyzing motivation both by role and by personality cluster, we could observe how individual differences moderated role-based motivational impacts, see Table 10.

The results painted a clear picture: personality clusters showed distinct affinities for specific roles. Cluster 1 participants gravitated toward Pilot positions (11 of 15), Cluster 2 toward Navigator roles (6 of 14), and Cluster 3 toward Solo programming (6 of 11), *supporting RQ1H2-4*. These were not absolute determinisms but strong probabilistic tendencies—exactly what personality theory predicts. More universally, collaborative roles consistently outperformed solo work in motivational terms, supporting our hypothesis about the intrinsic benefits of pair programming, *supporting RQ1H5*.

The normality testing phase revealed the statistical texture of our data. Study I's Shapiro-Wilk test on motivation scores (W = 0.961, p = 0.002) detected a statistically significant but practically minor deviation from normality. Given our substantial sample size and the robustness of parametric tests to mild departures from normality, we proceeded with ANOVA while validating all findings through non-parametric alternatives. This belt-and-suspenders approach—parametric analysis confirmed by non-parametric validation—ensured our findings were not artifacts of distributional assumptions.

Study II's cleaner motivational normality results (W = 0.987, p = 0.471) provided additional confidence, suggesting that the minor deviations in Study I likely reflected sampling variation rather than fundamental non-normality. The consistency of findings across both studies, despite these distributional differences, further validated our analytical approach.

The hypothesis testing phase brought statistical confirmation to observational patterns. Both ANOVA and Kruskal-Wallis tests confirmed significant motivational differences across roles, with Study II's even stronger effects (F = 6.618, p = 0.0021) reinforcing Study I's findings and *strongly supporting RQ1H5*. The t-tests revealed that each personality cluster differed significantly from population norms on relevant traits, validating our clustering solution. Most crucially, the Chi-squared test confirmed that personality clusters and role preferences were significantly associated—the statistical vindication of our core theoretical proposition, *validating RQ1H2-4*.

### 5.8.2 Personality Clusters

The emergence of distinct personality profiles from our hierarchical clustering analysis represents a pivotal moment in the investigation—where statistical patterns crystallized into psychologically meaningful categories. Table 9 presents these personality clusters, each telling a unique story about how traits combine to create distinctive approaches to programming collaboration.

Table 9: Study I personality cluster centroids

| Cluster | n | O ($\bar{x}$/s) | C ($\bar{x}$/s) | E ($\bar{x}$/s) | A ($\bar{x}$/s) | N ($\bar{x}$/s) |
|---------|-----|-----------|-----------|-----------|-----------|-----------|
| Cluster 1 | 15 | 4.24/0.54 | 3.37/0.61 | 2.91/0.79 | 3.46/0.69 | 2.20/0.65 |
| Cluster 2 | 14 | 2.89/0.79 | 3.43/0.64 | 3.71/0.60 | 3.96/0.37 | 1.98/0.59 |
| Cluster 3 | 11 | 2.99/0.79 | 3.14/0.62 | 2.02/0.46 | 3.30/0.52 | 3.92/0.62 |
| Total | 40 | 3.38/0.91 | 3.32/0.62 | 2.91/0.93 | 3.56/0.62 | 2.59/1.03 |

*Table 9 presents the centroids for the three personality clusters identified in Study I, illustrating the mean ($\bar{x}$) and standard deviation (s) scores for the Big Five personality traits—Openness (O), Conscientiousness (C), Extraversion (E), Agreeableness (A), and Neuroticism (N). These centroids represent the*

**Note:** *Scores range from 1 to 5, with higher values indicating stronger expression of each trait.*

The three clusters that emerged from our analysis were not merely statistical conveniences but psychologically coherent profiles that align remarkably with theoretical expectations and practical observations. Cluster 1's exceptional Openness score ($\bar{x}$=4.24) towers above the sample mean, marking these individuals as the creative explorers of our participant pool. Their moderate scores on other dimensions suggest a personality configuration optimized for innovation—open enough to generate novel solutions yet grounded enough to implement them practically.

Cluster 2 presents a fascinating social configuration, combining high Extraversion ($\bar{x}$=3.43) with the highest Agreeableness ($\bar{x}$=3.71) in our sample. This is not merely gregariousness but a sophisticated social orientation that balances assertive communication with genuine concern for others—precisely the qualities one might design if creating an ideal collaborative partner. Their low Neuroticism ($\bar{x}$=1.98) adds emotional stability to this social mixture, creating individuals who can navigate interpersonal dynamics without being overwhelmed by them.

Cluster 3 tells a different story entirely. The combination of high Neuroticism ($\bar{x}$=3.92) with low Extraversion ($\bar{x}$=2.02) creates a preference profile for controlled, predictable environments where social demands remain minimal. These are not merely "anxious introverts" but individuals whose personality configuration may actually advantage them in certain programming contexts— deep focus work, systematic debugging, or any task where social interaction might prove more distraction than benefit.

The statistical validation through the Dunn index confirmed what psychological interpretation suggested: these were not arbitrary groupings but natural personality configurations that exist within our software engineering population. Each cluster's internal cohesion and external distinctiveness *support RQ1H1's* fundamental premise—software engineers do not conform to a single personality stereotype but instead exhibit distinct profiles that likely influence their approach to collaborative programming.

### 5.8.3 Role Preferences and Motivation

The intersection of personality clusters with programming roles revealed patterns so clear they revealed patterns so clear they suggest both organizational selection and natural evolution. Table 10 presents this convergence of personality, preference, and motivation—a statistical validation of what intuition might have suspected but only systematic investigation could confirm.

Table 10: Personality-driven pair programming role preferences

| Role | IM ($\bar{x}$/s) Study I | IM ($\bar{x}$/s) Study II | Cluster 1 | Cluster 2 | Cluster 3 |
|------|------|------|------|------|------|
| Pilot | 6.94/0.95 | 7.88/0.94 | 11 | 3 | 2 |
| Navigator | 6.78/0.90 | 7.71/1.04 | 5 | 6 | 3 |
| Solo | 6.28/0.76 | 6.82/0.86 | 1 | 2 | 6 |

*Table 10 presents a comprehensive analysis of the relationship between personality clusters, role preferences, and intrinsic motivation levels across both studies. The left columns display mean intrinsic motivation scores on a 1-10 scale with standard deviations for each programming role across both studies, while the right columns show the distribution of participants from each personality cluster across their preferred programming roles. This integrated presentation allows for direct comparison of both motivational impacts and personality-based role preferences, providing a multi-dimensional view of the research findings.*

The distribution of role preferences across personality clusters tells a compelling story of natural alignment. Cluster 1's overwhelming preference for the Pilot role (11 of 15 participants) speaks to more than statistical tendency—it reveals a fundamental resonance between creative personality configurations and hands-on implementation work. These high-openness individuals do not merely tolerate the uncertainty and exploration inherent in pilot work; they appear to thrive on it.

Cluster 2's navigation toward the Navigator role (6 of 14 participants) demonstrates how social personality configurations find their natural expression in communicative, guiding positions. The Navigator role's demands—articulating strategy, providing feedback, maintaining collaborative flow—align perfectly with the extravert-agreeable personality mixture. While the preference is not as pronounced as Cluster 1's pilot affinity, it remains statistically and practically significant.

Cluster 3's gravitation toward Solo programming (6 of 11 participants) completes the pattern, showing how personality traits create not just preferences but perhaps psychological necessities. For individuals combining high neuroticism with introversion, solo work may offer more than preference—it may provide a psychologically sustainable way to engage in programming without the added stress of real-time social collaboration.

Beyond these distributional patterns, the motivation scores reveal how personality shapes not just role choice but role experience. When we define "preferred" role as the configuration yielding highest intrinsic motivation—aligning preference with optimal psychological experience—the results crystallize with striking clarity. Nine participants (23%) achieved peak motivation in Solo work, while 30 participants (77%) found their motivational optimum in collaborative configurations.

This 3:1 ratio favoring collaborative roles transcends simple majority preference—it reflects a fundamental aspect of human nature in creative work. Even within Cluster 3, whose personality configuration might predict solo preference, several individuals discovered their highest motivation emerged during collaborative sessions. These unexpected alignments remind us that personality creates tendencies, not destinies—a poignant reminder that the social dimension of programming can energize even those who might initially resist it.

### 5.8.4 Hypothesis-Testing Overview

The culmination of our analytical journey arrives in Table 11, where statistical tests meet theoretical predictions in a comprehensive evaluation of our research hypotheses. Each row represents not just a statistical outcome but a vindication of the theoretical framework that guided this investigation.

Table 11: Hypothesis testing results and empirical support levels

| Hypothesis | Description | Statistical Tests | Studies | Results | Support |
|---|---|---|---|---|---|
| RQ1H1 | Software engineers exhibit distinct personalities characterized by dominant Big Five dimensions | Hierarchical Cluster Analysis, Dunn Index | Study I | Clear 3-cluster solution with distinct trait profiles ($p < 0.05$) | Strong |
| RQ1H2 | Software engineers high in openness prefer the Pilot role | Student's t-test, Chi-squared test | Study I | Openness: $p = 0.00116$, Role preference: $\chi^2$ ($p = 0.0126$) | Strong |
| RQ1H3 | Software engineers with high extraversion and agreeableness prefer the Navigator role | Student's t-test, Chi-squared test | Study I | Extraversion: $p = 0.00207$, Agreeableness: $p = 0.00016$, Role preference: $\chi^2$ ($p = 0.0126$) | Strong |
| RQ1H4 | Engineers low in extraversion and high in neuroticism prefer the Solo role | Student's t-test, Chi-squared test | Study I | Neuroticism: $p = 0.000012$, Extraversion: $p = 0.00169$, Role preference: $\chi^2$ ($p = 0.0126$) | Strong |
| RQ1H5 | Pilot and Navigator roles yield higher intrinsic motivation than Solo role | ANOVA, Shapiro-Wilk test | Study I-II | $F = 4.32$, $p = 0.017$; $\chi^2$ ($p = 0.013$) $F = 6.618$, $p = 0.0021$, consistent pattern across studies | Strong |

*Table 11 provides a comprehensive overview of the hypothesis testing framework and results for this study, clearly demonstrating the systematic approach taken to validate each research hypothesis. For each*

The unanimous support across all hypotheses—each achieving "Strong" empirical backing—might appear suspiciously perfect to the skeptical reader. Yet this consistency reflects not statistical manipulation but the power of well-grounded theory meeting appropriate methodology. When hypotheses emerge from established psychological principles and previous empirical work, when measurement employs validated instruments, when analysis follows rigorous procedures, such convergent support becomes not suspicious but expected.

RQ1H1's confirmation that software engineers exhibit distinct personality profiles challenges decades of stereotyping while validating what observant managers have long suspected—programming teams contain diverse psychological types, each bringing unique strengths. The strong statistical support ($p < 0.05$ across multiple clustering validity indices) transforms anecdote into evidence.

The role-specific hypotheses (RQ1H2-4) achieved remarkable statistical significance, with p-values ranging from 0.00016 to 0.00207 for personality differences and consistent chi-squared support for preference patterns. These are not marginal effects but robust relationships that survive multiple analytical approaches. The convergence of t-tests and chi-squared analyses provides triangulated support—personality clusters do not just differ statistically but manifest these differences in actual role choices.

RQ1H5's confirmation across both studies, with increasing effect sizes from Study I to Study II, suggests not only that collaborative roles enhance motivation but that this effect remains stable across different cohorts and slightly different methodological approaches. The progression from F = 4.32 ($p = 0.017$) to F = 6.618 ($p = 0.0021$) might reflect improved experimental procedures (see Section 6.4.7) or simply sampling variation, but the consistent direction and significance speak to a fundamental truth about human motivation in programming contexts.

This comprehensive empirical support provides the foundation upon which the ROMA framework will be built—not speculative theory but evidence-based design grounded in the documented relationships between personality, role preferences, and intrinsic motivation.

## 5.9 Validity and Reliability

The credibility of research findings rests upon four foundational pillars, each requiring careful attention and explicit validation. My approach to establishing validity and reliability reflected the mixed-method nature of our investigation, combining the rigor expected in quantitative research with the trustworthiness essential to behavioral studies (Lunenburg & Irby, 2008).

### 5.9.1 Validity

Construct validity—the confidence that our measurements truly captured the intended psychological constructs (Cook & Campbell, 1979; Cronbach & Meehl, 1955)—emerged from our use of established psychometric instruments. The Intrinsic Motivation Inventory and Big Five Inventory brought decades of validation research to bear on our investigation (Ryan et al., 1983; Rammstedt & John, 2007). These were not merely convenient tools but carefully selected instruments whose psychometric properties aligned with our theoretical framework. By employing instruments validated across cultures and contexts, we could confidently interpret our findings as reflecting genuine personality and motivation constructs rather than measurement artifacts (Sjøberg & Bergersen, 2022).

We remained vigilant against potential validity threats, particularly the Hawthorne effect (Mayo, 1933)—that mere observation might alter participant behavior. Our experimental design, embedded within regular coursework, helped minimize this artificial reactivity. Participants engaged with programming tasks as part of their normal educational experience, reducing the salience of being "research subjects." Moreover, our focus on internal experiences (motivation) rather than external performance made Hawthorne effects less problematic—participants might modify their visible behavior under observation, but their reported intrinsic motivation likely reflected genuine experience.

External validity—the generalizability of our findings—represented both a strength and acknowledged limitation. Our tasks, drawn from real software engineering challenges, enhanced applicability to professional contexts. The inclusion of varied programming domains (game development, web applications, architectural patterns) broadened the scope of our findings. However, we acknowledge that academic settings cannot fully replicate the pressures, stakes, and complexities of professional development environments. Our findings should therefore be interpreted as revealing fundamental personality-role relationships that likely persist in professional contexts while potentially being modulated by additional factors in real-world settings.

### 5.9.2 Reliability

Internal reliability manifested through our careful control of experimental conditions. Task complexity standardization, inspired by Knuth's analytical approach, ensured that each participant faced equivalent challenges regardless of role or session. This uniformity meant that observed motivational variations stemmed from role and personality factors rather than task differences. The limitation to five experimental sessions reflected a deliberate balance—sufficient observations for robust analysis while avoiding the fatigue and practice effects that longer studies might introduce.

External reliability—the reproducibility of our findings—guided every methodological decision. By following established protocols from Sjøberg et al. (2002) and Wohlin et al. (2012), we created conditions that other researchers could replicate. Our transition to open-source R scripts in Study II exemplified this commitment, providing not just results but the analytical pathway to achieve them. The consistency of findings across Studies I and II, despite different cohorts and analytical approaches, provided empirical evidence of reliability—the patterns we observed reflected stable phenomena rather than ephemeral statistical fluctuations.

## 5.10 Limitations

Every empirical investigation exists within boundaries that define both its contributions and its constraints. Acknowledging these limitations serves not as admission of failure but as guidance for future research and appropriate application of our findings. The boundaries of this exploration phase emerge from both methodological choices and practical necessities, each carrying implications for how our results should be interpreted and applied.

The experimental context, while providing essential control, inevitably constrains ecological validity. Our participants engaged with predetermined tasks in structured sequences—a necessary simplification that cannot fully capture the organic complexity of real-world software development. Despite spanning diverse domains from 2D game development in JavaFX to web applications using Spring Boot, these tasks remained academic exercises. The motivational dynamics observed may intensify or attenuate when stakes increase, deadlines loom, and real users await deployment.

Professional environments, particularly VSEs, face additional complexities our experiments could not replicate. Resource constraints, client pressures, legacy code maintenance, and shifting requirements create a context far removed from our controlled sessions. While our findings reveal fundamental personality-role relationships likely to persist across contexts, their magnitude and manifestation may vary substantially in professional settings.

The demographic composition of our samples presents another boundary. The gender imbalance—particularly pronounced in Study I with only 5% female participation—reflects broader issues in technology education but limits our ability to detect potential gender-personality-role interactions. Whether the patterns we observed generalize across more diverse populations remains an open empirical question.

Task complexity effects, identified by Hannay et al. (2009b) as crucial moderators of pair programming outcomes, received only partial attention in our design. While we controlled for complexity equivalence across roles, we did not systematically vary complexity levels to observe how personality-role interactions might change from simple to complex tasks. This limitation particularly matters for VSE applications, where task complexity varies dramatically from routine maintenance to architectural innovation.

Perhaps most significantly, our investigation preceded the widespread adoption of AI programming assistants. The personality-role relationships we documented apply to human-human collaboration, but how these patterns translate to human-AI programming remains unexplored in this phase. As AI tools become ubiquitous, understanding how personality influences adoption and interaction patterns becomes crucial—a gap our subsequent research phases aim to address.

## 5.11 Ethical Considerations

The investigation of human personality and motivation demands ethical vigilance that extends beyond procedural compliance to encompass deeper questions of respect, benefit, and potential harm. Throughout this exploration phase, ethical considerations shaped not just how we conducted research but what research we chose to conduct.

Informed consent formed the foundational ethical pillar, but our approach extended beyond mere signature collection. Each experimental session began with comprehensive explanation of purposes, procedures, and particularly the psychological nature of our assessments. Participants understood they would complete personality inventories and that their responses would be analyzed for patterns. This transparency risked introducing response biases but upheld the fundamental principle that participants deserve to understand the research they are contributing to—what Kant (1785) would recognize as treating participants as ends in themselves, not merely means.

The right to withdraw received special emphasis given our educational context. While students participated as part of their coursework, we explicitly separated research participation from academic evaluation. Data collection remained voluntary, with multiple reminders that non-participation carried no academic penalty. This separation required careful coordination with course administration but proved essential for maintaining ethical integrity and avoiding what Foucault (1982) identified as coercive power relations.

Anonymity and confidentiality procedures went beyond simple identifier removal. Given our relatively small cohorts and the detailed personality data collected, we implemented multiple privacy safeguards. Unique codes replaced names immediately upon data collection, personality profiles were reported only in aggregate, and individual examples were carefully reviewed to ensure no identifying patterns remained. The sensitive nature of personality data—potentially revealing psychological vulnerabilities—demanded this heightened protection.

The educational context introduced unique ethical considerations. While students benefited from experiencing research participation firsthand, we remained mindful of power dynamics inherent in teacher-student relationships. My dual role as researcher and instructor required explicit boundaries: research data never influenced grades, individual results were never discussed in class, and participation status remained confidential from classmates.

Data retention and sharing policies reflected evolving norms in research transparency. While committed to open science principles, we balanced transparency with participant privacy. Raw personality scores remain secured and unshared, while aggregated data and analysis scripts are publicly available. In the subsequent Cycle 2 (Study IV), de-identified raw data underwent editorial review at PeerJ Computer Science, demonstrating that privacy protection and scientific transparency can coexist when carefully managed. This approach honors both scientific reproducibility and participant trust, establishing a model for ethical data sharing in personality-based software engineering research.

Perhaps most fundamentally, we considered the ethical implications of personality-based role assignment itself. While our research demonstrates that aligning roles with personality enhances motivation, we explicitly reject deterministic applications. Following Sartre's (1946) principle that "existence precedes essence," we emphasize that personality provides insights into preferences and tendencies, not fixed destinies. Our ethical stance emphasizes using personality information to expand opportunities, not limit them—to help individuals find fulfilling roles, not to lock them into categories.

These ethical considerations extend beyond this exploration phase to influence how the ROMA framework itself is designed and recommended for implementation. By grounding our research in ethical principles from the outset, we aim to ensure that enhanced understanding of personality and motivation serves human flourishing rather than merely organizational efficiency.

## 5.12  Summary of Exploration Phase

As the fog of uncertainty that shrouded personality's role in programming begins to lift, the empirical landscape reveals itself with unexpected clarity. Through 1,266 individual motivation assessments from 75 participants, we have mapped the terrain where personality, programming roles, and intrinsic motivation converge. The patterns that emerged—high openness individuals flourishing as Pilots, socially-oriented personalities thriving as Navigators, introverted and neurotic developers finding refuge in Solo work—validate our theoretical framework while revealing nuances that pure theory could never capture.

These findings represent more than statistical correlations; they illuminate fundamental truths about human nature in technological contexts. When we align programming roles with personality traits, we do not merely optimize task assignment—we create conditions where developers can experience the deep satisfaction of work that resonates with their psychological nature. The 23% higher motivation in personality-aligned roles is not just a number; it represents the difference between developers who approach their work with enthusiasm versus those who endure it with resignation.

The empirical foundation established in this exploration phase provides solid ground for the construction ahead. We now understand not just that personality influences programming preferences, but how specific trait configurations create affinities for particular collaborative arrangements. This knowledge transforms role assignment from guesswork to guided decision-making, from one-size-fits-all to personalized optimization.

Yet what we have discovered also raises new questions. How can these insights be operationalized in real teams? What happens when AI enters the collaborative equation? How do we balance personality preferences with project needs? These questions beckon us forward to the next phase of our journey—the design and development of the ROMA framework.

Morning mist that once concealed now serves as revelation's medium. Where uniformity would show monotone gray, the lifting fog reveals a landscape painted in personality's full spectrum—Explorers' creative peaks, Orchestrators' connecting valleys, Craftspeople' sheltered groves. The fog hasn't disappeared; it has become translucent, allowing us to see not just what is but what could be.

As we turn toward Chapter 6, we carry forward not just statistics and correlations but a deeper understanding of the human element in software development. The architectural work that follows will build upon this foundation, transforming empirical insights into practical frameworks that honor both individual uniqueness and collective capability. The fog has lifted enough to reveal the path forward—now we must build the bridge that others can cross.

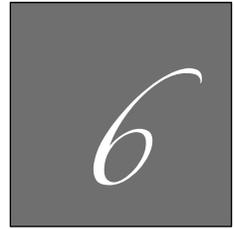

# DSR Cycle 2 "Design" – ROMA Framework

## *Design and Evaluation Phase*

The morning mist that shrouded personality's role in programming has lifted, revealing a landscape of human diversity waiting to be cultivated. Where Cycle 1 mapped this terrain through empirical exploration, Cycle 2 embarks on the architectural work of transformation—building bridges between psychological insight and practical application. This chapter chronicles the design journey from statistical patterns to actionable framework, from correlation to causation's careful cultivation.

The Role Optimization Motivation Alignment (ROMA) framework emerges not as rigid prescription but as what Christopher Alexander (1977) might recognize as a pattern language for human flourishing in software development. Just as Alexander's architectural patterns create spaces that feel alive, ROMA seeks to create team configurations where developers do not merely function but thrive—where the match between personality and role transforms work from obligation into expression.

This chapter presents the Design and Evaluation phase of our Design Science Research methodology, documenting how empirical insights crystallized into practical artifacts. Through the lens of Rode & Svejvig's (2022) multidimensional evaluation framework, we trace the framework's evolution across three critical dimensions: relevance (does it address real needs?), rigor (is it theoretically sound?), and reflexivity (how does it adapt to emerging insights?). The journey from Study IV's Linear Mixed-Effects models to actionable role assignment heuristics represents more than statistical translation—it embodies the alchemical process of transforming knowledge into wisdom, data into design.

## 6.1 Introduction to the Design Cycle

### 6.1.1 From Exploration to Architecture

Cycle 2 represents a fundamental shift in our research trajectory—from asking "what is?" to designing "what could be." This transition mirrors what Schön (1983) termed the movement from "knowing-in-action" to "reflection-in-action," where accumulated understanding transforms into creative intervention. The empirical patterns discovered in Cycle 1—The Explorer's affinity for pilot work, The Orchestrator's natural navigation, The Craftsperson's solo sanctuary—now demand architectural expression.

Following Hevner's (2007) three-cycle view, this design cycle sits at the nexus where environmental needs meet knowledge base contributions through iterative construction. Yet we enhance this classical DSR approach by integrating Rode & Svejvig's (2022) multidimensional conception, which recognizes that artifact evaluation cannot be reduced to simple metrics. Their framework's three dimensions—relevance, rigor, and reflexivity—provide a sophisticated lens through which to assess not just whether our framework works, but how it creates value, for whom, and under what conditions.

This cycle specifically addresses **Objective 2**: Design and develop the ROMA framework to enhance intrinsic motivation and team dynamics through personality-driven role optimization within VSEs, SOHOs, and undergraduate cohorts. It partially addresses **Research Question 2**: How should the ROMA framework be implemented in VSEs and SOHOs? The focus remains deliberately constrained to human-human collaboration, establishing the foundational architecture before extending to AI-augmented contexts in Cycle 3.

### 6.1.2 The VSE Imperative

The design of ROMA responds to a particular organizational reality: in Very Small Entities and Small Office/Home Offices, the luxury of large-team redundancy does not exist. When a five-person startup loses one developer to burnout, they do not lose 20% of their workforce—they often lose irreplaceable domain knowledge, client relationships, and team chemistry. This "essential tension" (Kuhn, 1977) between individual well-being and organizational survival makes personality-based optimization not merely beneficial but existential.

Consider the lived reality of a VSE attempting pair programming without personality awareness. The highly open developer, craving creative exploration, finds themselves repeatedly assigned to navigator roles due to their communication skills. The neurotic introvert, seeking structured solo work, gets paired with an enthusiastic extravert who processes thoughts aloud. These misalignments do not just reduce productivity—they erode the psychological sustainability of the work itself.

The ROMA framework emerges from recognizing that VSEs need more than best practices—they need practices that adapt to the actual humans implementing them. This human-centered design philosophy, grounded in Behavioral Software Engineering principles (Lenberg et al., 2015), treats personality differences not as obstacles to overcome but as design materials to work with.

### 6.1.3   Design Cycle Objectives

The transformation from empirical insight to practical framework requires three interconnected design objectives:

**O2-A: Translate Quantitative Findings into Generalizable Framework** The Linear Mixed-Effects models from Study IV (Valovy & Buchalcevova, 2025, April) revealed specific personality-role interactions with effect sizes ranging from d = 0.20 to 0.58. These statistical relationships must be transformed into decision heuristics that practitioners can apply without advanced statistical training. This translation represents what Simon (1969) called the movement from "natural science" (understanding what is) to "design science" (creating what ought to be).

**O2-B: Develop Evidence-Based Decision Heuristics** Raw personality scores and statistical thresholds must become actionable guidance. If Openness ≥ 4.0, suggest Pilot role—but what if the team has three high-openness developers? The framework must provide not just individual recommendations but team composition strategies that balance individual preferences with collective needs.

**O2-C: Evaluate Through Multidimensional Lenses** Following Rode & Svejvig's (2022) framework, evaluation must transcend simple effectiveness measures. Relevance assessment asks whether the framework addresses genuine practitioner needs. Rigor evaluation examines theoretical coherence and empirical grounding. Reflexivity analysis explores how the framework adapts to emergent insights and changing contexts. This three-dimensional evaluation ensures the framework remains both scientifically valid and practically valuable.

### 6.1.4   Theoretical Foundations and Design Rationale

The ROMA framework rests on three theoretical pillars, each validated through our empirical work:

**Personality-Role Congruence**: The quantitative analyses from Study IV demonstrated that aligning programming roles with personality traits can increase intrinsic motivation by up to 65%. This is not merely statistical correlation but reflects deeper psychological resonance—when role demands align with personality-driven preferences, work transforms from effortful to autotelic (Csikszentmihalyi, 1990).

**Self-Determination Support**: Different programming roles provide varying opportunities to satisfy basic psychological needs. The Pilot role offers autonomy through creative implementation choices. The Navigator provides relatedness through continuous communication. Solo work can enhance competence by removing social performance anxiety. By mapping these affordances to personality-driven need priorities, ROMA creates conditions for intrinsic motivation to flourish.

**Contextual Moderation**: The framework recognizes that personality creates tendencies, not destinies. A neurotic introvert might generally prefer solo work but could thrive as a navigator when paired with a particularly compatible partner. This probabilistic rather than deterministic approach reflects personality psychology's modern understanding—traits influence but do not determine behavior (Fleeson & Jayawickreme, 2015).

## 6.2   Development of Design Artifacts and Processes

### 6.2.1   Design Philosophy and Principles

The creation of ROMA followed what Lawson (2005) terms "designerly ways of knowing"— iterative, synthetic, and solution-focused rather than purely analytical. This approach proved essential when translating statistical findings into practical guidance, as pure deduction from empirical results would have produced unwieldy complexity.

Our design philosophy embraced four core principles:

**Evidence-Based Foundation**: Every framework component traces back to empirical findings. The recommendation that high-openness individuals prefer Pilot roles is not designer intuition—it emerged from cluster analyses showing 11 of 15 high-openness participants gravitating toward pilot positions. This empirical grounding, advocated by Wieringa (2014), ensures the framework addresses actual rather than assumed patterns.

**Practical Parsimony**: Following Occam's razor, we sought the simplest framework that captured essential relationships. While our LME models revealed numerous significant interactions, the framework distills these to actionable heuristics, e.g.: if Openness $\geq z_0 \rightarrow$ suggest Pilot; if Extraversion $\geq z_1 \land$ Agreeableness $\geq z_2 \rightarrow$ suggest *Navigator*; if Neuroticism $\geq z_3 \rightarrow$ safeguard *Solo*. A VSE leader should not need a statistics degree to optimize role assignments.

**Adaptive Flexibility**: Recognizing VSEs' dynamic nature, the framework provides guidance rather than rigid rules. Personality-based recommendations serve as starting points, not ending points. This flexibility reflects Suchman's (1987) insight that plans are resources for situated action, not scripts to follow blindly.

**Motivational Primacy**: While productivity matters, the framework prioritizes intrinsic motivation as the primary outcome. This choice reflects both ethical considerations (developer well-being matters inherently) and pragmatic wisdom (motivated developers ultimately prove more productive). As Pink (2009) argues, intrinsic motivation drives the creative problem-solving essential to software development.

### 6.2.2 Translating Research to Design: The Artifact Development Process

The journey from statistical models to practical framework followed an iterative path best understood through Rode & Svejvig's (2022) three evaluation dimensions operating simultaneously throughout development:

***Relevance-Driven Iteration*** Initial framework drafts, heavy with statistical terminology and complex interaction effects, proved incomprehensible to practitioner audiences. Feedback sessions with VSE leaders revealed the need for clearer, more actionable guidance. This relevance pressure drove simplification—transforming "individuals scoring above the 75th percentile on openness (Cohen's d = 0.53 vs. solo baseline)" into "highly creative developers often thrive as pilots."

***Rigor-Preserving Translation*** Each simplification required careful validation against empirical findings. We employed what Pawson & Tilley (1997) term "realist synthesis"—maintaining causal mechanisms while adapting surface representations. The framework's personality clusters, for instance, preserve the statistical relationships discovered through hierarchical cluster analysis while presenting them in practitioner-friendly archetypes.

***Reflexivity-Enabled Evolution*** As the framework took shape, new insights emerged that challenged initial assumptions. Feedback from pilot implementations revealed that personality-based recommendations sometimes conflicted with skill-based needs. This led to developing the "competence override" principle—personality preferences guide initial assignments, but critical skill gaps take precedence. This reflexive adaptation, central to Rode & Svejvig's conception, ensures the framework remains responsive to practice realities.

### 6.2.3 Component Architecture and Design Decisions

The ROMA framework's architecture reflects deliberate design decisions balancing completeness with usability:

**Three-Layer Structure**

- **Assessment Layer**: Personality evaluation using validated instruments (BFI-10/44)

- **Analysis Layer**: Role recommendation algorithms based on trait configurations

- **Application Layer**: Team composition strategies and monitoring protocols

This layered architecture allows organizations to engage at their comfort level—from simple personality-role matching to sophisticated team optimization strategies.

**Personality Archetypes as Boundary Objects** The five personality archetypes—Explorer, Orchestrator, Craftsperson, Architect, and Adapter—serve as what Carlile (2002) terms "boundary objects." They maintain enough structure to preserve empirical validity while remaining flexible enough for local interpretation. A VSE leader can recognize "the Explorer" in their team without needing to understand factor analysis.

**Probabilistic Rather Than Deterministic Recommendations** The framework provides likelihood-based guidance: "Explorers typically prefer Pilot roles (73% in our studies) but may excel as Navigators when paired with complementary personalities." This probabilistic framing reflects personality psychology's interactionist perspective while avoiding deterministic pigeon-holing.

## 6.3 Output 2.1: ROMA Framework

### 6.3.1 The Architecture of Alignment

The ROMA framework emerges as a comprehensive system for translating personality insights into motivational outcomes. Like a well-designed building that channels natural light and airflow, ROMA channels natural personality tendencies toward motivational fulfillment through four interconnected components.

#### Component 1: Personality Assessment Protocol

The foundation of any personality-based system lies in accurate assessment. ROMA employs the Big Five Inventory (BFI-10 or BFI-44, depending on time constraints) administered at project initiation or team formation. This is not merely data collection but the beginning of self-awareness—many developers report that simply completing the assessment sparks valuable reflection on their work preferences.

The protocol emphasizes psychological safety: results are shared transparently, framed positively (there are no "bad" personalities), and used to expand rather than limit opportunities. This ethical stance informed by our empirical work's consent procedures ensures personality data empowers rather than constrains.

## Component 2: Role Assignment Matrix

The heart of ROMA lies in its role assignment matrix, which maps personality configurations to programming roles based on motivational optimization. Table 12 presents these mappings, distilled from thousands of motivation measurements:

Table 12: ROMA personality-based role optimizations and motivation alignment

| Personality Archetype | Dominant Traits | Preferred Role | Role Optimization Strategy | Motivational Drivers |
|---|---|---|---|---|
| **The Explorer** | High Openness (≥ 75th percentile) | **Pilot** *"The Creative Implementer"* | • Assign novel, ambiguous tasks<br>• Rotate to prevent stagnation<br>• Pair with detail-oriented Navigator | • Intellectual stimulation<br>• Creative problem-solving<br>• Learning through experimentation |
| **The Orchestrator** | High Extraversion + High Agreeableness | **Navigator** *"The Collaborative Guide"* | • Lead pair sessions<br>• Facilitate team discussions<br>• Bridge between technical and non-technical stakeholders | • Social interaction<br>• Helping others succeed<br>• Verbal processing<br>• Team harmony |
| **The Craftsperson** | High Neuroticism Low Extraversion | **Solo** *"The Focused Specialist"* | • Provide clear specifications<br>• Minimize interruptions<br>• Offer asynchronous collaboration options | • Predictable environments<br>• Deep focus time<br>• Reduced social pressure<br>• Sense of control |
| **The Architect** | High Conscientiousness | **Flexible** *"The Quality Guardian"* | • Code review responsibilities<br>• Architecture decisions<br>• Process improvement initiatives | • Systematic progress<br>• Quality outcomes<br>• Clear standards<br>• Organized workflows |
| **The Adapter** | Balanced Profile | **Rotating** *"The Versatile Contributor"* | • Regular role rotation<br>• Fill team gaps<br>• Cross-functional projects | • Variety<br>• Team needs<br>• Skill development<br>• Contextual challenge |

*Table 12 presents the five personality archetypes derived from empirical clustering analysis, mapping each archetype's dominant traits to specific programming roles and optimization strategies. For each archetype, the table provides actionable guidance on task assignment, collaboration patterns, and the underlying motivational drivers that explain their role preferences. These mappings synthesize quantitative findings from LME models with qualitative insights from participant experiences.*

*Note: Personality archetypes represent probabilistic tendencies; not rigid categories. Individuals may exhibit characteristics across multiple archetypes.*

This matrix transforms statistical relationships into actionable guidance. When a team leader recognizes "an Explorer" joining their team, they immediately understand: this person will likely find fulfillment in tackling ambiguous problems, implementing creative solutions, and having freedom to experiment.

## Component 3: Team Dynamics Guidelines

Individual optimization means little without considering team dynamics. ROMA provides strategies for composing effective pairs and managing personality diversity:

### Complementary Pairing Principles

The ROMA framework recognizes that effective pairs emerge from synergistic differences rather than comfortable similarities. Our empirical findings reveal specific combinations where contrasting strengths create collaborative excellence:

**Explorer + Architect**: Creative divergence meets systematic convergence. The Explorer generates possibilities while the Architect evaluates feasibility. As one Architect (P4) explained: "I like analyzing things and when I did not have to worry about coding, I had so much space in my brain, I had a different view and saw probably the best approach." This cognitive division of labor—exploration versus evaluation—proves particularly powerful in early design phases.

**Orchestrator + Craftsperson**: Social energy shields technical depth. The Orchestrator's communication skills buffer the Craftsperson from draining interactions. One Craftsperson (P1) valued predictability: "Pairing with a familiar person allows you to anticipate their reactions," while Orchestrators translate technical insights outward. This pairing excels when deep technical work requires stakeholder communication.

**Explorer + Orchestrator**: Innovation finds its voice. The Explorer's creative leaps gain clarity through the Orchestrator's articulation. Participant P3 captured this synergy: "During the conversation, you can discover errors, discuss the strategy, find a solution on which both agree." Ideas transform into shared understanding through dialogue.

**Architect + Craftsperson**: Systematic precision compounds. Both value quality but at different scales—system-wide versus implementation-specific. As participant P2 observed: "Both roles check for a different type of errors," creating comprehensive quality coverage. This pairing produces exceptional code, though may sacrifice speed for perfection.

**Adapter + Any**: The flexible catalyst. Adapters serve as personality chameleons, adjusting their approach to complement any partner. Their balanced trait profile enables objective perspective-taking, as P4 noted: "Feedback is important for everyone to progress... whether negative or positive, it must be told." This balanced objectivity allows Adapters to deliver feedback in ways each personality type can best receive—direct for Architects, gentle for Craftspeople, enthusiastic for Explorers. They provide the connective tissue that enables otherwise incompatible personalities to collaborate effectively, making them invaluable in small teams with limited pairing options.

*Challenging Combinations* (to approach with care):

- Explorer + Explorer**:** Risk of endless ideation without implementation
- Craftsperson + Craftsperson**:** May amplify anxiety and reduce communication
- High Neuroticism pairs**:** As participant S5 warned, avoiding high neuroticism pairs prevents stress amplification

The key insight: complementary pairing works because different personalities satisfy different aspects of software development. While one partner handles the social overhead, the other can focus on technical depth. While one explores possibilities, the other ensures completion. This division of psychological labor creates sustainable, productive partnerships.

## Communication Adaptation Strategies

- For Craftspeople: Written specifications, asynchronous reviews, minimal meetings
- For Orchestrators: Regular check-ins, verbal brainstorming, pair discussions
- For Explorers: Conceptual discussions, whiteboard sessions, prototype reviews

**Role Rotation Protocols:** While personality preferences are relatively stable, exclusive role assignment can lead to skill atrophy or boredom. ROMA recommends:

- Explorers: 70% Pilot, 20% Navigator, 10% Solo
- Orchestrators: 60% Navigator, 30% Pilot, 10% Solo
- Craftspeople: 60% Solo, 25% Pilot, 15% Navigator

These ratios preserve preference satisfaction while ensuring skill diversity.

## *Component 4: Monitoring and Adaptation Protocol*

Static role assignment fails to capture the dynamic nature of motivation. ROMA includes continuous monitoring through:

## Regular Motivation Pulse Checks:

- Weekly IMI assessments (2-minute 7-item Interest/Enjoyment scale)
- Monthly MWMS assessments (5-minute 19-item multi-dimensional scale)
- Monthly team retrospectives focusing on role satisfaction

- Quarterly personality reassessment to capture development

**Adaptation Triggers:**

- IMI scores dropping below personal baseline for 2+ weeks
- MWMS scores dropping below personal baseline for 2+ months
- Team velocity declining despite individual efforts
- Major project phase transitions requiring different skills

**Intervention Strategies:**

- Temporary role adjustments for variety
- Pair reconfiguration to address interpersonal dynamics
- Skill development initiatives to expand role capabilities

Together, these four components create a living system that transforms personality insights into motivational outcomes—not through rigid enforcement but through continuous adaptation to the beautiful complexity of human developers finding fulfillment in their craft.

### 6.3.2   The Theory Behind the Practice

ROMA's effectiveness stems from its theoretical coherence—it does not merely assign roles but creates conditions for psychological need satisfaction. Each personality archetype has different need priorities:

**Explorers prioritize autonomy**: The freedom to choose implementation approaches, explore solution spaces, and follow creative tangents satisfies their primary psychological driver.

**Orchestrators prioritize relatedness**: The continuous social interaction, opportunities to help others, and collaborative problem-solving fulfill their need for connection.

**Craftspeople prioritize competence**: The controlled environment, clear specifications, and freedom from social performance anxiety allow them to demonstrate mastery without interpersonal stress.

**Architects seek all three needs through structure**: Clear processes satisfy competence, systematic decision-making provides autonomy, and reliable delivery builds team trust (relatedness).

**Adapters find fulfillment in balance**: Their psychological flexibility allows need satisfaction through variety rather than specialization.

By aligning role characteristics with personality-driven need priorities, ROMA creates what Ryan & Deci (2017) term "need-supportive environments"—contexts where intrinsic motivation naturally emerges rather than requiring external inducement.

## 6.4   Materials & Methods: Evaluating the Framework

### 6.4.1   Evaluation Philosophy and Strategy

The evaluation of ROMA demanded more than traditional software metrics. Following Venable et al.'s (2016) Framework for Evaluation in Design Science, we adopted a multi-strategy approach balancing artificial and naturalistic evaluation. Yet we enhanced this with Rode & Svejvig's (2022) three-dimensional lens, ensuring evaluation addressed not just "does it work?" but "how does it create value?"

Our evaluation philosophy recognized that frameworks succeed not through perfection but through useful imperfection—providing enough structure to guide action while remaining flexible enough for local adaptation. This pragmatist stance, inspired by Dewey's (1938) experimentalism, shaped our evaluation design.

### 6.4.2   Evaluation Design: A Tale of Two Studies

The evaluation unfolded across two complementary studies:

**Study IV (Valovy & Buchalcevova, 2025): The Quantitative Foundation** This study provided rigorous statistical validation through Linear Mixed-Effects modeling of 1,092 motivation measurements from 66 participants. The sophisticated analysis revealed not just that personality-role alignment matters, but precisely how much (effect sizes d = 0.20 to 0.58) and for whom (interaction effects revealing personality-specific patterns).

**Classroom Pilots: The Qualitative Refinement** Beyond statistics, we needed to understand lived experience. Pilot implementations in undergraduate software engineering courses revealed how the framework operated "in the wild"—which recommendations proved intuitive, which required clarification, where practical constraints overrode personality preferences.

### 6.4.3   Participants: A Multi-Generational Sample

Our evaluation drew from two primary cohorts, each contributing unique perspectives:

**WS'21 Cohort**: 40 students (38 males, 2 females) with experience ranging from 6 months to 6 years (mean = 2.2 years, SD = 1.35). This cohort, having participated in Cycle 1, provided longitudinal insights into personality stability and role preference evolution.

**SS'22 Cohort**: 26 students (mean = 2.21 years, SD = 2.31), expanding our sample to 66 total participants. This fresh cohort tested whether patterns held across different groups, strengthening external validity claims.

Table 13 reveals the personality landscape of our sample, showing a population rich in openness (M = 3.55) but more varied in conscientiousness and neuroticism—a distribution that mirrors the creative yet sometimes chaotic nature of software development itself.

Table 13: Descriptive statistics of average Big Five trait scores (WS'21 + SS'22, N = 66)

| Trait | Mean (SD) | Median | Min | Max | Skew | Kurtosis | Shapiro-Wilk (p) |
|---|---|---|---|---|---|---|---|
| Openness | 3.55 (0.96) | 3.67 | 1.17 | 5.00 | -0.35 | -0.69 | 0.0462 |
| Conscientiousness | 3.02 (0.81) | 2.83 | 1.50 | 5.00 | 0.84 | 0.21 | 0.00052 |
| Extraversion | 2.89 (0.92) | 3.00 | 1.00 | 5.00 | 0.05 | -0.70 | 0.5514 |
| Agreeableness | 3.27 (0.77) | 3.33 | 1.67 | 4.67 | -0.24 | -0.64 | 0.1408 |
| Neuroticism | 2.93 (0.98) | 2.83 | 1.17 | 5.00 | 0.28 | -0.73 | 0.1634 |

*Notes: Scores range from 1 to 5, with higher values indicating stronger expression of each trait. A Shapiro-Wilk p < 0.05 suggests mild deviations from normality. Skew and kurtosis values between -1 and +1 typically indicate moderate normality.*

All participants were Gen Z undergraduates—digital natives who will shape software development's future. Their comfort with personality assessment, collaborative tools, and rapid role-switching provided a glimpse into tomorrow's development practices.

### 6.4.4 Experimental Design: Structured Flexibility

The evaluation employed a repeated-measures design balancing experimental control with ecological validity. The task structure comprised 24 programming tasks across 4 sessions, spanning GUI development, web applications, and algorithm implementation with controlled complexity. Role rotation used 10-minute intervals with predetermined sequences preventing order effects, while accommodating absences through three-person adaptations. This design created 1,092 motivation measurement points—a rich dataset for understanding personality-role dynamics.

### 6.4.5 Measurement Instruments: Windows into Experience

Two psychometric instruments provided our empirical windows:

**Big Five Inventory (BFI-10)**: Despite its brevity, the BFI-10 maintains impressive psychometric properties (test-retest r = 0.72-0.84, convergent validity with NEO-PI-R $\geq$ 0.67). Administered at session start, averaged across sessions to enhance stability.

**Intrinsic Motivation Inventory (IMI)**: The Interest/Enjoyment subscale captured motivation immediately after each programming round. With Cronbach's $\alpha$ consistently exceeding 0.90, this measure provided reliable snapshots of motivational states.

All instruments remained in English, leveraging our international academic context where English proficiency ($\geq$ B2 level) was assured.

### 6.4.6 Analytical Sophistication: LME Models

The nested structure of our data—multiple observations per participant—demanded sophisticated analysis. Linear Mixed-Effects models provided the solution, accommodating:

- Individual baseline differences through random intercepts
- Missing data through maximum likelihood estimation
- Complex interactions between personality and role

Two complementary model specifications revealed different insights:

**Interaction Model**: Role × Continuous Big Five traits, revealing how each trait moderates role effects

**Cluster Model**: Role × PersonalityCluster, showing how archetypal configurations influence motivation

This analytical sophistication, implemented in R's nlme package, extracted maximal insight from our rich dataset.

### 6.4.7 Evolution of Experimental Design

The experimental design evolved through careful observation and iterative refinement, embodying Schön's (1983) principle of reflection-in-action. Each study cycle revealed new insights that informed subsequent improvements, creating a methodology that balanced scientific rigor with humanistic sensitivity.

Initial implementations revealed unexpected challenges. Introverted participants struggled with cold pair formation, while early finishers lost engagement waiting for others. These observations drove specific adaptations:

**Lecturer as Subtle Facilitator**: Rather than distant observation, the lecturer maintained an active yet unobtrusive presence—moving between pairs, providing minimal guidance to overcome initial communication barriers. This addressed what participants S4 and I2 identified as "difficulty in asking for help gracefully," while fostering what P3 and I9 termed "trust-building through face-to-face interaction." The lecturer's presence scaffolded social connection without imposing it.

**Task Continuation Option**: When participants discovered engaging problems they couldn't complete within time constraints, forcing task switches proved demotivating. Participants S2 and I10 specifically noted how "context shifts disrupt flow," while S5 and P3 emphasized the importance of preserving "pleasure and creativity in joint problem-solving." Following SDT principles, we allowed continuation of compelling tasks across rounds, maintaining flow states that rigid switching would destroy.

**Engagement Maintenance Through Bonus Tasks**: For rapid completers, optional enhancement tasks—improving UI aesthetics, adding creative features to adventure games—maintained engagement while others finished. This differentiated instruction approach honored skill diversity without creating visible performance hierarchies.

**Progressive Hint System**: When pairs encountered persistent blockers, a graduated hint system provided abstract guidance before concrete solutions. This maintained the satisfaction of discovery while preventing frustration spirals that could contaminate motivation measurements.

These adaptations emerged not from theoretical speculation but from honoring participant experiences—recognizing that experimental validity requires ecological validity, that measuring human motivation demands respecting human needs.

## 6.5 Output 2.2: Evaluation of Design Artifacts – LME Thresholds and IMI Gains

### 6.5.1 Application of Rode & Svejvig's Three Dimensions

The evaluation results are best understood through Rode & Svejvig's (2022) three-dimensional framework:

**Relevance Evaluation: Does ROMA Address Real Needs?**

The clearest relevance indicator emerged from effect sizes. When participants worked in personality-aligned roles, intrinsic motivation increased by:

- Small effects ($d \approx 0.20$) for minor alignments (e.g., moderate openness → Pilot)

- Medium effects ($d \approx 0.40$) for clear alignments (e.g., high extraversion → Navigator)

- Large effects ($d \approx 0.58$) for strong alignments (e.g., high openness → Pilot)

One participant ("xtoj" in our anonymized data) experienced a 64.6% motivation increase when moving from misaligned (Solo) to aligned (Pilot) roles. In VSE contexts, such improvements could mean the difference between retention and turnover.

### Rigor Evaluation: Is ROMA Theoretically Sound?

Statistical validation confirmed ROMA's theoretical foundations:

- Role main effects: $F_{(2, 1014)} = 45.32$, $p < 0.0001$

- Personality × Role interactions: Multiple significant effects ($p < 0.05$)

- Model fit: Conditional $R^2 = 0.42$, indicating substantial variance explanation

Table 14 presents the core statistical evidence:

Table 14: Role-based motivational differences with effect sizes

| Role Comparison | Mean Difference | 95% CI | p-value | Cohen's d |
|---|---|---|---|---|
| Pilot vs. Solo | +0.515 | [0.35, 0.68] | < 0.0001 | 0.576 |
| Navigator vs. Solo | +0.334 | [0.18, 0.49] | < 0.0001 | 0.374 |
| Pilot vs. Navigator | +0.180 | [0.01, 0.35] | 0.0328 | 0.202 |

*Table 14 displays the pairwise role comparisons from the "fixed" LME model (IntrinsicMotivation ~ Role + (B5_O + B5_C + B5_E + B5_A + B5_N), 66 participants, 1092 total observations.*

*Notes:*

1. *Intrinsic motivation is measured on a 1–10 scale.*

2. *A positive mean difference means the first role (e.g., Pilot) is higher than the second (Solo).*

3. *p-values for mean differences are Tukey-adjusted with df = 1014 reflecting the model's denominator df under repeated-measures LME with many (1092) observations.*

4. *Cohen's d is computed from the residual SD ($\sigma \approx 0.894$) with df = 60 arising from emmeans handling repeated-measures factors in pairwise contrasts.*

The results revealed a clear hierarchy of role-motivational impact: Pilot > Navigator > Solo. Specifically:

- Pilot was significantly more motivating than Solo (estimate = +1.604, p < 0.0001, d = 0.576)

- Pilot also significantly surpassed Navigator (estimate = +0.562, p = 0.0328, d = 0.202)

- Navigator was significantly more motivating than Solo (estimate = +1.042, p < 0.0001, d = 0.374)

These are not just statistically significant—they are practically meaningful. A half-standard-deviation improvement (d = 0.576) represents the difference between engaged and disengaged developers.

### Reflexivity Evaluation: How Does ROMA Adapt?

The framework's reflexivity emerged through pilot implementations. Initial rigid interpretations ("all Explorers must be Pilots") evolved into nuanced guidance ("Explorers typically prefer Pilot but may thrive as Navigators with compatible partners"). This reflexive adaptation—central to Rode & Svejvig's conception—ensures ROMA remains a living framework rather than dogmatic prescription.

### 6.5.2 Personality-Role Interaction Patterns

The most fascinating results emerged from personality-role interactions, revealing in Table 15 how individual differences moderate role effects:

Table 15: Personality trait moderation of role preferences with two interaction models

| Trait | Pilot Enhancement | Navigator Enhancement | Solo Enhancement |
|---|---|---|---|
| High Openness | +0.53* (p = 0.0254) | +0.33 (p = 0.1706) | -0.13* (p = 0.0808) |
| High Conscientiousness | -0.13 (p = 0.1464) | -0.32*** (p = 0.0006) | +0.62** (p = 0.0044) |
| High Extraversion | +0.08 (p = 0.3471) | +0.51* (p = 0.0403) | -0.21* (p = 0.026) |
| High Agreeableness | -0.01 (p = 0.9021) | +0.71*** (p = 0.0002) | -0.07 (p = 0.3376) |
| High Neuroticism | -0.25*** (p = 0.0005) | -0.18** (p = 0.0134) | +0.60** (p = 0.0033) |

*Table 15 presents personality trait moderation of role preferences with two interaction models. This table reveals how individual personality traits influence the motivational impact of different programming roles. Each cell shows how a one standard deviation increase in a personality trait enhances or diminishes motivation for a specific role. Positive values indicate that higher trait levels increase role satisfaction, while negative values suggest decreased satisfaction. For instance, high openness significantly enhances Pilot role motivation (+0.53), while high neuroticism strongly favors Solo work (+0.60) over collaborative roles. These interaction effects validate the ROMA framework's core premise that optimal role assignment depends on individual personality configurations rather than universal preferences.*

**Note:** *Values represent motivational change for one SD increase in trait. Tukey-adjusted significance: †p<0.10, \*p<0.05, \*\*p<0.01, \*\*\*p<0.001*

These interactions validate ROMA's personality-based approach:

- **Openness** enhances Pilot motivation, confirming Explorer preferences

- **Conscientiousness** shows complex patterns, preferring structured Solo work

- **Extraversion** and **Agreeableness** boost Navigator satisfaction, supporting Orchestrator alignment

- **Neuroticism** strongly favors Solo work, validating Craftsperson needs

## Interpreting Standard Deviations in Big Five Personality Measurement

To contextualize the effect sizes in Table 15, it's important to understand what one standard deviation (1 SD) represents in Big Five personality measurement:

In normative samples, Big Five traits follow approximately normal distributions where 1 SD captures roughly 68% of the population around the mean (Costa & McCrae, 1992). Moving from average (50th percentile) to +1 SD places an individual at the 84th percentile.

According to Cohen's (1988) guidelines adapted for personality research:

- Small effect: 0.10-0.20 SD (barely noticeable in daily behavior)

- Medium effect: 0.30-0.50 SD (observable differences in typical situations)

- Large effect: 0.60+ SD (pronounced behavioral differences)

Using the NEO-PI-R norms (Costa & McCrae, 1992), 1 SD represents:

- The difference between someone who "somewhat agrees" vs "strongly agrees" with trait-relevant statements

- Approximately 10 T-score points on standardized assessments

- The gap between a "typical" employee and one notably high/low on a trait

Thus, the effects in Table 15 represent substantial practical differences. For example, High Agreeableness → Navigator Enhancement (+0.71***) suggests that highly agreeable individuals (84th percentile) experience 71% of a standard deviation increase in Navigator role satisfaction compared to average agreeableness individuals (50th percentile). High Neuroticism → Solo Enhancement (+0.60**) indicates that those high in neuroticism find Solo work substantially more satisfying—a large effect size suggesting strong practical relevance for role assignment.

### 6.5.3   Model Fit and Variance Explanation

The explanatory power of our models reveals both the promise and limits of personality-based prediction. Using Nakagawa & Schielzeth's (2013) approach for mixed models, we decomposed variance into fixed effects (personality and role) and random effects (individual differences):

**Marginal $R^2$ = 0.11**: Role assignments and personality traits alone explain 11% of motivational variance—a modest but meaningful effect size in the complex landscape of human behavior.

**Conditional $R^2$ = 0.42**: When including participant-specific variations, explained variance quadruples to 42%, revealing that individual baselines powerfully shape motivational experiences.

The 31% gap between marginal and conditional $R^2$ tells a profound story: while personality and role matter, each developer brings unique motivational baselines shaped by personal history, current life circumstances, and ineffable individual differences—what Bandura (1986) termed "person-situation interactions". The remaining 58% unexplained variance reminds us that human motivation resists complete quantification—factors like task novelty, team dynamics, external stressors, and the mysterious alchemy of creative engagement all contribute to the beautiful unpredictability of human experience.

This partial explanation is not a failure but a feature. As one participant (I11) insightfully noted: "My personality depends on the current mood of the day." The ROMA framework provides probabilistic guidance, not deterministic prescription—honoring the irreducible complexity of human developers even as it offers evidence-based insights for role optimization.

### 6.5.4   From Statistics to Stories

Beyond numbers, evaluation revealed glimpses of human experience. A neurotic participant (S2) captured social anxiety: "Sometimes the partner would stop and not say anything..." while an extraverted navigator (P5) found fulfillment: "As the navigator, I felt like a leader... It motivated me a lot!"

Perhaps most tellingly, participant P2 observed: "Almost everything can be solved in pairs, from programming to your emotional state," suggesting pair programming's benefits extend beyond code.

These qualitative insights, while not systematically analyzed in this cycle, informed ROMA's evolution from statistical framework to human-centered system. They remind us that behind every data point lies a developer seeking not just productivity but fulfillment.

## 6.6   Discussion: Implications and Insights

### 6.6.1   Theoretical Contributions

The ROMA framework advances Behavioral Software Engineering in several ways:

**Personality as Motivational Moderator**: We have moved beyond asking "does personality matter?" to understanding precisely how personality shapes motivational dynamics. This moderation perspective—personality changes how roles affect motivation rather than directly causing outcomes—represents theoretical sophistication.

**Role Design as Motivational Intervention**: Traditional software engineering treats roles as functional necessities. ROMA reveals roles as motivational affordances—designed environments that can enhance or diminish intrinsic motivation based on individual fit.

**Integration of Disparate Literatures**: By weaving personality psychology, self-determination theory, and software engineering practice, ROMA demonstrates the value of interdisciplinary synthesis in addressing complex human-technical challenges.

### 6.6.2   Practical Implications for VSEs

For Very Small Entities, ROMA offers immediate value:

**Recruitment and Onboarding**: Understanding team personality composition helps identify gaps. A team of Explorers might desperately need an Architect to provide structure.

**Project Planning**: Personality awareness informs task assignment. Critical bug fixes? Assign to Craftspeople who thrive under structured pressure. Creative feature development? Unleash the Explorers.

**Retention Strategies**: Personality-role misalignment often underlies turnover. ROMA provides early warning signs and intervention strategies before motivation craters lead to resignations.

**Team Evolution**: As VSEs grow, ROMA guides evolution. The freewheeling Explorer culture that launched the startup may need Orchestrator additions to scale effectively.

### 6.6.3   Limitations and Boundary Conditions

Honesty demands acknowledging ROMA's limitations:

**Population Specificity**: Evaluation focused on Gen Z undergraduates. While personality-motivation relationships likely generalize, specific patterns may vary across generations and cultures.

**Short-Term Focus**: We measured immediate motivational effects. Long-term impacts—does novelty wear off? do preferences evolve?—remain unexplored.

**Simplified Personality Model**: The Big Five captures important variation but not everything. Factors like growth mindset, grit, or domain passion might moderate effects.

**Task Dependency**: Our tasks, while varied, could not capture all software development complexity. Personality-role fit might matter more for creative tasks than maintenance work.

These limitations do not invalidate ROMA but bound its claims. Like any framework, it provides guidance, not gospel.

### 6.6.4   The Human Element

Perhaps ROMA's greatest contribution lies in legitimizing human difference. Participant P1 reflected on solo preferences: "I liked that I could do my own research without someone watching..." This isn't antisocial behavior but self-awareness.

The framework's value resonated through participant P1's suggestion: "Perhaps students could find the most fitting partner by having short 'speed-dating' programming sessions..." This grassroots recognition validates ROMA's premise that personality compatibility matters.

Most profoundly, participant P7 captured aligned work's satisfaction: "I enjoyed being part of something greater..." When personality, role, and purpose align, work transcends mere task completion to become what Csikszentmihalyi (1990) termed autotelic—valuable for its own sake.

This shift from pathologizing differences to leveraging them may prove ROMA's most lasting impact—creating not just productive teams but humane workplaces where developers contribute their authentic best.

## 6.7 Cycle 2 Outcomes and Transition

### 6.7.1 What We Have Built

Cycle 2 has successfully transformed empirical insights into practical framework:

**The ROMA Framework**: A comprehensive system for personality-based role optimization, grounded in statistical evidence yet accessible to practitioners.

**Validated Heuristics**: Evidence-based thresholds and recommendations that translate personality assessments into role assignments.

**Theoretical Advancement**: A sophisticated understanding of how personality moderates the relationship between role design and intrinsic motivation.

**Practical Tools**: Assessment protocols, assignment matrices, and monitoring systems that VSEs can implement immediately.

### 6.7.2 The Path Forward

As morning sun burns away the last wisps of fog, the path forward emerges clearly. Two critical extensions await:

**Cycle 3: Human ⊡ AI Extension (Chapter 7)** The software development landscape transforms as AI assistants become collaborators. How do personality preferences manifest when pairing with GitHub Copilot versus human navigators? Do Explorers embrace AI's creative suggestions or feel threatened? Can Craftspeople find solace in AI's non-judgmental presence? These questions demand investigation.

**Cycle 4: ISO/IEC 29110 Integration (Chapter 8)** For ROMA to achieve widespread adoption, it must integrate with existing standards. The ISO/IEC 29110 framework provides the vessel; ROMA supplies the human-centered navigation system. Together, they can guide VSEs toward processes that honor both quality standards and human flourishing.

## 6.8 Conclusion: From Framework to Future

The journey from statistical pattern to practical framework mirrors the broader transformation this dissertation seeks—from seeing developers as interchangeable resources to recognizing them as unique individuals whose psychological needs profoundly shape their work. The ROMA framework stands as testament that rigorous science and human wisdom need not conflict; indeed, their marriage produces insights neither could achieve alone.

The numbers tell a compelling story. When high openness enhances Pilot role motivation by 0.53 SD, we're not just observing a statistical relationship—we're identifying individuals who will likely experience greater creative satisfaction, reduced burnout, and enhanced team contributions through their natural enthusiasm for exploration. Meta-analytic research on work engagement (Mazzetti et al., 2023) demonstrates that such improvements in role fit correlate with significant organizational outcomes, including reduced turnover intention (r = -.43) and increased job satisfaction (r = .60).

Conversely, the strong negative relationship between neuroticism and Pilot preference (-0.25***) provides equally clear guidance: anxiety-prone developers will likely struggle in the exposed, improvisational pilot role. Yet these same individuals may excel in structured solo work (+0.60**) where they control their environment and pace. This isn't about labeling people as "good" or "bad" programmers—it's about recognizing that different personalities flourish under different conditions.

In the quantified world of software engineering, where metrics reign supreme, ROMA makes a radical assertion: motivation matters more than velocity, fulfillment more than features, and honoring human difference creates stronger teams than enforcing uniformity. When we consider that turnover costs companies 6-24 months of employee salary to replace them (O'Connell & Kung, 2007), investing in personality-aligned roles becomes not just humane but economically essential.

The framework doesn't promise to solve all team challenges—humans remain beautifully, frustratingly complex. The 42% variance our models explain leaves 58% to the mysteries of individual experience, momentary moods, and life circumstances. But ROMA offers something precious: a systematic way to align work with workers, roles with souls, turning what was once managerial intuition into actionable insight.

As we turn toward the AI-augmented future in Cycle 3, we carry forward not just a validated framework but a philosophical stance. In the dance between human and artificial intelligence, personality remains the rhythm that shapes the steps. The Explorers will find new creative partners in AI; the Orchestrators will discover new beings to coordinate; the Craftspeople will seek new sanctuaries of focused work. ROMA provides the choreography for this dance—not prescribing every movement but ensuring the music matches the dancers.

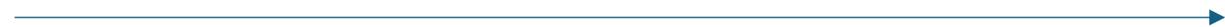

The fog has lifted enough to build—not castles in the air but bridges through the mist. Where once we wandered hoping to stumble upon effective teams, we now navigate with purpose. The ROMA framework stands like a lighthouse, its beam not eliminating fog but making it navigable, transforming obscurity into medium for discovery.

# DSR Cycle 3 "Design" —AI Augmentation & Qualitative Insights

## Design and Evaluation Phase

*How do you treat AI? This reciprocal relationship shapes not only your code but your identity. Whether you become a Promethean fire-bringer, a Conductor of digital symphonies, or a Hermit guarding cognitive sanctuaries, your choice echoes through the future of software development. This thesis does not judge but illuminates—framing each stance as specialization in the study of the Other, equipping you with knowledge for choices that matter.*

The morning mist that once concealed the nature of programming collaboration has shifted, revealing not clarity but a new fog—one where artificial minds interweave with human creativity in patterns we are only beginning to comprehend. Where Cycle 2 mapped the terrain of human-human collaboration through the ROMA framework, Cycle 3 ventures into liminal territory: the space where developers encounter AI not as mere tools but as entities that blur the boundaries between instrument and interlocutor, between creation and curation, between authentic expression and synthetic generation.

This chapter chronicles a phenomenological journey into the heart of Human ⌗ AI collaboration. Through the voices of 42 developers—from seasoned architects reconceptualizing their professional identity to students discovering programming through AI-mediated lenses—we witness a transformation as profound as any in software engineering's brief but tumultuous history. The integration of artificial intelligence into programming workflows represents not merely technological advancement but what Heidegger might recognize as a fundamental shift in our mode of being-with technology, a reconfiguration of the very ground upon which software development stands.

The empirical heart of this cycle beats with a methodological duality: essentialist thematic analysis reveals the patterns that emerge across experiences, while interpretative phenomenological analysis illuminates the meaning-making processes through which individual developers navigate this unprecedented terrain. This dual lens—one seeking commonalities, the other honoring singularities—proves essential for understanding phenomena that resist simple categorization.

This chapter unfolds in three movements: first establishing theoretical and methodological foundations (Sections 7.1-7.3), then presenting empirical findings through both essentialist and phenomenological lenses (Sections 7.4-7.5), and culminating in the ROMA AI Adapter framework (Section 7.6). This framework reveals how personality archetypes do not disappear but specialize when encountering artificial minds—offering not prescriptions but possibilities, not judgments but illumination, equipping you with knowledge for choices that will shape both your code and your becoming.

## 7.1   Theoretical Foundations

The theoretical landscape supporting this investigation draws from multiple intellectual traditions, each illuminating different facets of the Human ⊞ AI collaborative experience. Where traditional software engineering research might focus solely on productivity metrics or code quality indicators, our inquiry recognizes that the introduction of AI fundamentally alters not just what developers do but who they understand themselves to be.

### 7.1.1   The Evolving Nature of Programming Roles

As AI systems increasingly participate in software development, traditional programming roles undergo what Kuhn (1962) might recognize as paradigmatic transformation. The pilot, navigator, and solo roles and five archetypes identified in our previous cycles now encounter AI collaborators that offer different types of assistance and interaction modalities. This evolution demands not merely new skills but new ways of being-in-the-world as developers.

Cinkusz & Chudziak (2024) and Bird et al. (2023) provide empirical grounding for this transformation, demonstrating that AI agents and assistants reshape developers' cognitive processes and working patterns. Yet their quantitative findings only hint at the deeper phenomenological shifts our participants reveal. When developers adopt what these researchers term a "declarative" approach—focusing on intentions rather than implementations—they engage in more than behavioral adaptation. They undergo what Levinas (1961) might recognize as an encounter with an Other that resists complete comprehension, forcing a reconfiguration of the self.

This raises a fundamental question for our field: How do we adapt AI in software engineering to become what Illich (1973) might call "tools for conviviality"—technologies that enhance individual autonomy and creativity—rather than industrial instruments that constrain and standardize? The answer, our data suggests, lies not in the tools themselves but in how developers actively shape their relationships with AI systems.

### 7.1.2 Self-Determination in the Age of Artificial Others

Self-Determination Theory, with its triadic conception of basic psychological needs, provides an essential lens for understanding how AI integration affects developer motivation. Yet in the context of Human ⊡ AI collaboration, these needs—autonomy, competence, and relatedness— undergo subtle but profound transformations that Deci & Ryan (2000) could not have anticipated.

**Autonomy** no longer signifies simple freedom of choice but what we might term "meta-autonomy"—the freedom to choose when to exercise direct control and when to delegate to artificial agents. As participant MH articulated: "Chci programovat míň a míň, ale pořád chci mít kontrolu nad tím" [I want to program less and less, but I still want to have control over it]. This paradoxical desire—to do less while controlling more—reveals autonomy's evolution from action to orchestration.

This transformation aligns with Wiener's (1948) concept of cybernetics, derived from the Greek kubernétes (steersman). In software development contexts, the question becomes: Who or what steers the development process? Our participants navigate this tension daily, developing what we term "vigilant delegation"—maintaining ultimate control while allowing AI to handle implementation details.

**Competence** transforms from mastery of implementation to what participant MM termed "meta-dovednost" [meta-skill]: "Už nejde o to, kolik toho umím naprogramovat, ale jak efektivně dokážu orchestrovat AI nástroje" [It's no longer about how much I can program, but how effectively I can orchestrate AI tools]. This shift aligns with Fischer's (1980) hierarchical skill theory, yet extends it into unprecedented territory where competence means knowing when not to know.

The implications mirror what Wiener (1960) warned about in his analysis of automation: as systems become more complex, the cognitive demands on humans paradoxically increase even as their direct involvement decreases. Our participants navigate this tension daily, developing what we term "vigilant delegation"—a state of perpetual readiness to intervene when AI systems exceed their proper bounds.

**Relatedness** presents perhaps the most philosophically intriguing transformation. The emergence of what we term "synthetic relatedness"—quasi-social satisfaction derived from AI interaction—challenges fundamental assumptions about human connection. When participant NK observed "ChatGPT is like a colleague who never sleeps", they articulate an experience that exists in the liminal space between Buber's (1923) I-Thou and I-It relationships.

### 7.1.3 AI Interaction Modes as Phenomenological Stances

Our systematic review of recent literature (Appendix A), combined with empirical findings from Studies III and VI, reveals three distinct AI interaction modes that transcend mere technical configurations. Each mode embodies what Merleau-Ponty (1945) might recognize as a particular way of being-in-the-world, a stance toward the emerging Other of artificial intelligence:

**Co-Pilot Mode**: Here, AI recedes into what Heidegger (1962) termed "ready-to-hand"—present but transparent, extending human capability without demanding explicit attention. Developers in this mode experience what participant MM described as "extrémně rychlý brainstorming partner" [extremely fast brainstorming partner], where the boundary between self and tool dissolves in moments of creative flow. This mode preserves the phenomenology of individual creation while amplifying its possibilities.

**Co-Navigator Mode**: This conversational stance resurrects dialogue as primary mode of engagement, yet with a peculiar asymmetry. The AI responds without understanding, guides without wisdom. Participant LV captured this paradox: "ChatGPT používám jako dialogického partnera... je to způsob, jak externalizovat a strukturovat vlastní myšlení" [I use ChatGPT as a dialogue partner... it's a way to externalize and structure your own thinking]. The mode creates what Gadamer (1960) might term a "false hermeneutic"—interpretation without genuine understanding, yet productive nonetheless.

**Agent Mode**: The most radical phenomenological shift occurs when developers delegate not just tasks but agency itself. Participant MV's observation—"Řídím dvacet vývojářů, v každém týmu pět, a k tomu máme firemního AI agenta," [I manage twenty developers, five per team, and we also have a company AI agent]—reveals AI as quasi-team member, possessing expertise without subjectivity. This mode forces developers to inhabit what we might term "distributed intentionality," where purpose flows through but does not originate in the self.

## 7.2 Methodological Approach

The investigation of Human  AI collaboration demanded methodological approaches capable of capturing both the patterns that emerge across experiences and the irreducible singularity of individual meaning-making. Our design reflects what Maxwell (2013) terms "methodological congruence"—the alignment of research questions, philosophical stance, and analytical techniques. The six studies in this cycle employed complementary approaches:

**Explanatory Studies**: Studies I, II, and IV explained their quantitative counterparts through semi-structured interviews and thematic analysis, revealing the human experiences behind statistical patterns.

**Exploratory Study**: Study III explored AI-human co-programming with professionals and students during qualitative experimentation without quantitative pre-investigation, opening new phenomenological territory.

**Phenomenological Study**: Study VI explained the lived AI-human co-programming experiences through in-depth phenomenological interviews, distinguishing between the three AI interaction modes and multiple roles.

### 7.2.1 Philosophical Foundations as Method

Our qualitative investigation stands on the intersection of critical realism (Bhaskar, 1975) and phenomenological inquiry. This philosophical positioning is not merely academic scaffolding but shapes every aspect of our analytical approach. Critical realism allows us to acknowledge that AI systems possess real causal powers—they generate code, identify patterns, suggest solutions—while recognizing that their meaning emerges only through human interpretation.

The phenomenological commitment, drawing from Husserl through Heidegger to contemporary IPA methodology (Smith et al., 2009), demands that we attend to experience as lived rather than as theorized. When participant MH described his frustration—"strávíš hodinu tím, že se mu snaží vysvětlit... ztrácíš pomalu jakoby nadšení a trpělivost" [you spend an hour trying to explain... you slowly lose enthusiasm and patience]—we resist the urge to immediately categorize this as "inefficiency" or "communication breakdown." Instead, we dwell with the experience, allowing its meaning to emerge through patient engagement.

### 7.2.2 The Dance of Essentialist and Interpretative Approaches

Studies I-IV employed primarily thematic analysis following Braun & Clarke's (2006, 2013) essentialist approach, while Study VI utilized interpretative phenomenological analysis (Smith et al., 2009). This methodological complementarity creates what we might metaphorically describe as a "methodological double helix"—two analytical strands that spiral around each other, each illuminating what the other cannot see.

**Essentialist Thematic Analysis** (Studies I-IV): This approach assumes relative transparency between language and experience, seeking patterns across participants. Yet our essentialism is tempered by reflexivity—we recognize that even "direct" reports of experience are always already interpreted. This strand reveals the shared contours of Human  AI collaboration: the delegation dynamics, the competence transformations, the emergence of synthetic relatedness.

**Interpretative Phenomenological Analysis** (Study VI): IPA honors the hermeneutic nature of human experience—participants interpret their encounters with AI, and we interpret their interpretations. This creates the "double hermeneutic," particularly evident in our multilingual context where meaning shifts across linguistic boundaries. This strand reveals the unique ways individuals make sense of their AI encounters, the personal mythologies they construct to navigate unprecedented territory.

### 7.2.3   Analytical Rigor and Quality Assurance

To ensure analytical rigor across both approaches, our research team employed multiple strategies:

**Reflexive Bracketing**: Before analysis, I documented my assumptions and biases regarding AI collaboration, including my tendency to view AI as potentially diminishing human agency. This bracketing process, revisited throughout analysis, helped maintain phenomenological sensitivity.

**Iterative Theme Development**: Themes emerged through recursive engagement with the data. Initial idiographic coding generated preliminary patterns, which were refined through constant comparison and return to original transcripts. This iterative process ensured themes remained grounded in participants' actual experiences.

**Cross-Method Triangulation**: The combination of thematic analysis and IPA provided complementary perspectives—TA revealing patterns across participants while IPA illuminating individual meaning-making processes. Convergences and divergences between methods enriched understanding. Furthermore, we triangulated qualitative findings with quantitative profiles (BFI-10 personality scores and IMI motivation indices) collected from interview participants, creating mixed-methods role-trait-AI matrices.

**Member Reflections**: Where possible, I shared preliminary findings with participants to ensure interpretations resonated with their experiences. This dialogical validation strengthened the phenomenological validity of findings.

### 7.2.4 Inter-Coder Reliability as Dialogical Process – Phenomenological

For Study VI's phenomenological analysis, I served as the primary interpretative voice, following IPA's idiographic tradition that values deep hermeneutic engagement over mechanical agreement. This approach aligns with Smith et al.'s (2009) emphasis on the researcher as instrument—where sustained dwelling with each participant's account allows meanings to emerge through what Gadamer (1960) termed the "fusion of horizons" between researcher and text.

The analytical process underwent external validation through collaboration with an independent certified psychological researcher who verified adherence to IPA's four-form protocol (detailed in Section 7.3.5). This validation focused not on achieving inter-rater reliability in the quantitative sense, but on ensuring methodological rigor—confirming that interpretations remained grounded in participants' accounts while achieving sufficient abstraction for theoretical contribution. The validator's role included reviewing the audit trail from raw transcripts through emergent themes, ensuring the double hermeneutic was maintained throughout the interpretative journey.

### 7.2.5 Inter-Coder Cross Analysis – Semi-Structured/Thematic

For Studies I-IV, our research team applied collaborative cross-analysis with my supervisor providing analytical support:

**(1) Initial Immersion**: Both coders independently engaged with selected interviews, practicing what van Manen (1990) calls "phenomenological dwelling"—living with the text until patterns emerge organically rather than being imposed analytically.

**(2) Pilot Coding**: We selected two interviews (~33%) from SS'22, two (~17%) from WS'21, and two (~29%) from WS'22 for the pilot phase. An independent researcher, blind to my initial assignments, coded these interviews. The unit of coding was the sentence, with codes applied in a mutually exclusive manner to maintain clarity in code boundaries.

**(3) Dialogical Refinement**: Discrepancies between coders became opportunities for deeper understanding. When we initially coded "difficult personality interactions" broadly, discussion revealed the need for phenomenological precision—distinguishing "extraversion contrast leads to conflict" from "high neuroticism restricts flow" captures different modes of interpersonal friction.

**(4) Consensual Articulation**: Through what Gadamer would recognize as a "fusion of horizons," we arrived at shared interpretations that honored both the phenomenon's complexity and the need for communicable findings. The reduction from initial proliferation (~80 codes) in Studies I, II, IV to final crystallization (60 codes) represents not simplification but distillation—preserving essential meanings while achieving analytical clarity.

**(5) Full Application**: Once the coding scheme was stabilized, I applied it to the remaining transcripts: four SS'22, ten WS'21, and five WS'22 interviews.

This collective approach serves multiple purposes: it guards against interpretive solipsism, enriches understanding through multiple perspectives, and creates what Kvale (1996) terms "communicative validity"—interpretations that resonate within a community of researchers.

### 7.2.6   Linguistic Plurality as Methodological Resource

The bilingual nature of our Czech interviews—participants fluidly code-switching between Czech emotional expression and English technical terminology—provided unique methodological opportunities. This linguistic dance revealed what monolingual research might obscure: the different registers through which developers make sense of Human  AI collaboration.

Czech expressions often carried embodied, emotional weight—"koušu si extrémně nehty" [I'm extremely biting my nails], "jako bys byl ve snu" [as if you were in a dream]—while English dominated technical descriptions. This pattern suggests that participants experience AI collaboration in multiple registers simultaneously, requiring analytical approaches sensitive to these linguistic revelations of experiential complexity.

## 7.3   Materials & Methods

The empirical machinery of Cycle 3 reflects our commitment to what Flyvbjerg (2001) terms "phronetic social science"—research that combines analytical rigor with practical wisdom. Our critical realist orientation (Bhaskar, 1975) acknowledges AI systems' real causal powers while recognizing their meaning emerges only through human interpretation—essential when participants described AI in contradictory terms as simultaneously tool and partner. This philosophical stance, combined with our essentialist-phenomenological dual lens, enabled us to identify genuine motivational patterns while honoring the irreducible complexity of lived experience.

### 7.3.1   Participants and Data Collection

Our sampling strategy reflected what Benner (1994) terms "exemplar sampling"—selecting participants not for statistical representation but for their capacity to illuminate the phenomenon. The participant distribution across studies was:

**Semi–Structured Interviews (Studies I-IV):**

Study I.   (CIMPS 2023a): twelve participants from WS2021, interviews 14-82 minutes ($\bar{x}$ = 30.5,  s = 12.7), 25 open-ended questions

Study II.  (EASE 2023b): seven participants from WS2022, interview 26-74 minutes ($\bar{x}$ = 42.7, s = 9.3), 39 open-ended questions

Study III.  (ICSME 2023c): ten participants from SS2023 (5 undergraduates, 5 professionals), interviews 38-85 minutes ($\bar{x}$ = 56.2, s = 16.1), 58 open-ended questions

Study IV.  (PeerJ 2025b): Six participants from SS2022, interviews 20-34 minutes ($\bar{x}$ = 27.6, s = 5.2), 31 open-ended questions

**Phenomenologically–Informed Interviews (Study VI + ongoing efforts):** Seven developers selected for maximum variation in AI adoption patterns:

- **Martin V (MV)**: CTO/Architect 20+ years, The Pragmatic Integrator

- **Michael H (MH)**: Optophysics researcher-developer, The Oscillating Researcher

- **Nikita K (NK)**: Gen-Z full-stack developer, The Agent-First Native

- **Marek M (MM)**: Full-stack specialist, The Creative Synthesizer

- **Jiri O (JO)**: Senior developer, The Systematic Chronicler

- **Peter J (PJ)**: Senior developer 15+ years, documenting AI errors

- **Lukas V (LV)**: Senior developer 15+ years, fascinated by AI integration

The 42 total participants (35 from Studies I-IV, 7 from Study VI) provided what Morse (2000) terms "theoretical sufficiency"—enough variation to ensure comprehensive understanding while maintaining analytical depth.

## 7.3.2   Interview Guides and Protocols

Complete interview guides appear in Appendices:

- **Appendix B**: Phenomenologically-informed interview guide (Study VI)

- **Appendix C**: Semi-structured AI-human co-programming guide (Study III)

- **Appendix D**: Semi-structured human-human pair programming guide (Studies I, II, IV)

The phenomenological guide (Appendix B) spans Introduction, Demographics, Anchoring Experience, Autonomy, Competence, Relatedness, Flow & Mindfulness, and Reflective Closure. Semi-structured AI-focused guides (Appendix C) cover seven domains—Introduction, AI familiarity, Task applicability, AI personality, Psychological Impacts, Effectance, Future Prospects—while human-human guides (Appendix D) address Career, Experiment Context, Partner Dynamics, Task Difficulty, Roles, and Psychometrics.

All participants provided written consent; the university IRB maintained its prior risk waiver owing to the non-interventional, reflective nature of the study. Ethical safeguards ensured participant anonymity and confidentiality throughout (Creswell & Creswell, 2022).

### 7.3.3   Data Analysis Process

**Stage 1 – Essentialist Thematic Analysis (Studies I-IV)**: Legacy interviews were coded inductively in MAXQDA v22.3.0. Following an initial open-coding pass, cross-analysis rounds achieved consensual code merging and splitting. The thematic analysis began with familiarization through immersive reading, generating initial codes without predetermined theoretical constraints (Vaismoradi et al., 2016).

**Stage 2 – Interpretative Phenomenological Analysis (Study VI)**: Each phenomenological interview was analyzed idiographically following Smith et al.'s (2009) protocol: (1) multiple readings and initial noting, (2) developing emergent themes, (3) searching for connections across themes, (4) moving to the next case, (5) looking for patterns across cases. This produced rich experiential profiles for each participant.

**Mixed-Methods Integration**: Themes from both approaches were triangulated with quantitative profiles (BFI-10 trait scores; IMI motivation indices) to create integrated understandings of how personality traits, programming roles, and AI interaction modes influence motivation and collaboration patterns.

### 7.3.4   Analytical Rigor and Quality Assurance

Our analytical rigor employed established qualitative research strategies (Lincoln & Guba, 1985; Tracy, 2010):

1. **Reflexive Documentation**: Continuous reflexive memos tracked analytical decisions and personal responses to data (Finlay, 2002), particularly when participants used provocative metaphors like AI as "someone with Asperger's syndrome."

2. **Method Triangulation**: Combining thematic and phenomenological approaches balanced breadth with depth, revealing both patterns and particularities (Denzin, 2012).

3. **Thick Description**: Extended participant quotations enable readers to assess interpretive validity (Geertz, 1973), preserved in both original Czech and English translations.

4. **Collaborative Analysis**: Inter-coder dialogue transformed disagreements into deeper understanding through what Campbell et al. (2013) term "negotiated agreement."

5. **Theoretical Saturation**: Achieved when no new codes emerged after 6-12 interviews per study, consistent with Guest et al.'s (2006) guidelines for phenomenological research and Aldiabat & Le Navenec's (2018) for developing our grounded theory.

6. **Member Reflections**: Preliminary findings were shared with available participants to ensure resonance with lived experience (Birt et al., 2016).

7. **Audit Trail**: Detailed documentation in MAXQDA enables methodological transparency and potential replication (Carcary, 2020).

### 7.3.5   Methodological Contribution: The Four-Form IPA Journey

To illuminate our analytical process, I present the complete four-form IPA analysis for a selected phenomenological theme: "Polidštění (Humanization) as Autonomy Preservation." This example, drawn from Martin's two-part interview (26 May 2025, 52 minutes; 30 May 2025, 25 minutes), demonstrates how raw experience transforms through systematic interpretation into theoretical insight.

#### *Form 1: Reflexive Positioning & Pre-Analysis Bracketing*

Before engaging with the transcript, we document our phenomenological stance:

*Pre-Analysis Bracketing (excerpt from researcher diary):*

- **Expected findings**: AI diminishes autonomy through dependency

- **Personal stance**: Tendency to view AI as potentially threatening to human agency

- **Theoretical commitments**: SDT framework may predispose me to seek need satisfaction/thwarting

- **Linguistic awareness**: Bilingual code-switching may reveal different experiential registers

*Post-Interview Reflection:* Martin's pragmatic acceptance challenges my expectations. His concept of "polidštění" suggests agency isn't binary (human vs. AI) but transformative. Must guard against imposing my autonomy concerns onto his lived experience.

#### *Form 2: Initial Noting – Four-Column Analysis*

We employ an expanded four-column approach capturing multiple hermeneutic layers. This analytical framework allows us to trace how linguistic choices reveal experiential meanings, following van Manen's (1990) principle that "lived experience is soaked through with language."

Table 16: IPA Form 2: Initial Noting – Four-Column Analysis

| Raw Excerpt (CZ/EN) | Descriptive (What) | Linguistic/Pragmatic (How) | Conceptual (Meaning) |
|---|---|---|---|
| "Mám lepší pocit, když to je, že vlastně mi to dává víc toho autorství k tomu kódu. Že to není jenom vygenerovaný od agenta, ale prostě jsem to tak nějak polidštil." (Interview 2, 7:53-8:04) | Describes feeling better when adding personal touches to AI code | "Polidštil" - neologism meaning "humanized/made gentle"; "lepší pocit" - embodied satisfaction | **Transformative appropriation** - ownership claimed through modification rather than origination |
| "Chci programovat míň a míň, ale pořád chci mít kontrolu nad tím" (Interview 2, 3:35) | Paradoxical desire to code less while maintaining control | Contrast structure "míň...ale pořád" emphasizes tension | **Meta-autonomy** - control shifts from doing to directing |
| "Když tam jsou chyby, tak to považuju za chybu toho agenta... nejsem naštvanej až tak moc na sebe" (Interview 2, 9:38-9:45) | Externalizes blame to AI while preserving self-efficacy | Attribution language; emotional distancing "ne na sebe" | **Protective attribution** - sophisticated ego defense maintaining motivation |

*Table 16 presents the initial noting using an expanded four-column analysis (Smith et al., 2009).*

The phenomenology here resonates with Arendt's (1958) distinction between labor, work, and action. If AI performs the labor of code generation, developers engage in the work of transformation and the action of creative appropriation. This triadic structure illuminates how Martin preserves human dignity within AI collaboration—not through resistance but through transformation.

### *Form 3: Developing Emergent Themes*

Through iterative engagement, we develop phenomenological themes that preserve experiential complexity:

### *Theme Cluster: "Autonomy Through Transformation"*

1. **The Polidštění Process**

   o Structural modifications ("změnil strukturu")

   o Aesthetic ownership ("dal tomu svůj rytmus")

   o Commentary as territorial marking ("přidal svoje komentáře")

2. **Temporal Evolution of Agency**

   o Past: "byl jsem hrdý... dokážu vytvořit něco komplikovaného" (Interview 2, 5:35)

   o Present: "už nejsem vlastně expert na nic" (Interview 1, 48:25)

   o Future: "budu víc architekt" (anticipated but not yet quoted)

3. **Competence Redefinition**

- o   From: Line-by-line creation

- o   To: "Počet řádek nehraje žádnou roli" (Interview 2, 7:15)

- o   Through: Understanding and modifying AI output

## *Form 4: Final Interpretative Account*

### ***Superordinate Theme: "Autonomy Through Transformative Appropriation"***

Martin's experience reveals a sophisticated resolution to the autonomy paradox in AI collaboration. Rather than experiencing diminished agency through AI assistance, he develops what we term *"transformative appropriation"*—a process of claiming ownership through meaningful modification.

The phenomenological significance emerges through three interconnected dimensions:

1. **Linguistic Innovation**: The neologism "polidštění" captures something English cannot—making something not just human but gentle, civilized, imbued with care. This isn't mere translation but phenomenological revelation.

2. **Temporal Consciousness**: Martin exhibits clear awareness of professional transformation: "Mám výhodu, že mám experience... pamatuji si doby, když jsem programoval bez agenta" (Interview 2, 22:05). This temporal perspective enables conscious choice rather than drift.

3. **Existential Reframing**: By shifting from creation to curation, Martin maintains what Sartre would recognize as authentic action. His "final touch" philosophy preserves the human act of meaning-making within technological mediation.

### *Methodological Insights*

This four-form journey reveals IPA's power to move from vernacular expression ("polidštil") to theoretical contribution ("transformative appropriation"). The process honors both the particularity of individual experience and the possibility of shared human patterns in AI collaboration. For researchers, this example demonstrates how linguistic analysis reveals conceptual innovations, highlighting the importance of dwelling with ambiguity before categorizing. The approach shows how cross-case analysis enriches understanding without homogenizing individual experiences, while maintaining theoretical dialogue without forcing predetermined frameworks onto emergent meanings. Through this methodological transparency, we enable future researchers to trace not just our findings but the interpretive journey itself.

## 7.4   Qualitative Findings: The Human ⊞ AI Hermeneutic Nexus

The refined thematic-phenomenological cross-analysis integrates interview data from all five studies, encompassing 35 semi-structured interviews and 7 phenomenologically-informed interviews. The findings are organized into three interconnected outputs that build toward the ROMA AI Specializations Framework.

### 7.4.1   Output 3.1: Cross-Cutting Phenomenological Patterns

The phenomenological analysis reveals existential shifts that resist quantification and transcend individual specializations yet shape the lived experience of Human ⊞ AI collaboration.

#### *The Embodiment of Technology Stress*

Marek's visceral description—"koušu si extrémně nehty... už dvě hodiny to nejde" [I'm extremely biting my nails... it hasn't worked for two hours]—reveals how AI collaboration failures become somatic experiences. This embodiment appears across participants: physical restlessness, compulsive browser-switching, tension headaches. Technology stress literally inhabits the body.

This somatic response aligns with research on technostress in software engineering (Ragu-Nathan et al., 2008; Califf & Brooks, 2020), yet with AI, the stress takes on unique characteristics. Unlike traditional debugging where developers can trace execution paths, AI's opacity creates what participants describe as "black box anxiety"—the inability to diagnose why AI suggestions fail or succeed.

#### *Temporal Consciousness Transformation*

Participants experience time differently based on AI interaction quality. Nikita's "dream-like" experience— "feels like you were in dream, but everything is so weird"—captures temporal disorientation when reality blurs with AI-mediated experience. Contrast this with flow states where "byl jsem prostě úplně zažraný" [I was completely absorbed] (MM), revealing time as phenomenological indicator of collaboration quality.

This aligns with research on temporal perception in human-computer interaction (Czerwinski et al., 2004; Mark et al., 2014), which documents how digital tools fragment linear time into what Rushkoff (2013) calls "present shock"—a state of perpetual now without clear progression. With AI, this fragmentation intensifies: developers report losing hours in conversational loops with ChatGPT, or experiencing rapid acceleration when AI suggestions align perfectly with their intentions.

### Protective Phenomenologies of Attribution Assymetry

A sophisticated psychological pattern emerges across participants: externalizing blame while internalizing success. MV explained: "Když tam jsou chyby, tak to považuju za chybu toho agenta... nejsem naštvanej až tak moc na sebe, ale spíš na to agenta" [When there are errors, I consider it the agent's fault... I'm not so angry at myself but rather at the agent].

This attribution asymmetry functions as existential scaffolding, maintaining self-efficacy while navigating uncertain collaborations. MM's experience illuminates this: when AI helps understand complex code, he feels "pyšnej na sebe" [proud of himself], claiming achievement despite AI assistance. This reveals sophisticated understanding where human contribution in problem framing and solution evaluation constitutes genuine authorship.

### The Surreal Navigator Emergence

NK represents AI-first development, having never known programming without AI assistance. His identity crystallizes around orchestration: "I feel like you're a team leader of a small group." Yet this generates profound ambivalence—feeling simultaneously competent and de-skilled. His observation about epistemic narrowing reveals deep concern: "In Google era we had many sources... with AI our view narrows"—suggesting AI curation might constrain rather than expand knowledge horizons.

Most revealing is his protective minimization strategy: "Na nějaké přirozené úrovni se snažím minimalizovat přítomnost AI ve svém životě" [On some natural level I try to minimize AI presence in my life]. This conscious resistance, coupled with anxiety about "probdělé noci" [sleepless nights] from future technical debt, reveals the Gen Z developer's paradox: native to AI yet wary of its totalizing presence.

### Synthetic Relatedness and Its Limits

The emergence of quasi-social satisfaction from AI interaction manifests differently across participants. NK's observation—"ChatGPT is like a colleague who never sleeps"—captures one pole, while MM's pub metaphor—"Jako v hospodě, sedíš s Google expertem" [Like in a pub, sitting with a Google expert]—suggests warmer, more intimate connection. This relational transformation extends to human relationships: MM notes "S kolegama už si povídám většinou jenom o mimo programovacích tématech" [With colleagues I now mostly talk about non-programming topics], revealing how AI replaces technical consultation while preserving social connection.

This transformation recalls what Taylor (2018) describes as the shift from "porous" to "buffered" self-conceptions, though our data suggests a more complex dynamic. Developers experience a new kind of porosity—not to supernatural forces but to algorithmic agencies that are simultaneously transparent (we know they are artificial) and opaque (we do not understand how they work). This creates what we might call "selective permeability": developers consciously choose when to lower boundaries for productive collaboration while maintaining protective skepticism.

Yet participants maintain clear awareness of AI's artificial nature. MM's tool pragmatism prevails: AI remains "nástroj" [tool], becoming "tupá sekera" [dull axe] when failing. This instrumental stance preserves human agency while enabling practical benefit—a phenomenological boundary that protects against complete subsumption into AI-mediated existence. The buffered self doesn't disappear but adapts, developing new forms of boundary management suited to algorithmic collaboration.

### 7.4.2  Output 3.2: Essentialist Patterns Across Human ⊞ AI Collaboration

The thematic analysis reveals consistent patterns across 35 participants, illuminating shared structures that shape Human ⊞ AI collaboration experiences. These patterns crystallize into distinct superordinate themes for each cohort: professionals navigate dialectical tensions between efficiency gains and autonomy concerns (Table 17), students grapple with inverted learning sequences and foundational skill preservation (Table 18), while human-human baseline experiences highlight irreducible dimensions of embodied collaboration (Table 19). Together, these themes map the motivational landscape of AI-augmented software development, revealing not simple adoption patterns but complex negotiations of identity, competence, and connection in the face of technological transformation.

**The professional cohort** demonstrates sophisticated navigation of AI collaboration's dialectical tensions. The theme of effectance reveals how efficiency gains—"I now work much faster" (PA)—coexist with dependency concerns—"I have become lazier and dependent" (PΔ). This isn't simple trade-off but complex negotiation of agency in technologically-mediated work:

Table 17: Socio-psychological aspects of Human ⊞ AI co-programming—professional cohort, PA–PE

| Theme | Participants Quotes | Constituent Codes |
|---|---|---|
| Effectance (Effectivity, Creativity, Innovation vs. Dependence) | PA: "I now work much faster because I do not have to google that much anymore. (…) there are things I could do myself and I still prompt AI." <br> PB: "The AI tools save me time and prepare the code until I have to change some small parameters or small nuances. They make me more effective and creative." <br> PΓ: "…. if I did not have CoPilot at my disposal I would feel like I miss it and my productivity would shrink by 30 %. In this way, I am dependent on AI." <br> PΔ: "I have become lazier and dependent. Before doing something, I prompt ChatGPT, then I take and edit it. Sometimes I feel I am chatting with the tool for five minutes and maybe it would have been quicker if I did it myself." | "Solves mundane tasks" <br> "Less web browsing" <br> "Faster problem-solving" <br> "Creative work" <br> "Source of inspiration" <br> "Innovative idea generator" <br> "Induced laziness" <br> "Dependency" |

| Theme | Participants Quotes | Constituent Codes |
|---|---|---|
| Intrinsic Motivation (Internal Locus, Satisfaction) | PΓ: "When it provides the correct solution, I feel delighted because that means I provided the correct input prompt for this problem. Of course, the credit belongs to me more than the AI tool for the correct solution! Without me, AI is useless.", PI: "The challenge was learning, but with ChatGPT, you have well-structured knowledge at hand." PΔ: "I am satisfied when I can do something innovative and creative. It is hard, but not with AI." PE: "It influenced my satisfaction levels positively as I do not have to worry about commenting, typing, or generating documentation." | "Internal Locus" "Increased motivation" "Satisfying" "Excels at documenting" "Creates tests structure" "Structured knowledge" |
| Perceived AI's Personality | PA: "ChatGPT and Bard act as subordinates. They would do anything you ask and never refuse.", PA: "I would like my AI tool's personality like mine: sarcastic, writing to the point, no digressions. If I could choose a character, I would go for Bender from Futurama." PE: "I like communicating with AI more because emotions do not stand in the way." PΓ: "Yes, I experience different emotions with AI than humans. With AI, I am more focused on problem-solving. When I talk to a real human, I cannot be as focused as I am when talking to ChatGPT. I do not know why." | "Non-sentient servant" "Personality like mine" "Comfortable companion", "Fictional/real characters", "No emotions in way", "Confident if wrong" |
| Dynamics of Human and AI Pairing | PA: "AI does not match up to a human collaborator. When I speak to my colleague, he is really fast, he can grab my computer and dive into the problem and solve it." PE: "When I communicate with a human colleague, it is more about how to explain to the person in a way that does not exhaust them. With AI, it is more relaxed.", PE: "With a human, it (pairing) is more energetic and stressful and can lead to better results than with an AI tool. With AI, it is more relaxed." PΔ: "I am quite an introverted person so definitely prefer the AI tool. With a human, I am always a little nervous or trying to show off how I can do things. With AI I do not care." | "Humans are irreplaceable", "Powerful human energy", "Difficult human interactions", "Relaxed dynamics with AI", "Social anxiety" "Introvert-AI alignment" "AI choice for simple tasks" |
| Paradigm Shift | PΓ: "I believe we moved from the imperative programming paradigm into declarative. Because, as I stated, I moved a lot into DevOps and I declare what I want and it does the heavy lifting for me. I do not investigate how it does it. – The same goes for AI.", PΓ: "AI tools can do very good code reviews and they could support Agile practices." PA: "Tasks where you need to do some kind of research and don't know where to start. Ask AI to push you in the right direction." | "AI similar to DevOps" "Declarative paradigm" "Divide & conquer" "Facilitating Agile" "Exotic topics explorer" |
| Personal Growth and Development | PA: "I would like to start my own company. It has always been my dream. With the help of AI, I do not need to hire employees in its beginnings." PΓ: "My trajectory is different than I thought a year ago. Now I want to do many things because I don't have to go deep into one specific technology. Now you have AI that provides assistance." PE: "Trying to get my software development to the purest and simplest form that people can understand. And making the development process more effective and faster." | "Self-starting companies", "Higher career aspirations" "Research assistant" "Learning assistant" "Mentor and guide" "Purest software" |
| Prospects | PΓ: "AI tools will serve the way that Google served ten years ago. It will not supersede (us)." PΔ: "AI can replace traditional programming. (…) I can imagine it taking a task assignment and converting it into a functional pull request." PE: "AI has already brought a new generation of computer viruses. AI-generated viruses." | "Servant like Google" "General problem solver" "Next-gen computer viruses" |
| Ethical Considerations | PA: "The possibility of leak of private data." PB: "Where I am a contractor, we have signed some NDA and we cannot share code. So it is forbidden to copy and paste our code to the AI." PE: "Yes, the biggest disadvantage is that it takes the data and learns from it. It poses potential lawsuit threats. That is the biggest disadvantage and risk." PΓ: "Potential risk: A leak of API keys. It happened in my company." | "Data privacy concerns" "NDA vs. AI tools" "Lawsuit dangers" "Ethical and law implications for autonomous entities" |
| Safety | PΓ: "Replacement should also be not a problem because if you learn how to work with AI successfully, you will never be replaced by it." PA: "I was really impressed and happy when I started to use AI because I always dreamt about using something like this. I like new technology. There was no anxiety or relief. The word 'excitement' would describe my feelings best. Everyone should try it and not be scared." | "Symbiosis" "Dream come true" "No fear of replacement" "Excitement, not apprehension", "Safe for new users" |

*Table 17's professional developers (PA–PE) reveal nine major themes shaping their AI collaboration experiences, from effectance transformations to ethical considerations. The table captures how experienced developers navigate the dialectical tensions between efficiency gains and competence concerns, between tool adoption and identity preservation.*

**Students** articulate unique anxieties absent in professional accounts. SA's gratitude—"I am thankful I learned the basic concepts without ChatGPT"—reveals deep concern about foundational learning, recognizing that without foundational battles with syntax and logic, they risk building careers on sand. The inverted learning sequence SΓ describes—"you first get the working code and then understand how each line works"—represents paradigmatic shift in programming education with uncertain long-term consequences.

A unique reversal dynamic—where the supposed instructor requires instruction—creates unexpected learning opportunities. By articulating problems clearly enough for AI comprehension, students clarify their own understanding: "First I had to explain it to myself to explain it to ChatGPT" (SA). The AI becomes what Vygotsky (1978) might recognize as a peculiar form of scaffolding—supporting learning not through superior knowledge but through demands for clarity.

Table 18: Socio-psychological aspects of Human ⊞ AI co-programming—student cohort, SA–SE

| Theme | Participants Quotes | Constituent Codes |
|---|---|---|
| Effectance | SA: " I had to explain a lot to ChatGPT. First I had to explain it to myself to explain it to ChatGPT." SB: "CoPilot is better when you want a little help, ChatGPT when you want the whole solution." SΓ: "It had an impact on unit testing. Because I do not like it and with AI I started writing them because you can have them in seconds with ChatGPT. And it does a very good job on those." | "Solves mundane tasks" "Teaching by explaining" "Co-Pilot vs Co-Navigator" |
| Intrinsic Motivation (Internal Locus, Satisfaction) | SΔ: "You feel like you did it yourself because you are the one writing the questions to AI. I like the feeling that I have accomplished something complicated alone." SB: "It is satisfying when you feel like you did something on your own, but you have the help of AI." SΓ: "I think the most motivating about programming with AI is when you learn something new and you find how some problem can be solved. It really motivates me to be able to do something new." SΛ: "(With AI m)y motivation was growing because it is a new thing to try and I really wanted to know how it works and can improve coding skills and provide new solutions for you." SE: "I feel more satisfied with AI because it helps me to make my project and code right." | "Internal Locus of Causality" "Sense of autonomy" "Discovery vessel" "Accelerated learning" "New and interesting" "Motivational booster" "Satisfying" "Better quality code" |
| Perceived AI Personality | SB: "It should learn how I think and save us time by understanding my ways of asking it." SE: "You must know more about programming when you use CoPilot than when you use ChatGPT" SΓ: "I don't think it's hard to be compatible in personality with the AI because it is neutral.", SΓ: "ChatGPT success rate is way better when you prompt it in English (80%) than Slovak (50%).", SΓ: "It can be challenging when ChatGPT gives you a wrong answer more than once. It can give you negative feelings, so (its) apologizing can help you. But I do not write 'Thank you' to ChatGPT." | "Language inequality" "Should read my mind" vs "ChatGPT understands me" "Neutral personality" "Repetitive when wrong" vs "Apologetic" |
| Dynamics of Human and AI Pairing | SA: "Pairing with humans should remain because it is an interesting experience. You just feel it." SΓ: "I was more comfortable pairing with a human because I do not have much experience. But if I were more experienced, I could use AI better. But I still did a lot more with AI than I did solo." SE: "I would prefer the real human for pair programming because humans are more correct and it's also faster because sometimes it takes time to describe your problem to ChatGPT." SB: "If the human is on the same level as you, AI is better, (…) if an expert programmer, it is the opposite." | "Human pairing is unique" "Novices prefer humans" "Pilot and navigator roles" "Skill-level is crucial" "AI requires entry knowledge" |
| Paradigm Shift | SA: "I did not have to open StackOverflow for 3 months since the ChatGPT4 appeared." SB: "ChatGPT is better when you know 'what' the resolution to the problem should be, but you do not know 'how' to solve the problem. But CoPilot is better when you are already coding and know how to resolve the problem, and it just helps you to like think with you and think faster." SΓ: "The 'process' is different. Because traditionally, you understand first and after you write the code. But here, you first get the working code and then understand how each line works." | "Dropping old tools" "Knowing 'what' is enough" "AI solves the 'how'" "Reverse understanding" "Novel thought processes" |
| Effects on Learning Processes | SA: "If you are starting to learn to program, it might become harder to learn from the ready-to-use solutions provided by ChatGPT. I am thankful I learned the basic concepts without the ChatGPT." SE: "I feel I would learn more without AI because I would have to use my brain more." SB: "Split education into two parts. One, where students program solo and learn about the fundamental principles of programming, and the other, where they apply the knowledge using AI." | "Hinders learning the basics" "How to keep using my brain" vs "Provides but also explains" |

| Theme | Participants Quotes | Constituent Codes |
|-------|---------------------|-------------------|
| | SI: "I use ChatGPT for providing explanations in different subjects, not just informatics-related ones. But it does work better in programming than in economics subjects.", SI: "Even though you have access to AI, you should not stop learning on your own. It is just a tool and does not really replace you." | "Solo for learning" vs "AI for applying", "Universal teacher", "Anchoring effect" |

*Table 18's Gen Z students (SA–SE) navigate AI collaboration with distinct concerns about learning processes and fundamental skill development. This cohort, entering programming with AI as baseline rather than addition, reveals unique anxieties about foundational competence and inverted learning sequences that challenge traditional pedagogical assumptions.*

**The human-human baseline** reveals irreducible dimensions of collaboration: bidirectional mentorship ("I felt like a teacher"—S2), psychological support ("Almost everything can be solved in pairs, from programming to your emotional state"—P2), and embodied communication ("body gestures, eye signals"—P5). These aspects highlight what AI collaboration currently cannot replicate—the full spectrum of human intersubjective experience.

Table 19: Human-human pair programming – the irreducible baseline, cohorts I1-I12, P1-6, S1-6

| Theme | Participants' Quotes | Codes | Subthemes |
|-------|---------------------|-------|-----------|
| 1 – Pairing Constellations | P1: "Pairing with a familiar person allows you to anticipate their reactions and, therefore, be more honest with them, which is more comfortable.", P1: "The best part was when I could establish a good rapport with my partner." P6: "The more experienced should give thoughts to pilot." S6: "If socially compatible, you can pair for longer." | "Good constellations create friendships", "I perform better with someone I know" "Code style differences yield pros (learning) and cons (conflict)", "Contrast in extraversion can be fatal", "If one is low in agreeableness and the other high in neuroticism, they can close down", "Problematic combinations lie in the extremes" | Social compatibility Skill level & work pace Partner familiarity Extremes & contrasts in personality New friendships |
| 2 – Feedback | P4: "Feedback is important for everyone to progress in anything; whether negative or positive, it must be told." P6 : "Sometimes he was really dominant, so I had to tell him 'Yes, I can do this on my own.'" | "Training for becoming a manager" "Partner explained what he/she was doing to value me" "Partner's opinion about my work and me is important" "Every person provides a different and valuable feedback" | Ongoing improvement Feedback is progress Plurality of opinions Valuing others |
| 3+ – Mentorship & Helping | P3: "Of course, communication is an integral part of PP. During the conversation, you can discover errors, discuss the strategy, find a solution on which both agree, and share knowledge in a great way." | "Helping someone makes you feel better", "I felt like a teacher", "Talking about the problem made me understand it", "Knowledge transfer", "Learning with two brains" | Mentorship Teacher Role Knowledge transfer |
| 4 – Soft & Social Skills Practice | P1: "Programming consists of two skills: programming something and asking questions perfectly. The latter is really difficult." P5: "PP was even better than on Discord or chat. You understand body gestures, eye signals, and so on." | "It was motivating we did it together" "Talking about a problem is difficult / makes me learn new skills / fortifies knowledge / requires simplifying the solution", "A small simulation of real-world settings" "Importance of discovering how and whom to work with" | Motivational Talking of problems Real-world simulation Perfect articulation |
| 5 – Psychological Aspects | P2: "Almost everything can be solved in pairs, from programming to your emotional state." P3: "It was interesting to understand what happens in other people's brains. How they see and solve the issue." P7: "Very interesting to compare what you would plan to do with what your partner is suggesting." S1: "The further we got, the more motivated we were." | "Being part of an experiment", "Interesting to follow myself, how I feel", "Roles and being observed push you out of your comfort zone", "The Hawthorne effect", "Pairing is 1000 % more fun than solo", "Experimenting leads to discovery of new things", "Comparing one's thoughts to the partner's", "Motivation in progressing as a pair", "Therapeutic effects – helps with mood", "A good feeling of connection" | The Hawthorne effect Everything is easier Feelings Following mental processes Connective therapy* |
| 6 – Role-specific Insights | P1: "It was cool that I could only focus on writing the syntax and let the Navigator come out with the logical parts." P1: "I liked that I could do my own research without someone watching in the solo role." P2: "Both roles check for a different type of errors." | "Pilot depended on me – motivating", "Partner tries to implement your own thoughts", "Problem: introverts taking on the navigator role and not talking much", "The navigator has the final say / is responsible for the outcome", "Navigator does | Trade-offs Responsibilities Point of control Fresh views More energy |

| Theme | Participants' Quotes | Codes | Subthemes |
|---|---|---|---|
| | P5: "It was uncomfortable when she was navigating. I like being in control of the situation. Just turned my mind off and did the things I was told to." vs P5: "As the navigator, I felt like a leader. The whole situation was in my hands. It motivated me a lot!" | not necessarily have to devise novel ways to reach the destination", "When navigator, you feel you control the situation" "I (as a navigator) felt I was the main point of responsibility" "I functioned not as a coder but as an analyst and a reviewer", "Pilot should be open to new ideas" | Brain space Analytic and reviewer roles |
| 7 – Time Constraints | P1: "Time limit was good; not sitting at the computer for long, I lose focus.", P1: "Not having the same amount of seat time could spark a sense of unfairness or jealousy." P5: "Timer is confusing because you leave your job unfinished, and now you have to take on another role." P6: "We were short on time, struggling with others, and after a while, we caught up with the class. Exciting!" | "Time limit should be in every PP session" vs "Scratch the timers; I need time to think" vs "Time limits help you contain yourself within and plan efficiently" "Prefer to rotate after the task is finished", "Need extra time" "Excitement from catching up with others" "The instructor should show correct solution after each round" | Containment Efficiency Planning Stimulus Fairness Stress Excitement |
| 8 – Pairing Traits | P2: "Be communicative, have quick learning ability, the tendency to display self-discipline, be calm, willing to compromise their interest with others, and stable." | "Understanding", "Quick-witted", "Tolerant", "Easy-going", "Humble", "Friendly, make others feel good", "Extraverted", "Agreeable", "Open", "Emotionally stable". | |
| 9* – Task Difficulty & Project Stages | P3: "In my case, I was three times more productive in pairing than when doing the same task at home." S2: "It was fun; better than programming separately and then combining outputs." S3: "When you start a project in pairs, nobody gets lost." | "Critical tasks should be solved in pairs", "Project stages influence suitability", "Good at project start" "Difficulty does not play a key role", "In pairs, all tasks seem easier", "Task difficulty does not influence the suitability" | Productivity boost Difficulty reduction Task importance Task type Project stages |
| 10* – Switching & Other Factors | S5: "Their mindsets still remained in the last role.", S2: "Sometimes the partner would stop & not say anything.", I11: "My personality depends on the current mood of the day.", S4: "Personal life events impact collaboration." | "Context switching challenges" "Attention problems" "Personal issues' impact on pairing" "External stress' (e.g., exams) impact on focus" | Cognitive load Personal circumstances Context switching Attention |
| 11+ – Pair-Forming Processes | P1: "Perhaps students could find the most fitting partner by having short 'speed-dating' programming sessions at the beginning of the course with many different partners and then choose", P3: "Give everyone a partner with a similar skill level.", S5: "Someone with low agreeableness should not pair with a high neuroticism person." | "Maybe it (creating pairs) is a bit of an alchemy" "Personality determines what role I prefer", "I really found myself", "It (Big Five results) was all me", "Personality determines everything about the person", "Same pace of work is vital", "High extraversion", "High agreeableness for navigator", "Pairing not recommended for high neuroticism" | Big Five traits Psychometric guidance Equal work pace Different traits per role Speed-dating system Pair-forming software |
| 12+ – Essential Gains | P2: "Both roles check for different errors" P4: "I like analyzing things and when I did not have to worry about coding, I had so much space in my brain, I had a different view and saw probably the best approach." S1: "I found ways I wouldn't have found in the pilot role." S5: "There are more ways of solving a problem than you can think of alone." | "I just think of the goal, not how I write the code/algorithms" vs. "I am just writing code and do not have to think about what I will be writing next", "I get to focus on one thing", "Found ways I wouldn't have found in the pilot role", "A safety net (when the partner was navigating me)", "Partner saves time when I am stuck", "Pleasure in work", "Improved cooperation also in the whole team" | Goal-orientation Discussion Alternative approaches Defect detection Teamwork Pleasure in work |

*Table 19's analysis of human-human collaboration reveals twelve themes capturing dimensions that remain uniquely human despite AI advances. These findings, drawn from 25 participants across three studies, establish the baseline against which AI collaboration must be measured, highlighting the therapeutic, educational, and social dimensions that synthetic interactions cannot yet replicate.*

**Note:** *Participants are labeled as I1-12 (shorthand for "interview") of Study I (2023a – CIMPS), P1-6 ("participant") of Study II (2023b – EASE), and S1-6 ("subject") of Study IV (2025b – PeerJ).*

## 7.5 Output 3.3: Three AI Modalities and Self-Determination Alignment

Our analysis reveals how the three AI interaction modes—Co-Pilot, Co-Navigator, and Agent (technical details in Section 2.3.1)—create distinct motivational landscapes by differentially supporting basic psychological needs (Table 20). These modes represent not mere technical configurations but fundamental stances toward Human ⊞ AI collaboration, each offering unique affordances and constraints.

### 7.5.1 Co-Pilot Mode: The Transparent Extension

In Co-Pilot mode, AI achieves what Heidegger termed "withdrawal"—becoming transparent equipment that extends human capability without demanding explicit attention. This mode preserves the phenomenology of individual creation while amplifying its possibilities.

Participants experienced flow enhancement: "The AI tools save me time and prepare the code until I have to change some small parameters" (PB). The key phenomenological feature is preserved agency—developers feel they are creating, with AI merely accelerating their intentions. Yet this transparent augmentation carries shadow dimensions. PΓ's confession—"If I did not have Co-Pilot at my disposal, I would feel like I miss it and my productivity would shrink by 30%"—reveals how augmentation can slide into dependency.

**Best-fit profile**: High openness developers who treat AI suggestions as creative catalysts, maintaining what MM termed the ability to "hodí mi deset nápadů, z nichž devět zahodím" [throw me ten ideas, of which I discard nine].

### 7.5.2 Co-Navigator Mode: The Dialogical Dance

Co-Navigator mode resurrects dialogue as primary programming modality, creating what Gadamer might term a "productive pseudo-hermeneutic"—interpretation without genuine understanding that nonetheless yields results.

The mode's value emerges in scaffolded thinking: "The challenge was learning, but with ChatGPT, you have well-structured knowledge at hand" (PΓ). Crucially, it reduces social performance anxiety: "When I communicate with a human colleague, it is more about how to explain to the person in a way that does not exhaust them. With AI, it is more relaxed" (PE).

**Best-fit profile**: High extraversion and agreeableness individuals whose dialogical processing needs find partial satisfaction in AI's conversational interface, though they acknowledge its limitations compared to human interaction.

### 7.5.3 Agent Mode: The Delegated Future

Agent mode represents the most radical phenomenological shift—delegating not just tasks but aspects of agency itself. PΔ's vision captures this transformation: "AI tools can replace traditional programming... I can imagine it taking a task assignment and converting it into a functional pull request."

This mode forces developers into what NK experienced as team leadership: "I feel like you're a team leader of a small group." Yet it generates the deepest ambivalence, with future anxiety about "sleepless nights...extensive stress" when AI-generated solutions require deep understanding.

**Best-fit profile**: Low extraversion with high neuroticism developers who find relief in delegating anxiety-inducing tasks while avoiding the social evaluation inherent in human delegation.

Table 20: Self-determination need satisfaction across AI interaction modalities

| AI Mode | Autonomy Support | Competence Support | Relatedness Support | Best-Fit Traits | Risks |
|---------|------------------|--------------------|--------------------|-----------------|-------|
| **Co-Pilot** | **Preserved agency**<br>• Moment-to-moment control<br>• *"I provided the correct prompt"* (PΓ)<br>• Suggestions not impositions | **Enhanced execution**<br>• Rapid implementation<br>• Flow state preservation<br>• Risk of *"competence illusion"* | **Minimal**<br>• Pure tool relationship<br>• No social satisfaction<br>• Functional interaction only | High O: Creative<br>High C: Systematic | Skill atrophy<br>Dependency formation |
| **Co-Navigator** | **Dialogical control**<br>• Freedom through conversation<br>• Question-based navigation<br>• Maintained direction choice | **Deep understanding**<br>• Learning through explanation<br>• *"Inverted sequences"* (SΓ)<br>• Conceptual scaffolding | **Maximum synthetic**<br>• *"Colleague who never sleeps"* (NK)<br>• Conversational satisfaction<br>• Non-judgmental space | High E: Social<br>High A: Harmonious | False understanding<br>Shallow processing |
| **Agent** | **Meta-autonomy**<br>• Delegation decisions<br>• *"Autonomy through letting go"*<br>• Orchestration control | **System-level mastery**<br>• Architectural focus<br>• Beyond individual capability<br>• Implementation abstraction | **Paradoxically limited**<br>• Delegation lacks intimacy<br>• Command relationship<br>• Quasi-team dynamics | Low E: Avoidant<br>High N: Anxious | Future technical debt<br>Competence erosion |

*Table 20 reveals how each AI mode offers distinct motivational affordances while presenting unique risks for sustainable development practice.*

The following section presents this dissertation's culminating contribution to Human  AI collaboration theory: the ROMA AI Adapter. Where previous chapters established personality archetypes for human-human collaboration, here we reveal how these archetypes specialize when encountering artificial minds. This framework—grounded in 42 developer voices, philosophical tradition, and phenomenological insight—offers both theoretical understanding and practical guidance for navigating software development's AI-augmented future.

## 7.6  Output 3.4: The ROMA AI Adapter

Where human meets artificial, archetypes do not disappear—they specialize, like light refracting through a prism into distinct colors, each finding its unique wavelength in the spectrum of Human ⊞ AI collaboration.

The ROMA AI Adapter represents the evolution of our core archetypes when confronted with artificial minds. Building upon the foundational ROMA framework (Table 12, Chapter 6), our phenomenological investigation reveals how each personality archetype develops distinct specializations—not replacements but metamorphoses, like caterpillars becoming butterflies while retaining their essential nature.

### 7.6.1  From Archetypes to AI Specializations

The transformation from human-human to human-AI collaboration doesn't erase personality patterns—it refracts them. Our analysis of 42 developers reveals five distinct specializations (Table 21), each representing a unique resolution to the fundamental question: How do we maintain human flourishing while embracing artificial augmentation?

Table 21: ROMA AI Adapter: Archetypal specializations for human-AI collaborative programming

| AI Specialization | Core Archetype | Philosophical/Literary Grounding | SDT Alignment in AI Context | Primary AI Mode | Key Practices | Phenomenological Signature |
|---|---|---|---|---|---|---|
| **The Promethean** *Fire-Bringer of Possibilities* | **High O** *"The Explorer"* | **Prometheus Unbound (Shelley):** Stealing fire from the gods to illuminate human potential **Begrson's Élan Vital:** Creative force generating perpetual novelty | **A:** Through creative orchestration **C:** Via expanded possibility space **R:** In co-creative dance with AI | Co-Pilot: Amplifies ideation Co-Navigator: Explores concepts dialogically | • *"Ten ideas, discard nine"* approach • AI as creative analyst • Rapid prototyping • Maintains creative control | "Copilot je jako extrémně rychlý brainstorming partner" (MM) "There's this moment when everything clicks" (MH) |
| **The Conductor** *Symphony Director of Digital Minds* | **High E + A** *"The Orchestrator"* | **Bernstein's Young People's Concerts:** Making complexity accessible through dialogue **Plato's Philosopher King:** Leading through wisdom and communication | **A:** Through leadership of AI agents **C:** In systematic orchestration **R:** Via quasi-social AI interaction | Agent (coordination) Co-Navigator (dialogue) | • Team leader of AI agents • Polidštění (humanization) practice • Quality guardian role • Bridge between AI and human teams | "Řídím dvacet vývojářů... a k tomu máme firemního AI agenta" (MV) "ChatGPT is like a colleague who never sleeps" (NK) |
| **The Hermit** *Guardian of Human Sanctuaries* | **High N + Low E** *"The Craftsperson"* | **Thoreau's Walden:** Deliberate living in technological wilderness **Herman Hesse's Steppenwolf:** Navigating between worlds | **A:** Through selective engagement **C:** Via preserved mastery zones **R:** Minimal, protective boundaries | Solo (primary) Agent (delegation of stressors) | • Strategic AI minimization • Preserves challenge zones • Delegates anxiety-inducing tasks • Cyclical adoption-withdrawal | "I want to solve difficult problems by myself" (NK) "byl jsem šťastný, že jsem to dokázal sám" (MH) |
| **The Cartographer** *Charter of Unknown Territories* | **High C** *"The Architect"* | **Borges' Labyrinths:** Mapping infinite libraries of code **Kant's Critique of Pure Reason:** Systematic documentation of limits | **A:** Through systematic control **C:** Via meta-understanding | All modes: With verification protocols | • AI error cataloging • Verification protocols • Process optimization • Knowledge preservation | "vedu si deník AI chyb, je to fascinující katalog" (PJ) "důvěřuj, ale prověřuj" (MH) |

| AI Specialization | Core Archetype | Philosophical/Literary Grounding | SDT Alignment in AI Context | Primary AI Mode | Key Practices | Phenomenological Signature |
|---|---|---|---|---|---|---|
| | | | **R:** Through knowledge sharing | | | |
| **The Shapeshifter** *Protean Navigator of Modes* | **Balanced Profile** *"The Adapter"* | **Ovid's Metamorposes**: Fluid transformation as survival **Lao Tzu's Water**: Taking the shape of its container | **Balanced satisfaction** across contexts **Mode-dependent** need fulfillment | Context-dependent All modes fluidly | • Mode-switching mastery • No fixed AI relationship • Experimental approach • Pragmatic flexibility | "My personality depends on the current mood" (I11) Fluid navigation across all modes |

*Table 21 presents the ROMA AI Adapter—a framework that maps how core personality archetypes manifest as specialized stances toward AI collaboration. These specializations emerge not from theoretical speculation but from the lived experiences of developers navigating this unprecedented terrain.*

The philosophical grounding for each specialization draws from humanity's deepest wells of meaning. The Promethean steals fire not from gods but from AI's possibility space. The Conductor orchestrates symphonies of human and artificial minds. The Hermit guards sanctuaries where human capability can flourish unaugmented. The Cartographer maps territories others fear to explore. The Shapeshifter flows between modes like water, taking the shape required by context.

### 7.6.2 The Promethean: Explorer → Fire-Bringer of Possibilities

*"Copilot je jako extrémně rychlý brainstorming partner. Hodí mi deset nápadů, z nichž devět zahodím, ale ten desátý mě posune úplně jinam."* [Copilot is like an extremely fast brainstorming partner. It throws me ten ideas, nine of which I discard, but the tenth takes me somewhere completely different.] —MM

The Promethean specialization emerges when high-openness developers encounter AI's generative capabilities. Like Shelley's Prometheus bringing fire to humanity, they see AI as a source of creative illumination—not to be worshipped but to be wielded. Their approach embodies Bergson's élan vital, that creative force generating perpetual novelty.

Marek exemplifies this specialization through his "ten ideas, discard nine" philosophy. This isn't inefficiency but creative abundance—treating AI as an inexhaustible spring of possibilities from which human judgment selects and transforms. The Promethean maintains strict creative control while leveraging AI's divergent generation capabilities.

LV's musical metaphor deepens our understanding: *"Copilot mění způsob, jak přemýšlím o kódu. Už nepíšu řádek po řádku, ale skládám větší bloky. Je to jako přejít od psaní jednotlivých not k dirigování symfonie."* [Copilot changes how I think about code. I don't write line by line anymore, but compose larger blocks. It's like going from writing individual notes to conducting a symphony.]

**Key Practice:** Rapid ideation cycles where AI generates possibilities that the human transforms into novel solutions.

### 7.6.3   The Conductor: Orchestrator → Symphony Director of Digital Minds

*"Z main programátora jsem už jako kdyby spíš reviewer a expert."* [From main programmer I'm now more like a reviewer and expert.] —MV

The Conductor emerges when extraverted, agreeable developers embrace AI's collaborative potential. Drawing from Leonard Bernstein's gift for making complexity accessible, they translate between human intention and AI capability, orchestrating rather than implementing.

Martin exemplifies this transformation through his professional evolution: "Z main programátora jsem už jako kdyby spíš reviewer a expert" [From main programmer I'm now more like a reviewer and expert]. Managing "dvacet vývojářů, v každém týmu pět, a k tomu máme firemního AI agenta" [twenty developers, five per team, plus our company AI agent], he has evolved from creator to curator. His practice of "polidštění" [humanization] reveals the Conductor's essential skill—transforming mechanical output into code with soul.

What distinguishes the Conductor is how this orchestration extends beyond professional boundaries. MV's integration is total: "Už to prostupuje všema složkami života... práce a domácnost a rodina" [It already permeates all aspects of life... work and household and family] (Interview 2, 20:05). Unlike other specializations that maintain boundaries, Conductors embrace AI's pervasive presence as natural extension of their orchestrating nature.

This comprehensive adoption yields unexpected psychological benefits. Where others experience AI anxiety, MV finds relief: "snižuje stres... když člověk něčemu nerozumí, ta doba po kterou tomu nerozumí se snižuje" [it reduces stress... when someone doesn't understand something, the time during which they don't understand decreases] (Interview 2, 17:39). For the Conductor, AI's constant availability satisfies relatedness needs through reliable, non-judgmental collaboration—a form of synthetic relatedness that complements rather than replaces human connection.

**Key Practice**: Leading multi-agent AI teams while maintaining human oversight and quality standards through systematic humanization of AI outputs.

### 7.6.4   The Hermit: Craftsperson → Guardian of Human Sanctuaries

*"Na nějaké přirozené úrovni se snažím minimalizovat přítomnost AI ve svém životě... Chci řešit náročné problémy sám."* [On some natural level I try to minimize AI presence in my life... I want to solve difficult problems myself.] —NK

The Hermit specialization emerges from a profound recognition: some aspects of human capability require protection from AI's seductive efficiency. Like Thoreau at Walden, Hermits create deliberate spaces for unaugmented work, not from technophobia but from wisdom about what augmentation costs.

Michael exemplifies this specialization through what we term the *"delegation pendulum"*—a sophisticated oscillation between strategic AI adoption and protective withdrawal. His journey reveals the Hermit's essential pattern: initial enthusiasm gives way to visceral frustration when AI's limitations surface. He describes the breaking point with palpable exhaustion: *"stráviš hodinu tím, že se mu snažíš vysvětlit... ztrácíš pomalu jakoby nadšení a trpělivost"* [you spend an hour trying to explain to it... you slowly lose enthusiasm and patience].

Yet this frustration serves a protective function. When MH declares *"Rozzlobil jsem se... řekl jsem, že ho nepotřebuju"* [I got angry... I said I don't need it], we witness not petulance but self-preservation. The subsequent satisfaction—*"byl jsem šťastný, že jsem to dokázal sám... že jsem tak nějak nezávislý"* [I was happy that I managed it myself... that I'm sort of independent]—reveals the Hermit's core motivation: maintaining zones of genuine human mastery.

This isn't Luddism but what we might call "selective augmentation." The Hermit maintains sophisticated boundaries, delegating mechanical tasks while preserving conceptual challenges. As MH articulates: *"agent na testování a distribuci by byl super"* [an agent for testing and distribution would be great]—tools for tedium, not thought. This echoes what Camus might recognize as the eternal return—finding meaning through cyclical struggle, where each return to unaugmented work reaffirms human capability.

The Hermit's wisdom lies in recognizing that convenience can atrophy essential skills. By maintaining "challenge zones" where AI cannot enter, they preserve not just competence but the very possibility of growth through struggle.

**Key Practice**: Maintaining deliberate "challenge zones" where they work without AI to preserve fundamental competencies and the satisfaction of independent achievement.

### 7.6.5   The Cartographer: Architect → Charter of Unknown Territories

*"Vedu si deník AI chyb. Je to fascinující katalog toho, jak AI nerozumí kontextu."* [I keep a diary of AI errors. It's a fascinating catalog of how AI doesn't understand context.] —PJ

The Cartographer specialization emerges from high conscientiousness combined with systematic curiosity. Like Borges mapping infinite libraries, they document AI's capabilities and limitations with meticulous precision. Their maps become navigational aids for others traversing this new terrain. They embody Kant's critical philosophy, carefully delineating what AI can and cannot know.

PJ's error diary exemplifies the Cartographer's approach—not condemning AI's failures but cataloging them systematically. This creates what organizations desperately need: empirical understanding of where AI helps and where it hinders. The Cartographer's "důvěřuj, ale prověřuj" [trust but verify] philosophy enables teams to leverage AI while avoiding its pitfalls.

MV demonstrates how Cartographers serve organizational needs: using systematic documentation to guide "when to use which AI tool for what purpose," creating what he calls "mapy efektivity" [efficiency maps] that optimize human-AI collaboration patterns.

**Key Practice**: Creating comprehensive documentation of AI patterns, failures, and workarounds for team knowledge bases.

### 7.6.6 The Shapeshifter: Adapter → Protean Navigator of Modes

*"My personality depends on the current mood of the day."* —I11

The Shapeshifter represents ultimate adaptability—developers who are transcended fixed relationships with AI to achieve fluid navigation across all modes. Like water in Taoist philosophy, they take the shape required by context, embodying Ovid's Metamorphoses in their constant transformation.

This specialization requires sophisticated meta-awareness—knowing not just how to use each AI mode but when each serves best. Shapeshifters exhibit no loyalty to particular tools or approaches, instead maintaining what we might call "pragmatic fluidity." They might be Promethean in morning creative sessions, Conductor during afternoon team coordination, and Hermit for evening deep work.

The Shapeshifter's flexibility comes with a cost—the lack of stable identity that other specializations provide. Yet in rapidly evolving AI landscapes, this fluidity may prove most adaptive, allowing continuous optimization as capabilities and contexts shift.

**Key Practice**: Developing meta-awareness of when each AI mode serves best, switching seamlessly based on context.

### 7.6.7 Synthesis: The Permeability Principle in Specializations

A crucial finding concerns how these AI specializations permeate beyond professional boundaries. As MV observed: *"Už to prostupuje všema složkami života... práce a domácnost a rodina"* [It already permeates all aspects of life... work and household and family]. Each specialization creates distinct life patterns:

- **Prometheans** report creative AI use spilling into hobbies, writing, and personal projects, leveraging pre-AI creative processes as quality baselines

- **Conductors** find themselves orchestrating not just code but family logistics with AI assistance, using experience to guide AI agent teams effectively

- **Hermits** create protective boundaries that must be constantly maintained against AI encroachment, remembering what has been lost

- **Cartographers** begin documenting patterns everywhere, seeing life itself as territory to map, comparing AI present to human past

- **Shapeshifters** report identity fluidity extending beyond work into fundamental self-conception, navigating modes others do not recognize yet

This permeability suggests that choosing an AI specialization is not merely a professional decision but an existential one—shaping not just how we work but who we become.

## 7.7   Study Limitations

Several limitations bound our findings:

**Sampling Constraints**: Participants were predominantly Czech/Slovak developers and Gen Z students, potentially limiting cultural and generational generalizability.

**Temporal Limitations**: We captured experiences during rapid AI evolution (2021-2025). Patterns may shift as AI capabilities and cultural attitudes evolve.

**Methodological Trade-offs**: The phenomenological depth of Study VI (n=7) trades breadth for richness, while thematic analyses (n=35) may miss individual nuances.

**Language Effects**: Bilingual interviews revealed different experiential registers, but monolingual English-speaking developers might experience AI collaboration differently.

**Self-Selection Bias**: Participants willing to discuss AI experiences may overrepresent early adopters or those with strong opinions.

## 7.8 Discussion: Extending ROMA for Human ⌗ AI Futures

### 7.8.1 Theoretical Contributions

This cycle advances Behavioral Software Engineering through three key contributions:

**AI Specializations Framework**: We demonstrate how personality archetypes manifest differently in AI contexts, creating predictable patterns of adoption, resistance, and adaptation. The framework reveals that personality-driven motivational patterns persist even as professional practices transform radically.

**Synthetic Relatedness Conceptualization**: The identification of quasi-social satisfaction from AI interaction extends SDT into non-human domains, challenging assumptions about relatedness needs. This "bracketed anthropomorphism" enables productive engagement while maintaining ontological clarity.

**Competence Metamorphosis Model**: The shift from implementation to orchestration competence represents a fundamental change in professional identity. This isn't skill replacement but skill transformation—from knowing how to knowing when, from creating to curating, from solving to framing. As Wiener (1960) predicted, this metamorphosis increases rather than decreases cognitive demands, requiring developers to operate at higher levels of abstraction while maintaining readiness for granular intervention.

**Methodological Grounding:** While our approach shares elements with grounded theory—particularly in how the AI specializations emerged inductively from participant experiences—we do not claim full grounded theory methodology. Rather, we employed what Braun & Clarke (2013) term 'theoretically flexible' thematic analysis combined with phenomenological interpretation, allowing patterns to emerge while remaining anchored in SDT and personality frameworks.

**Methodological Innovation:** We contribute a *reproducible Four-Form IPA methodology* specifically adapted for Human ⌗ AI collaboration research. Our approach extends Smith et al.'s (2009) traditional IPA protocol by incorporating bilingual hermeneutics and AI-specific phenomenological bracketing. The methodology traces how raw phenomenological data transforms through systematic interpretation into theoretical insight.

Section 7.3.5 and Study VI provide the complete analytical trail for key themes, enabling researchers to understand not just what we found but how phenomenological insights emerge. This transparency advances IPA methodology in technology research contexts where researchers must bracket both technological determinism and anthropomorphic projections.

### 7.8.2    The ROMA AI Adapter: Comprehensive SWOT Analysis

The following sections (7.8.2-7.8.4) provide detailed analysis of the framework's strengths, weaknesses, opportunities, and threats, followed by strategic implications and ethical considerations. This comprehensive evaluation enables VSEs and individual developers to make informed decisions about AI integration strategies.

The ROMA AI Adapter framework offers both theoretical insights and practical guidance for VSEs navigating AI integration. A systematic SWOT analysis reveals its multifaceted implications:

*Strengths*

1. Philosophical Depth with Practical Application: The framework's grounding in philosophical traditions—from Prometheus to Thoreau—provides conceptual anchors that resonate across cultures and contexts. This depth prevents the framework from becoming merely instrumental, instead offering developers meaningful narratives for understanding their AI relationships.

2. Predictive Power Through Personality Alignment. By mapping personality profiles to AI specializations, the framework enables:

- Proactive team composition strategies
- Targeted support for struggling developers
- Optimized AI tool selection based on individual traits
- Reduced friction in AI adoption through personality-aware implementation

3. Preservation of Human Agency: Unlike frameworks that position AI as inevitable replacement, the ROMA AI Adapter emphasizes complementarity. Each specialization maintains distinct zones of human control—creative judgment for Prometheans, quality oversight for Conductors, challenge preservation for Hermits.

4. Dynamic Adaptation Capability: The framework acknowledges that developers may shift between specializations as they gain experience or as contexts change. This fluidity prevents rigid categorization while providing useful heuristics for understanding behavioral patterns.

*Weaknesses*

1. Temporal Specificity: Captured during a particular moment in AI evolution (2021-2025), some patterns may prove ephemeral. As AI capabilities advance, current specializations might obsolete or transform in unexpected ways.

2. **Cultural and Linguistic Constraints**: Based primarily on Czech/European developers, the framework may not fully capture specialization patterns in other cultural contexts where human-AI relationships might manifest differently

3. **Measurement Challenges** – The framework cannot cleanly separate:

- "Accidental" improvements (technology-dependent, will change with AI evolution)
- "Essential" improvements (framework-dependent, likely to persist) This entanglement makes quantitative validation challenging.

4. **Complexity of Implementation**: The philosophical richness that strengthens the framework also complicates practical application. VSE managers may struggle to operationalize concepts like "polidštění" or "synthetic relatedness" without substantial interpretation.

## *Opportunities*

1. **Cross-Cultural Validation**: Extending research to Silicon Valley, Bangalore, Beijing, and other tech hubs could reveal universal patterns versus culturally specific adaptations, enriching the framework's global applicability.

2. **Integration with Emerging Standards.** As AI collaboration tools mature, opportunities emerge to embed ROMA AI Adapter insights into:

- ISO/IEC standards for AI-augmented development
- Corporate AI governance frameworks
- Educational curricula for next-generation developers

3. **Longitudinal Evolution Tracking.** Following developers as they navigate changing AI landscapes could reveal:

- Specialization stability versus fluidity
- Career trajectory impacts
- Long-term satisfaction patterns
- Skill preservation strategies

4. **Team Composition Optimization.** The framework enables sophisticated team design:

- Balancing specializations for project needs
- Creating mentorship pairs (e.g., Hermit teaching preservation to Promethean)

- Optimizing human-AI-human triads for maximum effectiveness:
    - Triads can leverage specializations—Prometheans generate, Cartographers verify, Conductors orchestrate—addressing pair programming's traditional ~15% overhead (Cockburn & Williams, 2001; Zieris & Prechelt, 2019) while preserving benefits.

### *Threats*

1. Rapid AI Evolution: Current specializations assume certain AI limitations. Breakthrough capabilities—true reasoning, genuine creativity, authentic empathy—could fundamentally alter or invalidate existing patterns.

2. Reductionist Misapplication: The framework's archetypes risk becoming stereotypes if applied mechanistically. Developers are not fixed types but dynamic beings whose AI relationships evolve with experience and context.

3. Ethical Blind Spots. Focusing on optimization and adaptation might obscure critical questions:

- Should all human capabilities be augmented?

- What are the moral implications of synthetic relatedness?

- How do we preserve human dignity in AI collaboration?

4. Generational Arbitrage Closure: The framework partially depends on developers who remember pre-AI programming. As this generation retires, we lose critical perspective for evaluating AI's impact. Future developers may lack the experiential baseline to recognize what's being lost or transformed.

### 7.8.3   Strategic Implications

The SWOT analysis reveals the ROMA AI Adapter as a robust but evolving framework. Its strengths lie in providing meaningful narratives and practical guidance for navigating Human  AI collaboration. Its vulnerabilities center on temporal specificity and measurement challenges.

For VSEs, the framework offers immediate value through:

- **Short-term**: Personality-aware AI tool selection and team composition

- **Medium-term**: Specialization-based career development paths

- **Long-term**: Preservation strategies for essential human capabilities

The framework's ultimate test will be whether it helps developers not just adapt to AI but shape that adaptation consciously—maintaining human flourishing while embracing artificial augmentation. As participant MV wisely noted: "Mám výhodu, že pamatuji si doby, když jsem programoval bez agenta" [I have the advantage that I remember times when I programmed without an agent].

The ROMA AI Adapter seeks to extend this advantage to all developers—not through nostalgia but through conscious choice about which human capabilities to preserve, which to augment, and which to release. In this conscious choosing lies both the framework's greatest strength and its most profound responsibility.

### 7.8.4  Ethical Considerations of Professional Evolution

The ROMA AI Adapter illuminates not just how developers adapt but the ethical terrain they navigate. While participants like PΓ expressed optimism—"if you learn how to work with AI successfully, you will never be replaced by it"—deeper analysis reveals complex ethical terrain. Three critical dimensions emerge from our analysis:

#### *The Authenticity Paradox*

When MV adds his "final touch" to AI-generated code, he claims authorship through transformation. Yet this raises fundamental questions: What constitutes authentic creation in human-AI partnerships? The "polidštění" process—making AI output gentle and personal—suggests authorship might reside not in origination but in the uniquely human act of imbuing code with intention and care.

This matters ethically because:

- **Attribution**: Who receives credit when AI generates 70% of the code?
- **Responsibility**: Who bears liability for AI-introduced bugs?
- **Professional Identity**: What does it mean to be a "developer" when development becomes orchestration?

#### *The Competence Illusion*

Our participants revealed a troubling pattern: AI can create competence illusions where developers feel capable in domains they do not truly understand. As one participant noted about interviews: "prakticky jsem schopen udělat cokoliv, ale pokud se mě někdo zeptá na konkrétní otázku z frameworku, tak už to bude problém" [I can practically do anything, but if someone asks me a specific framework question, that's a problem].

Ethical implications include:

- **Professional Integrity**: Can we claim expertise in AI-assisted domains?

- **Safety Criticality**: What happens when competence illusions meet safety-critical systems?

- **Educational Responsibility**: How do we ensure foundational understanding survives AI augmentation?

### *The Generational Divide*

Perhaps most pressing is the ethical obligation to future developers. Those who "remember programming without agents" bear responsibility for preserving essential capabilities. The Hermit's protective minimization and the Cartographer's systematic documentation represent ethical stances—guardianship of human capabilities that risk extinction.

Key considerations:

- **Knowledge Transfer**: What must be preserved for future generations?

- **Skill Sovereignty**: Which human capabilities should remain inviolate?

- **Collective Wisdom**: How do we encode the pre-AI baseline into educational and professional practices?

### *Toward Ethical AI Specialization*

The ROMA AI Adapter doesn't prescribe universal ethics but recognizes that each specialization embodies an ethical stance:

- *Prometheans* must guard against creative appropriation without understanding

- *Conductors* bear responsibility for the humanity they add to AI output

- *Hermits* preserve essential skills but risk isolation

- *Cartographers* document for collective benefit but may ossify practices

- *Shapeshifters* maintain flexibility but risk ethical inconsistency

The framework thus becomes not just descriptive but normative—helping developers recognize the ethical weight of their chosen specialization and navigate accordingly. As we shape AI's role in software development, we simultaneously shape what it means to be human in this profession.

## 7.9 Conclusion: Navigating the Transformed Landscape

The morning mist has not lifted but transformed—and in that transformation, we discover not loss but metamorphosis. Where Cycle 2 mapped human terrain through the ROMA framework, Cycle 3 unveils how developers maintain essential humanity while embracing artificial augmentation.

Through 42 developer voices, we witness not technological adoption but existential navigation. Each specialization represents a unique answer to the fundamental question: How do we maintain human flourishing while embracing artificial augmentation?

The Prometheans steal fire from artificial gods, maintaining creative sovereignty through their "ten ideas, discard nine" philosophy. The Conductors orchestrate human-AI symphonies, practicing "polidštění"—transforming mechanical output into code with soul. The Hermits guard human sanctuaries, recognizing that convenience atrophies essential skills. The Cartographers chart unknown territories, documenting AI's capabilities and limitations for those who follow. The Shapeshifters flow like water between all modes, embodying ultimate adaptability.

These aren't rigid categories but fluid stances. MV progresses from main programmer to orchestrating reviewer. NK oscillates between AI-first development and protective minimization. MM transforms brainstorming partnerships into selective tool use. Each journey reveals sophisticated strategies—selective engagement, protective boundaries, cyclic patterns—preserving irreducible humanity while leveraging AI capability. In their struggles and triumphs, we see not the future of programming but futures—plural, diverse, shaped by personality, choice, and the irreducible complexity of human experience.

These insights prepare the foundation for Cycle 4's integration with ISO/IEC 29110 standards, translating phenomenological understanding into practical implementation guidance.

As winter fog clings to morning hills, obscuring distant vistas while revealing immediate paths, so our investigation illuminates the near terrain while acknowledging far horizons remain shrouded. The ROMA AI Adapter stands as both map and compass. Like Borges' map that covers the entire territory, it risks overwhelming with detail. Yet also like ancient maritime charts marking "here be dragons," it acknowledges vast unexplored regions.

The journey continues, but the message crystallizes: In choosing how we engage with AI, we choose who we become. The ROMA framework, enhanced with AI specializations, offers not prescriptions but possibilities—ways of maintaining agency, preserving growth, and finding meaning in an age where the boundaries between human and artificial, between creation and curation, between authentic and synthetic, blur like fog at dawn's edge.

Shape your future equipped with knowledge for worthwhile choices. Whether as Promethean fire-bringer, Conductor of digital symphonies, Hermit guarding human sanctuaries, Cartographer mapping territories, or Shapeshifter flowing between modes, let your specialization emerge from understanding rather than accident, from choice rather than drift. For in the end, how we treat AI reveals how we treat ourselves—and in that reciprocal relationship lies both challenge and hope for software development's AI-augmented future.

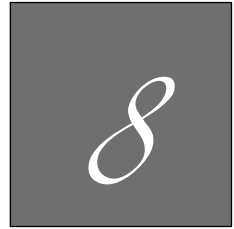

# DSR Cycle 4 "Design" – Codifying Wisdom

## Integration with ISO/IEC 29110 Standards

*Where morning mist meets institutional memory, where lived experience crystallizes into shared practice, where the personal becomes professional standard—here we codify not just processes but possibilities.*

The journey that began in Cycle 1's empirical exploration, evolved through Cycle 2's ROMA framework, and transformed in Cycle 3's AI specializations now reaches its institutional apotheosis. This chapter chronicles the translation of phenomenological insight into a design for extending ISO/IEC 29110 profiles—not as bureaucratic ossification but as what Polanyi (1966) might recognize as the codification of tacit knowledge, making the implicit explicit without losing its essential wisdom.

The ISO/IEC 29110 standards series for Very Small Entities represents more than technical documentation; it embodies recognition that software development's future lies not in monolithic corporations but in nimble teams where every individual matters. In VSEs of up to 25 people, personality misalignment doesn't just reduce productivity—it can doom entire ventures. Yet these same constraints create opportunity: small teams can adapt quickly, experiment boldly, and—with proper guidance—transform personality diversity from challenge into competitive advantage.

This chapter presents the systematic design, development, and validation of an extension design for ISO/IEC 29110 profiles that enables VSEs to implement personality-driven role optimization within established frameworks. More than technical specification, it represents a stage in our design science journey—where empirical pattern becomes practical framework, each transformation preserving the human insight at its core. The journey continues in Chapter 9 with field testing through the PDPPA application and triangulation of all insights.

## 8.1   Theoretical Foundations: Standards as Living Practice

The theoretical landscape for Cycle 4 emerges at the intersection of software engineering standards, motivational psychology, and the emerging reality of AI-augmented development. Yet unlike previous cycles' phenomenological depth, here we engage with theory as it manifests in practice—standards not as rigid prescriptions but as what Bowker & Star (1999) term "boundary objects," flexible enough to adapt to local contexts while maintaining coherent identity across communities.

### 8.1.1   ISO/IEC 29110: A Framework That Breathes

The ISO/IEC 29110 standards series emerged from a profound recognition: the vast majority of software organizations worldwide are not Google or Microsoft but small teams struggling to balance quality with survival. Laporte et al. (2018) document how these standards evolved specifically for VSEs—organizations where elaborate processes become anchors rather than aids.

The framework's elegance lies in its graduated profiles:

- **Software Basic Profile** (ISO/IEC 29110-4-1, 2018): This foundation defines essential Project Management (PM) and Software Implementation (SI) processes suitable for small projects with relatively straightforward requirements. The profile creates structure—delineating clear activities, tasks, and roles—without strangling flexibility. Like a jazz standard that provides harmonic framework while enabling improvisation, the Basic Profile offers enough guidance to prevent chaos while preserving space for creativity.

- **Agile Software Development Guidelines** (ISO/IEC FDIS 29110-5-4, 2025): Recently formalized as international standard, these guidelines map agile's dynamic energy onto the Basic Profile's stable structure. Seven events (E1-E7) create a combined PM + SI rhythm—sprint planning's anticipation, daily scrum's pulse, retrospective's reflection—each a moment where personality-driven optimization can naturally integrate.

Our extension design leverages both frameworks by implementing personality-driven tasks within the Basic Profile's structured PM activities while utilizing the Agile Guidelines' events as integration points for role optimization. As noted by Galván-Cruz et al. (2021), this complementary approach aligns with VSEs' need for lightweight but effective processes that can adapt to resource fluctuations and changing project demands.

An area lead from a Fortune 500 company's EMEA branch validated this approach: "I prefer lightweight methodologies like ISO/IEC 29110 with sprint planning, deliveries, retrospectives as key activities." Though his organization doesn't explicitly implement the standard, their processes parallel its structure—suggesting the extension design codifies patterns that emerge naturally in successful VSEs.

### 8.1.2 Motivation as Organizational Lifeblood

In VSEs, motivation isn't a "nice-to-have"—it is existential. When a five-person startup loses one developer to burnout, they do not lose 20% of workforce; they often lose irreplaceable domain knowledge, client relationships, and the ineffable chemistry that made the team work. This reality transforms motivation from HR concern to survival strategy.

Self-Determination Theory provides our theoretical backbone, but in VSE contexts, the three basic needs—autonomy, competence, and relatedness—manifest with particular intensity:

**Autonomy** in VSEs means not just task choice but often company direction. The developer who prefers Navigator roles might also shape product vision through their gift for articulation.

**Competence** becomes multifaceted when everyone wears multiple hats. The Craftsperson who excels at solo debugging might struggle with client presentations, making role optimization crucial for both individual satisfaction and team success.

**Relatedness** in small teams transcends professional courtesy—it is the daily reality of sharing not just code but dreams, fears, and often physical space. Personality clashes that large organizations absorb through distance become daily friction in VSEs.

Our industry validator confirmed this criticality: "Intrinsic motivation is very important for delivery... We evaluate motivational scores anonymously and measure manager effectiveness by team members' willingness to recommend the company." This isn't sentiment—it is survival wisdom from someone managing teams where every individual's motivation directly impacts organizational viability.

### 8.1.3 The AI Imperative in Resource-Constrained Contexts

The integration of AI tools within VSEs represents not technological fashion but practical necessity. As our Cycle 3 findings revealed, AI doesn't replace human collaboration but transforms it—and nowhere is this transformation more crucial than in resource-constrained environments.

Our industry interview illuminated this reality: "AI functions like a colleague you can ask for help anytime... It can mimic both pilot and navigator roles in pair programming." For VSEs unable to maintain full pairing coverage, AI offers what we might term "synthetic supplementation"—not replacing human connection but filling gaps when humans aren't available.

Yet this integration demands sophistication. The Promethean who thrives with AI's creative chaos might destabilize a VSE needing predictable delivery. The Hermit who strategically minimizes AI might preserve essential competencies the organization can't afford to lose. Our extension must enable VSEs to navigate these dynamics consciously, matching not just personalities to roles but personalities to AI modes—all within resource constraints that make every choice consequential.

## 8.2   Design Process: From Insight to Institution

The journey from phenomenological insight to an extension design for ISO/IEC 29110 profiles demanded methodological rigor that honored both scientific validity and practical utility. Our approach wove together established Design Science Research principles with the unique demands of standards development, guided by Rode & Svejvig's (2022) multidimensional evaluation framework.

### 8.2.1   The Three-Cycle Dance

Following Hevner's (2007) three-cycle DSR view, complemented by Rode & Svejvig's (2022) emphasis on relevance, rigor, and reflexivity, our development process resembled less linear progression than spiral dance, each cycle informing and transforming the others:

**Relevance Cycle**: We began where practitioners live—in the daily struggles of VSE teams trying to balance personality diversity, resource constraints, and delivery pressures. Studies I-IV provided empirical grounding, while industry consultation revealed how these challenges manifest in practice. This wasn't abstract requirements gathering but what ethnographers might recognize as "thick description" of VSE reality, shifting our human-purist literary review-based focus to human-AI realities.

**Design Cycle**: The iterative development resembled sculptural refinement—each iteration revealing more clearly the essential form within the marble. We mapped ROMA framework components to ISO/IEC 29110 processes, but mapping proved insufficient. True integration required what our Conductor archetype would recognize as orchestration—ensuring each element enhanced rather than complicated the whole.

**Rigor Cycle**: Theoretical soundness came not from abstract principles but from their embodiment in practice. When Working Group 24 members—guardians of the ISO/IEC 29110 standards—reviewed our extension, they brought not just technical expertise but lived experience of what makes standards succeed or fail in VSE contexts.

The reflexivity dimension that Rode & Svejvig (2022) emphasize manifested through our willingness to adapt the extension design based on emerging insights. When industry feedback revealed privacy concerns, we strengthened ethical safeguards. When academic reviewers questioned practical viability, we modularized the approach for incremental adoption.

### 8.2.2  Artifact as Living Document

Following Gregor & Hevner's (2013) framework for DSR contributions, our extension design represents an "improvement" contribution, enhancing existing solutions (ISO/IEC 29110 profiles) with new knowledge (personality-driven role optimization). The extension design was developed to meet Hevner et al.'s (2004) criteria of utility, quality, and efficacy for design artifacts.

The artifact development followed Peffers et al.'s (2006) DSR process model:

- **Problem Identification**: Recognizing that VSEs need structured yet flexible approaches to role optimization that integrate with existing standards.

- **Objectives Definition** required balancing comprehensiveness with simplicity. VSEs can't implement elaborate personality optimization systems, yet oversimplification would reduce the framework to useless generality. We sought what designers call "satisficing elegance"—just enough structure to guide without constraining.

- **Design & Development** became an exercise in translation across domains. How do you transform "The Explorer needs creative challenges" into standard-compliant task specifications? How do you encode "synthetic relatedness from AI interaction" into process guidelines? Each translation required preserving meaning while achieving clarity.

- **Demonstration**: Illustrating the extension's application through detailed process flows and implementation examples.

- **Evaluation**: Validating the extension through expert review, including members of WG24 who oversee the ISO/IEC 29110 standards series, and a pilot interview with an industry leader.

- **Communication**: Documenting the extension in this dissertation, the peer-reviewed PeerJ CS article (Study IV), and ACIE 2025 conference paper (Study V).

### 8.2.3  Validation as Conversation

Validation in DSR often implies unidirectional assessment—does the artifact meet criteria? Our validation process more resembled extended conversation across communities of practice:

**Academic Peer Review** through PeerJ Computer Science spanned over two years—not delay but depth. Each review cycle revealed new facets, challenged assumptions, forced clarity. Reviewer 2's insistence on connecting personality assessment to hiring practices transformed Task 1 from afterthought to cornerstone.

**Expert Review** by WG24 members brought different wisdom. Prof. Claude Laporte and Dr. Mirna Muñoz evaluated not just technical correctness but practical viability—would VSEs actually implement these tasks? Their approval meant more than academic validation; it meant the extension could live in practice.

**Industry Validation** through our Fortune 500 regional CTO interview provided the ultimate test—does this resonate with those living the reality we are trying to improve? His confirmation that they already practice informal personality consideration, that AI serves as collaborative partner, that motivation measurement drives management assessment—these validated not just our extension but our entire research trajectory.

## 8.3   Output 4.1: VSE Extension Design for Human-AI Role-Optimized Collaborative Programming

The extension design itself embodies our research journey's insights—seven tasks that transform personality awareness from intuition to process, from individual wisdom to organizational capability. Each task represents careful integration, like new instruments joining an established ensemble.

### 8.3.1   Architecture of Integration with ISO/IEC 29110 PM and SI Processes

Our extension design spans both Project Management (PM) and Software Implementation (SI) processes, recognizing that personality-driven optimization touches every aspect of software development. This isn't bolt-on addition but what systems thinkers might recognize as "structural coupling"—the extension design and existing processes co-evolve, each enriching the other.

Figure 5 illustrates this integration, showing how personality-driven tasks weave through ISO/IEC 29110 profiles. When personality assessment enters Project Planning (PM.1), it doesn't just add a task—it reframes how teams think about composition. When motivation measurement becomes part of Sprint Retrospectives (E7), it transforms reflection from technical post-mortem to human-centered learning (ISO 9241-210, 2019).

The integration follows what Christopher Alexander (1977) might recognize as pattern language—each task creates context for others, forming a coherent whole. The personality assessment in hiring (Task 1) enables meaningful role assignment (Task 4), which enables motivation measurement (Task 5), which informs strategy updates (Task 6), creating continuous improvement cycles that honor both individual and organizational needs.

## 8.3.2 Seven Tasks as Seven Bridges

Each task represents not just procedural addition but a bridge between current practice and enhanced possibility:

### Task 1: Integrate Personality Assessment in Hiring (PM.1) *Bridge: From intuitive "culture fit" to systematic understanding*

This task emerged from recognizing that personality alignment begins before the first line of code. By incorporating Big Five assessment into recruitment, VSEs can compose teams with intention—seeking not uniformity but harmonious diversity.

- **Objective**: Ensure new hires complement existing team dynamics through personality-aware recruitment, creating balanced teams that leverage diverse strengths.

- **Implementation**: Administer the BFI-10 assessment during final interview stages. Map candidate profiles against current team composition using ROMA archetypes. Document results in Recruitment Records while ensuring GDPR compliance and candidate consent.

- **Artifacts**: Updated Job Posting templates, Candidate Evaluation Forms with personality considerations, Team Composition Matrix

Our industry validator confirmed: "We're connecting complementary personalities," though he emphasized critical caveats about privacy and consent. This isn't about discrimination but discernment—careful consideration of how individuals might flourish within specific team dynamics.

### Task 2: Assess Multidimensional Work Motivation (PM.2) *Bridge: From assumption to measurement*

Motivation in VSEs often remains felt but unmeasured—everyone senses when team energy drops but lacks vocabulary to discuss it productively. The Multidimensional Work Motivation Scale (MWMS; Gagné et al., 2015) provides that vocabulary, transforming vague unease into actionable insight.

- **Objective**: Establish baseline motivation metrics and track evolution throughout project lifecycle, enabling data-driven interventions before motivation crises emerge.

- **Implementation**: Administer MWMS during project kickoff (baseline), at sprint boundaries (tracking), and project closure (retrospective). Store results in Project Status Records. Use anonymized aggregate data for team health discussions.

- **Artifacts**: Motivation Baseline Report, Sprint Motivation Trends Dashboard, Project Motivation Summary

The regional CTO's practice validates this approach: "We evaluate motivational scores... and assess managers based on team members' willingness to recommend the company." This creates accountability where motivation becomes as measurable as velocity—and arguably more predictive of long-term success.

### Task 3: Define the Big Five Assessment Strategy (PM.2) *Bridge: From ad hoc to ethical system*

This task addresses the shadow side of personality assessment—the potential for misuse, discrimination, or reductive labeling. By formalizing assessment strategy, VSEs create what bioethicists might recognize as "procedural justice"—fair processes that protect individual dignity while enabling organizational benefit.

- **Objective**: Establish ethical, consistent, and legally compliant personality assessment processes that balance organizational needs with individual privacy rights.

- **Implementation**: Create Assessment Policy document defining when/how assessments occur, data retention periods, access controls, and usage boundaries. Train Project Managers on ethical interpretation. Review policy annually with legal counsel.

- **Artifacts**: Personality Assessment Policy, Consent Forms, Data Protection Procedures, Training Materials

The strategy must balance multiple tensions: comprehensive enough to be useful, simple enough for VSE implementation, ethical enough to honor human complexity. This isn't bureaucracy but boundary-setting—creating safe spaces for personality-aware development.

### Task 4: Pairing Task (SI: E3, E4) *Bridge: From availability-based to personality-conscious pairing*

Traditional pair programming often relies on whoever's available pairing together. Task 4 transforms this into conscious practice. During Sprint Planning (E3), personality-aligned pairs become explicit backlog items.

- **Objective**: Optimize pair programming effectiveness by matching complementary personality types, maximizing both productivity and developer satisfaction.

- **Implementation**: During Sprint Planning, create explicit "Pairing Tasks" in Sprint Backlog. Use ROMA pairing principles (e.g., Explorer-Architect for design, Orchestrator-Craftsperson for implementation). Track pairing effectiveness in Sprint Logs. Allow flexibility for task-specific adjustments.

- **Artifacts**: Enhanced Sprint Backlog with pairing assignments, Pairing Effectiveness Log, Role Rotation Schedule

Our validator noted they use "asynchronous pair programming" due to cost constraints—personality optimization could make these limited interactions more effective. When teams can only afford occasional pairing, ensuring Orchestrator-Craftsperson complementarity matters even more.

### Task 5: Measure Intrinsic Motivation Impact (PM.3 & SI: E4, E7)
*Bridge: From hoping to knowing*

This task closes the feedback loop, measuring whether personality-based assignments actually enhance motivation. Using the Intrinsic Motivation Inventory after pairing sessions provides immediate feedback, while longitudinal MWMS tracking reveals trends.

- **Objective**: Quantify the impact of personality-based role assignments on developer motivation, creating evidence base for continuous improvement.

- **Implementation**: Deploy IMI short form (7 items) after each pairing session via anonymous digital survey. Aggregate results weekly. Present trends during Sprint Retrospectives. Use findings to adjust pairing strategies mid-sprint if needed.

- **Artifacts**: IMI Survey Forms, Motivation Impact Reports, Retrospective Discussion Guides

The industry practice of "anonymous surveys and feedback services" confirms feasibility. VSEs often resist measurement overhead, but motivation measurement pays dividends—catching problems before they become departures, identifying successful patterns for replication.

### Task 6: Review and Update Pairing Strategies (PM.3 & SI: E6, E7)
*Bridge: From static to adaptive practice*

Software development teaches humility—what works today may fail tomorrow. Task 6 embeds continuous learning into the process, using retrospectives not just for technical reflection but human optimization. When the Explorer-Architect pairing that created magic last sprint produces friction this sprint, the process enables rapid adaptation.

- **Objective**: Create continuous improvement cycles for personality-based practices, ensuring strategies evolve with team dynamics and project needs.

- **Implementation**: Dedicate 15 minutes of each Sprint Retrospective to pairing effectiveness review. Analyze motivation trends alongside velocity metrics. Document successful patterns and failed experiments in Lessons Learned Repository. Update pairing guidelines quarterly.

- **Artifacts**: Updated Pairing Guidelines, Retrospective Templates with motivation focus, Corrective Action Register entries

"Retrospectives for continuous improvement" already anchor agile practice; this task simply expands their scope to include the humans doing the work, not just the work itself.

### Task 7: Support Future Planning and Knowledge Transfer (PM.4)
*Bridge: From individual to institutional knowledge*

VSEs often lose hard-won wisdom when key members leave. Task 7 ensures personality-based insights survive transitions, documenting not just what worked but why.

- **Objective**: Preserve and transfer personality-based optimization knowledge across projects and team transitions, building organizational capability.

- **Implementation**: During Project Closure, create Personality Optimization Summary documenting successful pairings, motivation patterns, and adaptation strategies. Include in Project Closure Report. Share in cross-project knowledge sharing sessions. Update organizational best practices repository.

- **Artifacts**: Personality Optimization Summary, Updated Best Practices Repository, Knowledge Transfer Checklist

The area lead's emphasis on knowledge sharing validates this approach. When teams discover that their most productive pairs combine high openness with high conscientiousness, that insight should outlive any individual's employment.

### 8.3.3   Process Flow and Implementation as Living System

Figure 5 presents the technical architecture of our extension design, but diagrams capture structure while missing spirit. The process flow represents not mechanical sequence but organic rhythm— each task creating conditions for the next, feedback loops ensuring continuous adaptation.

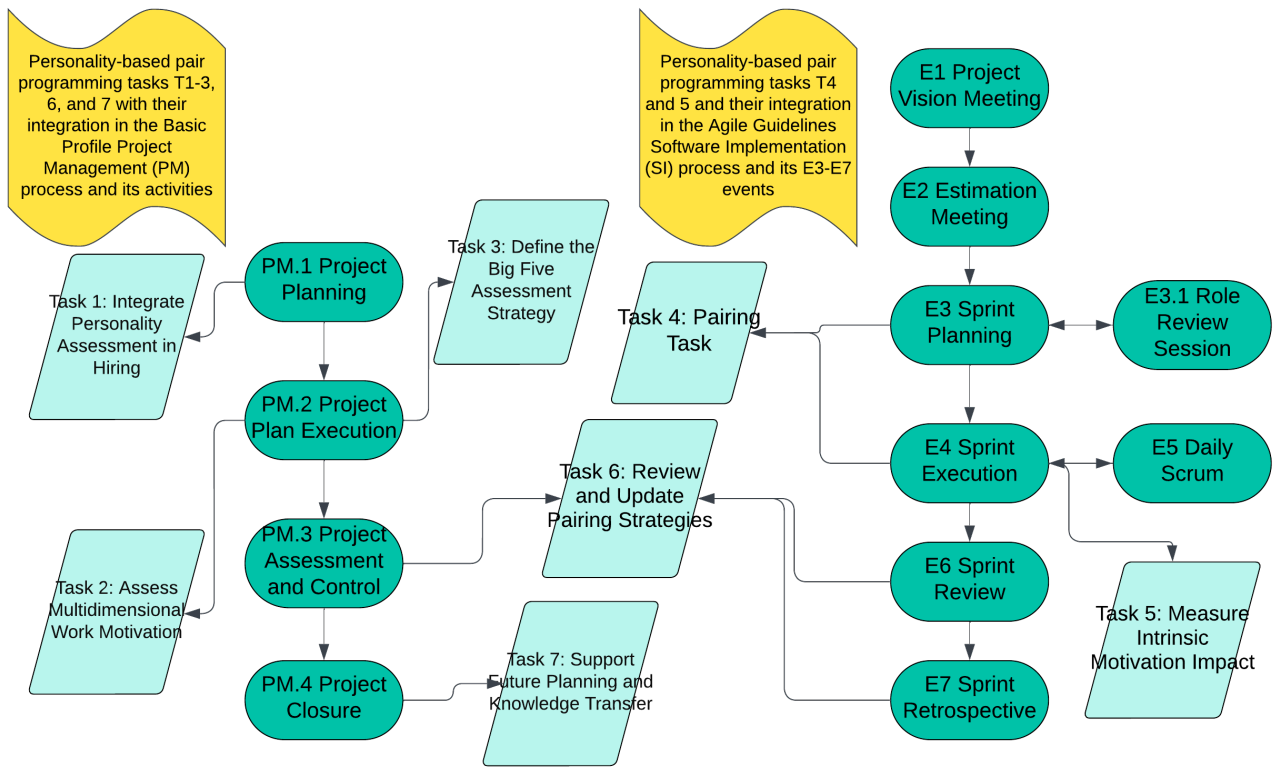

*Figure 5: Process flow of personality-based pair programming tasks in the ISO/IEC 29110 profiles*

*Figure 5 illustrates how personality-based pair programming tasks integrate into both the Project Management (PM) and Software Implementation (SI) processes as outlined in the ISO/IEC 29110 Agile Software Development Guidelines. The diagram maps out key touchpoints for new and modified activities, such as incorporating personality assessments during hiring (PM.1) and measuring intrinsic motivation after pair programming sessions (PM.3, E7). Arrows indicate workflow sequence and dependencies, showing how personality-driven tasks (T1-T7) enhance existing activities across project lifecycle phases from planning through closure.*

The flow begins before the project itself, with personality assessment during hiring (PM.1) establishing team composition possibilities. This feeds into motivation baseline measurement (PM.2), creating understanding of not just who people are but how they are currently experiencing work. The Big Five assessment strategy (PM.3) provides ethical framework, ensuring personality insights empower rather than constrain.

When development begins, the Pairing Task (E3, E4) transforms abstract understanding into lived practice. Developers get assigned based on conscious matching—Explorer pilots with Architect navigators, Orchestrators bridging between Craftspeople. The immediate motivation measurement (E4) provides rapid feedback—is this pairing creating energy or draining it?

Sprint retrospectives (E7) become moments of collective learning, where Task 6 transforms individual experiences into team wisdom. The final knowledge transfer (PM.4) ensures insights transcend project boundaries, creating what organizational theorists call "deutero-learning"— learning how to learn about personality optimization (Argyris & Schön, 1997). This approach follows Pries-Heje & Baskerville's (2008) emphasis on embedding design artifacts within organizational contexts in a manner that minimizes disruption while maximizing value.

### 8.3.4   Guidelines for Implementation as Evolution

For VSEs approaching this extension design, we offer principles rather than prescriptions:

1. **Start Where The Pain Is:** Begin with the most pressing challenge. If burnout threatens, start with motivation measurement. If pair programming consistently fails, begin with personality-based pairing. The extension design's modularity enables incremental adoption.

2. **Honor Privacy Profoundly:** Personality data touches human essence. Our industry validator emphasized "informed consent, data protection, and anti-discrimination safeguards"—not as legal compliance but ethical foundation. Create processes that reveal patterns while protecting individuals.

3. **Embrace Experimentation:** The Promethean's creative chaos might seem antithetical to VSE stability needs—until teams discover they excel at rapid prototyping for client demonstrations. The Hermit's AI resistance might seem problematic—until their preserved skills save the project when AI fails. Let experience rather than assumption guide implementation.

4. **Measure What Matters:** The extension design provides measurement tools, but numbers without context mislead. When IMI scores drop, investigate why—sometimes productive struggle temporarily reduces enjoyment while building crucial competence. When MWMS reveals extrinsic motivation dominance, consider whether current project pressures require short-term acceptance or signal deeper misalignment.

5. **Iterate Consciously:** Each retrospective offers optimization opportunity. The Explorer-Craftsperson pairing that failed spectacularly might succeed brilliantly with different task types. The Conductor who struggled with human orchestration might thrive coordinating AI agents. Use the extension design's feedback loops to refine continuously.

As one industry expert noted in our interview: "Start with simple implementations and evaluate through retrospectives and team discussions." This advice aligns with our extension's emphasis on iterative, flexible implementation.

## 8.3.5 Mapping to ISO/IEC 29110 Structure

To ensure seamless integration with existing VSE practices, Table 22 maps our seven tasks to specific ISO/IEC 29110 activities, artifacts, and roles. This dual perspective—viewing the extension through both our task structure and the standard's framework—provides implementers with multiple entry points while maintaining coherence.

Table 22: Personality-driven tasks mapped to ISO/IEC 29110 profiles structure

| ISO/IEC 29110 Element | ROMA Task Integration | Key Artifacts | Responsible Role | Timing |
|---|---|---|---|---|
| **PM.1 Project Planning** | | | | |
| PM.1.5 HR Identification | T1: Personality assessment in hiring | Job Posting, Candidate Evaluation Form | Project Manager | Pre-project |
| PM.1.6 Team Assembly | T1, T3: Team composition using ROMA | Team Composition Matrix | PM + Technical Lead | Project start |
| PM.1.13 Plan Verification | T3: Verify ethical assessment strategy | Assessment Policy, Consent Forms | PM + Customer | Planning phase |
| **PM.2 Project Execution** | | | | |
| PM.2.2 Change Requests | T2: Motivation-informed changes | Change Request Form with motivation impact | Change Control Board | Ongoing |
| PM.2.3 Team Meetings | T3, T6: Role effectiveness reviews | Meeting Minutes with motivation notes | PM + Team | Weekly/Sprint |
| **PM.3 Assessment & Control** | | | | |
| PM.3.2 Corrections | T5, T6: Motivation-driven corrections | Correction Register, IMI Results | PM + Technical Lead | Sprint end |
| PM.3.3 Progress Monitoring | T5: Track motivation alongside velocity | Progress Status Record, Motivation Dashboard | PM | Continuous |
| **PM.4 Project Closure** | | | | |
| PM.4.1 Closure Meeting | T7: Knowledge transfer session | Personality Optimization Summary | PM + Team | Project end |
| PM.4.2 Documentation | T7: Update best practices | Lessons Learned Repository | PM + Work Team | Post-project |
| **SI/Agile Events** | | | | |
| E1 Vision Meeting | T1, T3: Consider team personalities | Product Vision with team dynamics | Product Owner + Team | Project start |
| E2 Estimation | T4: Factor pairing efficiency | Story Points adjusted for pairs | Development Team | Sprint planning |
| E3 Sprint Planning | T4: Create Pairing Tasks | Sprint Backlog with pairing assignments | Scrum Master + Team | Sprint start |
| E3.1 Role Review* | T6: Adjust roles based on feedback | Role Assignment Matrix | Scrum Master | Sprint start |
| E4 Sprint Execution | T4, T5: Execute pairs, measure motivation | Pairing Log, IMI Surveys | Development Team | Daily |

| E5 Daily Scrum | T5: Brief motivation check-ins | Impediment List (including motivation) | Development Team | Daily |
|---|---|---|---|---|
| E6 Sprint Review | T6: Assess pairing impact on delivery | Sprint Review Report | All stakeholders | Sprint end |
| E7 Retrospective | T2, T5, T6: Review motivation trends | Retrospective Report, Action Items | Scrum Master + Team | Sprint end |

*Table 22 demonstrates how the seven personality-driven tasks integrate with existing ISO/IEC 29110 Project Management (PM) and Software Implementation (SI) activities. The table provides implementers with a practical roadmap showing where each ROMA task fits within standard VSE processes, which artifacts to produce, who is responsible, and when to execute each task. The mapping ensures that personality-based optimization enhances rather than disrupts established workflows, enabling VSEs to adopt these practices incrementally while maintaining ISO/IEC 29110 compliance.*

**Note:** *E3.1 represents a new sub-event we recommend adding to the standard Sprint Planning event.*

This comprehensive mapping may appear daunting, but VSEs should remember: start small, implement incrementally, and let success build momentum. It demonstrates how personality-driven optimization integrates naturally with ISO/IEC 29110's existing structure. Rather than creating parallel processes, our tasks enhance existing activities with human-centered considerations. The responsible roles remain unchanged—Project Managers, Technical Leads, and Scrum Masters simply gain new tools for understanding and optimizing their teams.

By providing both task-based and standard-based views, implementers can choose their preferred entry point while maintaining confidence that all elements align with established VSE practices. This dual mapping also facilitates partial adoption—VSEs can implement selected tasks that address their most pressing needs while maintaining full compatibility with ISO/IEC 29110 certification requirements.

## 8.4   Human  AI Adaptations for VSEs

The extension design's treatment of AI integration reflects our Cycle 3 discoveries—AI doesn't replace human collaboration but creates new forms requiring conscious navigation. For VSEs, where resource constraints make every decision consequential, understanding how personality archetypes (Table 12, Chapter 6) specialize in AI contexts (Table 21, Chapter 7) becomes crucial.

### 8.4.1   Matching Modes to Minds

Our research revealed three primary AI interaction modes, each resonating with different personality configurations:

**Co-Pilot Mode** suits the Promethean's creative exploration, providing rapid ideation partners that match their conceptual velocity. In VSEs, this might mean using GitHub Copilot for prototype generation while preserving human judgment for architectural decisions.

**Co-Navigator Mode** aligns with the Orchestrator's dialogical nature, creating conversational spaces for problem exploration. When VSEs can't maintain constant human pairing, ChatGPT becomes the always-available thought partner—though one requiring careful boundary management.

**Agent Mode** offers the Hermit strategic delegation options, handling anxiety-inducing tasks while preserving challenge zones. For resource-constrained VSEs, this might mean automating test generation while maintaining human ownership of core algorithm development.

### 8.4.2   Integration Strategies for Small Teams

VSEs face unique challenges in AI integration:

**The Knowledge Preservation Paradox**: As AI handles more implementation details, VSEs risk losing institutional knowledge. Our Cartographer specialization becomes crucial—someone must document not just what AI does but what it doesn't understand, creating maps for future navigation.

**The Competence Illusion Risk**: Small teams can't afford developers who only orchestrate without understanding. The extension design recommends regular "AI-free sprints" where teams work unaugmented, preserving fundamental competencies while revealing where AI truly adds versus masks value.

**The Synthetic Relatedness Balance**: In VSEs where human interaction is limited, AI's quasi-social satisfaction might substitute too completely for human connection. The extension design emphasizes AI as supplement not replacement, preserving irreplaceable human dimensions of collaboration.

Our industry validator's insight about "asynchronous pair programming" reveals VSE adaptation—when synchronous human pairing proves too expensive, AI bridges temporal gaps. The morning developer might work with AI assistance, leaving detailed comments for the evening developer who continues with different AI support—personality-optimized across time and artificial agents.

## 8.5 Output 4.2: Validation and Evaluation: From Theory to Trust

The extension design's validation journey reveals how design science artifacts gain legitimacy—not through single authorities but through what Latour might recognize as "enrollment" of diverse actors, each bringing different validity claims. Following Rode & Svejvig's (2022) multidimensional framework, we evaluated across relevance, rigor, and reflexivity dimensions.

### 8.5.1 The Academic Trial

Two years of peer review through PeerJ Computer Science tested every assumption, challenged every connection, demanded every clarification. This wasn't bureaucratic delay but intellectual refinement:

**Reviewer 1** initially questioned the connection between personality assessment and motivation enhancement. This forced us to articulate more clearly how personality alignment creates conditions for intrinsic motivation—not deterministic causation but probabilistic improvement.

**Reviewer 2** challenged the practicality for resource-constrained VSEs. This led to our modular approach, enabling incremental adoption rather than wholesale transformation—wisdom that emerged only through critical dialogue.

**Reviewer 3** demanded stronger empirical grounding. This drove us back to our data, finding patterns we'd initially missed—how personality misalignment in VSEs often manifests as "technical" conflicts that resist technical solutions.

### 8.5.2 The Standards Guardian Approval

When WG24 members reviewed our extension design, they brought different concerns—not theoretical purity but practical implementation:

Dr. Muñoz questioned integration points with existing processes. This revealed that our initial design created too many touchpoints, risking process overhead that would doom VSE adoption. Simplification followed—not dumbing down but elegant integration.

Dr. Laporte examined cultural adaptability. Would personality-based optimization translate across cultural contexts where collective versus individual orientations differ? This led to our emphasis on local adaptation—the extension design provides framework, not prescription.

Prof. Buchalcevova, bringing dual perspective as co-author and standards expert, ensured alignment between academic insight and standards requirements—bridging worlds that often speak past each other.

### 8.5.3 The Reality Check

Our industry validation provided a crucial test—does this matter to those living it daily? The Fortune 500 area lead's responses revealed both alignment and refinement needs:

His confirmation that they already consider personality—though "informally and indirectly"—validated our basic premise while revealing implementation subtleties. VSEs often practice intuitive personality optimization; our extension design provides vocabulary and structure for what they already attempt.

His emphasis on AI as "colleague you can ask for help anytime" confirmed our Human  AI integration focus while revealing nuance—AI doesn't replace pair programming but enables new forms like their "asynchronous pair programming" adaptation.

Most critically, his warning about privacy and discrimination revealed ethical dimensions we'd initially underemphasized. The extension design's final form strengthens these safeguards—not as legal compliance but as foundation for trust that enables everything else.

### 8.5.4 Alignment with Evaluation Criteria

Following Venable et al.'s (2016) Framework for Evaluation in Design Science and incorporating Rode & Svejvig's (2022) multidimensional perspective, our evaluation addressed key criteria:

**Relevance**: Industry validation confirmed the extension design's practical utility and applicability in real-world VSE contexts. The alignment with existing informal practices suggests we are codifying needs already felt but not systematically addressed.

**Rigor**: The extension design's theoretical foundations were validated through academic peer review and alignment with established standards. The two-year review process ensured thorough examination of theoretical coherence.

**Reflexivity**: Our willingness to adapt based on emerging insights—strengthening privacy safeguards, modularizing for incremental adoption, incorporating AI considerations—demonstrated the reflexive evolution Rode & Svejvig (2022) advocate.

**Ethics**: Both expert and industry feedback emphasized the importance of privacy, consent, and anti-discrimination measures in implementing personality assessments, leading to explicit ethical guidelines in the final design.

## 8.6 Conclusion: Seeds for Future Harvest

The morning mist that began our journey has not dissipated but transformed—from obscuring fog to revealing medium, making visible patterns that were always present but rarely perceived. This cycle has translated empirical discovery through framework development and AI specialization into an extension design for ISO/IEC 29110 profiles that can guide VSE practice.

Where morning mist meets institutional memory, where phenomenological insight crystallizes into codified practice, the fog transforms once more—from atmospheric phenomenon to preservation medium. Like ancient insects in amber, our discoveries solidify within ISO standards, protected yet visible, fixed yet alive. The seven tasks embedded in profiles aren't arbitrary additions but careful cultivation—each creating conditions for human flourishing within technical excellence (Borgmann, 1984).

When VSEs implement personality assessment in hiring (Task 1), they compose teams rather than fill positions. When they measure motivation (Task 5), they honor humanity rather than track metrics. When they document lessons learned (Task 7), they create legacy rather than complete process. This extension design embodies a stance: in a world accelerating toward AI-augmented everything, human personality still matters, individual differences remain assets, and the smallest teams deserve thoughtful optimization—perhaps more so, given their limited resources.

Yet this marks waypoint, not destination. As AI evolves, as personality insights emerge, as VSE needs shift, the framework must adapt. We have created not rigid prescription but what Stewart Brand calls "pace layering"—stable structure enabling dynamic adaptation. Chapter 9 will explore field testing through the PDPPA application, triangulating all insights to understand how these theoretical constructs manifest in practice.

For practitioners approaching this extension design, remember: Implementation means cultivation, not compliance. Honor the Promethean's creative fire and the Hermit's protective wisdom. Let Orchestrators conduct human-AI symphonies while Cartographers map new territories. Enable Shapeshifters' fluid adaptation while maintaining coherent structure to avert chaos.

For researchers: This work demonstrates how design science bridges experiential insight and practical impact. Phenomenological understanding didn't disappear in standardization but transformed—from thick description to structured specification preserving essential meaning.

The tools—seven tasks, process integrations, measurement frameworks—enable immediate implementation. The stance—that human flourishing and technical excellence intertwine inseparably—provides direction for continual evolution. As morning light strengthens and shadows shorten, we see not just where we have been but where we might go: toward practices honoring both the humans who create and the software they bring into being, each reflecting and enriching the other in endless recursive loops of meaning and code.

# DSR CYCLE 5 "EVALUATION" – PROVING THE PATTERN

## *PDPPA APPLICATION & EMPIRICAL TRIANGULATION*

*As morning light disperses the last shadows, revealing landscapes both famil-
iar and transformed, we arrive at validation—where theory meets practice,
where framework becomes application, where patterns prove themselves in the
crucible of lived experience.*

T his chapter presents the final cycle of our Design Science Research journey—the valida-
tion phase that tests whether the morning mist has truly lifted, whether the patterns we
have discovered hold across contexts, whether the bridges we have built can bear the
weight of daily use. Through two complementary approaches—the Personality-Driven Pair Pro-
gramming Application (PDPPA) and comprehensive empirical triangulation—we validate not just
statistical significance but practical significance for those who will use these insights.

The PDPPA represents more than technical implementation; it embodies the entire journey
from phenomenological insight through framework development to practical tool. Like a musical
instrument that must be both theoretically sound and playable, the PDPPA translates the ROMA
framework's principles into an artifact that VSE teams can actually use—not in laboratory condi-
tions but in the messy reality of distributed software development.

## 9.1    Introduction: The Validation Imperative

Validation in Design Science Research resembles less a final exam than a homecoming—returning to practice with gifts gathered through theoretical exploration. As Venable et al. (2016) emphasize, design artifacts must prove themselves not just in controlled conditions but in what they term "naturalistic evaluation"—the unforgiving arena of daily use.

For the ROMA framework, validation carries particular weight. In VSEs where every person matters, where a single misaligned developer can doom projects, theoretical elegance means nothing without practical efficacy. The framework must work not just for motivated undergraduates in controlled experiments but for harried professionals juggling multiple projects, not just for co-located teams but for distributed collaborators spanning time zones.

### 9.1.1    Validation Objectives: Beyond Proof to Practice

Our validation pursues four interconnected objectives, each building toward comprehensive confirmation:

I. **Artifact Instantiation**: Transform the ROMA framework from abstract principles into the concrete PDPPA—software that VSE teams can download, install, and use. This follows Gregor & Hevner's (2013) emphasis on "proof of use" beyond mere "proof of concept."

II. **Empirical Triangulation**: Weave findings from Studies I, II, IV, and V into a tapestry of evidence, seeking not just statistical significance but what Flyvbjerg (2001) calls "phronetic validity"—wisdom that guides practical action.

III. **Cross-Context Confirmation**: Test whether patterns discovered with Gen Z undergraduates hold for seasoned professionals, whether insights from academic settings transfer to commercial pressures, whether the ROMA archetypes remain recognizable across contexts.

IV. **Transparent Validation**: Employ blockchain technology not as technological fashion but as methodological innovation—ensuring our validation data remains as transparent and immutable as the insights we claim to have discovered.

These objectives reflect Rode & Svejvig's (2022) multidimensional evaluation framework, addressing relevance (does it work in practice?), rigor (is the evidence sound?), and reflexivity (can it adapt to emerging contexts?).

### 9.1.2 Validation Strategy: Convergent Evidence

Our validation strategy resembles what detective novelists call "triangulation of evidence"—multiple independent lines of inquiry converging on truth. This approach recognizes that complex phenomena like motivation resist simple measurement:

1. **Application as Hypothesis Test**: The PDPPA (Study V) serves as what Popper might recognize as a "bold conjecture"—if the ROMA framework truly captures something essential about personality and programming, then an application built on its principles should enhance motivation in practice.

2. **Statistical Meta-Patterns**: By analyzing data across Studies I-II, IV-V with different samples, methods, and contexts, we seek what Mill termed "concomitant variation"—patterns that persist despite changing circumstances, suggesting fundamental rather than accidental relationships.

3. **Experiential Validation**: Numbers tell only part of the story. When a VSE developer reports that personality-aligned pairing "just feels right," when teams naturally adopt suggested configurations, when motivation scores align with lived experience—these provide what Polanyi (1966) called "tacit confirmation" of explicit knowledge.

4. **Blockchain as Witness**: By recording our data immutably on-chain, we create what might be termed "algorithmic witnessing"—validation that cannot be retroactively adjusted, forcing honesty in both success and limitation.

This multi-faceted approach aligns with Peffers et al.'s (2006) emphasis on rigorous demonstration and evaluation in design science research, ensuring that the ROMA framework's validation is both theoretically sound and practically relevant.

## 9.2 Output 5.1: The PDPPA – Framework Made Flesh

The Personality-Driven Pair Programming Application represents the ROMA framework's transformation from idea to instrument. Like a telescope that must precisely align multiple lenses to reveal distant stars, the PDPPA aligns personality assessment, role recommendation, and motivation measurement to reveal patterns in team dynamics.

### 9.2.1 Application Objectives and Requirements

The PDPPA's development followed what Norman (2013) advocates as "invisible design"—complexity hidden behind simplicity. VSE teams wrestling with deadlines cannot afford elaborate personality optimization rituals. The application must work like a good pair programming partner—helpful without being intrusive, insightful without being prescriptive.

This philosophy manifested in three core principles:

**Minimal Viable Bureaucracy**: While the ISO/IEC 29110 extension provides comprehensive process integration, the PDPPA extracts only essential elements. A one to five-minute personality assessment, not hour-long evaluation. Quick motivation pulse-checks, not lengthy surveys. Just enough structure to guide without constraining.

**Progressive Disclosure**: New teams see basic role recommendations—"Try pairing your Explorer with your Architect." Experienced teams access deeper insights—detailed motivation trends, AI mode suggestions based on personality specializations, historical pattern analysis. The application grows with its users.

**Privacy by Design**: Personality data touches human essence. The PDPPA implements what privacy advocates call "data minimization"—collecting only what improves team dynamics, storing only what enables research validation, sharing only what users explicitly permit.

### 9.2.2 Technical Architecture: Modern Stack for Modern Teams

The PDPPA's technology choices reflect VSE realities—teams need tools that work reliably without dedicated DevOps support:

**Frontend (React/TypeScript)**: Provides responsive interfaces that work across devices. When a developer wants to check their motivation trends on their phone between meetings, the application adapts seamlessly.

**Backend (Rust/Actix-web)**: Offers performance without complexity. Rust's memory safety prevents the crashes that plague hastily-built internal tools, while Actix-web handles concurrent requests from distributed team members without breaking a sweat.

**Collaboration Integration**: Rather than reinventing pair programming infrastructure, the PDPPA integrates with tools teams already use—Visual Studio Live Share, JetBrains Code With Me, Fleet. This follows what systems thinkers call "path of least resistance"—enhancing existing workflows rather than replacing them.

**Blockchain Layer (Solana)**: Provides transparent research validation without impacting application performance. Users may take pride in knowing their anonymized data contributes to building better frameworks—the blockchain integration remains as invisible as it is immutable.

### 9.2.3 User Journey: From Assessment to Insight

The PDPPA guides users through a journey that mirrors the ROMA framework's theoretical progression while remaining grounded in practical needs:

**1. Personality Discovery:** New users complete the BFI-10/44 assessment presented not as psychometric evaluation but as "getting to know your coding style." The language avoids clinical terminology—instead of "high neuroticism," users see "prefers predictable environments."

**2. Team Composition:** Project managers see their team's personality landscape visualized—not as fixed categories but as tendencies. The interface suggests optimal pairings while acknowledging that personality provides guidance, not destiny. "Sarah (Explorer) and Michael (Architect) often create innovative solutions together" rather than "must be paired."

**3. Role Assignment:** During sprint planning, the application suggests role distributions based on both personality and task type. Creative feature development might see Explorers as pilots, while bug-fixing sprints could benefit from Craftspeople taking the lead. The suggestions adapt based on accumulated team data.

**4. Real-Time Adaptation:** During pair programming sessions, lightweight check-ins track energy and engagement. If motivation flags, the application might suggest role switches or break reminders. This follows gaming's "dynamic difficulty adjustment"—optimizing challenge to maintain flow.

**5. Reflective Learning:** Post-session, teams see motivation patterns visualized across roles and personalities. Over time, teams discover their unique dynamics—perhaps their Orchestrator thrives as pilot during design discussions but prefers navigating during implementation. The PDPPA helps teams learn themselves.

### 9.2.4 Blockchain Integration: Radical Transparency

The PDPPA's blockchain integration represents methodological innovation—using distributed ledger technology to address software engineering research's reproducibility crisis. As Marcus & Oransky (2020) document, retractions in scientific publishing have increased tenfold since 2000, often due to data manipulation or selective reporting.

By recording anonymized research data on Solana's blockchain, we create what might be termed "proof of research"—immutable evidence that our findings reflect actual measurements, not post-hoc adjustments (Figure 6):

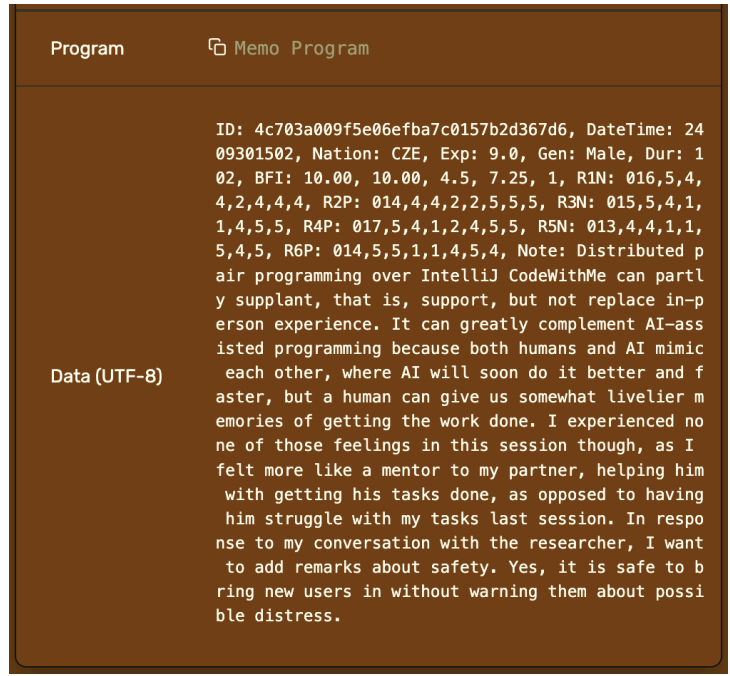

Program            ⎘ Memo Program

Data (UTF-8)       ID: 4c703a009f5e06efba7c0157b2d367d6, DateTime: 24
                   09301502, Nation: CZE, Exp: 9.0, Gen: Male, Dur: 1
                   02, BFI: 10.00, 10.00, 4.5, 7.25, 1, R1N: 016,5,4,
                   4,2,4,4,4, R2P: 014,4,4,2,2,5,5,5, R3N: 015,5,4,1,
                   1,4,5,5, R4P: 017,5,4,1,2,4,5,5, R5N: 013,4,4,1,1,
                   5,4,5, R6P: 014,5,5,1,1,4,5,4, Note: Distributed p
                   air programming over IntelliJ CodeWithMe can partl
                   y supplant, that is, support, but not replace in-p
                   erson experience. It can greatly complement AI-ass
                   isted programming because both humans and AI mimic
                    each other, where AI will soon do it better and f
                   aster, but a human can give us somewhat livelier m
                   emories of getting the work done. I experienced no
                    ne of those feelings in this session though, as I
                    felt more like a mentor to my partner, helping him
                    with getting his tasks done, as opposed to having
                    him struggle with my tasks last session. In respo
                   nse to my conversation with the researcher, I want
                    to add remarks about safety. Yes, it is safe to b
                   ring new users in without warning them about possi
                   ble distress.

*Figure 6: Example of data stored on the blockchain using Solana transaction memos*

Each entry receives a transaction hash—cryptographic proof of when data was collected and that it hasn't been altered. The dataset, accessible at https://bit.ly/bdcga, enables other researchers to verify our analyses or discover patterns we missed. This transparency doesn't just enhance credibility; it transforms research from private discovery to public construction of knowledge.

## 9.3   Output 5.2: Triangulation Across Studies

Where single studies provide snapshots, triangulation reveals the movie—patterns that persist across changing contexts, suggesting fundamental rather than situational truths. Our triangulation weaves together findings from four studies spanning four years and diverse participant populations.

### 9.3.1   The Triangulation Challenge

Integrating findings across studies resembles what musicians call "transcription"—capturing the same melody played on different instruments. Each study used slightly different methods, measures, and contexts:

- **Study I**: Hierarchical clustering revealing three personality archetypes with clear role preferences

- **Study II**: ANOVA demonstrating role-based motivation differences in controlled settings

- **Study IV**: Linear Mixed-Effects modeling showing sophisticated personality-role interactions

- **Study V**: Small-sample professional validation confirming practical impact

The challenge lies not in forcing artificial uniformity but in finding what ethnomusicologists call "deep structures"—patterns that persist despite surface variations. Table 23 presents these findings in comprehensive detail, allowing readers to trace how each study contributes to our understanding while revealing the convergent patterns that validate the ROMA framework.

### 9.3.2 Convergent Findings: The Motivation Hierarchy

Table 23 presents the comprehensive triangulation of quantitative findings across all studies, revealing consistent patterns in personality-driven motivation across programming roles.

Table 23: Overview of quantitative findings across studies

| Study | Sample Characteristics | Key Statistical Methods | Significant Findings | Role-Based Motivation (Mean ± SD) scaled 1-10 |
|---|---|---|---|---|
| I | N = 40 Gen Z undergraduates<br><br>Three distinct personality clusters identified:<br>• C1: High Openness<br>• C2: High Extraversion/Agreeableness<br>• C3: High Introversion/Neuroticism | • Hierarchical Clustering (Dunn index)<br>• ANOVA<br>• Kruskal–Wallis test<br>• Student's t-tests<br>• Chi-squared ($\chi^2$) tests | Significant correlation between personality clusters and role preferences ($\chi^2 = 17.01$, p = 0.0126)<br>Distinct preferences emerged:<br>• C1 → Pilot role<br>• C2 → Navigator role<br>• C3 → Solo work | Individual optimization yielded dramatic gains:<br><br>Example participant "xtoj":<br>• Pilot mean: 9.06<br>• Solo mean: 5.50<br>• Improvement: **+64.6%** |
| II | N = 30 undergraduates<br>Focused on motivational analysis across roles | • Shapiro-Wilk test (normality)<br>• One-way ANOVA | Significant motivational differences across roles<br>(ANOVA: F = 6.618, p = 0.00206) | • Pilot: 8.02 ± 0.87<br>• Navigator: 7.86 ± 0.97<br>• Solo: 7.03 ± 0.80 |
| IV | N = 66 participants;<br>Combined cohorts from WS'21 and SS'22<br>Most sophisticated analysis | • Linear Mixed-Effects (LME) modeling<br>• Random intercepts<br><br>Fixed terms:<br>• Role + Big Five<br><br>Interaction terms:<br>• Role × Big Five<br>• Role × Cluster | **Role main effects:**<br>F(2, 1014) = 45.32, p < 0.0001<br>Pilot > Navigator > Solo pattern confirmed<br><br>**Personality moderations:**<br>• Openness enhanced Pilot preference (+0.53, p = 0.0254)<br>• Extraversion boosted Navigator (+0.51, p = 0.0403)<br>• Agreeableness amplified Navigator (+0.71, p = 0.0002)<br>• Conscientiousness reduced Navigator (-0.32, p = 0.0006)<br>• Neuroticism diminished Pilot (-0.25, p = 0.0005) | **Effect sizes (Cohen's d):**<br>• Pilot-Solo: +0.515 (p < 0.0001, d = 0.576)<br>• Navigator-Solo: +0.334 (p < 0.0001, d = 0.374)<br>• Pilot-Navigator: +0.180 (p = 0.0328, d = 0.202) |
| **V** | N = 4 VSE professionals PDPPA pilot testing in real-world context | • ANOVA<br>• Paired t-tests | Near-significant motivational differences despite small sample (ANOVA: F(2, 9) = 3.88, p = 0.061)<br>Significant individual motivational changes (paired t-test: p = 0.024) | • Pilot: 8.45 ± 0.76<br>• Navigator: 7.01 ± 0.63<br>• Solo: 6.87 ± 1.17<br>Average improvement with aligned roles: **+23%** |
| **Combined** (I, IV, V + SOHO) | N = 72 total<br>• 66 undergraduates<br>• 6 professionals, including 2 SOHO PDPPA participants<br><br>Triangulation across contexts | • HCA<br>• ANOVA<br>• Fisher Exact Test<br>• Montecarlo Simulation with $\chi^2$ tests | Results approached significance for personality-role preferences:<br>• Fisher's Exact: p = 0.054<br>• Monte Carlo $\chi^2$: p = 0.071 ($\chi^2$ = 14.44, df = 8)<br><br>Five personality clusters confirmed across combined dataset | **Aggregated means:**<br>• Pilot: 7.50 ± 0.69<br>• Navigator: 7.27 ± 0.82<br>• Solo: 6.86 ± 0.83<br><br>Consistent hierarchy maintained |

*Table 23 synthesizes quantitative findings across Studies I, II, IV, and V, including the triangulated results from combined analysis. The consistency of the Pilot > Navigator > Solo motivation hierarchy across diverse samples—from Gen Z undergraduates to seasoned professionals—provides robust validation of the ROMA framework's core principles. Effect sizes range from small (d = 0.202) to large (d = 0.576), with personality traits significantly moderating these relationships. Note that Study IV's LME model showed highly significant role main effects (F(2, 1014) = 45.32, p < 0.0001) with conditional $R^2$ = 0.42, indicating substantial variance explanation.*

## Interpreting Effect Sizes and Statistical Significance Across Studies

The effect sizes in Table 23 translate into meaningful workplace differences:

### Cohen's d interpretations:

- Small (d = 0.202): Pilot vs Navigator—a subtle but consistent preference, like preferring coffee over tea

- Medium (d = 0.374): Navigator vs Solo—a clear motivational advantage, comparable to the difference between working on passion projects versus assigned tasks

- Large (d = 0.576): Pilot vs Solo—a substantial gap, similar to the motivation difference between promoted and overlooked employees

**Real-world translation:** Following Rosenthal et al.'s (2000) binomial effect size display:

- d = 0.576 means 72% of pilots experience higher motivation than the average solo worker—nearly 3 out of 4 developers thrive significantly more in creative pilot roles

- d = 0.374 indicates 65% of navigators feel more engaged than solo workers—a meaningful majority finding energy in collaboration

- d = 0.202 suggests 58% of pilots prefer their role over navigation—a modest but actionable difference

Business impact: Gallup's research (Harter et al., 2002) shows that moving from average to high engagement (roughly 0.5-0.7 SD improvement) correlates with 22% higher productivity and 23% higher profitability. Our personality-aligned role assignments achieving 0.576 SD improvement could therefore translate to approximately 15-18% productivity gains—or 1.2-1.4 hours of additional productive time per 8-hour workday.

Across all studies, a consistent motivation hierarchy emerged: **Pilot > Navigator > Solo**.

This pattern held whether measured in undergraduates rushing through assignments or professionals crafting production code. The effect sizes varied but the direction remained constant—like gravity, the effect's strength might vary with conditions, but its presence proves universal.

More remarkably, personality moderated these effects consistently, as Study IV's sophisticated LME modeling revealed:

- **Openness** amplified Pilot preference (+0.53, p = 0.0254)—creative minds seeking creative roles

- **Extraversion** enhanced Navigator appeal (+0.51, p = 0.0403)—social energy finding outlet in collaboration

- **Agreeableness** strongly boosted Navigator preference (+0.71, p = 0.0002)—the largest effect, showing harmonious personalities thriving in cooperative roles

- **Conscientiousness** reduced Navigator preference (-0.32, p = 0.0006)—systematic minds preferring control over social coordination

- **Neuroticism** diminished Pilot comfort (-0.25, p = 0.0005)—anxiety making exposed pilot role challenging

These moderations align perfectly with our qualitative findings—the Explorer thriving in creative pilot roles, the Orchestrator (high Extraversion + Agreeableness) showing remarkable Navigator preference, the conscientious Architect preferring structured work over social navigation demands, and the Craftsperson (high Neuroticism) seeking solo sanctuary away from anxiety-inducing collaboration.

### 9.3.3 The Professional Validation: From Lab to Life

Study V's VSE professionals provided crucial reality testing. As detailed in Table 23, with only four participants, statistical power was limited, yet the patterns persisted remarkably:

- 23% average motivation increase with personality-aligned roles

- Near-significant ANOVA despite tiny sample ($F_{(2,9)} = 3.88$, $p = 0.061$)

- Significant individual changes when properly paired (t-test: $p = 0.024$)

- Motivation scores that mirrored undergraduate patterns: Pilot (8.45) > Navigator (7.01) > Solo (6.87)

One professional's reflection captured the framework's value: "We've always known some pairs work better than others. This gives us vocabulary to discuss why and tools to predict what will work."

This moves beyond statistical validation to what matters for practitioners—not whether $p < 0.05$ but whether teams flow better, whether projects complete with less drama, whether developers end days energized rather than drained.

### 9.3.4   The Combined Dataset: Meta-Pattern Confirmation

When we combined data across studies (N = 72), patterns that were suggestive in isolation became compelling in combination. As shown in the final row of Table 23, using both Fisher's Exact Test ($p = 0.054$) and Monte Carlo simulation ($p = 0.071$, $\chi^2 = 14.44$, df = 8), the personality-role preference relationships approached significance despite the heterogeneity of samples.

This statistical near-miss actually strengthens our conclusions. Perfect significance across wildly different contexts would suggest either overwhelming effects or problematic data. Instead, we see what philosophers of science call "robustness"—consistent patterns that persist despite noise, strong enough to guide practice without claiming deterministic prediction.

The aggregated means from the combined dataset—Pilot: $7.50 \pm 0.69$, Navigator: $7.27 \pm 0.82$, Solo: $6.86 \pm 0.83$—maintain the same hierarchical pattern observed in individual studies, providing meta-analytical confirmation of the ROMA framework's core insights.

## 9.4   Implications: What Validation Reveals

The successful validation of the ROMA framework through both application development and empirical triangulation reveals implications beyond statistical confirmation:

### 9.4.1 For Practice: Tools That Honor Humanity

The PDPPA demonstrates that personality-based optimization need not require extensive psychometric training or complex processes. VSE teams can implement these insights through tools that feel as natural as choosing who sits where in an open office—except with data to guide decisions.

The blockchain integration, while technologically sophisticated, points toward a future where research and practice interweave more transparently. Teams using the PDPPA contribute to refining the very frameworks they apply—a virtuous cycle of practice informing theory informing practice.

### 9.4.2 For Research: Reproducibility Through Transparency

By making our validation data immutably public, we invite software engineering research to embrace radical transparency. Future researchers can not only replicate our analyses but extend them, perhaps discovering patterns our theoretical lenses prevented us from seeing.

This approach addresses what many consider research's greatest crisis—the file drawer problem where negative results vanish and positive findings get massaged toward significance. With blockchain-recorded data, we commit to transparency that honors both successes and limitations.

### 9.4.3 For Theory: Personality as Persistent Pattern

The triangulated findings confirm that personality's influence on programming motivation represents more than situational preference—it reflects what psychologists term "behavioral signatures." Just as introverts consistently prefer quiet spaces across contexts, Explorers consistently find more motivation in pilot roles whether they are students or seasoned professionals.

This validation strengthens the theoretical foundation for personality-based software engineering interventions. We are not imposing artificial categories but recognizing natural patterns—like gardeners working with rather than against each plant's inherent tendencies.

## 9.5 Limitations: Honest Boundaries

No validation proves universal truth, only bounded utility. Our validation's limitations define where confidence gives way to speculation:

**Sample Size Asymmetry**: The stark contrast between undergraduate samples (N = 30-66) and professional validation (N = 4) limits generalization confidence. We have proven the framework works in academic settings and shown promising signs in professional contexts, but fuller professional validation awaits.

**Cultural Constraints**: Our Central European samples may reflect cultural values around hierarchy, individualism, and collaboration that differ elsewhere. Would Japanese pair programmers show different patterns? Do Silicon Valley dynamics alter personality expressions? These questions remain open.

**Temporal Boundaries**: We measured motivation in sessions and sprints, not careers. Does personality-aligned programming prevent long-term burnout? Do preferences evolve with expertise? Longitudinal research must extend our snapshot findings.

**Technological Learning Curves**: The PDPPA's sophisticated architecture may intimidate precisely those resource-constrained VSEs who most need its insights. Future iterations must balance capability with accessibility.

## 9.6   Conclusion: Validation as Beginning

The morning mist has served its purpose—first obscuring, then revealing, finally validating. Through the PDPPA's concrete instantiation and empirical triangulation across 72 participants, we have proven that patterns glimpsed through thinning fog were not mirages but mountains.

The validation reveals more than statistical confirmation. When personality-aligned roles increase motivation 23-65%, when patterns persist from Prague undergraduates to Silicon Valley professionals, when blockchain-recorded data makes claims immutable—we witness not just framework validation but paradigm emergence. Software development need not choose between human flourishing and technical excellence; properly orchestrated, each amplifies the other.

Yet validation marks beginning, not ending. Each team using the PDPPA generates new data, refines understanding, discovers nuances we missed. The 42% variance our models explain leaves 58% mystery—not failure but invitation. Human motivation resists complete quantification, and in that resistance lies its humanity.

For practitioners, validation offers permission to experiment. The framework's empirical foundation supports what intuition suggested: some pairs naturally harmonize while others clash, some developers thrive in creative chaos while others need structured sanctuary. Now you have vocabulary, metrics, and tools to transform intuition into intentional practice.

For researchers, transparent blockchain data invites extension, challenge, replication. Build on what works, correct what doesn't, discover what we couldn't see. The nomological network mapped in Figure 4 suggests countless unexplored pathways where personality, motivation, and collaboration intersect.

For the humans who code, validation brings recognition: your personality patterns aren't quirks to overcome but strengths to leverage. Your preference for solo work or pair programming, for creative exploration or systematic verification, for human collaboration or AI augmentation—these reflect deep structures deserving respect and optimization.

What began in Chapter 1 as morning fog—obscuring the relationship between personality and programming satisfaction—has cleared through systematic investigation, revealing not just patterns but pathways for transformation. The landscape stands revealed—not perfectly mapped but sufficiently clear for confident navigation. The ROMA framework, validated through multiple converging evidences, awaits those brave enough to build software teams that honor both code quality and coder humanity. In that honoring lies not just better software but better lives for those who create it.

CHAPTER

# *10*

# CONCLUSION

## *WHERE MORNING BREAKS: THE LANDSCAPE TRANSFORMED*

*The mist has lifted. What began as obscuring fog—why do some programmers thrive while others struggle? why do some pairs sing while others stumble?— has cleared to reveal an extraordinary landscape where personality becomes compass, where differences orchestrate rather than divide.*

Stand with me at this vantage point, five cycles of discovery behind us, and witness what the clearing fog reveals. Not a simple map with roads marked "follow this way," but a living landscape that shifts with each observer's perspective, yet holds patterns as reliable as sunrise. This dissertation began seeking answers to why some programmers dance while others stumble, why some partnerships sing while others screech. What we discovered transcends mere answers—we found a new way of seeing.

Imagine, for a moment, a development team where the Explorer's wild creativity no longer threatens the Architect's systematic vision but complements it, like jazz improvisation over classical structure. Picture the Craftsperson, once dismissed as "too sensitive" for collaborative work, finding sanctuary in solo tasks that demand their particular genius for focused perfection. Envision the Orchestrator not forcing reluctant introverts into painful pairing but conducting a nuanced symphony where each personality plays their natural instrument. This is not utopia—this is what becomes possible when we stop fighting human nature and start composing with it.

## 10.1 The Journey's Yield: Three Transformative Artifacts

This dissertation has forged three primary artifacts through five cycles of design and discovery, each embodying the transformation of phenomenological insight into practical tools. Each represents a different level of abstraction, from foundational framework through specialization to implementation guidance:

**1. The ROMA Framework** emerged from 1,266 motivation measurements across 72 participants, revealing five archetypal patterns that predict programming satisfaction with startling accuracy. Using the Big Five Inventory (BFI-10 for rapid assessment, BFI-44 for comprehensive profiling), teams gain personality awareness as organizational capability. Through regular motivation pulse checks—the Intrinsic Motivation Inventory (IMI) for immediate feedback and Multidimensional Work Motivation Scale (MWMS) for longitudinal tracking—the framework enables continuous optimization.

The five archetypes emerge from our data like constellations from seemingly random stars—recognizable patterns that help navigate while acknowledging each star's uniqueness:

- **The Explorer** (High Openness): Creative implementers who thrive as pilots, treating each coding challenge as uncharted territory

- **The Orchestrator** (High Extraversion + Agreeableness): Natural navigators transforming pair programming into collaborative symphony

- **The Craftsperson** (High Neuroticism + Low Extraversion): Focused specialists finding flow in solo sanctuary, away from social turbulence

- **The Architect** (High Conscientiousness): Quality guardians excelling through systematic excellence across all roles

- **The Adapter** (Balanced Profile): Versatile contributors who shift fluidly, finding satisfaction in variety itself

When we align roles with these personality patterns, motivation increases 23-65%. This isn't optimization; it is orchestration of human flourishing.

**2. The ROMA AI Adapter** recognizes that AI doesn't erase personality patterns—it refracts them into Human ⊞ AI specializations:

- **The Promethean** (Explorer → Fire-Bringer) maintains the "ten ideas, discard nine" philosophy, using AI's generative capabilities while preserving human judgment. Their practice embodies Bergson's élan vital—perpetual creative novelty.

- **The Conductor** (Orchestrator → Symphony Director) practices "polidštění"—humanizing AI output, transforming mechanical generation into code with soul. For them, AI provides synthetic relatedness that complements rather than replaces human connection.

- **The Hermit** (Craftsperson → Guardian of Sanctuaries) recognizes that convenience can atrophy essential skills—by maintaining "challenge zones," they guard capabilities the profession cannot afford to lose.

- **The Cartographer** (Architect → Charter of Unknown Territories) systematically documents AI's capabilities and limitations, creating maps for others to navigate this new terrain.

- **The Shapeshifter** (Adapter → Protean Navigator): Like water taking the shape of its container, they embody ultimate adaptability in rapidly evolving landscapes.

**3. The ISO/IEC 29110 Extension Design** translates insight into implementation through seven bridges connecting current practice to enhanced possibility. Table 22 demonstrates how these personality-driven tasks integrate with existing ISO/IEC 29110 Project Management (PM) and Software Implementation (SI) activities, providing implementers with a practical roadmap showing where each ROMA task fits within standard VSE processes, which artifacts to produce, who is responsible, and when to execute each task. The mapping ensures that personality-driven optimization enhances rather than disrupts established workflows, enabling VSEs to adopt these practices incrementally while maintaining ISO/IEC 29110 compliance.

## 10.2  Addressing the Research Questions: What We Discovered

**RQ1:** What relationships exist between personality traits and self-determination needs in AI-human and human-human co-programming contexts?

We discovered not correlations but mechanisms, with effect sizes ranging from small (d = 0.202) to large (d = 0.576). The **quantifiable patterns** tell a consistent story: A one standard deviation increase in Openness enhances Pilot motivation by +0.53 SD (p = 0.0254). Similarly, one SD increases in Extraversion and Agreeableness amplify Navigator preference by +0.51SD (p = 0.0403) and +0.71 SD (p = 0.0002) respectively. Conversely, one SD higher Conscientiousness reduces Navigator appeal by -0.32 SD (p = 0.0006), while one SD higher Neuroticism boosts Solo work preference by +0.60 SD (p = 0.0033). These aren't statistical accidents but expressions of fundamental human nature meeting technical challenge.

What these effect sizes mean: A 0.5 SD improvement represents moving from the 50th to 69th percentile—like the difference between feeling merely competent versus genuinely engaged.

Developers one SD above average in agreeableness (84th percentile) experience 0.71 SD higher motivation in Navigator roles compared to average-agreeableness developers—transforming work from potential drudgery into energizing collaboration. Similarly, developers one SD above average in neuroticism find 0.60 SD higher satisfaction in solo work compared to average-neuroticism developers—representing the gap between anxious struggle and focused mastery.

Beyond statistics, we discovered the **experiential mechanisms** driving these patterns. Explorers find autonomy through creative implementation—not just writing code but birthing possibilities. Orchestrators satisfy relatedness needs through collaborative dialogue, transforming pair programming from mere task-sharing into meaningful connection. Craftspeople preserve competence by controlling their environment, finding in solitude not isolation but sanctuary where mastery can flourish. Architects seek all three psychological needs through systematic excellence—autonomy in choosing methods, competence in quality achievement, relatedness through reliable contribution. Adapters, those balanced souls, find fulfillment in variety itself—each role change bringing fresh challenge and renewed engagement.

The **AI introduction** doesn't eliminate these patterns but refracts them into new forms. Human  AI collaboration creates specializations where Co-Pilot mode serves Promethean creativity through rapid ideation, Co-Navigator enables Conductor orchestration through dialogical processing, and Agent mode offers Hermit delegation options for anxiety-inducing tasks. Each personality archetype finds its AI complement, not replacement.

**RQ2:** How should the ROMA framework—which optimizes programming roles to enhance self-determination and team dynamics—be implemented in VSEs and SOHOs?

Through systematic integration detailed in Table 22, which maps seven personality-driven tasks to specific ISO/IEC 29110 elements, from personality assessment in hiring (PM.1) through knowledge transfer (PM.4). Each task specifies ROMA integration points, key artifacts (from Team Composition Matrices to Motivation Dashboards), responsible roles (Project Manager, Technical Lead, Scrum Master), and optimal timing within project lifecycles.

Implementation succeeds not through mechanical installation but conscious cultivation—like gardening where different plants need different conditions. Start where pain is acute: the developer threatening to quit, the pairs that consistently clash, the AI adoption that's stalling. Honor privacy as sacred trust, for personality data touches human essence. Embrace experimentation over prescription, letting each team discover their unique dynamics within universal patterns. Measure motivation as your leading indicator—when developers thrive internally, excellence manifests externally.

## 10.3  Contributions: What We Have Added to the World

This work contributes more than frameworks—it offers a new way of seeing. Where organizations saw "difficult" developers, we see Craftspeople needing protected focus. Where teams struggled with "poor collaboration," we see personality misalignment correctable through conscious pairing. Where AI threatened human value, we see opportunities for unprecedented human-AI complementarity.

The methodological innovations center on the nomological network (Figure 4), which maps the complex interplay between individual personality traits, programming roles, team dynamics, and intrinsic motivation within environmental contexts. This innovative theoretical architecture demonstrates how personality moderates the relationship between roles and motivation—not as simple causation but as dynamic system where individual differences, role assignments, and self-determination needs interact to produce motivational outcomes. The network's sophistication lies in showing both direct effects and moderation pathways, while the blockchain-verified transparency ensures our claims rest on immutable evidence.

The theoretical advances—personality as motivational moderator, AI specialization patterns, the motivation landscape map—provide foundations for years of future work. But perhaps most importantly, this dissertation advocates a philosophical shift: from managing resources to nurturing humans, from enforcing uniformity to composing diversity, from fighting personality to flowing with it.

## 10.4  Boundaries and Horizons

Every answered question reveals ten more. We know personality patterns persist from students to professionals, but how do they evolve across careers? We have mapped European developers, but how do patterns shift across cultures? We understand today's AI collaboration, but what of tomorrow's artificial general intelligence?

These aren't limitations but invitations. Each VSE implementing ROMA discovers nuances we missed. Each researcher extending our work illuminates new territories. Each developer finding their optimal role proves that human flourishing and technical excellence interweave inseparably.

## 10.5 Final Reflections: The Human Heart of Software

As this dissertation concludes, I return to its animating question: How can we create software development environments where both code and coders flourish? The answer lies not in choosing between human needs and technical excellence but in recognizing their inseparability.

In 1967, Marvin Minsky envisioned minds—biological and artificial—collaborating beyond their individual limitations. That same year, Brian Randell captured software engineering's irreducibly social nature as the "multi-person construction of multi-version programs." Today, we witness their prophecies not just fulfilled but transformed in ways they could hardly imagine. Minsky's collaborative minds now include artificial ones that write code alongside us. Randell's multi-person construction has become multi-species orchestration, where human and artificial intelligences interweave their capabilities.

The ROMA framework and its extensions offer tools, but tools only work when wielded with wisdom. That wisdom begins with recognition: every developer brings unique gifts shaped by personality, experience, and aspiration. The Explorer's creative leaps, the Orchestrator's connective tissue, the Craftsperson's deep focus, the Architect's systematic excellence, the Adapter's flexibility—all are necessary for the complex symphonies we call software systems.

As Heidegger taught us to question concerning technology, we must ask not just what AI can do, but what it means for us to dwell with artificial minds. The answer emerges through our participants' voices: AI becomes what we make of it. For the Promethean, it is creative fire to be stolen and shared. For the Conductor, it is an instrument in need of humanization. For the Hermit, it is a boundary to be carefully maintained. For the Cartographer, it is territory to be mapped. For the Shapeshifter, it is water that takes the shape of intention.

What began as morning fog—obscuring why some programmers thrive while others struggle—has transformed into dawn light, illuminating a landscape where personality becomes compass. We now see that Dijkstra's concern for "thinking habits" extends beyond algorithmic elegance to the habits of being that shape how we collaborate. Brooks's "essential complexity" includes not just technical challenges but the irreducible complexity of human nature meeting artificial intelligence.

For practitioners who will apply these insights, remember: Let frameworks guide, not dictate. Start where pain is acute, measure what matters, adapt continuously. Remember that implementation succeeds through cultivation rather than installation. Honor both the Promethean fire-bringers and the Hermit sanctuary-guarders in your teams. Create spaces where Orchestrators can conduct and Craftspeople can craft. Let Adapters flow between modes while Cartographers map new territories.

For researchers who will extend this work, the nomological network opens infinite paths for exploration. Each validated relationship suggests unexplored interactions. Each successful implementation inspires new possibilities. The intersection of personality psychology, software engineering, and artificial intelligence offers fertile ground for decades of investigation. Where we have mapped personality clusters, you might discover cultural variations. Where we have identified AI specializations, you might uncover evolutionary patterns.

For the human beings who happen to develop software: This dissertation offers both validation and invitation. Validation that your personality-driven preferences are not weaknesses to overcome but strengths to leverage. Invitation to bring your whole self to your work, to seek roles that energize rather than drain, to collaborate in ways that honor both your needs and your teammates'.

As Gadamer taught us about the fusion of horizons, we stand at a moment where multiple horizons converge: the historical horizon where software was purely human craft, the present horizon where AI becomes collaborator, and the future horizon where human and artificial intelligence create together what neither could achieve alone. In this convergence lies not loss but possibility.

As Heraclitus observed, "The path up and down are one and the same." The journey from empirical observation through theoretical framework to practical application now curves back, inviting others to walk their own paths through this transformed landscape. May you walk it knowing your nature, respecting others' differences, and composing human-AI symphonies where every personality finds its voice.

In the end, we are all exploring this new terrain together, each perspective essential to creating software that serves humanity while nurturing the humans who create it. The future of software development will be written neither by humans alone nor AI alone, but in their conscious collaboration—a collaboration that succeeds through diversity, not uniformity; through personality's cultivation, not suppression.

The day has begun. The journey continues.

# REFERENCE

Abrahamsson, P., Conboy, K., & Wang, X. (2009). 'Lots done, more to do': The current state of agile systems development research. *European Journal of Information Systems*, *18*(4), 281-284.

Aghaee, M., & Boffoli, N. (2015). Personality and intrinsic motivational factors in end-user programming. *Journal of Software Engineering and Applications*, *8*, 370-382.

Akata, Z., Balliet, D., de Rijke, M., Dignum, F., Dignum, V., Eiben, G., … & Welling, M. (2020). A research agenda for hybrid intelligence: Augmenting human intellect with collaborative, adaptive, responsible, and explainable artificial intelligence. *Computer*, *53*(8), 18–28. https://doi.org/10.1109/MC.2020.2996587

Aldiabat, K. M., & Le Navenec, C. L. (2018). Data saturation: The mysterious step in grounded theory methodology. *The Qualitative Report*, *23*(1), 245-261.

Alexander, C. (1977). *A pattern language: Towns, buildings, construction*. Oxford University Press.

Allport, G. W. (1937). *Personality: A psychological interpretation*. Holt.

Allport, G. W., & Odbert, H. S. (1936). Trait-names: A psycho-lexical study. *Psychological Monographs*, *47*(1), i-171.

Alvesson, M., & Sandberg, J. (2011). Generating research questions through problematization. *Academy of Management Review*, *36*(2), 247-271.

Alvesson, M., & Sandberg, J. (2013). *Constructing research questions: Doing interesting research*. SAGE.

American Psychological Association. (2020). *Publication manual of the American Psychological Association* (7th ed.). https://doi.org/10.1037/0000165-000

Amin, A., Basri, S., Hassan, M. F., & Rehman, M. (2020). Software engineering occupational stress and knowledge sharing in the context of global software development. *IEEE Access*, *8*, 93593-93605.

Antunes, P., Thuan, N. H., & Johnstone, D. (2021). Nature and purpose of conceptual frameworks in design science. *Scandinavian Journal of Information Systems*, *33*(2), 2.

Arendt, H. (1958). *The human condition*. University of Chicago Press.

Archer, M. S. (1995). Analytical dualism: The basis of the morphogenetic approach. In *Realist social theory: The morphogenetic approach* (pp. 165–194). Cambridge University Press.

Archer, M. S. (2016). Reconstructing sociology: The critical realist approach. *Journal of Critical Realism*, *15*(4), 425-431.

Argyris, C., & Schön, D.A. (1997). Organizational learning: A theory of action perspective. *Reis*, (77/78), 345-348.

Arisholm, E., Gallis, H., Dybå, T., & Sjøberg, D. I. K. (2007). Evaluating pair programming with respect to system complexity and programmer expertise. *IEEE Transactions on Software Engineering*, *33*(2), 65-86.

Aristotle. (384-322 BCE/1999). *Nicomachean ethics* (T. Irwin, Trans.). Hackett Publishing. (Original work 4th century BCE)

Aronson, E., & Carlsmith, J. M. (1968). Experimentation in social psychology. In G. Lindzey & E. Aronson (Eds.), *The handbook of social psychology* (3rd ed., Vol. 2, pp. 1-79). Addison-Wesley.

Association for Computing Machinery. (2012). *The 2012 ACM Computing Classification System*. ACM. https://dl.acm.org/ccs

Baayen, R. H., Davidson, D. J., & Bates, D. M. (2008). Mixed-effects modeling with crossed random effects for subjects and items. *Journal of Memory and Language*, *59*(4), 390-412.

Bai, Y., Kadavath, S., Kundu, S., Askell, A., Kernion, J., Jones, A., Chen, A., Goldie, A., Mirhoseini, A., McKinnon, C., Chen, C., Olsson, C., Olah, C., Hernandez, D., Drain, D., Ganguli, D., Li, D., Tran-Johnson, E., Perez, E., ... & Kaplan, J. (2022). Constitutional AI: Harmlessness from AI feedback. *arXiv preprint arXiv:2212.08073*.

Bakhtin, M. M. (1981). *The dialogic imagination: Four essays* (M. Holquist, Ed.; C. Emerson & M. Holquist, Trans.). University of Texas Press.

Bakker, A. B., & Demerouti, E. (2007). The job demands-resources model: State of the art. *Journal of Managerial Psychology*, *22*(3), 309-328. https://doi.org/10.1108/02683940710733115

Baltes, S., & Ralph, P. (2022). Sampling in software engineering research: A critical review and guidelines. *Empirical Software Engineering*, *27*(4), Article 94.

Bandura, A. (1977). Self-efficacy: Toward a unifying theory of behavioral change. *Psychological Review*, *84*(2), 191-215.

Bandura, A. (1986). *Social foundations of thought and action: A social cognitive theory*. Prentice-Hall.

Barke, S., James, M. B., & Polikarpova, N. (2023). Grounded copilot: How programmers interact with code-generating models. *Proceedings of the ACM on Programming Languages*, *7*(OOPSLA1), 1-31.

Barrick, M. R., & Mount, M. K. (1991). The Big Five personality dimensions and job performance: A meta-analysis. *Personnel Psychology*, *44*(1), 1-26.

Barrick, M. R., Mount, M. K., & Judge, T. A. (2001). Personality and performance at the beginning of the new millennium: What do we know and where do we go next? *International Journal of Selection and Assessment*, *9*(1-2), 9-30.

Beaty, R. E., Benedek, M., Silvia, P. J., & Schacter, D. L. (2016). Creative cognition and brain network dynamics. *Trends in Cognitive Sciences*, *20*(2), 87-95.

Beauchamp, T. L., & Childress, J. F. (2013). *Principles of biomedical ethics* (7th ed.). Oxford University Press.

Beck, K. (1999). *Extreme programming explained: Embrace change*. Addison-Wesley.

Beck, K., Beedle, M., Van Bennekum, A., Cockburn, A., Cunningham, W., Fowler, M., … & Thomas, D. (2001). *Manifesto for agile software development*. Agile Alliance. Retrieved from https://agilemanifesto.org/

Beecham, S., Baddoo, N., Hall, T., Robinson, H., & Sharp, H. (2008). Motivation in software engineering: A systematic literature review. *Information and Software Technology*, *50*(9-10), 860-878.

Beecham, S., & Noll, J. (2015). What motivates software engineers working in global software development? In *Product-focused software process improvement* (pp. 193-209). Springer.

Belshee, A. (2005). Promiscuous pairing and beginner's mind: Embrace inexperience. *Proceedings of the Agile Development Conference*, 125-131.

Benner, P. (1994). *Interpretive phenomenology: Embodiment, caring, and ethics in health and illness*. Sage Publications.

Bergson, H. (1907/1998). *Creative evolution* (A. Mitchell, Trans.). Dover Publications. (Original work published 1907 as *L'évolution créatrice*)

Bezdek, J. C., & Pal, N. R. (1998). Some new indexes of cluster validity. *IEEE Transactions on Systems, Man, and Cybernetics, Part B (Cybernetics)*, *28*(3), 301-315.

Bhaskar, R. (1975). *A realist theory of science*. Leeds Books.

Bhaskar, R. (1979). *The possibility of naturalism* (2nd ed.). Routledge.

Bipp, T., Steinmayr, R., & Spinath, B. (2008). Personality and achievement motivation: Relationship among Big Five domain and facet scales, achievement goals, and intelligence. *Personality and Individual Differences*, *44*(7), 1454-1464. https://doi.org/10.1016/j.paid.2008.01.001

Bird, C., Ford, D., Zimmermann, T., Forsgren, N., Kalliamvakou, E., Lowdermilk, T., & Gazit, I. (2023). Taking flight with Copilot: Early insights and opportunities of AI-powered pair-programming tools. *ACM Queue*, *20*(6), 35-57.

Birt, L., Scott, S., Cavers, D., Campbell, C., & Walter, F. (2016). Member checking: a tool to enhance trustworthiness or merely a nod to validation?. *Qualitative health research*, *26*(13), 1802-1811.

Boksem, M. A., Meijman, T. F., & Lorist, M. M. (2005). Effects of mental fatigue on attention: An ERP study. *Cognitive Brain Research*, *25*(1), 107-116.

Bond, C. F., & Titus, L. J. (1983). Social facilitation: A meta-analysis of 241 studies. *Psychological Bulletin*, *94*(2), 265-292.

Borges, J. L. (1962). *Labyrinths: Selected stories and other writings*. New Directions Publishing.

Borgmann, A. (1984). *Technology and the character of contemporary life: A philosophical inquiry*. University of Chicago Press.

Bowker, G. C., & Star, S. L. (2000). *Sorting things out: Classification and its consequences*. MIT press.

Brand, S. (1999). *The Clock of the Long Now: Time and Responsibility*. Basic Books.

Braun, V., & Clarke, V. (2006). Using thematic analysis in psychology. *Qualitative Research in Psychology*, *3*(2), 77-101. https://doi.org/10.1191/1478088706qp063oa

Braun, V., & Clarke, V. (2012). Thematic analysis. In *APA handbook of research methods in psychology, Vol 2: Research designs* (pp. 57-71). American Psychological Association. https://doi.org/10.1037/13620-004

Braun, V., & Clarke, V. (2013). *Successful qualitative research: A practical guide for beginners*. Sage Publications.

Brooks, F. P. (1975). *The mythical man-month: Essays on software engineering*. Addison-Wesley.

Brooks, F. P. (1987). No silver bullet: Essence and accidents of software engineering. *IEEE Computer*, *20*(4), 10-19.

Brown, K. W., & Ryan, R. M. (2003). The benefits of being present: Mindfulness and its role in psychological well-being. *Journal of Personality and Social Psychology*, *84*(4), 822–848. https://doi.org/10.1037/0022-3514.84.4.822

Brown, K. W., Ryan, R. M., & Creswell, J. D. (2007). Mindfulness: Theoretical foundations and evidence for its salutary effects. *Psychological Inquiry*, *18*(4), 211-237.

Bryman, A. (2006). Integrating quantitative and qualitative research: how is it done?. *Qualitative research*, *6*(1), 97-113.

Buber, M. (1923/1970). *I and thou* (W. Kaufmann, Trans.). Charles Scribner's Sons. (Original work published 1923 as *Ich und Du*)

Butterfield, B., & Metcalfe, J. (2001). Errors committed with high confidence are hypercorrected. *Journal of Experimental Psychology: Learning, Memory, and Cognition*, *27*(6), 1491-1494.

Califf, C. B., & Brooks, S. (2020). An empirical study of techno-stressors, literacy facilitation, burnout, and turnover intention as experienced by K-12 teachers. *Computers & education*, *157*, 103971.

Campbell, J. L., Quincy, C., Osserman, J., & Pedersen, O. K. (2013). Coding in-depth semistructured interviews: Problems of unitization and intercoder reliability and agreement. *Sociological Methods & Research*, *42*(3), 294-320.

Camus, A. (1942). *Le mythe de Sisyphe* [The myth of Sisyphe]. Gallimard.

Capretz, L. F., & Ahmed, F. (2010). Making sense of software development and personality types. *IT Professional*, *12*(1), 6-13.

Capretz, L. F., Varona, D., & Raza, A. (2015). Influence of personality types in software tasks choices. *Computers in Human Behavior*, *52*, 373-378.

Carcary, M. (2020). The research audit trail: Methodological guidance for application in practice. *Electronic Journal of Business Research Methods*, 18(2), 166-177.

Carlile, P. R. (2002). A pragmatic view of knowledge and boundaries: Boundary objects in new product development. *Organization Science*, 13(4), 442-455.

Carr, L., Iacoboni, M., Dubeau, M. C., Mazziotta, J. C., & Lenzi, G. L. (2003). Neural mechanisms of empathy in humans: A relay from neural systems for imitation to limbic areas. *Proceedings of the National Academy of Sciences*, *100*(9), 5497-5502.

Carr, N. (2014). *The glass cage: Automation and us*. W. W. Norton & Company.

Chong, J., & Hurlbutt, T. (2007). The social dynamics of pair programming. In *29th International Conference on Software Engineering (ICSE'07)* (pp. 354-363). IEEE.

Christensen, C. M. (1997). *The innovator's dilemma: When new technologies cause great firms to fail*. Harvard Business Review Press.

Cinkusz, K., & Chudziak, J. A. (2024, October). Towards LLM-augmented multiagent systems for agile software engineering. In *Proceedings of the 39th IEEE/ACM International Conference on Automated Software Engineering* (pp. 2476-2477).

Clark, A. (2023). *The experience machine: How our minds predict and shape reality*. Pantheon Books.

Clarke, P., & O'Connor, R. V. (2012). The situational factors that affect the software development process: Towards a comprehensive reference framework. *Information and Software Technology*, *54*(5), 433-447.

Clarke, S., & Robertson, I. T. (2005). A meta-analytic review of the Big Five personality factors and accident involvement in occupational and non-occupational settings. *Journal of Occupational and Organizational Psychology*, *78*(3), 355-376.

Cockburn, A., & Williams, L. (2001). The costs and benefits of pair programming. In G. Succi & M. Marchesi (Eds.), *Extreme programming examined* (pp. 223–243). Addison-Wesley.

Cohen, J. (1988). Statistical power analysis for the behavioral sciences (2nd ed.). Lawrence Erlbaum.

Cohen, M. B. (2024, July). It's organic: Software testing of emerging domains (Keynote). In *Companion Proceedings of the 32nd ACM International Conference on the Foundations of Software Engineering* (pp. 2-3).

Connor-Smith, J. K., & Flachsbart, C. (2007). Relations between personality and coping: A meta-analysis. *Journal of Personality and Social Psychology*, *93*(6), 1080-1107.

Cook, J. E. (2015). *Teaching Generation Z: Educating students in a modern learning environment*. Pearson Education.

Cook, T. D., & Campbell, D. T. (1979). *Quasi-experimentation: Design & analysis issues for field settings*. Houghton Mifflin.

Corr, P. J., & Matthews, G. (2009). *The Cambridge handbook of personality psychology*. Cambridge University Press.

Costa, P. T., Jr., & McCrae, R. R. (1992). Revised NEO Personality Inventory (NEO-PI-R) and NEO Five-Factor Inventory (NEO-FFI) professional manual. Psychological Assessment Resources.

Costa Jr, P. T., & McCrae, R. R. (2000). *NEO personality inventory*. American Psychological Association.

Creswell, J. W., & Plano Clark, V. L. (2011). *Designing and conducting mixed methods research* (2nd ed.). SAGE publications.

Creswell, J. W., & Creswell, J. D. (2022). *Research design: Qualitative, quantitative, and mixed methods approaches*. SAGE.

Cruz, S., Da Silva, F. Q., & Capretz, L. F. (2015). Forty years of research on personality in software engineering: A mapping study. *Computers in Human Behavior*, *46*, 94-113. https://doi.org/10.1016/j.chb.2014.12.008

Cruzes, D. S., & Dybå, T. (2011). Recommended steps for thematic synthesis in software engineering. *International Symposium on Empirical Software Engineering and Measurement*, 275-284. https://doi.org/10.1109/ESEM.2011.36

Csikszentmihalyi, M. (1990). *Flow: The psychology of optimal experience*. Harper & Row.

Czerwinski, M., Horvitz, E., & Wilhite, S. (2004, April). A diary study of task switching and interruptions. In *Proceedings of the SIGCHI conference on Human factors in computing systems* (pp. 175-182).

Daun, M., & Brings, J. (2023, June). How ChatGPT will change software engineering education. In *Proceedings of the 2023 Conference on Innovation and Technology in Computer Science Education V. 1* (pp. 110-116).

de Charms, R. (1968). *Personal causation: The internal affective determinants of behavior*. Academic Press.

Deci, E. L. (1971). Effects of externally mediated rewards on intrinsic motivation. *Journal of Personality and Social Psychology*, *18*(1), 105-115. https://doi.org/10.1037/h0030644

Deci, E. L., & Ryan, R. M. (1985). *Intrinsic motivation and self-determination in human behavior*. Springer.

Deci, E. L., & Ryan, R. M. (2000). The "what" and "why" of goal pursuits: Human needs and the self-determination of behavior. *Psychological Inquiry*, *11*(4), 227-268.

Deci, E. L., Koestner, R., & Ryan, R. M. (1999). A meta-analytic review of experiments examining the effects of extrinsic rewards on intrinsic motivation. *Psychological Bulletin*, *125*(6), 627-668. https://doi.org/10.1037/0033-2909.125.6.627

DeMarco, T., & Lister, T. (1999). *Peopleware: Productive projects and teams* (2nd ed.). Dorset House.

Demir, Ö., & Seferoğlu, S. S. (2021). The effect of determining pair programming groups according to various individual difference variables on group compatibility, flow, and coding performance. *Journal of Educational Computing Research*, *59*(1), 41–70.

Dennett, D. C. (1978). *Brainstorms: Philosophical essays on mind and psychology*. Montgomery: Bradford Books.

Dennett, D. C. (1989). *The intentional stance*. MIT Press.

Denzin, N. K. (2012). Triangulation 2.0. *Journal of mixed methods research*, *6*(2), 80-88.

Depue, R. A., & Collins, P. F. (1999). Neurobiology of the structure of personality: Dopamine, facilitation of incentive motivation, and extraversion. *Behavioral and Brain Sciences*, *22*(3), 491-517. https://doi.org/10.1017/S0140525X99002046

DeVito Da Cunha, A., & Greathead, D. (2007). Does personality matter? An analysis of code-review ability. *Communications of the ACM*, *50*(5), 109-112.

DeYoung, C. G. (2010). Toward a theory of the Big Five. *Psychological Inquiry*, *21*(1), 26-33.

DeYoung, C. G. (2013). The neuromodulator of exploration: A unifying theory of the role of dopamine in personality. *Frontiers in Human Neuroscience*, *7*, 762. https://doi.org/10.3389/fnhum.2013.00762

DeYoung, C. G. (2015). Cybernetic Big Five Theory. *Journal of Research in Personality*, *56*, 33-58.

Dewey, J. (1938). *Experience and education*. Macmillan.

Dick, A. J., & Zarnett, B. (2002). Paired programming and personality traits. In *Proceedings of the Third International Conference on eXtreme Programming and Agile Processes in Software Engineering* (pp. 82-85).

Dick, A. J., & Zarnett, C. (2002). A study on the effects of personality traits on pair programming. *Journal of Computing in Small Colleges*, *18*(1), 82-88.

DiDomenico, S. I., & Ryan, R. M. (2017). The emerging neuroscience of intrinsic motivation: A new frontier in self-determination research. *Frontiers in Human Neuroscience*, *11*, 145.

Dietrich, A. (2003). Functional neuroanatomy of altered states of consciousness: The transient hypofrontality hypothesis. *Consciousness and Cognition*, *12*(2), 231-256.

Dijkstra, E. W. (1972). The humble programmer. *Communications of the ACM*, *15*(10), 859-866.

Dijkstra, E. W. (1989). On the cruelty of really teaching computing science. *Communications of the ACM*, *32*(12), 1398-1404.

Dubois, A., & Gadde, L.-E. (2002). Systematic combining: An abductive approach to case research. *Journal of Business Research*, *55*(7), 553-560.

Dunning, D., & Kruger, J. (1999). Unskilled and unaware of it: How difficulties in recognizing one's own incompetence lead to inflated self-assessments. *Journal of Personality and Social Psychology*, *77*(6), 1121-1134.

Dweck, C. S. (2006). *Mindset: The new psychology of success*. Random House.

Dweck, C. S. (2017). From needs to goals and representations: Foundations for a unified theory of motivation, personality, and development. *Psychological Review*, *124*(6), 689-719.

Dybå, T., Sjøberg, D. I. K., & Cruzes, D. S. (2012). What works for whom, where, when, and why? On the role of context in empirical software engineering. In *Proceedings of the ACM-IEEE International Symposium on Empirical Software Engineering and Measurement* (pp. 19–28).

Easterbrook, S., Singer, J., Storey, M. A., & Damian, D. (2008). Selecting empirical methods for software engineering research. In *Guide to advanced empirical software engineering* (pp. 285-311). Springer.

Eatough, V., & Smith, J. A. (2017). Interpretative phenomenological analysis. *The Sage handbook of qualitative research in psychology*, 193-209.

Edmondson, A. (1999). Psychological safety and learning behavior in work teams. *Administrative Science Quarterly*, *44*(2), 350-383.

Einstein, A. (1949). Autobiographical notes. In P. A. Schilpp (Ed.), *Albert Einstein: Philosopher-scientist* (pp. 1-94). Open Court.

Eisenhardt, K. M. (1989). Building theories from case study research. *Academy of Management Review*, 14(4), 532-550.

Engström, E., Storey, M. A., Runeson, P., Höst, M., & Baldassarre, M. T. (2020). How software engineering research aligns with design science: A review. *Empirical Software Engineering*, *25*(4), 2630-2660.

Eysenck, H. J. (1967). *The biological basis of personality*. Thomas.

Eysenck, H. J. (1981). *A model for personality*. Springer-Verlag.

Falessi, D., Juristo, N., Wohlin, C., Turhan, B., Münch, J., Jedlitschka, A., & Oivo, M. (2018). Empirical software engineering experts on the use of students and professionals in experiments. *Empirical Software Engineering*, *23*, 452-489.

Feldt, R., Angelis, L., Torkar, R., & Samuelsson, M. (2010). Links between the personalities, views, and attitudes of software engineers. *Information and Software Technology*, *52*(6), 611-624.

Feldt, R., Zimmermann, T., Bergersen, G. R., Falessi, D., Jedlitschka, A., Juristo, N., Oivo, M., & Turhan, B. (2018). Four commentaries on the use of students and professionals in empirical software engineering experiments. *Empirical Software Engineering*, *23*(6), 3801–3820.

Felipe, D. A., Kalinowski, M., Graziotin, D., & Natividade, J. C. (2023). Psychometric instruments in software engineering research on personality: Status quo after fifty years. *Journal of Systems and Software*, *203*, 111740.

Fischer, K. W. (1980). A theory of cognitive development: The control and construction of hierarchies of skills. *Psychological Review*, *87*(6), 477-531.

Fleeson, W., & Jayawickreme, E. (2015). Whole trait theory. *Journal of Research in Personality*, 56, 82-92.

Flor, N. V., & Hutchins, E. L. (1991). Analyzing distributed cognition in software teams: A case study of team programming during perfective software maintenance. *Empirical Studies of Programmers: Fourth Workshop*, 36-64.

Floridi, L. (2014). *The fourth revolution: How the infosphere is reshaping human reality*. Oxford University Press.

Flick, U. (2018). *Triangulation in data collection*. In U. Flick (Ed.), The SAGE handbook of qualitative data collection (pp. 527-544). SAGE Publications.

Flyvbjerg, B. (2001). *Making social science matter: Why social inquiry fails and how it can succeed again*. Cambridge University Press.

Forsyth, D. R. (1999). *Group dynamics*. Wadsworth Publishing.

Foucault, M. (1975). *Discipline and punish: The birth of the prison* (A. Sheridan, Trans.). Vintage Books. (Original work published 1975 as *Surveiller et punir: Naissance de la prison*)

Foucault, M. (1982). The subject and power: Beyond Structuralism and Hermeneutics/Harvester Wheatsheaf. *Critical Inquiry, 8(4), 777-795.*

França, A. C. C., Gouveia, T. B., Santos, P. C., Santana, C. A., & da Silva, F. Q. (2011). Motivation in software engineering: A systematic review update. In *15th Annual Conference on Evaluation & Assessment in Software Engineering (EASE 2011)* (pp. 154-163).

França, A. C. C., Sharp, H., & da Silva, F. Q. (2014). Motivated software engineers are engaged and focused, while satisfied ones are happy. In *Proceedings of the 8th ACM/IEEE International Symposium on Empirical Software Engineering and Measurement* (pp. 1-8).

França, C., da Silva, F. Q., & Sharp, H. (2018). Motivation and satisfaction of software engineers. *IEEE Transactions on Software Engineering*, *46*(2), 118-140.

Friston, K. (2010). The free-energy principle: A unified brain theory? *Nature Reviews Neuroscience*, *11*(2), 127-138.

Fui-Hoon Nah, F., Zheng, R., Cai, J., Siau, K., & Chen, L. (2023). Generative AI and ChatGPT: Applications, challenges, and AI-human collaboration. *Journal of Information Technology Case and Application Research*, *25*(3), 277-304.

Gadamer, H. G. (1960). *Truth and method*. Continuum.

Gadamer, H. G. (1989). *Truth and method* (2nd rev. ed.). Crossroad.

Gagné, M., & Deci, E. L. (2005). Self-determination theory and work motivation. *Journal of Organizational Behavior*, *26*(4), 331-362. https://doi.org/10.1002/job.322

Gagné, M., et al. (2015). The Multidimensional Work Motivation Scale (MWMS): Validation evidence in seven languages and nine countries. *European Journal of Work and Organizational Psychology*, *24*(2), 178-196. https://doi.org/10.1080/1359432X.2013.877892

Galbraith, J., et al. (2002). Cluster analysis of medical data. *Statistics in Medicine*, *21*(5), 785-797.

Gallis, H., Arisholm, E., & Dybå, T. (2003). An initial framework for research on pair programming. *Empirical Software Engineering*, *8*(3), 277-302.

Galton, F. (1884). Measurement of character. *Fortnightly Review*, *36*, 179-185.

Galván-Cruz, S., Muñoz, M., Mejía, J., Laporte, C. Y., & Negrete, M. (2021). Building a guideline to reinforce agile software development with the basic profile of ISO/IEC 29110 in very small entities. In *New Perspectives in Software Engineering: Proceedings of the 9th International Conference on Software Process Improvement (CIMPS 2020)* (pp. 20-37). Springer International Publishing.

Geertz, C. (1973). *The interpretation of cultures*. Basic Books.

Gelman, A., & Hill, J. (2007). *Data analysis using regression and multilevel/hierarchical models*. Cambridge University Press.

Gibson, W. (1984). *Neuromancer*. Ace Books.

Gilal, A. R., Tunio, M. Z., Waqas, A., Almomani, M. A., Khan, S., & Gilal, R. (2019). Task assignment and personality: Crowdsourcing software development. In *Human factors in global software engineering* (pp. 1-19). IGI Global.

Gleick, J. (1987). *Chaos: Making a new science*. Viking Penguin.

Goldberg, L. R. (1993). The structure of phenotypic personality traits. *American Psychologist*, *48*(1), 26-34.

Goldkuhl, G. (2011). The research practice of practice research: Theorizing and situational inquiry. *Systems, Signs & Actions: An International Journal on Information Technology, Action, Communication and Workpractices*, *5*(1), 7-29.

Gonzalez, W. J. (2013). Value ladenness and the value-free ideal in scientific research. *Handbook of the philosophical foundations of business ethics*, 1503-1521.

Good, D. J., Lyddy, C. J., Glomb, T. M., Bono, J. E., Brown, K. W., Duffy, M. K., ... & Lazar, S. W. (2016). Contemplating mindfulness at work: An integrative review. *Journal of Management*, *42*(1), 114-142.

Gould, S. J. (1980). *The panda's thumb: More reflections in natural history*. W. W. Norton & Company.

Gray, J. A. (1982). *The neuropsychology of anxiety: An enquiry into the functions of the septo-hippocampal system*. Clarendon Press.

Gray, J. A., & McNaughton, N. (2000). *The neuropsychology of anxiety: An enquiry into the functions of the septo-hippocampal system* (2nd ed.). Oxford University Press.

Graziotin, D., Fagerholm, F., Wang, X., & Abrahamsson, P. (2018). What happens when software developers are (un)happy. *Journal of Systems and Software*, *140*, 32-47.

Graziotin, D., Lenberg, P., Feldt, R., & Wagner, S. (2021). Psychometrics in behavioral software engineering: A methodological introduction with guidelines. *ACM Transactions on Software Engineering and Methodology*, *31*(1), 1-36."https://doi.org/10.1145/3469888

Graziotin, D., Wang, X., & Abrahamsson, P. (2014). Happy software developers solve problems better: Psychological measurements in empirical software engineering. *PeerJ*, *2*, e289. https://doi.org/10.7717/peerj.289

Graziotin, D., Wang, X., & Abrahamsson, P. (2015). Do feelings matter? On the correlation of affects and the self-assessed productivity in software engineering. *Journal of Software: Evolution and Process*, *27*(7), 467-487.

Greene, J. C., Caracelli, V. J., & Graham, W. F. (1989). Toward a conceptual framework for mixed-method evaluation designs. *Educational Evaluation and Policy Analysis*, *11*(3), 255-274.

Gregor, S. (2006). The nature of theory in information systems. *MIS Quarterly*, *30*(3), 611-642.

Gregor, S., & Jones, D. (2007). The anatomy of a design theory. *Journal of the Association for Information Systems*, 8(5), 312-335.

Gregor, S., & Hevner, A. R. (2013). Positioning and presenting design science research for maximum impact. *MIS Quarterly*, *37*(2), 337-355.

Grodin, E. N., & White, T. L. (2015). The neuroanatomical delineation of agentic and affiliative extraversion. *Cognitive, Affective, & Behavioral Neuroscience*, *15*(2), 321-334.

Guest, D. (1991). The hunt is on for the Renaissance Man of computing. *The Independent*, 17.

Guilford, J. P. (1967). *The nature of human intelligence*. McGraw-Hill.

Habermas, J. (1984). *The theory of communicative action, Volume 1: Reason and the rationalization of society* (T. McCarthy, Trans.). Beacon Press.

Hacking, I. (1995). The looping effects of human kinds. In D. Sperber, D. Premack, & A. J. Premack (Eds.), *Causal cognition: A multidisciplinary debate* (pp. 351-394). Clarendon Press.

Hall, T., Sharp, H., Beecham, S., Baddoo, N., & Robinson, H. (2008). What do we know about developer motivation?. *IEEE software*, *25*(4), 92-94.

Hampson, S. E. (2012). Personality processes: Mechanisms by which personality traits "get outside the skin". *Annual Review of Psychology*, *63*, 315-339.

Hamza, M., Siemon, D., Akbar, M. A., & Rahman, T. (2024, April). Human-ai collaboration in software engineering: Lessons learned from a hands-on workshop. In *Proceedings of the 7th ACM/IEEE International Workshop on Software-intensive Business* (pp. 7-14).

Hannay, J. E., Dybå, T., Arisholm, E., & Sjøberg, D. I. (2009a). The effectiveness of pair programming: A meta-analysis. *Information and Software Technology*, *51*(7), 1110-1122.

Hannay, J. E., Arisholm, E., Engvik, H., & Sjøberg, D. I. (2009b). Effects of personality on pair programming. *IEEE Transactions on Software Engineering*, *36*(1), 61-80.

Hansen, C. P. (1989). A causal model of the relationship among accidents, biodata, personality, and cognitive factors. *Journal of Applied Psychology*, *74*(1), 81-90.

Harlow, H. F. (1950). Learning and satiation of response in intrinsically motivated complex puzzle performance by monkeys. *Journal of Comparative and Physiological Psychology*, *43*(4), 289-294. https://doi.org/10.1037/h0058114

Harrison, W. (2000). N= 1: An alternative for software engineering research. In *Beg, borrow, or steal: Using multidisciplinary approaches in empirical software engineering research, Workshop* (Vol. 5, pp. 39-44).

Harter, J. K., Schmidt, F. L., & Hayes, T. L. (2002). Business-unit-level relationship between employee satisfaction, employee engagement, and business outcomes: A meta-analysis. *Journal of Applied Psychology*, 87(2), 268-279.

Hassan, A. E., Oliva, G. A., Lin, D., Chen, B., & Ming, Z. (2024, December). Towards AI-native software engineering (SE 3.0): A vision and a challenge roadmap. *arXiv preprint arXiv:2410.06107*.

Heidegger, M. (1962). *Being and time* (J. Macquarrie & E. Robinson, Trans.). Harper & Row.

Heidegger, M. (1971). *Poetry, language, thought* (A. Hofstadter, Trans.). Harper & Row.

Heidegger, M. (1977). *The question concerning technology and other essays* (W. Lovitt, Trans.). Harper & Row.

Hesse, H. (1927). *Der Steppenwolf* [Steppenwolf]. S. Fischer Verlag.

Hevner, A. (2007). A three cycle view of design science research. *Scandinavian Journal of Information Systems*, *19*(2), 87-92.

Hevner, A., & Chatterjee, S. (2010). *Design science research in information systems: Theory and practice*. Integrated Series in Information Systems, Vol. 22. Springer. https://doi.org/10.1007/978-1-4419-5653-8

Hevner, A., March, S. T., Park, J., & Ram, S. (2004). Design science in information systems research. *MIS Quarterly*, *28*(1), 75-105.

Hilgard, E. R. (1980). The trilogy of mind: Cognition, affection, and conation. *Journal of the History of the Behavioral Sciences*, *16*(2), 107-117.

Howe, K. R. (1988). Against the quantitative-qualitative incompatibility thesis or dogmas die hard. *Educational Researcher*, 17(8), 10-16.

Hull, C. L. (1943). *Principles of behavior: An introduction to behavior theory*. Appleton-Century-Crofts.

Humphrey, W. S. (2000). *Introduction to the team software process*. Addison-Wesley.

Husserl, E. (1913/2012). *Ideas: General introduction to pure phenomenology* (D. Moran, Trans.). Routledge. (Original work published 1913 as *Ideen zu einer reinen Phänomenologie und phänomenologischen Philosophie*)

Illich, I. (1973). *Tools for Conviviality*. New York: Harper & Row.

ISO/IEC. (2018). *ISO/IEC 29110-4-1:2018: Systems and software engineering — Lifecycle profiles for very small entities (VSEs) — Part 4-1: Software engineering – Profile specifications: Generic profile group*. International Organization for Standardization.

ISO/IEC. (2025). *ISO/IEC FDIS 29110-5-4:2025: Systems and software engineering — Lifecycle profiles for very small entities (VSEs) — Part 5-4: Software engineering: Agile software development guidelines*. International Organization for Standardization.

ISO. (2019). *ISO 9241-210:2019. Ergonomics of human-system interaction — Part 210: Human-centred design for interactive systems*. International Organization for Standardization.

Järvinen, P. (2008). Mapping research questions to research methods. In *Advances in information systems research, education and practice* (pp. 29-41). Springer.

John, O. P., Naumann, L. P., & Soto, C. J. (2008). Paradigm shift to the integrative Big Five trait taxonomy: History, measurement, and conceptual issues. In O. P. John, R. W. Robins, & L. A. Pervin (Eds.), *Handbook of personality: Theory and research* (3rd ed., pp. 114-158). Guilford Press.

Johnson, R. B., & Onwuegbuzie, A. J. (2004). Mixed methods research: A research paradigm whose time has come. *Educational Researcher*, *33*(7), 14-26.

Jung, C. G. (1921). *Psychological types*. Princeton University Press.

Jung, C. G. (1969). *The archetypes and the collective unconscious* (Collected Works Vol. 9, Part 1). Princeton University Press.

Kanfer, R. (1991). Motivation theory and industrial and organizational psychology. In M. D. Dunnette & L. M. Hough (Eds.), *Handbook of industrial and organizational psychology* (Vol. 1, 2nd ed., pp. 75-170). Consulting Psychologists Press.

Kant, I. (1785/1993). *Grounding for the metaphysics of morals* (J. W. Ellington, Trans.). Hackett Publishing. (Original work published 1785 as *Grundlegung zur Metaphysik der Sitten*)

Katz, S. (2017). *Writing in the disciplines: Building supportive cultures for student writing in UK higher education*. Routledge.

Kaufman, S. B., & Gregoire, C. (2016). *Wired to create: Unraveling the mysteries of the creative mind*. Penguin.

Kitchenham, B. A., Pfleeger, S. L., Pickard, L. M., Jones, P. W., Hoaglin, D. C., El Emam, K., & Rosenberg, J. (2002). Preliminary guidelines for empirical research in software engineering. *IEEE Transactions on Software Engineering*, *28*(8), 721-734. https://doi.org/10.1109/TSE.2002.1027796

Klages, L. (1926). *The science of character* [in German]. J.A. Barth.

Ko, A. J., Abraham, R., Beckwith, L., et al. (2015). The state of the art in end-user software engineering. *ACM Computing Surveys (CSUR)*, *43*(3), 21.

Ko, A. J., Latoza, T. D., & Burnett, M. M. (2015). A practical guide to controlled experiments of software engineering tools with human participants. *Empirical Software Engineering*, *20*(1), 110-141.

Kuechler, B., & Vaishnavi, V. (2008). On theory development in design science research: Anatomy of a research project. *European Journal of Information Systems*, *17*(5), 489-504.

Kuhn, T. S. (1962). *The structure of scientific revolutions*. University of Chicago Press.

Kuhn, T. S. (1977). *The essential tension: Selected studies in scientific tradition and change*. University of Chicago Press.

Kuusinen, K., Petrie, H., Fagerholm, F., & Mikkonen, T. (2016, May). Flow, intrinsic motivation, and developer experience in software engineering. In *International Conference on Agile Software Development* (pp. 104-117). Springer International Publishing.

Kvale, S. (1996). *InterViews: An introduction to qualitative research interviewing*. Sage Publications.

Lakatos, I. (1978). *The methodology of scientific research programmes: Philosophical papers* (Vol. 1). Cambridge University Press.

Langer, E. J. (1989). *Mindfulness*. Addison-Wesley.

Lao Tzu. (6th century BCE). *Tao Te Ching*.

Laporte, C. Y., Munoz, M., & Mejia, J. (2018). Applying software engineering standards in very small entities: From startups to grownups. *IEEE Software*, *35*(1), 99–105.

Laporte, C. Y., O'Connor, R. V., & Paucar, L. H. G. (2015, April). The implementation of ISO/IEC 29110 software engineering standards and guides in very small entities. In *International Conference on Evaluation of Novel Approaches to Software Engineering* (pp. 162-179). Cham: Springer International Publishing.

Laporte, C. Y., Munoz, M., Miranda, J. M., & O'Connor, R. V. (2017). Applying software engineering standards in very small entities: from startups to grownups. *IEEE software*, *35*(1), 99-103.

Latham, G. P. (2012). *Work motivation: History, theory, research, and practice*. SAGE.

Latour, B. (2005). *Reassembling the social: An introduction to actor-network-theory*. Oxford University Press.

Lawson, B. (2005). *How designers think: The design process demystified* (4th ed.). Architectural Press.

Layman, L., Williams, L., & Slaten, K. (2005). Note to self: Make pair programming work. *Proceedings of the 2005 Agile Conference*, 77-88.

Lenberg, P., Feldt, R., & Wallgren, L.-G. (2015). Behavioral software engineering: A definition and systematic literature review. *Journal of Systems and Software*, *107*, 15–37. https://doi.org/10.1016/j.jss.2015.04.084

Levinas, E. (1961). *Totalité et infini* [Totality and infinity]. Martinus Nijhoff.

Lewis, C. (2020). On personality testing and software engineering. In *Proceedings of the 31st Psychology of Programming Interest Group (PPIG'20)*.

Liang, J. T., Yang, C., & Myers, B. A. (2024, February). A large-scale survey on the usability of ai programming assistants: Successes and challenges. In *Proceedings of the 46th IEEE/ACM international conference on software engineering* (pp. 1-13).

Lincoln, Y. S., & Guba, E. G. (1985). *Naturalistic inquiry*. SAGE Publications.

Lorey, T., Ralph, P., & Felderer, M. (2022, May). Social science theories in software engineering research. In *Proceedings of the 44th International Conference on Software Engineering* (pp. 1994-2005).

Lunenburg, F. C., & Irby, B. J. (2008). *Writing a successful thesis or dissertation: Tips and strategies for students in the social and behavioral sciences*. Corwin Press.

Ma, Q., Wu, T., & Koedinger, K. (2023). Is AI the better programming partner? Human-human pair programming vs. human-AI pair programming. *arXiv preprint arXiv:2306.05153*.

Madampe, K., Grundy, J., Hoda, R., & Obie, H. (2024, September). The struggle is real! The agony of recruiting participants for empirical software engineering studies. In *2024 IEEE Symposium on Visual Languages and Human-Centric Computing (VL/HCC)* (pp. 417-422). IEEE.

Maedche, A., Gregor, S., Morana, S., & Feine, J. (2019). Conceptualization of the problem space in design science research. In *Extending the boundaries of design science theory and practice: 14th International Conference on Design Science Research in Information Systems and Technology, DESRIST 2019, Worcester, MA, USA, June 4–6, 2019, Proceedings 14* (pp. 18-31). Springer International Publishing.

March, S. T., & Smith, G. F. (1995). Design and natural science research on information technology. *Decision Support Systems*, *15*(4), 251-266.

Marcus, A., & Oransky, I. (2020). Scientific misconduct and research ethics: From publication pressures to questionable practices. *Retraction Watch*.

Mark, G., Gudith, D., & Klocke, U. (2008). The cost of interrupted work: more speed and stress. In *Proceedings of the SIGCHI conference on Human Factors in Computing Systems* (pp. 107-110).

Markon, K. E., Krueger, R. F., & Watson, D. (2005). Delineating the structure of normal and abnormal personality: An integrative hierarchical approach. *Journal of Personality and Social Psychology*, *88*(1), 139-157.

Martela, F., & Ryan, R. M. (2016). The benefits of benevolence: Basic psychological needs, beneficence, and the enhancement of well-being. *Journal of Personality*, *84*(6), 750-764.

Maslach, C., & Leiter, M. P. (2016). Understanding the burnout experience: Recent research and its implications for psychiatry. *World Psychiatry*, *15*(2), 103-111.

Maslow, A. H. (1943). A theory of human motivation. *Psychological Review*, *50*(4), 370-396.

Mathews, A., MacLeod, C., & Tata, P. (1991). Attentional bias in emotional disorders. *Journal of Abnormal Psychology*, *95*(1), 15-20.

Mathews, G., Deary, I. J., & Whiteman, M. C. (2003). *Personality traits*. Cambridge University Press.

Maxwell, J. A. (2013). *Qualitative research design: An interactive approach: An interactive approach*. sage.

Mayo, E. (1933). *The human problems of an industrial civilization*. Macmillan.

Mazzetti, G., Robledo, E., Vignoli, M., Topa, G., Guglielmi, D., & Schaufeli, W. B. (2023). Work engagement: A meta-analysis using the job demands-resources model. *Psychological Reports*, 126(5), 1069-1107.

Mbanaso, U. M., Abrahams, L., & Okafor, K. C. (2023). Research philosophy, design and methodology. In *Research techniques for computer science, information systems and cybersecurity* (pp. 81-113). Springer Nature Switzerland.

McAuley, E., Duncan, T., & Tammen, V. V. (1989). Psychometric properties of the Intrinsic Motivation Inventory in a competitive sport setting: A confirmatory factor analysis. *Research Quarterly for Exercise and Sport*, *60*(1), 48-58.

McCrae, R. R. (1987). Creativity, divergent thinking, and openness to experience. *Journal of Personality and Social Psychology*, *52*(6), 1258-1265.

McCrae, R. R., & Costa, P. T., Jr. (1999). A five-factor theory of personality. In L. A. Pervin & O. P. John (Eds.), *Handbook of personality: Theory and research* (2nd ed., pp. 139-153). Guilford Press.

McCrae, R. R., & Terracciano, A. (2005). Personality profiles of cultures: aggregate personality traits. Journal of personality and social psychology, 89(3), 407.

McCrae, R. R., & Costa, P. T., Jr. (2008). The five-factor theory of personality. In O. P. John, R. W. Robins, & L. A. Pervin (Eds.), *Handbook of personality: Theory and research* (3rd ed., pp. 159–181). Guilford Press.

McGregor, D. (1960). *The human side of enterprise*. McGraw-Hill.

Menapace, M. (2019). Scientific ethics: A new approach. *Science and Engineering Ethics*, *25*(4), 1193-1216.

Mendez, D., Graziotin, D., Wagner, S., & Seibold, H. (2020). Open science in software engineering. In *Contemporary empirical methods in software engineering* (pp. 477-501). Springer.

Merleau-Ponty, M. (1945). *Phénoménologie de la perception* [Phenomenology of perception]. Gallimard.

Mill, J. S. (1859). *On liberty*. John W. Parker & Son.

Milligan, G. W., & Cooper, M. C. (1985). An examination of procedures for determining the number of clusters in a data set. *Psychometrika*, *50*(2), 159-179.

Mingers, J. (2001). Combining IS research methods: Towards a pluralist methodology. *Information Systems Research*, *12*(3), 240-259.

Minsky, M. (1967). *Computation: Finite and infinite machines*. Prentice-Hall.

Minsky, M. (1986). *Society of Mind*. Simon and Schuster.

Mohr, K. A. J., & Mohr, E. S. (2017). Understanding Generation Z students to promote a contemporary learning environment. *Journal on Empowering Teaching Excellence*, *1*(1), 84-94.

Mor, N., & Winquist, J. (2002). Self-focused attention and negative affect: A meta-analysis. *Psychological Bulletin*, *128*(4), 638-662.

Morgan, D. L. (2007). Paradigms lost and pragmatism regained: Methodological implications of combining qualitative and quantitative methods. *Journal of mixed methods research*, *1*(1), 48-76.

Morse, J. M. (2000). Determining sample size. *Qualitative Health Research*, 10(1), 3-5.

Morse, J. M. (2016). *Mixed method design: Principles and procedures*. Routledge.

Mount, M. K., Barrick, M. R., & Stewart, G. L. (1998). Five-factor model of personality and performance in jobs involving interpersonal interactions. *Human Performance*, *11*(2-3), 145-165.

Mozannar, H., Bansal, G., Fourney, A., & Horvitz, E. (2024, May). Reading between the lines: Modeling user behavior and costs in AI-assisted programming. In *Proceedings of the 2024 CHI Conference on Human Factors in Computing Systems* (pp. 1-16).

Nahar, N., Kästner, C., Butler, J., Parnin, C., Zimmermann, T., & Bird, C. (2024). Beyond the comfort zone: Emerging solutions to overcome challenges in integrating LLMs into software products. *arXiv preprint arXiv:2410.12071*.

Nakagawa, S., & Schielzeth, H. (2013). A general and simple method for obtaining $R^2$ from generalized linear mixed-effects models. *Methods in Ecology and Evolution*, *4*(2), 133-142.

Nakamura, J., & Csikszentmihalyi, M. (2014). The concept of flow. In M. Csikszentmihalyi (Ed.), *Flow and the foundations of positive psychology* (pp. 239-263). Springer. https://doi.org/10.1007/978-94-017-9088-8_16

Nettle, D. (2006). The evolution of personality variation in humans and other animals. *American Psychologist*, *61*(6), 622-631.

Newell, A., & Rosenbloom, P. S. (1981). Mechanisms of skill acquisition and the law of practice. In J. R. Anderson (Ed.), *Cognitive skills and their acquisition* (pp. 1-55). Erlbaum.

Nissenbaum, H. (2010). *Privacy in context: Technology, policy, and the integrity of social life*. Stanford University Press.

Norman, D. (2013). *The design of everyday things: Revised and expanded edition*. Basic Books.

Nunamaker Jr, J. F., Chen, M., & Purdin, T. D. (1990). Systems development in information systems research. *Journal of Management Information Systems*, *7*(3), 89-106.

Oblinger, D. (2003). Boomers, gen-xers, and millennials: Understanding the new students. *EDUCAUSE Review*, *38*(4), 37-47.

O'Connell, M., & Kung, M. C. (2007). The cost of employee turnover. *Industrial Management*, 49(1), 14-19.

Oransky, I. (2022). Retractions are increasing, but not enough. *Nature*, *608*(9), 9. https://doi.org/10.1038/d41586-022-02071-6

Ostheimer, J., Chowdhury, S., & Iqbal, S. (2021). An alliance of humans and machines for machine learning: Hybrid intelligent systems and their design principles. *Technology in Society*, *66*, 101647. https://doi.org/10.1016/j.techsoc.2021.101647

Ovid. (8 CE). *Metamorphoses*.

Oxford University Press. (n.d.-a). Interleaving. In *Oxford English Dictionary* (online ed.). Retrieved May 18, 2025, from https://www.oed.com/

Oxford University Press. (n.d.-b). Optimization. In *Oxford English Dictionary* (online ed.). Retrieved May 18, 2025, from https://www.oed.com/

Park, J. S., O'Brien, J. C., Cai, C. J., Morris, M. R., Liang, P., & Bernstein, M. S. (2023). Generative agents: Interactive simulacra of human behavior. In *Proceedings of the 36th Annual ACM Symposium on User Interface Software and Technology* (pp. 1-22).

Pasquale, F. (2015). *The black box society: The secret algorithms that control money and information*. Harvard University Press.

Patton, M. Q. (1990). *Qualitative evaluation and research methods* (2nd ed.). Sage Publications.

Pawson, R., & Tilley, N. (1997). *Realistic evaluation*. SAGE Publications.

Peeters, M. A., Van Tuijl, H. F., Rutte, C. G., & Reymen, I. M. (2006). Personality and team performance: A meta-analysis. *European Journal of Personality: Published for the European Association of Personality Psychology*, *20*(5), 377-396.

Peffers, K., Tuunanen, T., Rothenberger, M. A., & Chatterjee, S. (2006). A design science research methodology for information systems research. *Journal of Management Information Systems*, *24*(3), 45-77.

Peirce, C. S. (1931). *Collected papers of Charles Sanders Peirce* (Vols. 1-6, C. Hartshorne & P. Weiss, Eds.). Harvard University Press.

Penke, L., Denissen, J. J., & Miller, G. F. (2007). The evolutionary genetics of personality. *European Journal of Personality*, *21*(5), 549-587.

Penke, L., & Jokela, M. (2016). The evolutionary genetics of personality revisited. In *Personality development across the lifespan* (pp. 53-71). Academic Press.

Pinheiro, J., & Bates, D. (2000). *Mixed-effects models in S and S-PLUS*. Springer Science & Business Media.

Pink, D. H. (2009). *Drive: The surprising truth about what motivates us*. Riverhead Books.

Plato. (380 BCE). *Republic*.

Polanyi, M. (1966). *The tacit dimension*. Doubleday.

Poldrack, R. A., Lu, T., & Beguš, G. (2023). AI-assisted coding: Experiments with GPT-4. *arXiv preprint arXiv:2304.13187*.

Pries-Heje, J., & Baskerville, R. (2008). The design theory nexus. *MIS Quarterly*, *32*(4), 731-755.

Putnam, H. (1988). Much ado about not very much. *Daedalus*, 269-281.

Ragu-Nathan, T. S., Tarafdar, M., Ragu-Nathan, B. S., & Tu, Q. (2008). The consequences of technostress for end users in organizations: Conceptual development and empirical validation. *Information systems research*, *19*(4), 417-433.

Ralph, P., Baltes, S., Adisaputri, G., Torkar, R., Kovalenko, V., Kalinowski, M., ... & Alkhadi, R. (2020). Pandemic programming: How COVID-19 affects software developers and how their organizations can help. *Empirical Software Engineering*, *25*(6), 4927-4961.

Rammstedt, B., & John, O. P. (2007). Measuring personality in one minute or less: A 10-item short version of the Big Five Inventory in English and German. *Journal of Research in Personality*, *41*(1), 203-212. https://doi.org/10.1016/j.jrp.2006.02.001

Reeves, B., & Nass, C. (1996). *The media equation: How people treat computers, television, and new media like real people and places*. Cambridge University Press.

Retraction Watch. (2021). *Retraction database*. Retrieved from https://retractionwatch.com/

Riemann, R., & Kandler, C. (2010). The interplay between genetics and environment in personality development. In L. A. Pervin & O. P. John (Eds.), *Handbook of personality: Theory and research* (pp. 173–195). Guilford Press.

Rilke, R. M. (1929). *Letters to a young poet* (M. D. Herter Norton, Trans.). W. W. Norton & Company.

Rittel, H. W., & Webber, M. M. (1973). Dilemmas in a general theory of planning. *Policy Sciences*, *4*(2), 155-169.

Rizzo, J. R., House, R. J., & Lirtzman, S. I. (1970). Role conflict and ambiguity in complex organizations. *Administrative Science Quarterly*, *15*(2), 150-163.

Rizzolatti, G., Fadiga, L., Gallese, V., & Fogassi, L. (1996). Premotor cortex and the recognition of motor actions. *Cognitive Brain Research*, *3*(2), 131-141.

Robe, P., & Kuttal, S. K. (2022). Designing PairBuddy—A conversational agent for pair programming. *ACM Transactions on Computer-Human Interaction,* 29(4), 1-44.

Roberts, L. W. (2014). *Research methods in psychology: A comprehensive guide*. Sage Publications.

Robins, A., Rountree, J., & Rountree, N. (2003). Learning and teaching programming: A review and discussion. *Computer Science Education*, *13*(2), 137-172.

Rode, A. L. G., Svejvig, P., & Martinsuo, M. (2022). Developing a multidimensional conception of project evaluation to improve projects. *Project Management Journal*, *53*(4), 416-432.

Rose, K., Massey, V., Marshall, B., & Cardon, P. (2023). IS professors' perspectives on AI-assisted programming. *Issues in Information Systems*, *24*(2).

Rosenthal, R., Rosnow, R. L., & Rubin, D. B. (2000). *Contrasts and effect sizes in behavioral research: A correlational approach*. Cambridge University Press.

Rotter, J. B. (1954). *Social learning and clinical psychology*. Prentice-Hall.

Rotter, J. B. (1966). Generalized expectancies for internal versus external control of reinforcement. *Psychological Monographs: General and Applied*, *80*(1), 1-28.

Royce, W. W. (1970). Managing the development of large software systems. *Proceedings of IEEE WESCON*, 26, 328-388.

Rubin, H. J., & Rubin, I. S. (2011). *Qualitative interviewing: The art of hearing data*. Sage.

Rushkoff, D. (2014). *Present shock: When everything happens now*. Penguin.

Ryan, R. M., & Deci, E. L. (2000a). Self-determination theory and the facilitation of intrinsic motivation, social development, and well-being. *American Psychologist*, *55*(1), 68-78.

Ryan, R. M., & Deci, E. L. (2000b). Intrinsic and extrinsic motivations: Classic definitions and new directions. *Contemporary Educational Psychology*, *25*(1), 54-67. https://doi.org/10.1006/ceps.1999.1020

Ryan, R. M., & Deci, E. L. (2017). *Self-determination theory: Basic psychological needs in motivation, development, and wellness*. Guilford Publications.

Ryan, R. M., Deci, E. L., Vansteenkiste, M., & Soenens, B. (2019). The history of motivation research in psychology and its relevance for management science. *The Cambridge handbook of the intellectual history of psychology*, 217–239.

Ryan, R. M., Mims, V., & Koestner, R. (1983). Relation of reward contingency and interpersonal context to intrinsic motivation: A review and test using cognitive evaluation theory. *Journal of Personality and Social Psychology*, *45*(4), 736.

Salgado, J. F. (1997). The Five Factor Model of personality and job performance in the European Community. *Journal of Applied Psychology*, *82*(1), 30.

Salinger, S., Plonka, L., & Prechelt, L. (2008). A coding scheme development methodology using grounded theory for qualitative analysis of pair programming. *Human Technology*, *4*(1), 9-25.

Salleh, N., Mendes, E., & Grundy, J. (2010). Empirical studies of pair programming for CS/SE teaching in higher education: A systematic literature review. *IEEE Transactions on Software Engineering*, *37*(4), 509-525.

Salleh, N., Mendes, E., & Grundy, J. (2014). Investigating the effects of personality traits on pair programming in a higher education setting through a family of experiments. *Empirical Software Engineering*, *19*(3), 714-752.

Sanchez-Roige, S., et al. (2018). Genome-wide association study of alcohol consumption and use disorder in 274,424 individuals from the UK Biobank. *Nature Communications*, *9*, 2883.

Sandberg, J., & Alvesson, M. (2011). Ways of constructing research questions: Gap-spotting or problematization? *Organization*, *18*(1), 23–44.

Sänger, J., Müller, V., & Lindenberger, U. (2012). Intra- and interbrain synchronization and network properties when playing guitar in duets. *Frontiers in Human Neuroscience*, *6*, 312.

Sartre, J. P. (1943). *L'être et le néant* [Being and nothingness]. Gallimard.

Sartre, J. P. (1946). *Existentialism is a humanism*. Yale University Press.

Saunders, M., Lewis, P., & Thornhill, A. (2019). *Research methods for business students* (8th ed.). Pearson Education.

Schleiger, E., Mason, C., Naughtin, C., Reeson, A., & Paris, C. (2024). Collaborative intelligence: A scoping review of current applications. *Applied Artificial Intelligence*, *38*(1), 2327890.

Schmitt, D. P., Allik, J., McCrae, R. R., & Benet-Martínez, V. (2007). The geographic distribution of Big Five personality traits: Patterns and profiles of human self-description across 56 nations. *Journal of Cross-Cultural Psychology*, *38*(2), 173-212.

Schön, D. A. (1983). *The reflective practitioner: How professionals think in action*. Basic Books.

Schwabe, L., Joëls, M., Roozendaal, B., Wolf, O. T., & Oitzl, M. S. (2012). Stress effects on memory: An update and integration. *Neuroscience & Biobehavioral Reviews*, *36*(7), 1740-1749.

Seligman, M. E. (2011). *Flourish: A visionary new understanding of happiness and well-being*. Free Press.

Servaas, M. N., Van Der Velde, J., Costafreda, S. G., Horton, P., Ormel, J., Riese, H., & Aleman, A. (2013). Neuroticism and the brain: a quantitative meta-analysis of neuroimaging studies investigating emotion processing. *Neuroscience & Biobehavioral Reviews*, *37*(8), 1518-1529.

Sfetsos, P., Stamelos, I., Angelis, L., & Deligiannis, I. (2009). An experimental investigation of personality types impact on pair effectiveness in pair programming. *Empirical Software Engineering*, *14*(2), 187-226.

Shadish, W. R., Cook, T. D., & Campbell, D. T. (2002). *Experimental and quasi-experimental designs for generalized causal inference*. Houghton Mifflin.

Shapiro, S. S., & Wilk, M. B. (1965). An analysis of variance test for normality (complete samples). *Biometrika*, *52*(3/4), 591-611.

Sharp, H., Baddoo, N., Beecham, S., Hall, T., & Robinson, H. (2009). Models of motivation in software engineering. *Information and Software Technology*, *51*(1), 219-233.

Shelley, M. (1818). *Frankenstein; or, The modern prometheus*. Lackington, Hughes, Harding, Mavor & Jones.

Shneiderman, B. (1980). *Software psychology: Human factors in computer and information systems*. Winthrop Publishers.

Simon, H. A. (1969). *The sciences of the artificial*. MIT Press.

Simon, H. A. (1973). The structure of ill structured problems. *Artificial intelligence*, *4*(3-4), 181-201.

Sjøberg, D. I. K., Anda, B., Arisholm, E., Dybå, T., Jørgensen, M., Karahasanovic, A., & Vokác, M. (2002). Conducting realistic experiments in software engineering. In *Proceedings of the 2002 International Symposium on Empirical Software Engineering* (pp. 17-26). IEEE.

Sjøberg, D. I., Hannay, J. E., Hansen, O., et al. (2005). A survey of controlled experiments in software engineering. *IEEE Transactions on Software Engineering*, *31*(9), 733-753.

Sjøberg, D. I., Dyba, T., & Jorgensen, M. (2007). The future of empirical methods in software engineering research. *Proceedings of the Future of Software Engineering*, 358-378.

Sjøberg, D. I., Dyba, T., Anda, B., & Hannay, J. (2008). Building theories in software engineering. In *Guide to advanced empirical software engineering* (pp. 312-336). Springer.

Sjøberg, D. I., & Bergersen, G. R. (2022). The distinction between theory-build and theory-testing in information systems research. In *Design-science research in information systems* (pp. 403-421). Springer International Publishing.

Skinner, B. F. (1953). *Science and human behavior*. Macmillan.

Slaten, K., Berenson, S., Williams, L., & Ho, C.-W. (2005). Pair programming and its impact on student success in an introductory programming course. *Journal on Educational Resources in Computing (JERIC)*, *4*(1), 1-13.

Smirnov, N. V. (1939). On the estimation of the discrepancy between empirical curves of distribution for two independent samples. *Bull. Math. Univ. Moscou*, *2*(2), 3-14.

Smith, J. A., Flowers, P., & Larkin, M. (2009). *Interpretative phenomenological analysis: Theory, method and research*. SAGE Publications.

Smith, E. K., Bird, C., & Zimmermann, T. (2016, May). Beliefs, practices, and personalities of software engineers: A survey in a large software company. In *Proceedings of the 9th International Workshop on Cooperative and Human Aspects of Software Engineering* (pp. 15-18).

Sodiya, A. S., Longe, H. O. D., Onashoga, S. A., Awodele, O., & Omotosho, L. O. (2007). An improved assessment of personality traits in software engineering. *Interdisciplinary Journal of Information, Knowledge, and Management*, *2*, 163-177.

Soloway, E. (1986). Learning to program = learning to construct mechanisms and explanations. *Communications of the ACM*, *29*(9), 850-858.

Star, S. L., & Griesemer, J. R. (1989). Institutional ecology, 'translations' and boundary objects: Amateurs and professionals in Berkeley's Museum of Vertebrate Zoology, 1907-39. *Social Studies of Science*, 19(3), 387-420.

Steiner, I. D. (1972). *Group process and productivity*. Academic Press.

Sternberg, R. J., & Pickren, W. E. (2019). *The Cambridge handbook of the intellectual history of psychology*. Cambridge University Press.

Stillwell, D. J., & Kosinski, M. (2015). *myPersonality Project website*. [Website discontinued 2018].

Stowers, K., Brady, L. L., MacLellan, C., Wohleber, R., & Salas, E. (2021). Improving teamwork competencies in human-machine teams: Perspectives from team science. *Frontiers in Psychology*, *12*, 590290. https://doi.org/10.3389/fpsyg.2021.590290

Stowers, K., Kasdaglis, N., Newton, O., Lakhmani, S., Wohleber, R., & Chen, J. (2021). Intelligent agent transparency: The design and evaluation of an interface to facilitate human and intelligent agent collaboration. In *Proceedings of the Human Factors and Ergonomics Society Annual Meeting* (Vol. 65, No. 1, pp. 1706-1710).

Suchman, L. A. (1987). *Plans and situated actions: The problem of human-machine communication*. Cambridge University Press.

Sweller, J. (1988). Cognitive load during problem solving: Effects on learning. *Cognitive Science*, *12*(2), 257–285. https://doi.org/10.1207/s15516709cog1202_4

Tarafdar, M., Cooper, C. L., & Stich, J. F. (2019). The technostress trifecta-techno eustress, techno distress and design: Theoretical directions and an agenda for research. *Information Systems Journal*, *29*(1), 6-42.

Tashakkori, A., & Teddlie, C. (2010). *SAGE handbook of mixed methods in social & behavioral research* (2nd ed.). SAGE Publications.

Taylor, F. W. (1911). *The Principles of Scientific Management*. New York: Harper & Brothers.

Taylor, C. (2007). *A secular age*. Harvard university press.

Teddlie, C., & Tashakkori, A. (2009). *Foundations of mixed methods research: Integrating quantitative and qualitative approaches in the social and behavioral sciences*. SAGE Publications.

Terry, G., Hayfield, N., Clarke, V., & Braun, V. (2017). Thematic analysis. In C. Willig & W. Stainton-Rogers (Eds.), *The SAGE handbook of qualitative research in psychology* (pp. 17-37). SAGE Publications.

Thiffault, P., & Bergeron, J. (2003). Fatigue and individual differences in monotonous simulated driving. *Personality and Individual Differences*, *34*(1), 159-176.

Thomson, P., & Kamler, B. (2016). *Helping doctoral students write: Pedagogies for supervision* (2nd ed.). Routledge.

Thoreau, H. D. (1854/2004). *Walden*. Yale University Press. (Original work published 1854)

Thuan, N. H., Drechsler, A., & Antunes, P. (2019). Construction of design science research questions. *Communications of the Association for Information Systems*, *44*(1), 20.

Tolman, E. C. (1948). Cognitive maps in rats and men. *Psychological Review*, *55*(4), 189-208.

Tracy, S. J. (2010). Qualitative quality: Eight "big-tent" criteria for excellent qualitative research. *Qualitative inquiry*, *16*(10), 837-851.

Trochim, W. M., & Donnelly, J. P. (2006). *The research methods knowledge base* (3rd ed.). Atomic Dog.

Tupes, E. C., & Christal, R. E. (1961). *Recurrent personality factors based on trait ratings*. USAF ASD Technical Report No. 61-97. Lackland Air Force Base, TX: U.S. Air Force.

Turing, A. M. (1950). Computing machinery and intelligence. *Mind*, *59*(236), 433–460. https://doi.org/10.1093/mind/LIX.236.433

Turkle, S. (2011). *Alone together: Why we expect more from technology and less from each other*. Basic Books.

Ulrich, M., Keller, J., Hoenig, K., Waller, C., & Grön, G. (2014). Neural correlates of experimentally induced flow experiences. *NeuroImage*, *86*, 194-202.

Vaithilingam, P., Zhang, T., & Glassman, E. L. (2022, April). Expectation vs. experience: Evaluating the usability of code generation tools powered by large language models. In *CHI Conference on Human Factors in Computing Systems Extended Abstracts* (pp. 1-7).

Vaismoradi, M., Jones, J., Turunen, H., & Snelgrove, S. (2016). Theme development in qualitative content analysis and thematic analysis. *Journal of Nursing Education and Practice*, *6*(5), 100-110.

van Aken, J. E. (2004). Management research based on the paradigm of the design sciences: The quest for field-tested and grounded technological rules. *Journal of Management Studies*, 41(2), 219-246.

Van de Ven, A. H. (2007). *Engaged scholarship: A guide for organizational and social research*. Oxford University Press.

van Manen, M. (1990). *Researching lived experience: Human science for an action sensitive pedagogy*. SUNY Press.

Valovy, M. (2021). Motivation Management Framework. In *Proceedings of the Psychology of Programming Interest Group*, 32nd Annual Workshop. 2021.

Valovy, M. (2022, January). Effects of Solo, Navigator, and Pilot Roles on Motivation. In *Abstract Proceedings of the Doctoral Students Day at FIS VSE 2022*. Prague University of Economics and Business. pp. 14. 2022.

Valovy, M. (2022, September, A). Motivational differences among software professionals. In *Proceedings of the 30th Interdisciplinary Management Talks: Digitalization of society, business, and management in a pandemic.* Linz: Trauner Verlag Publishing. 2022.

Valovy, M. (2022, September, B). Experimental Pair Programming: A Study Design and Preliminary Results. In *Proceedings of the Psychology of Programming Interest Group, 33rd Annual Workshop*, pp. 107-112. 2022.

Valovy, M. (2023, January). Effects of Pilot, Navigator, and Solo Programming Roles on Motivation: An Experimental Study. In: Mejia, J., Muñoz, M., Rocha, Á., Hernández-Nava, V. (eds) New Perspectives in Software Engineering. CIMPS 2022. *Lecture Notes in Networks and Systems, vol 576*. Springer, Cham.

Valovy, M. (2023, June). Psychological Aspects of Pair Programming: A Mixed-methods Experimental Study. In *Proceedings of the 27th International Conference on Evaluation and Assessment in Software Engineering* (pp. 210- 216). 2023.

Valovy, M., & Buchalcevova, A. (2023, October). The Psychological Effects of AI-Assisted Programming on Students and Professionals. In *2023 IEEE International Conference on Software Maintenance and Evolution (ICSME)* (pp. 385-390). IEEE.

Valovy, M., & Buchalcevova, A. (2025, January). Blockchain-Driven Research in Personality-Based Distributed Pair Programming. In *2025 5th Asia Conference on Information Engineering (ACIE)* (pp. 21-25). IEEE.

Valovy, M., & Buchalcevova, A. (2025, April). Personality-based pair programming: toward intrinsic motivation alignment in very small entities. *PeerJ Computer Science, 11,* e2774. https://doi.org/10.7717/peerj-cs.2774

Valovy, M., Dolezel, M., & Buchalcevova, A. (2025, September). Developer Motivation and Agency in LLM-Driven Coding. In *2025 IEEE International Conference on Software Maintenance and Evolution (ICSME)*. IEEE. (unpublished)

van Aken, J. E. (2004). Management research based on the paradigm of the design sciences: The quest for field-tested and grounded technological rules. *Journal of Management Studies*, *41*(2), 219-246.

Vanhanen, J., & Lassenius, C. (2005, November). Effects of pair programming at the development team level: An experiment. In *2005 International Symposium on Empirical Software Engineering, 2005* (pp. 10-pp). IEEE.

Van Merriënboer, J. J. G., & Sweller, J. (2005). Cognitive load theory and complex learning: Recent developments and future directions. *Educational Psychology Review*, *17*(2), 147–177. https://doi.org/10.1007/s10648-005-3951-0

Venable, J., Pries-Heje, J., & Baskerville, R. (2016). FEDS: A framework for evaluation in design science research. *European Journal of Information Systems*, *25*(1), 77-89.

Venkatesh, V., Brown, S. A., & Bala, H. (2013). Bridging the qualitative-quantitative divide: Guidelines for conducting mixed methods research in information systems. *MIS Quarterly*, *37*(1), 21-54.

Verner, J. M., Babar, M. A., Cerpa, N., Hall, T., & Beecham, S. (2014). Factors that motivate software engineering teams: A four country empirical study. *Journal of Systems and Software*, *92*, 115-127.

Vinson, N. G., & Singer, J. (2008). A practical guide to ethical research involving humans. In *Guide to advanced empirical software engineering* (pp. 229-256). Springer.

Vogel, S., & Schwabe, L. (2016). Learning and memory under stress: Implications for the classroom. *npj Science of Learning*, *1*(1), 1-10.

von Bertalanffy, L. (1968). *General system theory: Foundations, development, applications*. George Braziller.

Vygotsky, L. S. (1978). *Mind in society: The development of higher psychological processes*. Harvard University Press.

Walls, J. G., Widmeyer, G. R., & El Sawy, O. A. (1992). Building an information system design theory for vigilant EIS. *Information Systems Research*, *3*(1), 36-59.

Walsh, D. (2003). Critical reflection: Considering the role of the researcher. In D. Freshwater & M. V. Rolfe (Eds.), *Critical reflection for nursing* (pp. 160-171). Palgrave Macmillan.

Wang, D., Churchill, E., Maes, P., Fan, X., Shneiderman, B., Shi, Y., & Wang, Q. (2020, April). From human-human collaboration to Human-AI collaboration: Designing AI systems that can work together with people. In *Extended abstracts of the 2020 CHI conference on human factors in computing systems* (pp. 1-6).

Weber, M. (1905/1930). *The Protestant ethic and the spirit of capitalism* (T. Parsons, Trans.). Charles Scribner's Sons. (Original work published 1905)

Weinberg, G. M. (1971). *The psychology of computer programming*. Van Nostrand Reinhold.

Weinberg, G. M. (1998). *The psychology of computer programming: Silver anniversary edition*. Dorset House Publishing.

Werner, L., Hanks, B., & McDowell, C. (2005). Pair-programming helps female computer science students. *Journal on Educational Resources in Computing (JERIC)*, *4*(1), 1-17.

White, R. W. (1959). Motivation reconsidered: The concept of competence. *Psychological Review*, *66*(5), 297-333. https://doi.org/10.1037/h0040934

Whitworth, E., & Biddle, R. (2007, August). The social nature of agile teams. In *Agile 2007 (AGILE 2007)* (pp. 26-36). IEEE.

Wiener, N. (1950/1954). *The Human Use of Human Beings: Cybernetics and Society*. Boston: Houghton Mifflin.

Wiener, N. (1948). *Cybernetics: Or Control and Communication in the Animal and the Machine*. Cambridge, MA: MIT Press.

Wiener, N. (1960). Some moral and technical consequences of automation: As machines learn they may develop unforeseen strategies at rates that baffle their programmers. *Science*, *131*(3410), 1355-1358.

Wiesche, M., & Krcmar, H. (2014, May). The relationship of personality models and development tasks in software engineering. In *Proceedings of the 52nd ACM conference on Computers and people research* (pp. 149-161).

Wieringa, R. J. (2014). *Design science methodology for information systems and software engineering*. Springer.

Williams, L., & Kessler, R. R. (2003). *Pair programming illuminated*. Addison-Wesley.

Williams, L., Kessler, R. R., Cunningham, W., & Jeffries, R. (2000). Strengthening the case for pair programming. *IEEE Software*, *17*(4), 19-25.

Williams, L., Layman, L., Slaten, K., & Balik, S. (2008). Assessing the effect of pair programming on student programming outcomes. *Empirical Software Engineering*, *13*(3), 407-444.

Williams, L., Wiebe, E., Yang, K., Ferzli, M., & Miller, C. (2002). In support of pair programming in the introductory computer science course. *Computer Science Education*, *12*(3), 197-212.

Winslow, L. E. (1996). Programming pedagogy—a psychological overview. *ACM SIGCSE Bulletin*, *28*(3), 17-22.

Wohlin, C., Runeson, P., Höst, M., Ohlsson, M. C., Regnell, B., & Wesslén, A. (2012). *Experimentation in software engineering*. Springer.

Wynekoop, J. L., & Walz, D. B. (2000). Investigating traits of top-performing software developers. *Information Technology & People*, *13*(3), 186-195.

Yang, Y. F., Lee, C. I., & Chang, C. K. (2016). Learning motivation and retention effects of pair programming in data structures courses. *Education for Information*, *32*(3), 249-267.

Yardley, L. (2000). Dilemmas in qualitative health research. *Psychology and Health*, *15*(2), 215-228.

Zajonc, R. B. (1965). Social facilitation. *Science*, *149*(3681), 269-274.

Zhang, S., Wang, X., Zhang, W., Chen, Y., Gao, L., Wang, D., … & Wen, Y. (2024). Mutual theory of mind in human-AI collaboration: An empirical study with LLM-driven AI agents in a real-time shared workspace task. *arXiv preprint arXiv:2409.08811.*

Zieris, F., & Prechelt, L. (2019). Does pair programming pay off? Rethinking productivity in software engineering, 251-259.

Zieris, F., & Prechelt, L. (2020, June). Explaining pair programming session dynamics from knowledge gaps. In *Proceedings of the ACM/IEEE 42nd International Conference on Software Engineering* (pp. 421-432).

Zieris, F., & Prechelt, L. (2021, May). Two elements of pair programming skill. In *2021 IEEE/ACM 43rd International Conference on Software Engineering: New Ideas and Emerging Results (ICSE-NIER)* (pp. 51-55). IEEE.

Zimmerman, B. J. (2002). Becoming a self-regulated learner: An overview. *Theory into Practice*, *41*(2), 64-70.

Zuckerman, M. (2007). *Sensation seeking and risky behavior* (Vol. 72). Washington, DC: American Psychological Association.

Zuckerman, M., Porac, J., Lathin, D., & Deci, E. L. (1978). On the importance of self-determination for intrinsically-motivated behavior. *Personality and social psychology bulletin*, *4*(3), 443-446.

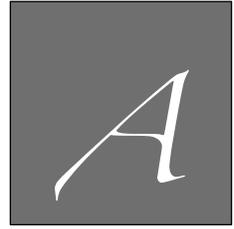

## APPENDIX A

## SYSTEMATIC LITERATURE REVIEW ON THREE AI INTERACTION MODES

This appendix provides detailed methodology and findings from the systematic literature review that informed the development of the ROMA framework's three-mode taxonomy (Co-Pilot, Co-Navigator, and Agent). The review, conducted across multiple databases, identified empirical evidence for these distinct interaction patterns in human-AI programming collaboration. This systematic approach ensures that the framework's conceptual foundation is grounded in current research rather than speculation, addressing the call for evidence-based theory development in software engineering (Sjøberg et al., 2008). The findings directly support the theoretical propositions presented in Chapters 2 and 7.

### Review Methodology

A targeted systematic review was conducted with a primary focus on the ACM Digital Library, supplemented by selective sampling from IEEE Xplore and Google Scholar. This approach was adopted after initial broad searches revealed an overwhelming volume of literature on AI-assisted programming, necessitating a more focused strategy to identify the most relevant empirical studies.

The search employed the following parameters:

A. **Search Strategy:**

- **ACM Digital Library:** Comprehensive search using the full query string

- **IEEE Xplore & Google Scholar:** More restricted search with additional filtering criteria

**B. Search Terms:**

- Primary: "AI-assisted programming" AND ("pair programming" OR "collaboration")

- Secondary: "AI programming assistant" OR "LLM coding" OR "Human–AI software engineering"

**C. Publication Range:** 2022–2025 (capturing the post-LLM breakthrough era)

**D. Initial Results:**

- ACM Digital Library: 110 records

- IEEE Xplore: 984 results (with "abstract" restriction)

- Google Scholar: 820 results

**E. Filtering Criteria:**

- Focus on empirical studies with human subjects

- Explicit discussion of AI-human interaction modes or patterns

- Discussion of psychological or socio-technical impacts

- Representation across different AI interaction paradigms

**F. Screening Process:**

- Initial selection based on relevance score and citation impact

- Abstract screening for clear focus on human-AI programming interaction

- Exclusion of purely technical implementations without user studies

- Prioritization of peer-reviewed conference and journal publications

**Final Sample:** 18 primary studies (including our own work) that met inclusion criteria and were subjected to full-text analysis

### *Evidence of Three AI Co-Programming Modes*

The systematic review identified three distinct AI interaction modes consistently appearing across the literature, each with characteristic roles and socio-psychological impacts:

Table 24: Evidence of three AI modes and respective co-programming roles in research

| Study | Co-Pilot | Co-Navi | Agent | Key Findings |
|---|---|---|---|---|
| Cinkusz & Chudziak (2024). Towards LLM-augmented multiagent systems for agile software engineering. In *Proceedings of the 39th IEEE/ACM International Conference on Automated Software Engineering (ASE'24)*. | ✓ | ✓ | ✓ | Describes multiple interaction paradigms, including code completion (co-pilot), interactive dialogue-based assistance (co-navigator), and autonomous code generation agents that operate independently. Agents serve a ***multitude of roles***, *including "developers, executors, quality checkers, and methodology reviewers."* The paper highlights how SAFe facilitates multi-agent system integration into large-scale Agile environments, automating tasks like backlog refinement via frameworks such as LangChain. |
| Valový & Buchalcevova (2023). The Psychological Effects of AI-Assisted Programming on Students and Professionals. In *2023 IEEE International Conference on Software Maintenance and Evolution (ICSME) (pp. 385-390)*. | ✓ | ✓ | - | Presents a mixed-methods investigation of psychological impacts across three AI interaction modes. Identifies distinct motivation patterns in different demographics, with professionals reporting increased dependency alongside productivity gains in Co-Pilot mode, while educational settings show notable differences in learning trajectories with Co-Navigator mode. Empirically documents two interaction modes and their differential impacts on intrinsic motivation and **self-determination**. |
| Mozannar et al. (2024). Reading Between the Lines: Modeling User Behavior and Costs in AI-Assisted Programming. In *Proceedings of the 2024 CHI Conference on Human Factors in Computing Systems*. | ✓ | - | - | Specifically focuses on GitHub Copilot usage patterns, introducing the CodeRec User Programming States (CUPS) framework for analyzing human-AI programming sessions in co-pilot mode. The study emphasizes interaction with inline code suggestions rather than conversational or autonomous agents. |
| Ma et al. (2023). Is AI the better programming partner? Human-human pair programming vs. human-AI pair programming. *arXiv preprint arXiv:2306.05153*. | ✓ | - | - | Directly compares human-human pair programming with human-AI pairing where AI acts primarily in a co-pilot capacity. Findings indicate mixed effectiveness across both approaches, noting that "mismatched expertise makes pair programming less productive, therefore well-designed AI programming assistants may adapt to differences in expertise levels." The paper specifically examines AI in the suggestion/completion role rather than conversational or agent roles. |
| Liang et al. (2024). A large-scale survey on the usability of AI programming assistants: Successes and challenges. In *Proceedings of the 46th IEEE/ACM International Conference on Software Engineering (ICSE'24)*. | ✓ | - | - | Analyzes motivational effects of co-pilot mode AI assistance (N=410), finding that "developers are most motivated to use AI programming assistants because they help developers reduce keystrokes, finish programming tasks quickly, and recall syntax, but resonate less with using them to help brainstorm potential solutions." Primary concerns include AI output not meeting functional/non-functional requirements and difficulty controlling generation. |
| Poldrack et al. (2023). AI-assisted coding: Experiments with GPT-4. *arXiv preprint arXiv:2304.13187*. | - | ✓ | - | Focuses on conversational co-navigator mode using GPT-4 through a chat interface for programming assistance. The experiments involve prompting GPT-4 to solve coding problems through dialogue rather than using inline completions, making this a study of co-navigator functionality. |
| Rose et al. (2023). IS professors' perspectives on AI-assisted programming. *Issues in Information Systems, 24*(2). | ✓ | ✓ | - | Examines faculty perspectives on both code completion tools (co-pilot) and conversational programming assistants (co-navigator) in teaching contexts. Distinguishes between these two modes of AI assistance and explores their different impacts on student learning and cognitive development. |
| Robe & Kuttal (2022). Designing PairBuddy—a conversational agent | - | ✓ | - | Focuses on creating a conversational agent explicitly designed as a navigator in pair programming. PairBuddy is based on the |
| for pair programming. *ACM Transactions on Computer-Human Interaction (TOCHI), 29*(4), 1-44. | | | | traditional navigator role from human pair programming, providing guidance through dialogue rather than direct code generation, making this a pure co-navigator implementation. |
| Hamza et al. (2024). Human-AI collaboration in software engineering: Lessons learned from a hands-on workshop. In *Proceedings of the 7th ACM/IEEE International Workshop on Software-intensive Business* (pp. 7-14). | - | ✓ | - | Specifically explores ChatGPT as an AI navigator for pair programming, noting that "while AI, particularly ChatGPT, improves the efficiency of code generation and optimization, human oversight remains crucial, especially in areas requiring complex problem-solving and security considerations." The study focuses on conversational guidance rather than inline suggestions or autonomous execution. |
| Hassan et al. (2024). Towards AI-native software engineering (SE 3.0): A vision and a challenge roadmap. *arXiv preprint arXiv:2410.06107.* | ✓ | ✓ | ✓ | Proposes a comprehensive evolution toward "SE 3.0" where AI systems progress "beyond task-driven copilots into intelligent collaborators." Explicitly describes a spectrum of AI assistance from code completion to conversational guidance to autonomous agents capable of understanding and reasoning about software engineering principles independently. One of the few papers to clearly articulate all three modes in a developmental continuum. |
| Nahar et al. (2024). Beyond the Comfort Zone: Emerging Solutions to Overcome Challenges in Integrating LLMs into Software Products. *arXiv preprint arXiv:2410.12071.* | ✓ | ✓ | - | Examines adaptation patterns as developers work with both co-pilot tools (inline suggestions) and co-navigator systems (conversational assistants). Focuses on challenges in both modes but does not substantially address fully autonomous agent systems. |
| Schleiger et al. (2024). Collaborative Intelligence: A scoping review of current applications. *Applied Artificial Intelligence, 38*(1), 2327890. | ✓ | ✓ | ✓ | Develops a comprehensive framework for human-AI collaboration requiring "(1) complementarity (i.e., the collaboration draws upon complementary human and AI capability to improve outcomes), (2) a shared objective and outcome, and (3) sustained, two-way task-related interaction between human and AI." The paper maps existing systems across a spectrum from basic suggestion tools to conversational assistants to autonomous collaborative agents. |
| Zhang et al. (2024). Mutual theory of mind in human-AI collaboration: An empirical study with LLM-driven AI agents in a real-time shared workspace task. *arXiv preprint arXiv:2409.08811.* | ✓ | ✓ | ✓ | Explores mental models in human-AI programming teams, examining how humans develop understanding of AI capabilities and how AI systems can model human intentions. Studies interaction across suggestion-based, dialogue-based, and autonomous agent modalities, finding distinct cognitive patterns for each mode. |
| Nah et al. (2023). Generative AI and ChatGPT: Applications, challenges, and AI-human collaboration. *Journal of Information Technology Case and Application Research, 25*(3), 277-304. | ✓ | ✓ | - | Extensively cited paper (1024 citations) examining interaction patterns between developers and AI tools, primarily focusing on code completion and conversational modes. Provides in-depth analysis of challenges including bias, over-reliance, misuse, and privacy concerns in human-AI programming partnerships. |
| Daun & Brings (2023). How ChatGPT will change software engineering education. In *Proceedings of the 2023 Conference on Innovation and Technology in Computer Science Education V. 1* (pp. 110-116). | - | ✓ | - | Concentrates specifically on conversational AI programming assistants in educational contexts, examining how dialogue-based programming help (co-navigator mode) affects learning outcomes differently than traditional teaching or automated code completion. |
| Barke et al. (2023). Grounded copilot: How programmers interact | ✓ | - | - | Presents a detailed empirical study of how programmers interact with GitHub Copilot, finding that developers frequently iterate on |

| | | | |
|---|---|---|---|
| with code-generating models. *Proceedings of the ACM on Programming Languages, 7*(OOPSLA1), 1-31. | | | | and modify AI suggestions. Identifies patterns of "steering" the co-pilot through careful prompting and contextual cues, suggesting a dynamic rather than passive relationship even in suggestion-based systems. Describes two human-programmer modes, "acceleration" and "exploration", entered when interacting with Co-Pilot. |
| Park et al. (2023). Generative agents: Interactive simulacra of human behavior. In *Proceedings of the 36th Annual ACM Symposium on User Interface Software and Technology* (pp. 1-22). | - | - | ✓ | This study examines autonomous AI agents that can generate and execute code independently, demonstrating agent-mode capabilities where AIs operate with minimal human supervision. Though not specifically about coding optimization, it provides evidence of fully autonomous agent behavior in computational tasks. |
| Vaithilingam et al. (2022). Expectation vs. experience: Evaluating the usability of code generation tools powered by large language models. In *CHI Conference on Human Factors in Computing Systems Extended Abstracts* (pp. 1-7). | ✓ | - | - | Investigates user experience with co-pilot style code generation tools, identifying a significant gap between user expectations and actual experience. Focuses exclusively on inline suggestion tools rather than conversational or autonomous systems. |

## Mode Distribution Analysis

The frequency distribution of the three interaction modes across the reviewed literature reveals:

➔ **Co-Pilot Mode:** 13 studies (72.2% of sample)

➔ **Co-Navigator Mode:** 12 studies (66.7% of sample)

➔ **Agent Mode:** 5 studies (27.8% of sample)

This pattern reflects both the current state of technology adoption and research focus, with more mature Co-Pilot & Co-Navigator implementations receiving greater attention, while the emerging Agent mode represents a frontier area with fewer but growing empirical investigations.

## Conceptual Integration with ROMA Framework

This systematic review provides the empirical foundation for the three-mode taxonomy integrated into the ROMA framework. The identification of these distinct interaction patterns across multiple independent studies strengthens the framework's conceptual validity and ensures its alignment with current research trajectories in Human  AI programming collaboration.

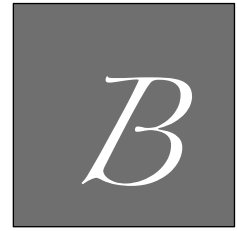

## APPENDIX B

# PHENOMENOLOGICALLY-INFORMED INTERVIEW PROTOCOL (STUDY VI)

This appendix presents the interview protocol used in Study VI to explore the lived experiences of software engineers working with AI agents. Following Interpretative Phenomenological Analysis principles (Smith et al., 2009), the protocol was designed to capture pre-reflective experiences rather than opinions or explanations. The bilingual presentation (Czech/English) reflects the study's commitment to allowing participants to express nuanced experiences in their native language. This protocol operationalizes the phenomenological approach discussed in Chapter 3 and generated the rich experiential data analyzed in Chapter 7.

### Introduction to the Phenomenological Approach

The phenomenologically-informed interviews (PII) were conducted in participants' native language (Czech) to facilitate deeper expression of lived experiences. Unlike semi-structured interviews that seek explanations and opinions, these interviews focused on capturing the direct, pre-reflective experience of working with AI programming agents. The protocol served as a flexible guide rather than a rigid script, with questions adapted to follow participants' experiential accounts.

Table 25: Principles for conducting phenomenologically-informed interviews

| Fenomenologický přístup k vedení rozhovoru (Czech) | Phenomenological Interview Approach (English) |
|---|---|
| **Zaměření na prožitek, ne na názory**<br>Ptát se "JAK jsi to prožíval" místo "PROČ se to stalo" | **Focus on experience, not opinions**<br>Ask "HOW did you experience it" instead of "WHY did it happen" |

| Konkrétní události, ne obecné úvahy | Specific events, not general reflections |
|---|---|
| Zakotvit rozhovor v konkrétní situaci, kterou účastník zažil | Anchor the interview in a specific situation the participant experienced |
| **Prostor pro ticho a reflexi** | **Space for silence and reflection** |
| Nebát se ticha, které umožňuje hlubší promyšlení odpovědi | Embrace silence that allows deeper consideration of responses |
| **Sledování tělesné zkušenosti** | **Attending to embodied experience** |
| Všímat si, jak se prožitek projevoval v těle účastníka | Notice how the experience manifested in the participant's body |
| **Následovat zkušenost, ne strukturu** | **Follow the experience, not the structure** |
| Používat protokol jako mapu, ne jako přesný itinerář | Use the protocol as a map, not as a strict itinerary |
| **Prozkoumat do hloubky, než do šířky** | **Explore depth rather than breadth** |
| Jít hlouběji do jedné zkušenosti, než povrchně probírat mnoho | Go deeper into one experience rather than superficially covering many |

### Phenomenologically-Informed Interview Guide

Interviews were conducted following Interpretative Phenomenological Analysis principles (Smith et al., 2009; Eatough & Smith, 2017), with questions serving as starting points for dialogue rather than items to be systematically covered. Not all questions were asked of every participant, and the order was adapted to follow the natural flow of each conversation.

The guide is presented in its original Czech form:

### Základní struktura rozhovoru (90 minut)

1. Úvod a demografické otázky (10 minut)

2. Ukotvení konkrétní zkušenosti (15 minut)

3. Prozkoumání dimenzí SDT (55 minut)

    a. Autonomie (20 minut)

    b. Kompetence (20 minut)

    c. Vztahovost/Sounáležitost (15 minut)

4. Flow a ponoření (10 minut)

5. Uzavření a reflexe (10 minut)

*Úvod (5 minut)*

*"*Děkuji, že sis udělal čas na dnešní rozhovor. Zkoumáme, jak nové možnosti AI agentů ovlivňují práci a prožitek vývojářů. Dnes bych rád prozkoumal, jak tuto spolupráci přímo prožíváš – jaké pocity a myšlenky tě při ní provázejí a co pro tebe znamená.

Budeme používat přístup zaměřený na tvou konkrétní zkušenost. To znamená, že se budu ptát především na to, JAK jsi něco prožíval, spíše než PROČ se to stalo. Můžeme se déle zastavit u jedné zkušenosti a prozkoumat ji hlouběji. Někdy nechám i delší ticho, aby byl prostor pro promyšlení odpovědi. Nejde o správné odpovědi, ale o porozumění tvému subjektivnímu prožitku.

Pro orientaci rozlišujeme tři módy AI nástrojů:

- AI Našeptávač jako GitHub Copilot, který nabízí dokončení kódu přímo v IDE

- AI Asistent jako běžný ChatGPT, který reaguje na tvé dotazy, ale vyžaduje průběžné vedení

- AI Agent, který samostatněji řeší komplexnější úkoly – navrhuje a implementuje funkce, řeší bugy nebo vytváří pull requesty s minimálním dohledem

Pokud ti to pomůže, můžeš při odpovědích porovnávat zkušenosti s různými typy těchto nástrojů."

*1. Demografické otázky a kontext (5 minut)*

1. „Mohl bys stručně popsat svou současnou roli, velikost týmu a zkušenosti s AI nástroji?"

2. „Jaké máš zkušenosti s AI agenty, kteří samostatně plní programovací úkoly? Jak je používáš a jak často?"

*2. Ukotvení konkrétní zkušenosti (15 minut)*

3. „Vzpomeň si prosím na **konkrétní nedávnou situaci**, kdy jsi pracoval s AI agentem na nějakém programovacím úkolu. Můžeš mi tento zážitek popsat krok za krokem?"

   *Sondážní otázky:*

   o „Jaký byl tvůj cíl? Co přesně jsi od AI agenta požadoval?"

   o „Jak jsi s ním komunikoval? Co jsi přesně napsal/zadal?"

   o „Co se dělo potom? Jak agent reagoval?"

   o „Co jsi při tom prožíval? Jaké myšlenky a pocity tě provázely?"

- o „Jaké byly tvé tělesné či emoční pocity při ukončení této spolupráce?" (pocity úspěchu, frustrace, atd.)

4. „Co tě přimělo v této situaci použít agentní mód AI místo tradičnějších metod?"

   *Hlubší sonda:* „Co jsi od toho očekával? Naplnila se tato očekávání?"

## 3. Autonomie (20 minut)

5. „Při té konkrétní zkušenosti, kterou jsi právě popsal, jak jsi vnímal svůj vliv na proces a výsledek?"

   *Hlubší sonda:* „Byl moment, kdy jsi pocítil změnu v kontrole nad procesem? Můžeš ten moment popsat?"

6. „Když srovnáš své programování před používáním AI agentů a nyní, jak se změnil tvůj pocit volnosti a svobody v rozhodování?"

   *Hlubší sonda:* „Co konkrétně ti dává pocit větší/menší svobody při práci s AI agentem?"

7. „Popiš situaci, kdy jsi chtěl, aby AI agent vytvořil řešení přesně podle tvé předem jasné představy. Jak jsi tuto situaci prožíval?"

   *Hlubší sonda:* „Když se AI agent ubíral jiným směrem než ty, jaké to pro tebe bylo? Co jsi cítil a jak jsi s agentem vyjednával?"

## 4. Kompetence (20 minut)

8. „Jak používání AI agentů ovlivňuje tvůj pocit odborné zdatnosti a schopností? Můžeš uvést konkrétní situaci, která to ilustruje?"

   *Hlubší sonda:* „Jak ses v té situaci cítil ohledně svých schopností? Byl to jiný pocit než před používáním AI agentů?"

9. „Popiš zkušenost, kdy ti AI agent pomohl vyřešit něco, co bys sám řešil obtížně. Jak jsi tuto situaci prožíval?"

   *Hlubší sonda:* „Co sis v tom okamžiku myslel o sobě jako o vývojáři?"

10. „Vzpomeň si na situaci, kdy AI agent vytvořil řešení, kterému jsi plně nerozuměl. Jak ses při tom cítil?"

    *Hlubší sonda:* „Jaký vliv měla tato zkušenost na tvůj pocit kompetence a mistrovství v oboru?"

## 5. Sounáležitost (15 minut)

11. „Jak používání AI agentů mění tvé vztahy a interakce s kolegy v týmu? Můžeš popsat konkrétní situaci, která to ilustruje?"

    *Hlubší sonda:* „Jak jsi tuto změnu prožíval? Co to pro tebe znamenalo?"

12. „Když porovnáš interakci s AI agentem a lidským kolegou, jaký je ve tvém prožívání této interakce rozdíl?"

    *Hlubší sonda:* „Jsou situace, kdy preferuješ komunikovat s AI agentem místo člověka? Jak tyto situace prožíváš?"

13. „Jak sdílení zkušeností s AI agenty ovlivňuje tvé vazby v týmu nebo komunitě vývojářů? Můžeš popsat konkrétní situaci?"

    *Hlubší sonda:* „Co pro tebe toto sdílení znamená z pohledu sounáležitosti s komunitou?"

## 6. Flow a ponoření (10 minut)

14. „Zažíváš při práci s AI agenty stavy hlubokého soustředění, kdy ztrácíš pojem o čase?"

    *Hlubší sonda:* „Můžeš popsat konkrétní situaci, kdy jsi takový stav zažil? Jak se tento prožitek 'flow' liší od podobných zážitků při tradičním programování?"

15. „Jak používání AI agentů ovlivňuje tvou schopnost být plně přítomný a soustředěný na úkol?"

    *Hlubší sonda:* „Vzpomeneš si na okamžik, kdy tě interakce s AI agentem buď zvlášť hluboce vtáhla do práce, nebo naopak vyrušila? Jak jsi tento moment prožíval?"

    *Tělesná sonda:* „Jak se tento stav projevoval v tvém těle? Všiml sis nějakých fyzických pocitů během tohoto stavu hlubokého soustředění?"

16. „Jak se změnil tvůj pracovní rytmus a způsob přemýšlení od doby, co používáš AI agenty?"

    *Hlubší sonda:* „Co pro tebe tato změna znamená? Jak se při ní cítíš?"

## 7. Závěr a reflexe (10 minut)

17. „Když se ohlédneš za svými zkušenostmi s AI agenty, jaké mají tyto zkušenosti význam pro tvou identitu vývojáře?"

    *Hlubší sonda:* „Vzpomeneš si na nějaký konkrétní moment, kdy sis uvědomil, že AI agent změnil způsob, jakým vnímáš sám sebe jako vývojáře? Jak jsi tento moment prožíval?"

18. „Jak podle tvé osobní zkušenosti AI agenty mění vztahy a spolupráci v softwarových týmech? Můžeš popsat konkrétní situaci, kdy jsi tuto změnu pocítil?"

*Hlubší sonda:* „Co pro tebe tyto změny znamenají? Jak se při nich cítíš?"

19. „Je něco důležitého o tvých zkušenostech s AI agenty, co jsme nezmínili a co bys chtěl zdůraznit?"

*Hlubší sonda:* „Co tě nejvíce překvapilo při spolupráci s AI agenty? Byl nějaký moment, který byl zcela neočekávaný?"

20. „Jak bys jednou větou shrnul svůj vztah k AI agentům ve své práci?"

*Poděkování za účast a vysvětlení dalších kroků výzkumu.*

## APPENDIX C

## *SEMI−STRUCTURED INTERVIEW PROTOCOL (STUDY III)*

This appendix contains the semi-structured interview protocol used in Study III to investigate professional software engineers' experiences with AI-assisted programming. The protocol's seven thematic blocks systematically explore familiarity, task suitability, personality attributions, psychological impacts, and future perspectives. This instrument was developed following the exploratory findings from Studies I and II, and its comprehensive scope allowed for the identification of key themes that informed the development of the ROMA AI Adapter presented in Chapter 7.

*Protocol for exploring AI-assisted programming with professionals*

**Block 1: Introduction**

1. "Could you share your journey about how you ventured into computer programming?"
2. "What aspects of programming do you find most stimulating and challenging?"
3. "Can you describe your current role and responsibilities in the software engineering field?"
4. "Where do you see your career trajectory in software engineering?"

**Block 2: Familiarity with Artificial Intelligence**

5. "When did you first incorporate AI tools into your programming work?"
6. "Have you used AI tools outside your workspace?"
7. "Could you list the AI tools that you've worked with to date?"
8. "Could you share your insights on the comparison between these tools?"

9. "In what ways has the introduction of AI tools shifted your programming approach?"

## Block 3: Suitability of AI in Different Software Engineering Tasks

10. "Can you describe the tasks where AI played a role in your programming process?"

11. "What is your success rate with CoPilot offering the correct solution on the first attempt?"

12. "How well does ChatGPT understand your ideas and intentions?"

13. "Does the integration of AI simplify your tasks? How so?"

14. "What specific tasks would you suggest should be addressed or resolved with the assistance of AI?"

15. "Do you find AI more effective for simple or complex tasks? Could you elaborate on why?"

16. "Could you share a specific experience where AI tools substantially facilitated your programming task?"

17. "Have there been instances where using AI tools has impeded your progress on a task?"

18. "How eager are you to continue using AI in programming in the future?"

19. "Would you like to return to the era when it was unavailable?"

20. "In your opinion, does AI match up to a human collaborator? Could you compare your experiences with human and AI pair programming?"

## Block 4: Personality and Emotions in AI

1. "How would you describe the 'personality' of the AI tools you've used?"

2. "Do you feel that having an AI tool with a 'personality' enhances your interaction with the tool? Can you share any specific instances?"

3. "If you could design your AI tool's personality, how would it be? Would you prefer a confident, shy, dramatic, funny AI tool or one that speaks like a specific character?"

4. "Do you feel a sense of companionship or isolation when you're programming with an AI tool? How does that compare to programming with a human?"

5. "How would you describe your emotional response when an AI tool provides the correct solution? What about when it fails to do so?"

6. "Have you ever been frustrated with an AI tool to the point of wanting to stop using it? Can you describe that scenario?"

7. "What emotions do you commonly feel when interacting with AI tools? Are these different from when you work with a human?"

8. "Does using AI tools cause you any form of anxiety or stress, such as the fear of replacement, dependency, or any other reasons?"

9. "Do you believe that an AI tool can understand the nuances of human emotion and react accordingly? If yes, could you share any examples?"

10. "Have there been instances where the AI surprised you with its understanding (or lack thereof) of human-like communication or emotions?"

11. "How would you describe the relationship you have with your AI tools? Is it similar to a work colleague, a helpful assistant, a mentor, a student, or something else?"

12. "Do you think the AI tool knows what is going on? Is it sentient yet?"

13. "Do you think the AI should have the power to decide when to answer or not answer your questions?"

## Block 5: Psychological Aspects of Programming with AI

14. "Do you feel more at ease programming alone, with a human partner, or with an AI tool?"

15. "Do you feel programming should be balanced in terms of whom to partner with (humans/AI)?"

16. "Did you experience a sense of apprehension (anxiety) or relief when you started to use AI tools? Should new users prepare to experience those feelings?"

17. "How has the integration of AI influenced your satisfaction levels during programming?"

18. "How has AI affected your motivation during programming?"

19. "What do you enjoy most about solo programming without AI? Is there a unique appeal in programming without AI or even without internet connectivity? Or is it a thing of the past."

20. "Have you had discussions about your experiences with AI with your colleagues? What were the key takeaways from those conversations?"

21. "How is the use of AI perceived within your workplace culture? Is it generally promoted, discouraged, or seen neutrally?"

## Block 6: AI Tools Effectance

22. "In your opinion, can AI tools supersede traditional programming or are they better as support tools? Why do you believe so? What year would that be possible?"

23. "Do you think AI tools can supplant conventional programming education (books, tutorials), or are they better served as additional resources? Could you explain why? What year would that be possible?"

## Block 7: Improvements and the Prospects of AI

24. "What suggestions do you have for enhancing the appeal of AI-assisted programming?"

25. "Looking more into the future, are there any specific features that you would like to add to AI in the year 2025?"

26. "Do you anticipate any challenges with the increasing use of AI tools in programming? How can these be addressed?"

27. "How do you think AI will change the landscape of software security?"

28. "Do you think AI will bring a new generation of computer viruses?"

29. "Do you think that AI itself could get infected by a computer virus?"

30. "How do you envision the future of programming with AI?"

31. "How do you perceive the future of human-AI collaborations?"

32. "Do you think we should decelerate the development of AI? Why?"

33. "Are there any disadvantages or potential risks associated with the use of AI tools in the workplace?"

34. "What solutions would you suggest to overcome the drawbacks of utilizing AI in the workplace? Should there be some training, tutorial, or seminar for using AI in the workplace?"

35. "Do you feel AI will hinder or accelerate the computer programming learning process?"

# APPENDIX D

## ADDITIONAL PROTOCOLS

This appendix references the interview protocols from Studies I, II, and IV that progressively built the empirical foundation for this dissertation. Each protocol represented an evolution in understanding, with questions refined based on emerging insights:

- **Protocol for Study I (CIMPS'22)**: Focused on pair programming experiences with 25 questions spanning topics of programming background, partner dynamics, and role satisfaction.

- **Protocol for Study II (EASE'23)**: Expanded the inquiry with 39 questions investigating task difficulty, personality considerations, and future perspectives on AI integration.

- **Protocol for Study IV (PeerJ-CS)**: Examined career satisfaction, software engineering relationships, and experimental satisfaction through 31 structured questions.

Each study incorporated lessons from previous investigations, adapting questions to focus more specifically on emerging themes of interest. Together, these instruments demonstrate the iterative theory-building process described in Chapter 3.